\documentclass[longauth]{aa}  
\usepackage{natbib,twoopt}
\usepackage[dvipsnames]{xcolor}
\usepackage[breaklinks=true]{hyperref} 
\bibpunct{(}{)}{;}{a}{}{,}             
\makeatletter
  \newcommandtwoopt{\citeads}[3][][]{\href{http://adsabs.harvard.edu/abs/#3}%
    {\def\hyper@linkstart##1##2{}%
     \let\hyper@linkend\@empty\citealp[#1][#2]{#3}}}
  \newcommandtwoopt{\citepads}[3][][]{\href{http://adsabs.harvard.edu/abs/#3}%
    {\def\hyper@linkstart##1##2{}%
     \let\hyper@linkend\@empty\citep[#1][#2]{#3}}}
  \newcommandtwoopt{\citetads}[3][][]{\href{http://adsabs.harvard.edu/abs/#3}%
    {\def\hyper@linkstart##1##2{}%
     \let\hyper@linkend\@empty\citet[#1][#2]{#3}}}
  \newcommandtwoopt{\citeyearads}[3][][]%
    {\href{http://adsabs.harvard.edu/abs/#3}
    {\def\hyper@linkstart##1##2{}%
     \let\hyper@linkend\@empty\citeyear[#1][#2]{#3}}}
\makeatother
\hypersetup{colorlinks=true,linkcolor=blue,citecolor=blue,urlcolor=blue}

    \newcommand{\NNOGGCNS}{431}
     \newcommand{\NNOGTOT}{3016}
     \newcommand{\NRAVERV}{2520}
    \newcommand{\NSIMBADRV}{12\,852}
 \newcommand{\NGCNSWITHRVS}{82\,358}
    \newcommand{\NDRTWORV}{125\,354}
\newcommand{\NGCNSRVS}{135\,790}
     \newcommand{\NREJECT}{880\,428}
      \newcommand{\NFINAL}{331\,312}
  \newcommand{\NDISTFIFTY}{300\,526}
    \newcommand{\NDISTTEN}{331\,312}
\newcommand{\NINEIGHTMAS}{1\,211\,740}
\newcommand{\orcit}[1]{\protect\href{https://orcid.org/#1}{\protect\includegraphics[width=8pt]{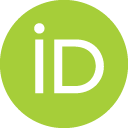}}}

\newcommand{\TENPC}{316}
\newcommand{\problimit}{0.38}%

\newcommand\kms{\ensuremath{\text{\,km\,s}^{-1}}}

\newcommand\gbp{\ensuremath{G_\mathrm{BP}}}
\newcommand\grp{\ensuremath{G_\mathrm{RP}}}

\newcommand{\G}{\textit{Gaia}}
\newcommand\gaia{\textit{Gaia}}
\newcommand\hip{\textsc{Hipparcos\xspace}}

\newcommand\dmed{\texttt{dist\_50}}
\newcommand\dtol{\texttt{dist\_1}}

\newcommand\gdrone{\textit{Gaia}~DR1\xspace}
\newcommand\gdrtwo{\textit{Gaia}~DR2\xspace}
\newcommand\gdrthree{\textit{Gaia}~EDR3\xspace}
\newcommand{\USun}{$U_\odot$}
\newcommand{\VSun}{$V_\odot$}
\newcommand{\WSun}{$W_\odot$}

\newcommand{\Vsun}{$V_\odot$}

\usepackage{graphicx}
\usepackage{txfonts}

\usepackage{colortbl}
\usepackage{textcomp}

\usepackage{gensymb}
\usepackage{listings}
\usepackage{ulem}

\definecolor{dkgreen}{rgb}{0,0.6,0}
\definecolor{gray}{rgb}{0.5,0.5,0.5}
\definecolor{mauve}{rgb}{0.58,0,0.82}
\definecolor{golden}{rgb}{0.86,0.65,0.01}

\makeatletter
\renewcommand*\maketitle{%
  \thispagestyle{firstpage}
\begingroup
    \if@wideboxfn
    \setlength\bibindent{1.4\parindent}
    \else
    \setlength\bibindent{\parindent}
    \fi
    \renewcommand*\thefootnote{\@fnsymbol\c@footnote}%
    \renewcommand\@makefntext[1]{%
    \ifaa@longfn\hsize\textwidth\fi
    \noindent
    \hb@xt@\bibindent{\hss\@makefnmark\enspace}##1}
  \ifaa@twocolumn
  \begingroup
    \begin{aa@strip}
          \aa@maketitle
    \end{aa@strip}
    \@thanks
  \endgroup
  \else
    \begingroup
      \let\thanks\footnote
      \aa@maketitle
    \endgroup
  \fi
\endgroup
  \setcounter{footnote}{0}%
}
\makeatother

\begin{document} 

 
\title{\G\ Early Data Release 3:\\The \G\ Catalogue of Nearby
  Stars\footnote{Tables 2,3, the distance probability density function,
    the healpixel magnitude limits, the missing object list, the white
    dwarf probability list and a white dwarf comparison and are only
    available in electronic form at the CDS via
    \url{http://cdsweb.u-strasbg.fr/cgi-bin/qcat?J/A+A/}}}

\author{
{\it Gaia} Collaboration
\and R.L.      ~Smart                         \orcit{0000-0002-4424-4766}\inst{\ref{inst:0001}}\thanks{Contact: richard.smart@inaf.it}
\and L.M.      ~Sarro                         \orcit{0000-0002-5622-5191}\inst{\ref{inst:0002}}
\and J.        ~Rybizki                       \orcit{0000-0002-0993-6089}\inst{\ref{inst:0003}}
\and C.        ~Reyl\'{e}                     \orcit{0000-0003-2258-2403}\inst{\ref{inst:0004}}
\and A.C.      ~Robin                         \orcit{0000-0001-8654-9499}\inst{\ref{inst:0004}}
\and N.C.      ~Hambly                        \orcit{0000-0002-9901-9064}\inst{\ref{inst:0006}}
\and U.        ~Abbas                         \orcit{0000-0002-5076-766X}\inst{\ref{inst:0001}}
\and M.A.      ~Barstow                       \orcit{0000-0002-7116-3259}\inst{\ref{inst:0008}}
\and J.H.J.    ~de Bruijne                    \orcit{0000-0001-6459-8599}\inst{\ref{inst:0009}}
\and B.        ~Bucciarelli                   \orcit{0000-0002-5303-0268}\inst{\ref{inst:0001}}
\and J.M.      ~Carrasco                      \orcit{0000-0002-3029-5853}\inst{\ref{inst:0011}}
\and W.J.      ~Cooper                        \orcit{0000-0003-3501-8967}\inst{\ref{inst:0012},\ref{inst:0001}}
\and S.T.      ~Hodgkin                       \orcit{0000-0002-5470-3962}\inst{\ref{inst:0014}}
\and E.        ~Masana                        \orcit{0000-0002-4819-329X}\inst{\ref{inst:0011}}
\and D.        ~Michalik                      \orcit{0000-0002-7618-6556}\inst{\ref{inst:0009}}
\and J.        ~Sahlmann                      \orcit{0000-0001-9525-3673}\inst{\ref{inst:0017}}
\and A.        ~Sozzetti                      \orcit{0000-0002-7504-365X}\inst{\ref{inst:0001}}
\and A.G.A.    ~Brown                         \orcit{0000-0002-7419-9679}\inst{\ref{inst:0019}}
\and A.        ~Vallenari                     \orcit{0000-0003-0014-519X}\inst{\ref{inst:0020}}
\and T.        ~Prusti                        \orcit{0000-0003-3120-7867}\inst{\ref{inst:0009}}
\and C.        ~Babusiaux                     \orcit{0000-0002-7631-348X}\inst{\ref{inst:0022},\ref{inst:0023}}
\and M.        ~Biermann                      \inst{\ref{inst:0024}}
\and O.L.      ~Creevey                       \orcit{0000-0003-1853-6631}\inst{\ref{inst:0025}}
\and D.W.      ~Evans                         \orcit{0000-0002-6685-5998}\inst{\ref{inst:0014}}
\and L.        ~Eyer                          \orcit{0000-0002-0182-8040}\inst{\ref{inst:0027}}
\and A.        ~Hutton                        \inst{\ref{inst:0028}}
\and F.        ~Jansen                        \inst{\ref{inst:0009}}
\and C.        ~Jordi                         \orcit{0000-0001-5495-9602}\inst{\ref{inst:0011}}
\and S.A.      ~Klioner                       \orcit{0000-0003-4682-7831}\inst{\ref{inst:0031}}
\and U.        ~Lammers                       \orcit{0000-0001-8309-3801}\inst{\ref{inst:0032}}
\and L.        ~Lindegren                     \orcit{0000-0002-5443-3026}\inst{\ref{inst:0033}}
\and X.        ~Luri                          \orcit{0000-0001-5428-9397}\inst{\ref{inst:0011}}
\and F.        ~Mignard                       \inst{\ref{inst:0025}}
\and C.        ~Panem                         \inst{\ref{inst:0036}}
\and D.        ~Pourbaix                      \orcit{0000-0002-3020-1837}\inst{\ref{inst:0037},\ref{inst:0038}}
\and S.        ~Randich                       \orcit{0000-0003-2438-0899}\inst{\ref{inst:0039}}
\and P.        ~Sartoretti                    \inst{\ref{inst:0023}}
\and C.        ~Soubiran                      \orcit{0000-0003-3304-8134}\inst{\ref{inst:0041}}
\and N.A.      ~Walton                        \orcit{0000-0003-3983-8778}\inst{\ref{inst:0014}}
\and F.        ~Arenou                        \orcit{0000-0003-2837-3899}\inst{\ref{inst:0023}}
\and C.A.L.    ~Bailer-Jones                  \inst{\ref{inst:0003}}
\and U.        ~Bastian                       \orcit{0000-0002-8667-1715}\inst{\ref{inst:0024}}
\and M.        ~Cropper                       \orcit{0000-0003-4571-9468}\inst{\ref{inst:0046}}
\and R.        ~Drimmel                       \orcit{0000-0002-1777-5502}\inst{\ref{inst:0001}}
\and D.        ~Katz                          \orcit{0000-0001-7986-3164}\inst{\ref{inst:0023}}
\and M.G.      ~Lattanzi                      \orcit{0000-0003-0429-7748}\inst{\ref{inst:0001},\ref{inst:0050}}
\and F.        ~van Leeuwen                   \inst{\ref{inst:0014}}
\and J.        ~Bakker                        \inst{\ref{inst:0032}}
\and J.        ~Casta\~{n}eda                 \orcit{0000-0001-7820-946X}\inst{\ref{inst:0053}}
\and F.        ~De Angeli                     \inst{\ref{inst:0014}}
\and C.        ~Ducourant                     \orcit{0000-0003-4843-8979}\inst{\ref{inst:0041}}
\and C.        ~Fabricius                     \orcit{0000-0003-2639-1372}\inst{\ref{inst:0011}}
\and M.        ~Fouesneau                     \orcit{0000-0001-9256-5516}\inst{\ref{inst:0003}}
\and Y.        ~Fr\'{e}mat                    \orcit{0000-0002-4645-6017}\inst{\ref{inst:0058}}
\and R.        ~Guerra                        \orcit{0000-0002-9850-8982}\inst{\ref{inst:0032}}
\and A.        ~Guerrier                      \inst{\ref{inst:0036}}
\and J.        ~Guiraud                       \inst{\ref{inst:0036}}
\and A.        ~Jean-Antoine Piccolo          \inst{\ref{inst:0036}}
\and R.        ~Messineo                      \inst{\ref{inst:0063}}
\and N.        ~Mowlavi                       \inst{\ref{inst:0027}}
\and C.        ~Nicolas                       \inst{\ref{inst:0036}}
\and K.        ~Nienartowicz                  \orcit{0000-0001-5415-0547}\inst{\ref{inst:0066},\ref{inst:0067}}
\and F.        ~Pailler                       \inst{\ref{inst:0036}}
\and P.        ~Panuzzo                       \orcit{0000-0002-0016-8271}\inst{\ref{inst:0023}}
\and F.        ~Riclet                        \inst{\ref{inst:0036}}
\and W.        ~Roux                          \inst{\ref{inst:0036}}
\and G.M.      ~Seabroke                      \inst{\ref{inst:0046}}
\and R.        ~Sordo                         \orcit{0000-0003-4979-0659}\inst{\ref{inst:0020}}
\and P.        ~Tanga                         \orcit{0000-0002-2718-997X}\inst{\ref{inst:0025}}
\and F.        ~Th\'{e}venin                  \inst{\ref{inst:0025}}
\and G.        ~Gracia-Abril                  \inst{\ref{inst:0076},\ref{inst:0024}}
\and J.        ~Portell                       \orcit{0000-0002-8886-8925}\inst{\ref{inst:0011}}
\and D.        ~Teyssier                      \orcit{0000-0002-6261-5292}\inst{\ref{inst:0079}}
\and M.        ~Altmann                       \orcit{0000-0002-0530-0913}\inst{\ref{inst:0024},\ref{inst:0081}}
\and R.        ~Andrae                        \inst{\ref{inst:0003}}
\and I.        ~Bellas-Velidis                \inst{\ref{inst:0083}}
\and K.        ~Benson                        \inst{\ref{inst:0046}}
\and J.        ~Berthier                      \orcit{0000-0003-1846-6485}\inst{\ref{inst:0085}}
\and R.        ~Blomme                        \orcit{0000-0002-2526-346X}\inst{\ref{inst:0058}}
\and E.        ~Brugaletta                    \orcit{0000-0003-2598-6737}\inst{\ref{inst:0087}}
\and P.W.      ~Burgess                       \inst{\ref{inst:0014}}
\and G.        ~Busso                         \orcit{0000-0003-0937-9849}\inst{\ref{inst:0014}}
\and B.        ~Carry                         \orcit{0000-0001-5242-3089}\inst{\ref{inst:0025}}
\and A.        ~Cellino                       \orcit{0000-0002-6645-334X}\inst{\ref{inst:0001}}
\and N.        ~Cheek                         \inst{\ref{inst:0092}}
\and G.        ~Clementini                    \orcit{0000-0001-9206-9723}\inst{\ref{inst:0093}}
\and Y.        ~Damerdji                      \inst{\ref{inst:0094},\ref{inst:0095}}
\and M.        ~Davidson                      \inst{\ref{inst:0006}}
\and L.        ~Delchambre                    \inst{\ref{inst:0094}}
\and A.        ~Dell'Oro                      \orcit{0000-0003-1561-9685}\inst{\ref{inst:0039}}
\and J.        ~Fern\'{a}ndez-Hern\'{a}ndez   \inst{\ref{inst:0099}}
\and L.        ~Galluccio                     \orcit{0000-0002-8541-0476}\inst{\ref{inst:0025}}
\and P.        ~Garc\'{i}a-Lario              \inst{\ref{inst:0032}}
\and M.        ~Garcia-Reinaldos              \inst{\ref{inst:0032}}
\and J.        ~Gonz\'{a}lez-N\'{u}\~{n}ez    \orcit{0000-0001-5311-5555}\inst{\ref{inst:0092},\ref{inst:0104}}
\and E.        ~Gosset                        \inst{\ref{inst:0094},\ref{inst:0038}}
\and R.        ~Haigron                       \inst{\ref{inst:0023}}
\and J.-L.     ~Halbwachs                     \orcit{0000-0003-2968-6395}\inst{\ref{inst:0108}}
\and D.L.      ~Harrison                      \orcit{0000-0001-8687-6588}\inst{\ref{inst:0014},\ref{inst:0110}}
\and D.        ~Hatzidimitriou                \orcit{0000-0002-5415-0464}\inst{\ref{inst:0111}}
\and U.        ~Heiter                        \orcit{0000-0001-6825-1066}\inst{\ref{inst:0112}}
\and J.        ~Hern\'{a}ndez                 \inst{\ref{inst:0032}}
\and D.        ~Hestroffer                    \orcit{0000-0003-0472-9459}\inst{\ref{inst:0085}}
\and B.        ~Holl                          \orcit{0000-0001-6220-3266}\inst{\ref{inst:0027},\ref{inst:0066}}
\and K.        ~Jan{\ss}en                    \inst{\ref{inst:0117}}
\and G.        ~Jevardat de Fombelle          \inst{\ref{inst:0027}}
\and S.        ~Jordan                        \orcit{0000-0001-6316-6831}\inst{\ref{inst:0024}}
\and A.        ~Krone-Martins                 \orcit{0000-0002-2308-6623}\inst{\ref{inst:0120},\ref{inst:0121}}
\and A.C.      ~Lanzafame                     \orcit{0000-0002-2697-3607}\inst{\ref{inst:0087},\ref{inst:0123}}
\and W.        ~L\"{ o}ffler                  \inst{\ref{inst:0024}}
\and A.        ~Lorca                         \inst{\ref{inst:0028}}
\and M.        ~Manteiga                      \orcit{0000-0002-7711-5581}\inst{\ref{inst:0126}}
\and O.        ~Marchal                       \inst{\ref{inst:0108}}
\and P.M.      ~Marrese                       \inst{\ref{inst:0128},\ref{inst:0129}}
\and A.        ~Moitinho                      \orcit{0000-0003-0822-5995}\inst{\ref{inst:0120}}
\and A.        ~Mora                          \inst{\ref{inst:0028}}
\and K.        ~Muinonen                      \orcit{0000-0001-8058-2642}\inst{\ref{inst:0132},\ref{inst:0133}}
\and P.        ~Osborne                       \inst{\ref{inst:0014}}
\and E.        ~Pancino                       \orcit{0000-0003-0788-5879}\inst{\ref{inst:0039},\ref{inst:0129}}
\and T.        ~Pauwels                       \inst{\ref{inst:0058}}
\and A.        ~Recio-Blanco                  \inst{\ref{inst:0025}}
\and P.J.      ~Richards                      \inst{\ref{inst:0139}}
\and M.        ~Riello                        \orcit{0000-0002-3134-0935}\inst{\ref{inst:0014}}
\and L.        ~Rimoldini                     \orcit{0000-0002-0306-585X}\inst{\ref{inst:0066}}
\and T.        ~Roegiers                      \inst{\ref{inst:0142}}
\and C.        ~Siopis                        \inst{\ref{inst:0037}}
\and M.        ~Smith                         \inst{\ref{inst:0046}}
\and A.        ~Ulla                          \inst{\ref{inst:0145}}
\and E.        ~Utrilla                       \inst{\ref{inst:0028}}
\and M.        ~van Leeuwen                   \inst{\ref{inst:0014}}
\and W.        ~van Reeven                    \inst{\ref{inst:0028}}
\and A.        ~Abreu Aramburu                \inst{\ref{inst:0099}}
\and S.        ~Accart                        \inst{\ref{inst:0150}}
\and C.        ~Aerts                         \orcit{0000-0003-1822-7126}\inst{\ref{inst:0151},\ref{inst:0152},\ref{inst:0003}}
\and J.J.      ~Aguado                        \inst{\ref{inst:0002}}
\and M.        ~Ajaj                          \inst{\ref{inst:0023}}
\and G.        ~Altavilla                     \orcit{0000-0002-9934-1352}\inst{\ref{inst:0128},\ref{inst:0129}}
\and M.A.      ~\'{A}lvarez                   \orcit{0000-0002-6786-2620}\inst{\ref{inst:0158}}
\and J.        ~\'{A}lvarez Cid-Fuentes       \orcit{0000-0001-7153-4649}\inst{\ref{inst:0159}}
\and J.        ~Alves                         \orcit{0000-0002-4355-0921}\inst{\ref{inst:0160}}
\and R.I.      ~Anderson                      \orcit{0000-0001-8089-4419}\inst{\ref{inst:0161}}
\and E.        ~Anglada Varela                \orcit{0000-0001-7563-0689}\inst{\ref{inst:0099}}
\and T.        ~Antoja                        \orcit{0000-0003-2595-5148}\inst{\ref{inst:0011}}
\and M.        ~Audard                        \orcit{0000-0003-4721-034X}\inst{\ref{inst:0066}}
\and D.        ~Baines                        \orcit{0000-0002-6923-3756}\inst{\ref{inst:0079}}
\and S.G.      ~Baker                         \orcit{0000-0002-6436-1257}\inst{\ref{inst:0046}}
\and L.        ~Balaguer-N\'{u}\~{n}ez        \orcit{0000-0001-9789-7069}\inst{\ref{inst:0011}}
\and E.        ~Balbinot                      \orcit{0000-0002-1322-3153}\inst{\ref{inst:0168}}
\and Z.        ~Balog                         \orcit{0000-0003-1748-2926}\inst{\ref{inst:0024},\ref{inst:0003}}
\and C.        ~Barache                       \inst{\ref{inst:0081}}
\and D.        ~Barbato                       \inst{\ref{inst:0027},\ref{inst:0001}}
\and M.        ~Barros                        \orcit{0000-0002-9728-9618}\inst{\ref{inst:0120}}
\and S.        ~Bartolom\'{e}                 \orcit{0000-0002-6290-6030}\inst{\ref{inst:0011}}
\and J.-L.     ~Bassilana                     \inst{\ref{inst:0150}}
\and N.        ~Bauchet                       \inst{\ref{inst:0085}}
\and A.        ~Baudesson-Stella              \inst{\ref{inst:0150}}
\and U.        ~Becciani                      \orcit{0000-0002-4389-8688}\inst{\ref{inst:0087}}
\and M.        ~Bellazzini                    \orcit{0000-0001-8200-810X}\inst{\ref{inst:0093}}
\and M.        ~Bernet                        \inst{\ref{inst:0011}}
\and S.        ~Bertone                       \orcit{0000-0001-9885-8440}\inst{\ref{inst:0182},\ref{inst:0183},\ref{inst:0001}}
\and L.        ~Bianchi                       \inst{\ref{inst:0185}}
\and S.        ~Blanco-Cuaresma               \orcit{0000-0002-1584-0171}\inst{\ref{inst:0186}}
\and T.        ~Boch                          \orcit{0000-0001-5818-2781}\inst{\ref{inst:0108}}
\and A.        ~Bombrun                       \inst{\ref{inst:0188}}
\and D.        ~Bossini                       \orcit{0000-0002-9480-8400}\inst{\ref{inst:0189}}
\and S.        ~Bouquillon                    \inst{\ref{inst:0081}}
\and A.        ~Bragaglia                     \orcit{0000-0002-0338-7883}\inst{\ref{inst:0093}}
\and L.        ~Bramante                      \inst{\ref{inst:0063}}
\and E.        ~Breedt                        \orcit{0000-0001-6180-3438}\inst{\ref{inst:0014}}
\and A.        ~Bressan                       \orcit{0000-0002-7922-8440}\inst{\ref{inst:0194}}
\and N.        ~Brouillet                     \inst{\ref{inst:0041}}
\and A.        ~Burlacu                       \inst{\ref{inst:0196}}
\and D.        ~Busonero                      \orcit{0000-0002-3903-7076}\inst{\ref{inst:0001}}
\and A.G.      ~Butkevich                     \inst{\ref{inst:0001}}
\and R.        ~Buzzi                         \orcit{0000-0001-9389-5701}\inst{\ref{inst:0001}}
\and E.        ~Caffau                        \orcit{0000-0001-6011-6134}\inst{\ref{inst:0023}}
\and R.        ~Cancelliere                   \orcit{0000-0002-9120-3799}\inst{\ref{inst:0201}}
\and H.        ~C\'{a}novas                   \orcit{0000-0001-7668-8022}\inst{\ref{inst:0028}}
\and T.        ~Cantat-Gaudin                 \orcit{0000-0001-8726-2588}\inst{\ref{inst:0011}}
\and R.        ~Carballo                      \inst{\ref{inst:0204}}
\and T.        ~Carlucci                      \inst{\ref{inst:0081}}
\and M.I       ~Carnerero                     \orcit{0000-0001-5843-5515}\inst{\ref{inst:0001}}
\and L.        ~Casamiquela                   \orcit{0000-0001-5238-8674}\inst{\ref{inst:0041}}
\and M.        ~Castellani                    \orcit{0000-0002-7650-7428}\inst{\ref{inst:0128}}
\and A.        ~Castro-Ginard                 \orcit{0000-0002-9419-3725}\inst{\ref{inst:0011}}
\and P.        ~Castro Sampol                 \inst{\ref{inst:0011}}
\and L.        ~Chaoul                        \inst{\ref{inst:0036}}
\and P.        ~Charlot                       \inst{\ref{inst:0041}}
\and L.        ~Chemin                        \orcit{0000-0002-3834-7937}\inst{\ref{inst:0213}}
\and A.        ~Chiavassa                     \orcit{0000-0003-3891-7554}\inst{\ref{inst:0025}}
\and M.-R. L.  ~Cioni                         \orcit{0000-0002-6797-696x}\inst{\ref{inst:0117}}
\and G.        ~Comoretto                     \inst{\ref{inst:0216}}
\and T.        ~Cornez                        \inst{\ref{inst:0150}}
\and S.        ~Cowell                        \inst{\ref{inst:0014}}
\and F.        ~Crifo                         \inst{\ref{inst:0023}}
\and M.        ~Crosta                        \orcit{0000-0003-4369-3786}\inst{\ref{inst:0001}}
\and C.        ~Crowley                       \inst{\ref{inst:0188}}
\and C.        ~Dafonte                       \orcit{0000-0003-4693-7555}\inst{\ref{inst:0158}}
\and A.        ~Dapergolas                    \inst{\ref{inst:0083}}
\and M.        ~David                         \orcit{0000-0002-4172-3112}\inst{\ref{inst:0224}}
\and P.        ~David                         \inst{\ref{inst:0085}}
\and P.        ~de Laverny                    \inst{\ref{inst:0025}}
\and F.        ~De Luise                      \orcit{0000-0002-6570-8208}\inst{\ref{inst:0227}}
\and R.        ~De March                      \orcit{0000-0003-0567-842X}\inst{\ref{inst:0063}}
\and J.        ~De Ridder                     \orcit{0000-0001-6726-2863}\inst{\ref{inst:0151}}
\and R.        ~de Souza                      \inst{\ref{inst:0230}}
\and P.        ~de Teodoro                    \inst{\ref{inst:0032}}
\and A.        ~de Torres                     \inst{\ref{inst:0188}}
\and E.F.      ~del Peloso                    \inst{\ref{inst:0024}}
\and E.        ~del Pozo                      \inst{\ref{inst:0028}}
\and A.        ~Delgado                       \inst{\ref{inst:0014}}
\and H.E.      ~Delgado                       \orcit{0000-0003-1409-4282}\inst{\ref{inst:0002}}
\and J.-B.     ~Delisle                       \orcit{0000-0001-5844-9888}\inst{\ref{inst:0027}}
\and P.        ~Di Matteo                     \inst{\ref{inst:0023}}
\and S.        ~Diakite                       \inst{\ref{inst:0239}}
\and C.        ~Diener                        \inst{\ref{inst:0014}}
\and E.        ~Distefano                     \orcit{0000-0002-2448-2513}\inst{\ref{inst:0087}}
\and C.        ~Dolding                       \inst{\ref{inst:0046}}
\and D.        ~Eappachen                     \inst{\ref{inst:0243},\ref{inst:0152}}
\and B.        ~Edvardsson                    \inst{\ref{inst:0245}}
\and H.        ~Enke                          \orcit{0000-0002-2366-8316}\inst{\ref{inst:0117}}
\and P.        ~Esquej                        \orcit{0000-0001-8195-628X}\inst{\ref{inst:0017}}
\and C.        ~Fabre                         \inst{\ref{inst:0248}}
\and M.        ~Fabrizio                      \orcit{0000-0001-5829-111X}\inst{\ref{inst:0128},\ref{inst:0129}}
\and S.        ~Faigler                       \inst{\ref{inst:0251}}
\and G.        ~Fedorets                      \inst{\ref{inst:0132},\ref{inst:0253}}
\and P.        ~Fernique                      \orcit{0000-0002-3304-2923}\inst{\ref{inst:0108},\ref{inst:0255}}
\and A.        ~Fienga                        \orcit{0000-0002-4755-7637}\inst{\ref{inst:0256},\ref{inst:0085}}
\and F.        ~Figueras                      \orcit{0000-0002-3393-0007}\inst{\ref{inst:0011}}
\and C.        ~Fouron                        \inst{\ref{inst:0196}}
\and F.        ~Fragkoudi                     \inst{\ref{inst:0260}}
\and E.        ~Fraile                        \inst{\ref{inst:0017}}
\and F.        ~Franke                        \inst{\ref{inst:0262}}
\and M.        ~Gai                           \orcit{0000-0001-9008-134X}\inst{\ref{inst:0001}}
\and D.        ~Garabato                      \orcit{0000-0002-7133-6623}\inst{\ref{inst:0158}}
\and A.        ~Garcia-Gutierrez              \inst{\ref{inst:0011}}
\and M.        ~Garc\'{i}a-Torres             \orcit{0000-0002-6867-7080}\inst{\ref{inst:0266}}
\and A.        ~Garofalo                      \orcit{0000-0002-5907-0375}\inst{\ref{inst:0093}}
\and P.        ~Gavras                        \orcit{0000-0002-4383-4836}\inst{\ref{inst:0017}}
\and E.        ~Gerlach                       \orcit{0000-0002-9533-2168}\inst{\ref{inst:0031}}
\and R.        ~Geyer                         \orcit{0000-0001-6967-8707}\inst{\ref{inst:0031}}
\and P.        ~Giacobbe                      \inst{\ref{inst:0001}}
\and G.        ~Gilmore                       \orcit{0000-0003-4632-0213}\inst{\ref{inst:0014}}
\and S.        ~Girona                        \orcit{0000-0002-1975-1918}\inst{\ref{inst:0159}}
\and G.        ~Giuffrida                     \inst{\ref{inst:0128}}
\and R.        ~Gomel                         \inst{\ref{inst:0251}}
\and A.        ~Gomez                         \orcit{0000-0002-3796-3690}\inst{\ref{inst:0158}}
\and I.        ~Gonzalez-Santamaria           \orcit{0000-0002-8537-9384}\inst{\ref{inst:0158}}
\and J.J.      ~Gonz\'{a}lez-Vidal            \inst{\ref{inst:0011}}
\and M.        ~Granvik                       \orcit{0000-0002-5624-1888}\inst{\ref{inst:0132},\ref{inst:0280}}
\and R.        ~Guti\'{e}rrez-S\'{a}nchez     \inst{\ref{inst:0079}}
\and L.P.      ~Guy                           \orcit{0000-0003-0800-8755}\inst{\ref{inst:0066},\ref{inst:0216}}
\and M.        ~Hauser                        \inst{\ref{inst:0003},\ref{inst:0285}}
\and M.        ~Haywood                       \orcit{0000-0003-0434-0400}\inst{\ref{inst:0023}}
\and A.        ~Helmi                         \orcit{0000-0003-3937-7641}\inst{\ref{inst:0168}}
\and S.L.      ~Hidalgo                       \orcit{0000-0002-0002-9298}\inst{\ref{inst:0288},\ref{inst:0289}}
\and T.        ~Hilger                        \orcit{0000-0003-1646-0063}\inst{\ref{inst:0031}}
\and N.        ~H\l{}adczuk                   \inst{\ref{inst:0032}}
\and D.        ~Hobbs                         \orcit{0000-0002-2696-1366}\inst{\ref{inst:0033}}
\and G.        ~Holland                       \inst{\ref{inst:0014}}
\and H.E.      ~Huckle                        \inst{\ref{inst:0046}}
\and G.        ~Jasniewicz                    \inst{\ref{inst:0295}}
\and P.G.      ~Jonker                        \orcit{0000-0001-5679-0695}\inst{\ref{inst:0152},\ref{inst:0243}}
\and J.        ~Juaristi Campillo             \inst{\ref{inst:0024}}
\and F.        ~Julbe                         \inst{\ref{inst:0011}}
\and L.        ~Karbevska                     \inst{\ref{inst:0027}}
\and P.        ~Kervella                      \orcit{0000-0003-0626-1749}\inst{\ref{inst:0301}}
\and S.        ~Khanna                        \orcit{0000-0002-2604-4277}\inst{\ref{inst:0168}}
\and A.        ~Kochoska                      \orcit{0000-0002-9739-8371}\inst{\ref{inst:0303}}
\and M.        ~Kontizas                      \orcit{0000-0001-7177-0158}\inst{\ref{inst:0111}}
\and G.        ~Kordopatis                    \orcit{0000-0002-9035-3920}\inst{\ref{inst:0025}}
\and A.J.      ~Korn                          \orcit{0000-0002-3881-6756}\inst{\ref{inst:0112}}
\and Z.        ~Kostrzewa-Rutkowska           \inst{\ref{inst:0019},\ref{inst:0243}}
\and K.        ~Kruszy\'{n}ska                \orcit{0000-0002-2729-5369}\inst{\ref{inst:0309}}
\and S.        ~Lambert                       \orcit{0000-0001-6759-5502}\inst{\ref{inst:0081}}
\and A.F.      ~Lanza                         \orcit{0000-0001-5928-7251}\inst{\ref{inst:0087}}
\and Y.        ~Lasne                         \inst{\ref{inst:0150}}
\and J.-F.     ~Le Campion                    \inst{\ref{inst:0313}}
\and Y.        ~Le Fustec                     \inst{\ref{inst:0196}}
\and Y.        ~Lebreton                      \orcit{0000-0002-4834-2144}\inst{\ref{inst:0301},\ref{inst:0316}}
\and T.        ~Lebzelter                     \orcit{0000-0002-0702-7551}\inst{\ref{inst:0160}}
\and S.        ~Leccia                        \orcit{0000-0001-5685-6930}\inst{\ref{inst:0318}}
\and N.        ~Leclerc                       \inst{\ref{inst:0023}}
\and I.        ~Lecoeur-Taibi                 \orcit{0000-0003-0029-8575}\inst{\ref{inst:0066}}
\and S.        ~Liao                          \inst{\ref{inst:0001}}
\and E.        ~Licata                        \orcit{0000-0002-5203-0135}\inst{\ref{inst:0001}}
\and H.E.P.    ~Lindstr{\o}m                  \inst{\ref{inst:0001},\ref{inst:0324}}
\and T.A.      ~Lister                        \orcit{0000-0002-3818-7769}\inst{\ref{inst:0325}}
\and E.        ~Livanou                       \inst{\ref{inst:0111}}
\and A.        ~Lobel                         \inst{\ref{inst:0058}}
\and P.        ~Madrero Pardo                 \inst{\ref{inst:0011}}
\and S.        ~Managau                       \inst{\ref{inst:0150}}
\and R.G.      ~Mann                          \orcit{0000-0002-0194-325X}\inst{\ref{inst:0006}}
\and J.M.      ~Marchant                      \inst{\ref{inst:0331}}
\and M.        ~Marconi                       \orcit{0000-0002-1330-2927}\inst{\ref{inst:0318}}
\and M.M.S.    ~Marcos Santos                 \inst{\ref{inst:0092}}
\and S.        ~Marinoni                      \orcit{0000-0001-7990-6849}\inst{\ref{inst:0128},\ref{inst:0129}}
\and F.        ~Marocco                       \orcit{0000-0001-7519-1700}\inst{\ref{inst:0336},\ref{inst:0337}}
\and D.J.      ~Marshall                      \inst{\ref{inst:0338}}
\and L.        ~Martin Polo                   \inst{\ref{inst:0092}}
\and J.M.      ~Mart\'{i}n-Fleitas            \orcit{0000-0002-8594-569X}\inst{\ref{inst:0028}}
\and A.        ~Masip                         \inst{\ref{inst:0011}}
\and D.        ~Massari                       \orcit{0000-0001-8892-4301}\inst{\ref{inst:0093}}
\and A.        ~Mastrobuono-Battisti          \orcit{0000-0002-2386-9142}\inst{\ref{inst:0033}}
\and T.        ~Mazeh                         \orcit{0000-0002-3569-3391}\inst{\ref{inst:0251}}
\and P.J.      ~McMillan                      \orcit{0000-0002-8861-2620}\inst{\ref{inst:0033}}
\and S.        ~Messina                       \orcit{0000-0002-2851-2468}\inst{\ref{inst:0087}}
\and N.R.      ~Millar                        \inst{\ref{inst:0014}}
\and A.        ~Mints                         \orcit{0000-0002-8440-1455}\inst{\ref{inst:0117}}
\and D.        ~Molina                        \orcit{0000-0003-4814-0275}\inst{\ref{inst:0011}}
\and R.        ~Molinaro                      \orcit{0000-0003-3055-6002}\inst{\ref{inst:0318}}
\and L.        ~Moln\'{a}r                    \orcit{0000-0002-8159-1599}\inst{\ref{inst:0351},\ref{inst:0352},\ref{inst:0353}}
\and P.        ~Montegriffo                   \inst{\ref{inst:0093}}
\and R.        ~Mor                           \orcit{0000-0002-8179-6527}\inst{\ref{inst:0011}}
\and R.        ~Morbidelli                    \orcit{0000-0001-7627-4946}\inst{\ref{inst:0001}}
\and T.        ~Morel                         \inst{\ref{inst:0094}}
\and D.        ~Morris                        \inst{\ref{inst:0006}}
\and A.F.      ~Mulone                        \inst{\ref{inst:0063}}
\and D.        ~Munoz                         \inst{\ref{inst:0150}}
\and T.        ~Muraveva                      \orcit{0000-0002-0969-1915}\inst{\ref{inst:0093}}
\and C.P.      ~Murphy                        \inst{\ref{inst:0032}}
\and I.        ~Musella                       \orcit{0000-0001-5909-6615}\inst{\ref{inst:0318}}
\and L.        ~Noval                         \inst{\ref{inst:0150}}
\and C.        ~Ord\'{e}novic                 \inst{\ref{inst:0025}}
\and G.        ~Orr\`{u}                      \inst{\ref{inst:0063}}
\and J.        ~Osinde                        \inst{\ref{inst:0017}}
\and C.        ~Pagani                        \inst{\ref{inst:0008}}
\and I.        ~Pagano                        \orcit{0000-0001-9573-4928}\inst{\ref{inst:0087}}
\and L.        ~Palaversa                     \inst{\ref{inst:0370},\ref{inst:0014}}
\and P.A.      ~Palicio                       \orcit{0000-0002-7432-8709}\inst{\ref{inst:0025}}
\and A.        ~Panahi                        \orcit{0000-0001-5850-4373}\inst{\ref{inst:0251}}
\and M.        ~Pawlak                        \orcit{0000-0002-5632-9433}\inst{\ref{inst:0374},\ref{inst:0309}}
\and X.        ~Pe\~{n}alosa Esteller         \inst{\ref{inst:0011}}
\and A.        ~Penttil\"{ a}                 \orcit{0000-0001-7403-1721}\inst{\ref{inst:0132}}
\and A.M.      ~Piersimoni                    \orcit{0000-0002-8019-3708}\inst{\ref{inst:0227}}
\and F.-X.     ~Pineau                        \orcit{0000-0002-2335-4499}\inst{\ref{inst:0108}}
\and E.        ~Plachy                        \orcit{0000-0002-5481-3352}\inst{\ref{inst:0351},\ref{inst:0352},\ref{inst:0353}}
\and G.        ~Plum                          \inst{\ref{inst:0023}}
\and E.        ~Poggio                        \orcit{0000-0003-3793-8505}\inst{\ref{inst:0001}}
\and E.        ~Poretti                       \orcit{0000-0003-1200-0473}\inst{\ref{inst:0385}}
\and E.        ~Poujoulet                     \inst{\ref{inst:0386}}
\and A.        ~Pr\v{s}a                      \orcit{0000-0002-1913-0281}\inst{\ref{inst:0303}}
\and L.        ~Pulone                        \orcit{0000-0002-5285-998X}\inst{\ref{inst:0128}}
\and E.        ~Racero                        \inst{\ref{inst:0092},\ref{inst:0390}}
\and S.        ~Ragaini                       \inst{\ref{inst:0093}}
\and M.        ~Rainer                        \orcit{0000-0002-8786-2572}\inst{\ref{inst:0039}}
\and C.M.      ~Raiteri                       \orcit{0000-0003-1784-2784}\inst{\ref{inst:0001}}
\and N.        ~Rambaux                       \inst{\ref{inst:0085}}
\and P.        ~Ramos                         \orcit{0000-0002-5080-7027}\inst{\ref{inst:0011}}
\and M.        ~Ramos-Lerate                  \inst{\ref{inst:0396}}
\and P.        ~Re Fiorentin                  \orcit{0000-0002-4995-0475}\inst{\ref{inst:0001}}
\and S.        ~Regibo                        \inst{\ref{inst:0151}}
\and V.        ~Ripepi                        \orcit{0000-0003-1801-426X}\inst{\ref{inst:0318}}
\and A.        ~Riva                          \orcit{0000-0002-6928-8589}\inst{\ref{inst:0001}}
\and G.        ~Rixon                         \inst{\ref{inst:0014}}
\and N.        ~Robichon                      \orcit{0000-0003-4545-7517}\inst{\ref{inst:0023}}
\and C.        ~Robin                         \inst{\ref{inst:0150}}
\and M.        ~Roelens                       \orcit{0000-0003-0876-4673}\inst{\ref{inst:0027}}
\and L.        ~Rohrbasser                    \inst{\ref{inst:0066}}
\and M.        ~Romero-G\'{o}mez              \orcit{0000-0003-3936-1025}\inst{\ref{inst:0011}}
\and N.        ~Rowell                        \inst{\ref{inst:0006}}
\and F.        ~Royer                         \orcit{0000-0002-9374-8645}\inst{\ref{inst:0023}}
\and K.A.      ~Rybicki                       \orcit{0000-0002-9326-9329}\inst{\ref{inst:0309}}
\and G.        ~Sadowski                      \inst{\ref{inst:0037}}
\and A.        ~Sagrist\`{a} Sell\'{e}s       \orcit{0000-0001-6191-2028}\inst{\ref{inst:0024}}
\and J.        ~Salgado                       \orcit{0000-0002-3680-4364}\inst{\ref{inst:0079}}
\and E.        ~Salguero                      \inst{\ref{inst:0099}}
\and N.        ~Samaras                       \orcit{0000-0001-8375-6652}\inst{\ref{inst:0058}}
\and V.        ~Sanchez Gimenez               \inst{\ref{inst:0011}}
\and N.        ~Sanna                         \inst{\ref{inst:0039}}
\and R.        ~Santove\~{n}a                 \orcit{0000-0002-9257-2131}\inst{\ref{inst:0158}}
\and M.        ~Sarasso                       \orcit{0000-0001-5121-0727}\inst{\ref{inst:0001}}
\and M.        ~Schultheis                    \orcit{0000-0002-6590-1657}\inst{\ref{inst:0025}}
\and E.        ~Sciacca                       \orcit{0000-0002-5574-2787}\inst{\ref{inst:0087}}
\and M.        ~Segol                         \inst{\ref{inst:0262}}
\and J.C.      ~Segovia                       \inst{\ref{inst:0092}}
\and D.        ~S\'{e}gransan                 \orcit{0000-0003-2355-8034}\inst{\ref{inst:0027}}
\and D.        ~Semeux                        \inst{\ref{inst:0248}}
\and S.        ~Shahaf                        \orcit{0000-0001-9298-8068}\inst{\ref{inst:0251}}
\and H.I.      ~Siddiqui                      \orcit{0000-0003-1853-6033}\inst{\ref{inst:0426}}
\and A.        ~Siebert                       \orcit{0000-0001-8059-2840}\inst{\ref{inst:0108},\ref{inst:0255}}
\and L.        ~Siltala                       \orcit{0000-0002-6938-794X}\inst{\ref{inst:0132}}
\and E.        ~Slezak                        \inst{\ref{inst:0025}}
\and E.        ~Solano                        \inst{\ref{inst:0431}}
\and F.        ~Solitro                       \inst{\ref{inst:0063}}
\and D.        ~Souami                        \orcit{0000-0003-4058-0815}\inst{\ref{inst:0301},\ref{inst:0434}}
\and J.        ~Souchay                       \inst{\ref{inst:0081}}
\and A.        ~Spagna                        \orcit{0000-0003-1732-2412}\inst{\ref{inst:0001}}
\and F.        ~Spoto                         \orcit{0000-0001-7319-5847}\inst{\ref{inst:0186}}
\and I.A.      ~Steele                        \orcit{0000-0001-8397-5759}\inst{\ref{inst:0331}}
\and H.        ~Steidelm\"{ u}ller            \inst{\ref{inst:0031}}
\and C.A.      ~Stephenson                    \inst{\ref{inst:0079}}
\and M.        ~S\"{ u}veges                  \inst{\ref{inst:0066},\ref{inst:0442},\ref{inst:0003}}
\and L.        ~Szabados                      \orcit{0000-0002-2046-4131}\inst{\ref{inst:0351}}
\and E.        ~Szegedi-Elek                  \orcit{0000-0001-7807-6644}\inst{\ref{inst:0351}}
\and F.        ~Taris                         \inst{\ref{inst:0081}}
\and G.        ~Tauran                        \inst{\ref{inst:0150}}
\and M.B.      ~Taylor                        \orcit{0000-0002-4209-1479}\inst{\ref{inst:0448}}
\and R.        ~Teixeira                      \orcit{0000-0002-6806-6626}\inst{\ref{inst:0230}}
\and W.        ~Thuillot                      \inst{\ref{inst:0085}}
\and N.        ~Tonello                       \orcit{0000-0003-0550-1667}\inst{\ref{inst:0159}}
\and F.        ~Torra                         \orcit{0000-0002-8429-299X}\inst{\ref{inst:0053}}
\and J.        ~Torra$^\dagger$               \inst{\ref{inst:0011}}
\and C.        ~Turon                         \orcit{0000-0003-1236-5157}\inst{\ref{inst:0023}}
\and N.        ~Unger                         \orcit{0000-0003-3993-7127}\inst{\ref{inst:0027}}
\and M.        ~Vaillant                      \inst{\ref{inst:0150}}
\and E.        ~van Dillen                    \inst{\ref{inst:0262}}
\and O.        ~Vanel                         \inst{\ref{inst:0023}}
\and A.        ~Vecchiato                     \orcit{0000-0003-1399-5556}\inst{\ref{inst:0001}}
\and Y.        ~Viala                         \inst{\ref{inst:0023}}
\and D.        ~Vicente                       \inst{\ref{inst:0159}}
\and S.        ~Voutsinas                     \inst{\ref{inst:0006}}
\and M.        ~Weiler                        \inst{\ref{inst:0011}}
\and T.        ~Wevers                        \orcit{0000-0002-4043-9400}\inst{\ref{inst:0014}}
\and \L{}.     ~Wyrzykowski                   \orcit{0000-0002-9658-6151}\inst{\ref{inst:0309}}
\and A.        ~Yoldas                        \inst{\ref{inst:0014}}
\and P.        ~Yvard                         \inst{\ref{inst:0262}}
\and H.        ~Zhao                          \orcit{0000-0003-2645-6869}\inst{\ref{inst:0025}}
\and J.        ~Zorec                         \inst{\ref{inst:0469}}
\and S.        ~Zucker                        \orcit{0000-0003-3173-3138}\inst{\ref{inst:0470}}
\and C.        ~Zurbach                       \inst{\ref{inst:0471}}
\and T.        ~Zwitter                       \orcit{0000-0002-2325-8763}\inst{\ref{inst:0472}}
}
\institute{
     INAF - Osservatorio Astrofisico di Torino, via Osservatorio 20, 10025 Pino Torinese (TO), Italy\relax                                                                                                                                                                                                       \label{inst:0001}
\and Dpto. de Inteligencia Artificial, UNED, c/ Juan del Rosal 16, 28040 Madrid, Spain\relax                                                                                                                                                                                                                     \label{inst:0002}
\and Max Planck Institute for Astronomy, K\"{ o}nigstuhl 17, 69117 Heidelberg, Germany\relax                                                                                                                                                                                                                     \label{inst:0003}
\and Institut UTINAM CNRS UMR6213, Universit\'{e} Bourgogne Franche-Comt\'{e}, OSU THETA Franche-Comt\'{e} Bourgogne, Observatoire de Besan\c{c}on, BP1615, 25010 Besan\c{c}on Cedex, France\relax                                                                                                               \label{inst:0004}
\and Institute for Astronomy, University of Edinburgh, Royal Observatory, Blackford Hill, Edinburgh EH9 3HJ, United Kingdom\relax                                                                                                                                                                                \label{inst:0006}
\and School of Physics and Astronomy, University of Leicester, University Road, Leicester LE1 7RH, United Kingdom\relax                                                                                                                                                                                          \label{inst:0008}
\and European Space Agency (ESA), European Space Research and Technology Centre (ESTEC), Keplerlaan 1, 2201AZ, Noordwijk, The Netherlands\relax                                                                                                                                                                  \label{inst:0009}
\and Institut de Ci\`{e}ncies del Cosmos (ICCUB), Universitat  de  Barcelona  (IEEC-UB), Mart\'{i} i  Franqu\`{e}s  1, 08028 Barcelona, Spain\relax                                                                                                                                                              \label{inst:0011}
\and Centre for Astrophysics Research, University of Hertfordshire, College Lane, AL10 9AB, Hatfield, United Kingdom\relax                                                                                                                                                                                       \label{inst:0012}
\and Institute of Astronomy, University of Cambridge, Madingley Road, Cambridge CB3 0HA, United Kingdom\relax                                                                                                                                                                                                    \label{inst:0014}
\and RHEA for European Space Agency (ESA), Camino bajo del Castillo, s/n, Urbanizacion Villafranca del Castillo, Villanueva de la Ca\~{n}ada, 28692 Madrid, Spain\relax                                                                                                                                          \label{inst:0017}
\and Leiden Observatory, Leiden University, Niels Bohrweg 2, 2333 CA Leiden, The Netherlands\relax                                                                                                                                                                                                               \label{inst:0019}
\and INAF - Osservatorio astronomico di Padova, Vicolo Osservatorio 5, 35122 Padova, Italy\relax                                                                                                                                                                                                                 \label{inst:0020}
\and Univ. Grenoble Alpes, CNRS, IPAG, 38000 Grenoble, France\relax                                                                                                                                                                                                                                              \label{inst:0022}
\and GEPI, Observatoire de Paris, Universit\'{e} PSL, CNRS, 5 Place Jules Janssen, 92190 Meudon, France\relax                                                                                                                                                                                                    \label{inst:0023}
\and Astronomisches Rechen-Institut, Zentrum f\"{ u}r Astronomie der Universit\"{ a}t Heidelberg, M\"{ o}nchhofstr. 12-14, 69120 Heidelberg, Germany\relax                                                                                                                                                       \label{inst:0024}
\and Universit\'{e} C\^{o}te d'Azur, Observatoire de la C\^{o}te d'Azur, CNRS, Laboratoire Lagrange, Bd de l'Observatoire, CS 34229, 06304 Nice Cedex 4, France\relax                                                                                                                                            \label{inst:0025}
\and Department of Astronomy, University of Geneva, Chemin des Maillettes 51, 1290 Versoix, Switzerland\relax                                                                                                                                                                                                    \label{inst:0027}
\and Aurora Technology for European Space Agency (ESA), Camino bajo del Castillo, s/n, Urbanizacion Villafranca del Castillo, Villanueva de la Ca\~{n}ada, 28692 Madrid, Spain\relax                                                                                                                             \label{inst:0028}
\and Lohrmann Observatory, Technische Universit\"{ a}t Dresden, Mommsenstra{\ss}e 13, 01062 Dresden, Germany\relax                                                                                                                                                                                               \label{inst:0031}
\and European Space Agency (ESA), European Space Astronomy Centre (ESAC), Camino bajo del Castillo, s/n, Urbanizacion Villafranca del Castillo, Villanueva de la Ca\~{n}ada, 28692 Madrid, Spain\relax                                                                                                           \label{inst:0032}
\and Lund Observatory, Department of Astronomy and Theoretical Physics, Lund University, Box 43, 22100 Lund, Sweden\relax                                                                                                                                                                                        \label{inst:0033}
\and CNES Centre Spatial de Toulouse, 18 avenue Edouard Belin, 31401 Toulouse Cedex 9, France\relax                                                                                                                                                                                                              \label{inst:0036}
\and Institut d'Astronomie et d'Astrophysique, Universit\'{e} Libre de Bruxelles CP 226, Boulevard du Triomphe, 1050 Brussels, Belgium\relax                                                                                                                                                                     \label{inst:0037}
\and F.R.S.-FNRS, Rue d'Egmont 5, 1000 Brussels, Belgium\relax                                                                                                                                                                                                                                                   \label{inst:0038}
\and INAF - Osservatorio Astrofisico di Arcetri, Largo Enrico Fermi 5, 50125 Firenze, Italy\relax                                                                                                                                                                                                                \label{inst:0039}
\and Laboratoire d'astrophysique de Bordeaux, Univ. Bordeaux, CNRS, B18N, all{\'e}e Geoffroy Saint-Hilaire, 33615 Pessac, France\relax                                                                                                                                                                           \label{inst:0041}
\vfill\break
\and Mullard Space Science Laboratory, University College London, Holmbury St Mary, Dorking, Surrey RH5 6NT, United Kingdom\relax                                                                                                                                                                                \label{inst:0046}
\and University of Turin, Department of Physics, Via Pietro Giuria 1, 10125 Torino, Italy\relax                                                                                                                                                                                                                  \label{inst:0050}
\and DAPCOM for Institut de Ci\`{e}ncies del Cosmos (ICCUB), Universitat  de  Barcelona  (IEEC-UB), Mart\'{i} i  Franqu\`{e}s  1, 08028 Barcelona, Spain\relax                                                                                                                                                   \label{inst:0053}
\and Royal Observatory of Belgium, Ringlaan 3, 1180 Brussels, Belgium\relax                                                                                                                                                                                                                                      \label{inst:0058}
\and ALTEC S.p.a, Corso Marche, 79,10146 Torino, Italy\relax                                                                                                                                                                                                                                                     \label{inst:0063}
\and Department of Astronomy, University of Geneva, Chemin d'Ecogia 16, 1290 Versoix, Switzerland\relax                                                                                                                                                                                                          \label{inst:0066}
\and Sednai S\`{a}rl, Geneva, Switzerland\relax                                                                                                                                                                                                                                                                  \label{inst:0067}
\and Gaia DPAC Project Office, ESAC, Camino bajo del Castillo, s/n, Urbanizacion Villafranca del Castillo, Villanueva de la Ca\~{n}ada, 28692 Madrid, Spain\relax                                                                                                                                                \label{inst:0076}
\and Telespazio Vega UK Ltd for European Space Agency (ESA), Camino bajo del Castillo, s/n, Urbanizacion Villafranca del Castillo, Villanueva de la Ca\~{n}ada, 28692 Madrid, Spain\relax                                                                                                                        \label{inst:0079}
\and SYRTE, Observatoire de Paris, Universit\'{e} PSL, CNRS,  Sorbonne Universit\'{e}, LNE, 61 avenue de l’Observatoire 75014 Paris, France\relax                                                                                                                                                              \label{inst:0081}
\and National Observatory of Athens, I. Metaxa and Vas. Pavlou, Palaia Penteli, 15236 Athens, Greece\relax                                                                                                                                                                                                       \label{inst:0083}
\and IMCCE, Observatoire de Paris, Universit\'{e} PSL, CNRS, Sorbonne Universit{\'e}, Univ. Lille, 77 av. Denfert-Rochereau, 75014 Paris, France\relax                                                                                                                                                           \label{inst:0085}
\and INAF - Osservatorio Astrofisico di Catania, via S. Sofia 78, 95123 Catania, Italy\relax                                                                                                                                                                                                                     \label{inst:0087}
\and Serco Gesti\'{o}n de Negocios for European Space Agency (ESA), Camino bajo del Castillo, s/n, Urbanizacion Villafranca del Castillo, Villanueva de la Ca\~{n}ada, 28692 Madrid, Spain\relax                                                                                                                 \label{inst:0092}
\and INAF - Osservatorio di Astrofisica e Scienza dello Spazio di Bologna, via Piero Gobetti 93/3, 40129 Bologna, Italy\relax                                                                                                                                                                                    \label{inst:0093}
\and Institut d'Astrophysique et de G\'{e}ophysique, Universit\'{e} de Li\`{e}ge, 19c, All\'{e}e du 6 Ao\^{u}t, B-4000 Li\`{e}ge, Belgium\relax                                                                                                                                                                  \label{inst:0094}
\and CRAAG - Centre de Recherche en Astronomie, Astrophysique et G\'{e}ophysique, Route de l'Observatoire Bp 63 Bouzareah 16340 Algiers, Algeria\relax                                                                                                                                                           \label{inst:0095}
\and ATG Europe for European Space Agency (ESA), Camino bajo del Castillo, s/n, Urbanizacion Villafranca del Castillo, Villanueva de la Ca\~{n}ada, 28692 Madrid, Spain\relax                                                                                                                                    \label{inst:0099}
\and ETSE Telecomunicaci\'{o}n, Universidade de Vigo, Campus Lagoas-Marcosende, 36310 Vigo, Galicia, Spain\relax                                                                                                                                                                                                 \label{inst:0104}
\and Universit\'{e} de Strasbourg, CNRS, Observatoire astronomique de Strasbourg, UMR 7550,  11 rue de l'Universit\'{e}, 67000 Strasbourg, France\relax                                                                                                                                                          \label{inst:0108}
\and Kavli Institute for Cosmology Cambridge, Institute of Astronomy, Madingley Road, Cambridge, CB3 0HA\relax                                                                                                                                                                                                   \label{inst:0110}
\and Department of Astrophysics, Astronomy and Mechanics, National and Kapodistrian University of Athens, Panepistimiopolis, Zografos, 15783 Athens, Greece\relax                                                                                                                                                \label{inst:0111}
\and Observational Astrophysics, Division of Astronomy and Space Physics, Department of Physics and Astronomy, Uppsala University, Box 516, 751 20 Uppsala, Sweden\relax                                                                                                                                         \label{inst:0112}
\and Leibniz Institute for Astrophysics Potsdam (AIP), An der Sternwarte 16, 14482 Potsdam, Germany\relax                                                                                                                                                                                                        \label{inst:0117}
\vfill\break
\and CENTRA, Faculdade de Ci\^{e}ncias, Universidade de Lisboa, Edif. C8, Campo Grande, 1749-016 Lisboa, Portugal\relax                                                                                                                                                                                          \label{inst:0120}
\and Department of Informatics, Donald Bren School of Information and Computer Sciences, University of California, 5019 Donald Bren Hall, 92697-3440 CA Irvine, United States\relax                                                                                                                              \label{inst:0121}
\and Dipartimento di Fisica e Astronomia ""Ettore Majorana"", Universit\`{a} di Catania, Via S. Sofia 64, 95123 Catania, Italy\relax                                                                                                                                                                             \label{inst:0123}
\and CITIC, Department of Nautical Sciences and Marine Engineering, University of A Coru\~{n}a, Campus de Elvi\~{n}a S/N, 15071, A Coru\~{n}a, Spain\relax                                                                                                                                                       \label{inst:0126}
\and INAF - Osservatorio Astronomico di Roma, Via Frascati 33, 00078 Monte Porzio Catone (Roma), Italy\relax                                                                                                                                                                                                     \label{inst:0128}
\and Space Science Data Center - ASI, Via del Politecnico SNC, 00133 Roma, Italy\relax                                                                                                                                                                                                                           \label{inst:0129}
\and Department of Physics, University of Helsinki, P.O. Box 64, 00014 Helsinki, Finland\relax                                                                                                                                                                                                                   \label{inst:0132}
\and Finnish Geospatial Research Institute FGI, Geodeetinrinne 2, 02430 Masala, Finland\relax                                                                                                                                                                                                                    \label{inst:0133}
\and STFC, Rutherford Appleton Laboratory, Harwell, Didcot, OX11 0QX, United Kingdom\relax                                                                                                                                                                                                                       \label{inst:0139}
\and HE Space Operations BV for European Space Agency (ESA), Keplerlaan 1, 2201AZ, Noordwijk, The Netherlands\relax                                                                                                                                                                                              \label{inst:0142}
\and Applied Physics Department, Universidade de Vigo, 36310 Vigo, Spain\relax                                                                                                                                                                                                                                   \label{inst:0145}
\and Thales Services for CNES Centre Spatial de Toulouse, 18 avenue Edouard Belin, 31401 Toulouse Cedex 9, France\relax                                                                                                                                                                                          \label{inst:0150}
\and Instituut voor Sterrenkunde, KU Leuven, Celestijnenlaan 200D, 3001 Leuven, Belgium\relax                                                                                                                                                                                                                    \label{inst:0151}
\and Department of Astrophysics/IMAPP, Radboud University, P.O.Box 9010, 6500 GL Nijmegen, The Netherlands\relax                                                                                                                                                                                                 \label{inst:0152}
\and CITIC - Department of Computer Science and Information Technologies, University of A Coru\~{n}a, Campus de Elvi\~{n}a S/N, 15071, A Coru\~{n}a, Spain\relax                                                                                                                                                 \label{inst:0158}
\and Barcelona Supercomputing Center (BSC) - Centro Nacional de Supercomputaci\'{o}n, c/ Jordi Girona 29, Ed. Nexus II, 08034 Barcelona, Spain\relax                                                                                                                                                             \label{inst:0159}
\and University of Vienna, Department of Astrophysics, T\"{ u}rkenschanzstra{\ss}e 17, A1180 Vienna, Austria\relax                                                                                                                                                                                               \label{inst:0160}
\and European Southern Observatory, Karl-Schwarzschild-Str. 2, 85748 Garching, Germany\relax                                                                                                                                                                                                                     \label{inst:0161}
\and Kapteyn Astronomical Institute, University of Groningen, Landleven 12, 9747 AD Groningen, The Netherlands\relax                                                                                                                                                                                             \label{inst:0168}
\and Center for Research and Exploration in Space Science and Technology, University of Maryland Baltimore County, 1000 Hilltop Circle, Baltimore MD, USA\relax                                                                                                                                                  \label{inst:0182}
\and GSFC - Goddard Space Flight Center, Code 698, 8800 Greenbelt Rd, 20771 MD Greenbelt, United States\relax                                                                                                                                                                                                    \label{inst:0183}
\and EURIX S.r.l., Corso Vittorio Emanuele II 61, 10128, Torino, Italy\relax                                                                                                                                                                                                                                     \label{inst:0185}
\and Harvard-Smithsonian Center for Astrophysics, 60 Garden St., MS 15, Cambridge, MA 02138, USA\relax                                                                                                                                                                                                           \label{inst:0186}
\and HE Space Operations BV for European Space Agency (ESA), Camino bajo del Castillo, s/n, Urbanizacion Villafranca del Castillo, Villanueva de la Ca\~{n}ada, 28692 Madrid, Spain\relax                                                                                                                        \label{inst:0188}
\and CAUP - Centro de Astrofisica da Universidade do Porto, Rua das Estrelas, Porto, Portugal\relax                                                                                                                                                                                                              \label{inst:0189}
\and SISSA - Scuola Internazionale Superiore di Studi Avanzati, via Bonomea 265, 34136 Trieste, Italy\relax                                                                                                                                                                                                      \label{inst:0194}
\and Telespazio for CNES Centre Spatial de Toulouse, 18 avenue Edouard Belin, 31401 Toulouse Cedex 9, France\relax                                                                                                                                                                                               \label{inst:0196}
\vfill\break
\and University of Turin, Department of Computer Sciences, Corso Svizzera 185, 10149 Torino, Italy\relax                                                                                                                                                                                                         \label{inst:0201}
\and Dpto. de Matem\'{a}tica Aplicada y Ciencias de la Computaci\'{o}n, Univ. de Cantabria, ETS Ingenieros de Caminos, Canales y Puertos, Avda. de los Castros s/n, 39005 Santander, Spain\relax                                                                                                                 \label{inst:0204}
\and Centro de Astronom\'{i}a - CITEVA, Universidad de Antofagasta, Avenida Angamos 601, Antofagasta 1270300, Chile\relax                                                                                                                                                                                        \label{inst:0213}
\and Vera C Rubin Observatory,  950 N. Cherry Avenue, Tucson, AZ 85719, USA\relax                                                                                                                                                                                                                                \label{inst:0216}
\and University of Antwerp, Onderzoeksgroep Toegepaste Wiskunde, Middelheimlaan 1, 2020 Antwerp, Belgium\relax                                                                                                                                                                                                   \label{inst:0224}
\and INAF - Osservatorio Astronomico d'Abruzzo, Via Mentore Maggini, 64100 Teramo, Italy\relax                                                                                                                                                                                                                   \label{inst:0227}
\and Instituto de Astronomia, Geof\`{i}sica e Ci\^{e}ncias Atmosf\'{e}ricas, Universidade de S\~{a}o Paulo, Rua do Mat\~{a}o, 1226, Cidade Universitaria, 05508-900 S\~{a}o Paulo, SP, Brazil\relax                                                                                                              \label{inst:0230}
\and M\'{e}socentre de calcul de Franche-Comt\'{e}, Universit\'{e} de Franche-Comt\'{e}, 16 route de Gray, 25030 Besan\c{c}on Cedex, France\relax                                                                                                                                                                \label{inst:0239}
\and SRON, Netherlands Institute for Space Research, Sorbonnelaan 2, 3584CA, Utrecht, The Netherlands\relax                                                                                                                                                                                                      \label{inst:0243}
\and Theoretical Astrophysics, Division of Astronomy and Space Physics, Department of Physics and Astronomy, Uppsala University, Box 516, 751 20 Uppsala, Sweden\relax                                                                                                                                           \label{inst:0245}
\and ATOS for CNES Centre Spatial de Toulouse, 18 avenue Edouard Belin, 31401 Toulouse Cedex 9, France\relax                                                                                                                                                                                                     \label{inst:0248}
\and School of Physics and Astronomy, Tel Aviv University, Tel Aviv 6997801, Israel\relax                                                                                                                                                                                                                        \label{inst:0251}
\and Astrophysics Research Centre, School of Mathematics and Physics, Queen's University Belfast, Belfast BT7 1NN, UK\relax                                                                                                                                                                                      \label{inst:0253}
\and Centre de Donn\'{e}es Astronomique de Strasbourg, Strasbourg, France\relax                                                                                                                                                                                                                                  \label{inst:0255}
\and Universit\'{e} C\^{o}te d'Azur, Observatoire de la C\^{o}te d'Azur, CNRS, Laboratoire G\'{e}oazur, Bd de l'Observatoire, CS 34229, 06304 Nice Cedex 4, France\relax                                                                                                                                         \label{inst:0256}
\and Max-Planck-Institut f\"{ u}r Astrophysik, Karl-Schwarzschild-Stra{\ss}e 1, 85748 Garching, Germany\relax                                                                                                                                                                                                    \label{inst:0260}
\and APAVE SUDEUROPE SAS for CNES Centre Spatial de Toulouse, 18 avenue Edouard Belin, 31401 Toulouse Cedex 9, France\relax                                                                                                                                                                                      \label{inst:0262}
\and \'{A}rea de Lenguajes y Sistemas Inform\'{a}ticos, Universidad Pablo de Olavide, Ctra. de Utrera, km 1. 41013, Sevilla, Spain\relax                                                                                                                                                                         \label{inst:0266}
\and Onboard Space Systems, Lule\aa{} University of Technology, Box 848, S-981 28 Kiruna, Sweden\relax                                                                                                                                                                                                           \label{inst:0280}
\and TRUMPF Photonic Components GmbH, Lise-Meitner-Stra{\ss}e 13,  89081 Ulm, Germany\relax                                                                                                                                                                                                                      \label{inst:0285}
\and IAC - Instituto de Astrofisica de Canarias, Via L\'{a}ctea s/n, 38200 La Laguna S.C., Tenerife, Spain\relax                                                                                                                                                                                                 \label{inst:0288}
\and Department of Astrophysics, University of La Laguna, Via L\'{a}ctea s/n, 38200 La Laguna S.C., Tenerife, Spain\relax                                                                                                                                                                                        \label{inst:0289}
\and Laboratoire Univers et Particules de Montpellier, CNRS Universit\'{e} Montpellier, Place Eug\`{e}ne Bataillon, CC72, 34095 Montpellier Cedex 05, France\relax                                                                                                                                               \label{inst:0295}
\and LESIA, Observatoire de Paris, Universit\'{e} PSL, CNRS, Sorbonne Universit\'{e}, Universit\'{e} de Paris, 5 Place Jules Janssen, 92190 Meudon, France\relax                                                                                                                                                 \label{inst:0301}
\and Villanova University, Department of Astrophysics and Planetary Science, 800 E Lancaster Avenue, Villanova PA 19085, USA\relax                                                                                                                                                                               \label{inst:0303}
\and Astronomical Observatory, University of Warsaw,  Al. Ujazdowskie 4, 00-478 Warszawa, Poland\relax                                                                                                                                                                                                           \label{inst:0309}
\and Laboratoire d'astrophysique de Bordeaux, Univ. Bordeaux, CNRS, B18N, all\'{e}e Geoffroy Saint-Hilaire, 33615 Pessac, France\relax                                                                                                                                                                           \label{inst:0313}
\vfill\break
\and Universit\'{e} Rennes, CNRS, IPR (Institut de Physique de Rennes) - UMR 6251, 35000 Rennes, France\relax                                                                                                                                                                                                    \label{inst:0316}
\and INAF - Osservatorio Astronomico di Capodimonte, Via Moiariello 16, 80131, Napoli, Italy\relax                                                                                                                                                                                                               \label{inst:0318}
\and Niels Bohr Institute, University of Copenhagen, Juliane Maries Vej 30, 2100 Copenhagen {\O}, Denmark\relax                                                                                                                                                                                                  \label{inst:0324}
\and Las Cumbres Observatory, 6740 Cortona Drive Suite 102, Goleta, CA 93117, USA\relax                                                                                                                                                                                                                          \label{inst:0325}
\and Astrophysics Research Institute, Liverpool John Moores University, 146 Brownlow Hill, Liverpool L3 5RF, United Kingdom\relax                                                                                                                                                                                \label{inst:0331}
\and IPAC, Mail Code 100-22, California Institute of Technology, 1200 E. California Blvd., Pasadena, CA 91125, USA\relax                                                                                                                                                                                         \label{inst:0336}
\and Jet Propulsion Laboratory, California Institute of Technology, 4800 Oak Grove Drive, M/S 169-327, Pasadena, CA 91109, USA\relax                                                                                                                                                                             \label{inst:0337}
\and IRAP, Universit\'{e} de Toulouse, CNRS, UPS, CNES, 9 Av. colonel Roche, BP 44346, 31028 Toulouse Cedex 4, France\relax                                                                                                                                                                                      \label{inst:0338}
\and Konkoly Observatory, Research Centre for Astronomy and Earth Sciences, MTA Centre of Excellence, Konkoly Thege Mikl\'{o}s \'{u}t 15-17, 1121 Budapest, Hungary\relax                                                                                                                                        \label{inst:0351}
\and MTA CSFK Lend\"{ u}let Near-Field Cosmology Research Group,
Konkoly Observatory, CSFK, Konkoly Thege Mikl\'os \'ut 15-17, H-1121
Budapest, Hungary\relax\label{inst:0352}
\and ELTE E\"{ o}tv\"{ o}s Lor\'{a}nd University, Institute of Physics, 1117, P\'{a}zm\'{a}ny P\'{e}ter s\'{e}t\'{a}ny 1A, Budapest, Hungary\relax                                                                                                                                                               \label{inst:0353}
\and Ru{\dj}er Bo\v{s}kovi\'{c} Institute, Bijeni\v{c}ka cesta 54, 10000 Zagreb, Croatia\relax                                                                                                                                                                                                                   \label{inst:0370}
\and Institute of Theoretical Physics, Faculty of Mathematics and Physics, Charles University in Prague, Czech Republic\relax                                                                                                                                                                                    \label{inst:0374}
\and INAF - Osservatorio Astronomico di Brera, via E. Bianchi 46, 23807 Merate (LC), Italy\relax                                                                                                                                                                                                                 \label{inst:0385}
\and AKKA for CNES Centre Spatial de Toulouse, 18 avenue Edouard Belin, 31401 Toulouse Cedex 9, France\relax                                                                                                                                                                                                     \label{inst:0386}
\and Departmento de F\'{i}sica de la Tierra y Astrof\'{i}sica, Universidad Complutense de Madrid, 28040 Madrid, Spain\relax                                                                                                                                                                                      \label{inst:0390}
\and Vitrociset Belgium for European Space Agency (ESA), Camino bajo del Castillo, s/n, Urbanizacion Villafranca del Castillo, Villanueva de la Ca\~{n}ada, 28692 Madrid, Spain\relax                                                                                                                            \label{inst:0396}
\and Department of Astrophysical Sciences, 4 Ivy Lane, Princeton University, Princeton NJ 08544, USA\relax                                                                                                                                                                                                       \label{inst:0426}
\and Departamento de Astrof\'{i}sica, Centro de Astrobiolog\'{i}a (CSIC-INTA), ESA-ESAC. Camino Bajo del Castillo s/n. 28692 Villanueva de la Ca\~{n}ada, Madrid, Spain\relax                                                                                                                                    \label{inst:0431}
\and naXys, University of Namur, Rempart de la Vierge, 5000 Namur, Belgium\relax                                                                                                                                                                                                                                 \label{inst:0434}
\and EPFL - Ecole Polytechnique f\'{e}d\'{e}rale de Lausanne, Institute of Mathematics, Station 8 EPFL SB MATH SDS, Lausanne, Switzerland\relax                                                                                                                                                                  \label{inst:0442}
\and H H Wills Physics Laboratory, University of Bristol, Tyndall Avenue, Bristol BS8 1TL, United Kingdom\relax                                                                                                                                                                                                  \label{inst:0448}
\and Sorbonne Universit\'{e}, CNRS, UMR7095, Institut d'Astrophysique de Paris, 98bis bd. Arago, 75014 Paris, France\relax                                                                                                                                                                                       \label{inst:0469}
\and Porter School of the Environment and Earth Sciences, Tel Aviv University, Tel Aviv 6997801, Israel\relax                                                                                                                                                                                                    \label{inst:0470}
\and Laboratoire Univers et Particules de Montpellier, Universit\'{e} Montpellier, Place Eug\`{e}ne Bataillon, CC72, 34095 Montpellier Cedex 05, France\relax                                                                                                                                                    \label{inst:0471}
\and Faculty of Mathematics and Physics, University of Ljubljana, Jadranska ulica 19, 1000 Ljubljana, Slovenia\relax                                                                                                                                                                                             \label{inst:0472}
}

   \date{Received ; accepted }

 
  \abstract
   {}
   {We produce a clean and well-characterised catalogue of
     objects within 100\,pc of the Sun from the \G\ Early Data
     Release 3. We characterise the catalogue through comparisons to the
     full data release, external catalogues, and simulations. We carry
     out a first analysis of the science that is possible with this sample to
     demonstrate its potential and best practices for its use. }
   {The selection of objects within 100\,pc from the full catalogue
     used selected training sets, machine-learning procedures,
     astrometric quantities, and solution quality indicators to determine a
     probability that the astrometric solution is reliable. The
     training set construction exploited the astrometric data, quality
     flags, and external photometry.  For all candidates we calculated
     distance posterior probability densities using Bayesian
     procedures and mock catalogues to define
     priors. Any object with reliable astrometry and a non-zero
     probability of being within 100\,pc is included in the catalogue.
   }
   {We have produced a catalogue of \NFINAL\ objects that we estimate
     contains at least 92\% of stars of stellar type  M9 within 100\,pc of the Sun. We
     estimate that 9\% of the stars in this catalogue probably lie outside 100\,pc, but
     when the distance probability function is used, a correct
     treatment of this contamination is possible. We produced luminosity functions with a high signal-to-noise ratio for the main-sequence stars,
     giants, and white dwarfs. We examined in detail the Hyades
     cluster, the white dwarf population, and wide-binary systems
     and produced candidate lists for all three samples. We detected
     local manifestations of several streams, superclusters, and halo
     objects, in which we identified 12 members of \G\ Enceladus. We
     present the first direct parallaxes of five objects in multiple
     systems within 10\,pc of the Sun. }
   {We provide the community with a large, well-characterised
     catalogue of objects in the solar neighbourhood. This is a
     primary benchmark for measuring and understanding fundamental
     parameters and descriptive functions in astronomy. }


   \titlerunning{The \textit{Gaia} Catalogue of Nearby Stars}
   \authorrunning{Gaia Collaboration} 

   \maketitle
%
\section{Introduction}

The history of astronomical research is rich with instances in which
improvements in our observational knowledge have led to breakthroughs
in our theoretical understanding.  The protracted astronomical timescales
have required astronomers to employ
significant ingenuity to extrapolate today's snapshot in time to understanding the
history and evolution of even the local part of our Galaxy.  This is
hampered by the fact that our knowledge and census of the Galaxy, 
including the local region, is incomplete. The
difficulty has primarily been in the resources required to determine distances and
the lack of a sufficiently deep and complete census of nearby objects,
both of which will be resolved by the ESA \G\
mission.  \G\ will determine distances, motions, and colours of all the
stars, except for the very brightest, in the solar neighbourhood.

The solar neighbourhood has been considerably studied since the
beginning of the past century when astronomers began to routinely
measure stellar parallaxes. In 1957 this effort was formalised with
the publication of 915 known stars within 20\,pc
\citep{1957MiABA...8....1G}.  Various updates and extensions
to larger distances produced what became the Catalogue of Nearby Stars,
including all known stars, 3803, within 25\,pc released in 1991
\cite[CNS,][]{1991adc..rept.....G}. The \hip\ mission increased
the quantity and quality of the CNS content; however, the magnitude
limit of \hip\ resulted in an incompleteness for faint objects. In 1998
the CNS dataset was moved
online\footnote{\url{https://wwwadd.zah.uni-heidelberg.de/datenbanken/aricns/}}
and currently has 5835 entries, but it is no longer updated.  The most
recent update of the CNS by \citet{2010PASP..122..885S} was to provide
accurate coordinates and near-infrared magnitudes taken from the Two
Micron Sky Survey \citep[2MASS,][]{2006AJ....131.1163S}.

The CNS has been used in various investigations, gathering over 300
citations from the studies of wide-binary systems
\citep{2010A&A...514A..98C, 2002ApJ...572L..79L, 1994RMxAA..28...43P,
  1991AJ....101..625L}, searches for solar twins
\citep{1993A&A...274..825F}, statistics for extra-solar planet hosts
\citep{2007ApJS..173..143B, 2007ApJ...670..833J, 2006ApJ...649..389P},
the local luminosity function \citep{2002AJ....124.2721R,
  1999AJ....117..508G, 1998AJ....116.2513M,
  1983nssl.conf..163W,1997AJ....113.2246R}, the mass-luminosity
  relation \citep{1999ApJ...512..864H}, to galactic and local
  kinematics \citep{1997A&A...323..781B, 1974HiA.....3..395W}. The
  utility of the CNS has been limited by its incompleteness and the
  lack of high-precision parallaxes. Other compilations of nearby
  objects have either limited the type of objects to, for example,
  ultra-cool dwarfs and 25\,pc \cite{2019ApJ...883..205B}, cooler T/Y
  dwarfs and 20\,pc \cite{2019ApJS..240...19K}, complete spectral
  coverage but limited volume, such as the REsearch Consortium On
  Nearby Stars 10\,pc sample \cite{2018AJ....155..265H}, or, with the
  inclusion of substellar objects and an 8\,pc
  volume \cite{2012ApJ...753..156K}. However, these catalogues have
   by necessity all been based on multiple observational sources and
  astrometry of limited precision. The high astrometric precision and
  faint magnitude survey mode of \G\ will provide a census that will
  be more complete, in a larger volume, and homogeneous. It is therefore
  easier to characterise.

In this contribution we present the \G\ Catalogue of Nearby Stars
(GCNS), a first attempt to make a census of all stars in the solar
neighbourhood using the \G\ results.  In the GCNS we define the solar
neighbourhood to be a sphere of radius 100\,pc centred on  the Sun. This will be volume-complete for all
objects earlier than M8 at the nominal
$G$=20.7 magnitude limit of \G. Later type objects will be too faint
for \G\ at 100\,pc, resulting in progressively smaller complete
volumes with increasing spectral type. In section 2 we discuss
the generation of the GCNS, in Section 3 we present an overview of
the catalogue contents and availability, in Section 4 we carry out
some quality assurance tests, and in Section 5 we report an
example for a scientific exploitation of the GCNS.

\section{GCNS generation \label{sec:cat-gen}}
\label{sec:2Cataloggeneration}

In this section we describe the process by which we have generated the GCNS starting from a selection of all
sources in the \gdrthree archive with measured parallaxes
$\hat{\varpi} > 8$\,mas (we use $\varpi$ for true
parallaxes and $\hat{\varpi}$ for measured parallaxes). 
The process is composed of two phases: in the first phase
(Sect. \ref{sec:randomforest}), we attempt to remove sources with
spurious astrometric solutions using a random forest
classifier \citep{breiman2001}; and in the second phase
(Sect. \ref{sec:distance_estimation}), we infer posterior probability
densities for the true distance of each source. The GCNS is then
defined based on the classifier probabilities and the properties of
the distance posterior distribution according to criteria specified
below. These procedures are critical for the catalogue generation, and the details pertain to the area of machine-learning. 

\begin{figure}[!htb]
\center{\includegraphics[width=0.45\textwidth]{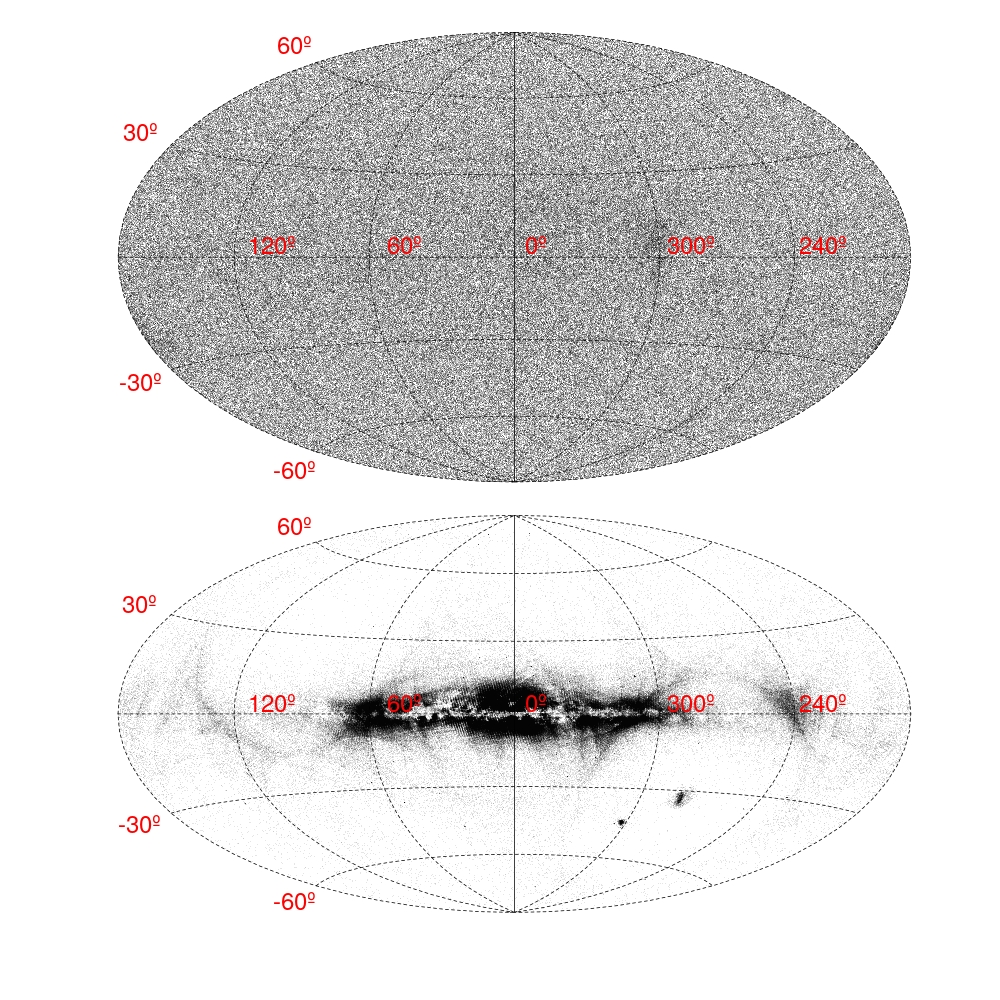}}  
\caption{\label{fig:lbgoodbad}
Distribution of selected (top panel) and rejected (bottom panel)
sources according to the random forest classifier in Galactic
coordinates in an Aitoff projection.}
\end{figure}

\subsection{Removal of spurious sources}
\label{sec:randomforest}

In order to generate the first selection of sources inside 100\,pc, we
constructed a classifier to identify poor astrometric solutions that
result in observed parallaxes greater than 10\,mas from true sources
within the 100\,pc radius.  For objects with \G, $G$=20, the median
uncertainty of \gdrthree parallaxes is~0.5\,mas \citep{EDR3-DPACP-121},
and the global zero-point is between {-20} to
{-40}\,$\mu$as \citep{EDR3-DPACP-132}, therefore the 10\,mas
boundary is extremely well defined.  We started by selecting a
sample with $\hat{\varpi} \ge$ 8\,mas to minimise the sample
size and avoid introducing a large loss of sources due to the
parallax measurement uncertainty. Using the GeDR3mock catalogue
\citep[cf. Sec:\,\ref{sec:distance_estimation}]{2020PASP..132g4501R},
we estimate that about 55 sources lie truly within 100\,pc but
are lost in the primary selection at 8\,mas.  {We find a total
of 1211740 sources with measured parallaxes $\hat{\varpi} \ge
8$\,mas}.

Spurious astrometric solutions can be due to a number of reasons, but the causes
that produce such large parallaxes are mostly related to the inclusion
of outliers in the measured positions because close pairs are only
resolved for certain transits and scan directions \citep[see Section
  7.9 of][]{2018A&A...616A...1G}. This is more likely to occur in
regions of high surface density of sources or for close binary systems
(either real or due to perspective effects). {Parallax errors of
  smaller magnitude are more likely due to the presence of more than
  one object in the astrometric window or to binary orbital motion
  that is not accounted for}.

We aim at classifying sources into two categories based solely on
astrometric quantity and quality indicators. We
explicitly leave photometric measurements out of the selection in
order to avoid biases from preconceptions relative to the loci in the
colour-absolute magnitude diagram (CAMD) where sources are expected.
A classifier that uses
the position of sources in the CAMD, and is therefore trained with
examples from certain regions in this diagram, such as the main
sequence, red clump, or white dwarf (hereafter WD) sequences, might yield an
incomplete biased catalogue in the sense that sources out of these
classical loci would be taken for poor astrometric solutions. In contrast, we aim at separating the two categories (loosely speaking,
good and poor astrometric solutions) based on predictive variables
other than those arising from the photometric measurements, and use the
resulting CAMD as external checks of the selection
procedure. This will allow us to identify true nearby objects with
problematic photometry, as we show in subsequent sections.


\begin{figure*}[!htb]
    \sidecaption
    \includegraphics[width=12cm]{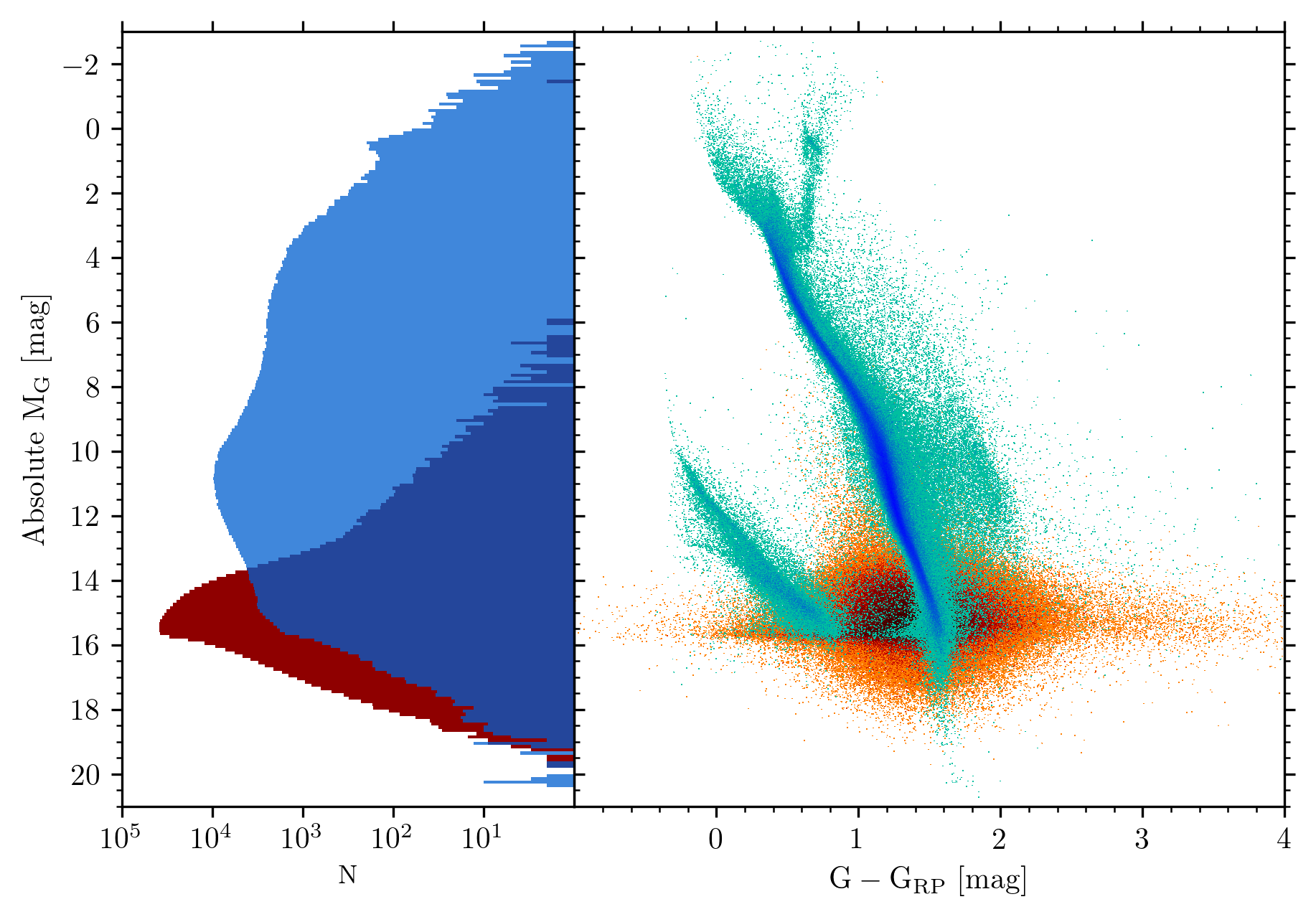}
    \caption{\label{fig:CAMDGminRP_KDE-MG}
    Left panel: Distribution of absolute $G$ magnitudes for the full  \gdrthree
    $\hat{\varpi} \ge$ 8\,mas sample. The blue distribution is for selected sources and the red one for rejected sources using a bin
    size of $\sigma_{M_G} = 0.1$\,mag. The slight bump in the
    distribution of selected sources at ${M_G} = 15$\,mag that coincides
    with the maximum of the rejected sources is probably indicative of
    contamination. \newline ~ \newline
    Right panel: CAMD diagram for the full sample. The blue points are good solutions and the red poor ones. The strip of source with nominally good
    solutions connecting the main and white dwarf sequence at $M_G\sim
    15$ is unexpected and due to contamination of the GCNS by faint
    objects at distances of 80-120\,pc, as discussed in
    Sec.~\ref{sec:contamandcompl}. \newline ~ \newline}
\end{figure*}

In order to construct the classification model, we created a training
set with examples in both categories as follows. For the set of poor
astrometric solutions, we queried the \gdrthree archive for sources with
parallaxes $\hat{\varpi} < -8$ mas. The query returned 512\,288 sources. We
assumed that the mechanism by which large (in absolute value) spurious
parallaxes are produced is the same regardless of the sign and that
the distribution of astrometric quantities that the model infers
from this set of large negative parallaxes is therefore equivalent (i.e. unbiased with respect) to that of the set of large spurious
parallaxes. We include in Appendix\,\ref{app:tsbad} a series of
histograms with the distributions of the predictive variables in both
the training set and the resulting classification. The latter is
inevitably a consequence of the former (the training set), but the good
match of the distributions for the $\hat{\varpi} < -8$\,mas (training
set) and $\hat{\varpi} > 8$\,mas (sources classified as poor astrometric
solutions) is reassuring.

Sources with poor astrometric solutions are
expected to have small true parallaxes (we estimate their mean true
parallax to be 0.25\,mas, as justified below) and are scattered towards
high absolute values due to data reduction problems, as
those described above. By using the large negative parallax sample as
training set for the class of poor astrometric solutions, we avoided
potential contamination by sources that lie truly within the 125 pc radius or
the incompleteness (and therefore bias) associated with the selection of
only very clear cases of poor astrometry. 

The set of examples of good astrometric solutions within the
8\,mas limit was constructed as follows. We first selected sources in low-density regions of the sky (those with absolute values of the Galactic
latitudes greater than 25\degree\ and at angular distances from the centres of the Large and Small Magellanic Clouds greater than 12 and 9 degrees, respectively)
and kept only sources with a positive cross-match in the 2MASS
catalogue. As a result, we assembled a set of 291\,030 sources with
photometry in five bands: $G$, \grp, $J, H,$ and $K$. We avoided the use of \gbp\ magnitudes because they have known limits for faint red objects \citep[see Section 8 of][]{EDR3-DPACP-117}.

From these we constructed a representation space with one colour index
($G-J$) and four absolute magnitudes ($M_G, M_{\rm RP}, M_H,\text{and } M_K$). We
fit models of the source distribution
in the loci of WDs, the red clump and giant branch, and the main
sequence. The models for the WDs, giant branch, and red clump stars are
Gaussian mixture models, while the main-sequence model is based on the
5D principal curve \citep{doi:10.1080/01621459.1989.10478797}. We used
these models to reject sources with positions in representation space
far from these high-density loci (presumably due to incorrect
cross-matches or poor astrometry). As a result, we obtained a set of
274\,108 sources with consistent photometry in the \G\ and 2MASS
bands. This is less than half the number of sources with parallaxes
more negative than -8\,mas. We recall that the selection of
this set of examples of good astrometric solutions is based on
photometric measurements and parallaxes, but we only required that the
photometry in the five bands is consistent. The photometric
information is not used  later on, and the subsequent classification of all sources
into the two categories of good and spurious astrometric measurements
is based only on the astrometric quantities described below. This selection would therefore only bias the resulting catalogue if it excluded
sources with good astrometric solutions whose astrometric properties
were significantly different from those of the training examples.


\begin{table}
\begin{center}
\begin{tabular}{ r|rr} 
    &     1  &    2 \\
 \hline
  1 & 89706   & 128   \\
  2  &  83 & 90119\\
\end{tabular}
\caption{\label{tab:confMatSel} Confusion matrix of the classifier evaluated in the test set. Class 1
represents good astrometric solutions (positives), and class 2
represents poor solutions (negatives). The first row shows the number of
class 1 examples classified as good astrometric solutions (true
positives, first column) and as poor solutions (false negatives, second
column). The second row shows the number of class 2 examples
classified as class 1 (false positives, first column) and class 2
(true negatives; second columns). The total number of
misclassifications for the set of test examples is 0.1\%.
}
\end{center}
\end{table}

The classification model consists of a random
forest \citep{breiman2001} trained on predictor variables selected
from a set of 41 astrometric features listed in
Table\,\ref{tab:importance-all}. Table\,\ref{tab:importance-all}
includes the feature names as found in the \G\ archive and its
importance measured with the mean decrease in accuracy \citep[two
leftmost columns]{Breiman2002} or Gini index \citep[two rightmost
columns]{gini1912variabilita}.  We selected features (based on the Gini
index) even though random forests inherently down-weight the effect
of unimportant features. We did this for the sake of efficiency. The
selected features are shaded in grey in Table\,\ref{tab:importance-all},
and we shade in red one particular variable
(\texttt{astrometric\_params\_solved}) that can only take two values and was
not selected despite the nominal relevance. 
The set of $2\times 274\,108$ 
examples (we selected exactly the same number of examples in the two categories
and verify the validity of this balanced training set choice below)
was divided into a training set (67\%) and a test set (33\%) in order
to assess the accuracy of the classifier and determine the
probability threshold that optimises completeness and
contamination. We find the optimum probability in the corresponding receiver
operating curve (ROC), which is $p = $\problimit,\ yielding a
sensitivity of 0.9986 (the fraction of correctly classified good
examples in the test set) and a specificity of 0.9991 (the same
fraction, but for the poor category). The random forest consists of 5000
decision trees built by selecting amongst three randomly selected
predictors at each split. Variations in the number of trees or
candidate predictors did not produce better results, as evaluated on
the test set. These can be summarised by the confusion matrix shown in
Table\,\ref{tab:confMatSel}.

Figure \ref{fig:lbgoodbad} shows the distribution in the sky of
selected (top) and rejected (bottom) sources. The distribution of
selected sources looks uniform, as expected, with the exception of the
the slight over-density at $l,b\approx(300,10)$ that is probably part of the
Lower Centaurus Crux subgroup of the Sco OB2 association at 115\,pc
\citep{2018A&A...620A.172Z}. 
The bottom panel highlights problematic sky areas
related to high surface density regions and/or specificities of the
scanning law. 
In order to detect signs of incompleteness and/or
contamination, we inspected the 
distribution of absolute $G$ magnitudes for both sets of sources
(Fig. \ref{fig:CAMDGminRP_KDE-MG} left panel). The distribution 
of spurious sources shows a main component centred at $M_{G}\sim 15$;
this coincides with a local bump in the distribution function of selected sources, which may be indicative of contamination.

\begin{figure}[!htb]
\center{\includegraphics[width=0.45\textwidth]{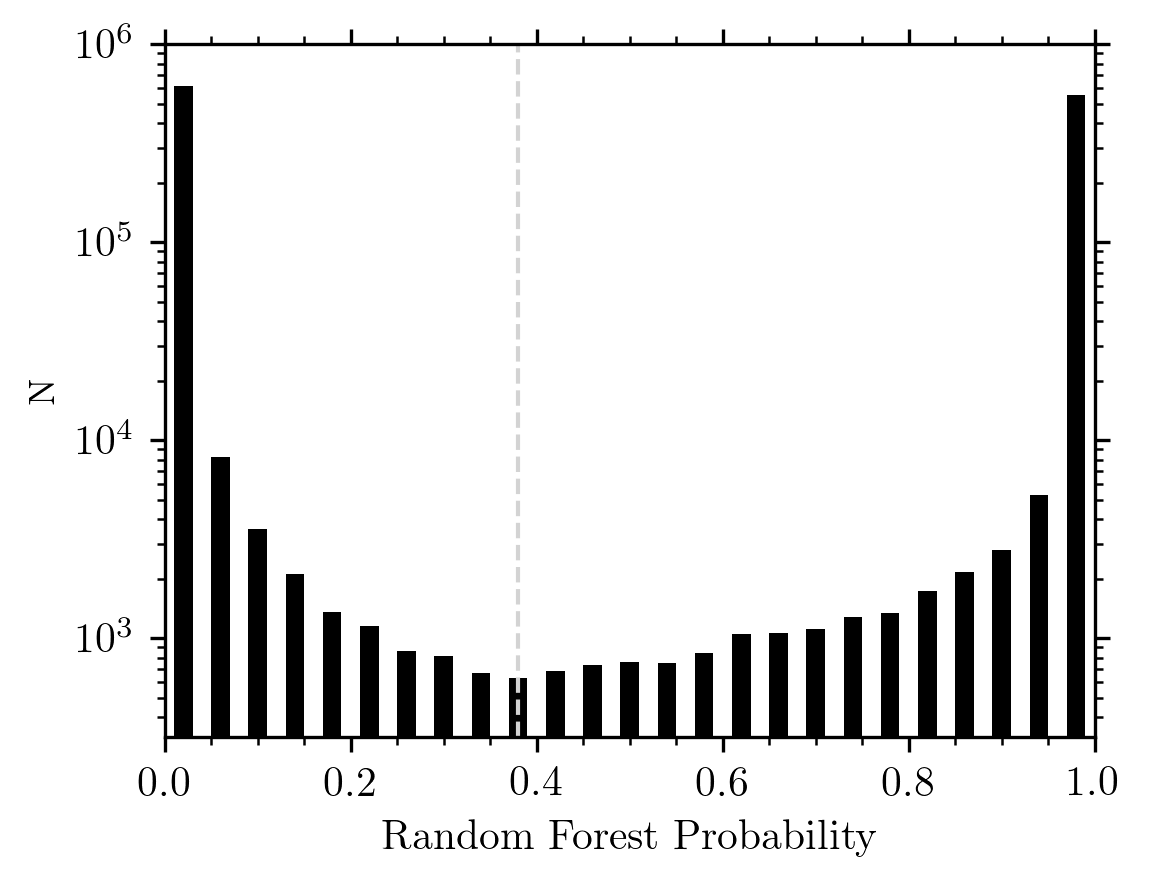}}  
\caption{\label{fig:hist} Histogram of the classification probabilities (in the category of good astrometric solutions) produced by the random forest. The vertical axis is in logarithmic scale.}
\end{figure}

Figure \ref{fig:hist} shows a (logarithmic) histogram with the derived
membership probabilities. Neither the number of sources with 
$\hat\varpi < -8$\,mas or the comparison of the numbers of sources 
classified as good and poor provide evidence for a significant imbalance 
in the true proportions of the classes. We therefore discarded the revision 
of the training set proportions or the inclusion of additional actions 
to recalibrate the classification probabilities due to a class imbalance. Finally,
the right panel of Fig.\,\ref{fig:CAMDGminRP_KDE-MG} shows a
colour-absolute magnitude diagram (CAMD) for the full sample colour-coded by probability $p$. The rejected sources 
are predominantly in areas of the CAMD that are usually empty, consistent with our hypothesis that the parallaxes are unreliable.

As final confirmation for the assumptions underlying the training set
definition we attempted to estimate the mean true parallax of the
poor astrometric solution by determining the negative value of the observed
parallax that results in approximately the same number of sources 
as those classified as poor astrometric solutions by our
random forest classifier. We find 638\,796 sources classified as poor
astrometric solutions, which is similar to the number of sources
with $\hat{\varpi} \le -7.5$\,mas (639\,058). If the
distribution of true parallaxes of sources with poor astrometric
solutions were symmetric { (which is not necessarily
true),} then its mode
could be estimated as $(8-7.5)/2 = 0.25$\,mas or 4\,kpc.


The random forest classifier described above is a
solution for the particular problem of separating good and poor
astrometric solutions in the solar neighbourhood, but it is not applicable
at larger distances. Good and poor astrometric solutions are well
separated in the space of input variables because the former are of
exquisite quality. As the measured parallax decreases, the proportions
of both classes change in the input parameter space and the degree of
overlap between the two increases. We therefore expect misclassifications
to increase for smaller observed parallaxes, also because the
fraction of sources in each category varies and increases more
steeply for the poor astrometric solutions. Finally, we would like to emphasise that a probability below the selection threshold does not necessarily mean that the source does not lie within 100\,pc. The astrometric solution of a source can be problematic (and the source therefore rejected by the random forest) even if it is located within 100\,pc. 


\subsection{Simple Bayesian distance estimation}
\label{sec:distance_estimation}
\begin{figure}[!htb]
\center{\includegraphics[width=0.5\textwidth]{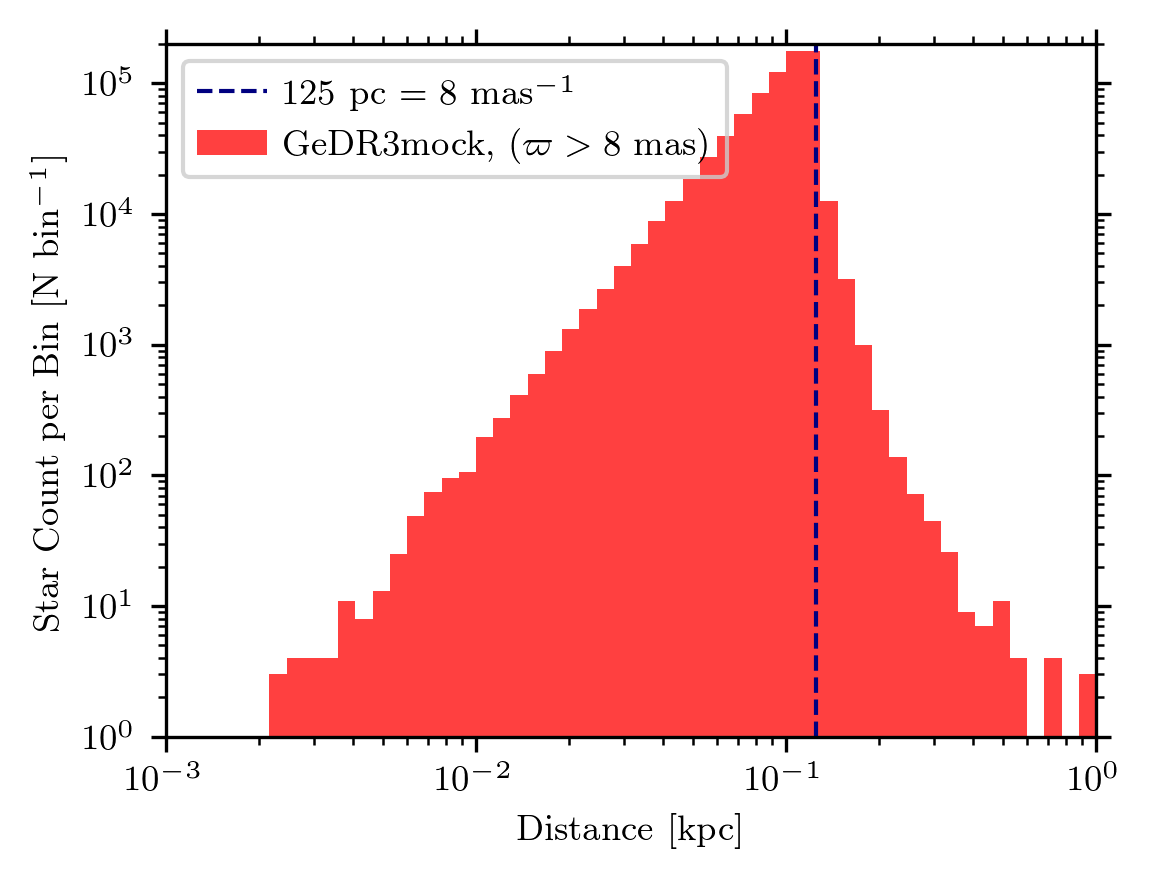}}  
\caption{Distance distribution of stars in GeDR3mock selected on observed parallax $>$ 8\,mas. We use this distribution as a prior for our simple Bayesian distance estimation.}
\label{fig:prior_distance}

\end{figure}

In order to infer distances from the observed parallaxes, we need an
expected distance distribution (prior) for the sources in our sample
selection ($\hat{\varpi} \ge 8$\,mas). We assumed that we have
removed all poor solutions. The simplest prior is a single
distribution that does not depend on sky position or type of star
(e.g. colour). We defined an empirical prior based on synthetic samples
using the GeDR3mock, which includes all
the stars down to $G$ = 20.7\,mag. The parallax uncertainty for GeDR3mock was empirically trained on \gdrtwo\ data and was lowered according to the
longer time baseline of \gdrthree. The mock \texttt{parallax\_error}
distribution is narrower than that of the empirical \gdrthree, therefore we artificially
increased the spread in $\log({\texttt{parallax\_error}})$, 
see the query below. Because the catalogue only contains the true
parallaxes, we selected observed parallaxes through the following query,
which can be performed on the GAVO TAP
service\footnote{\url{http://dc.g-vo.org/tap}}:

\begin{lstlisting}
SELECT * FROM(
SELECT parallax, GAVO_RANDOM_NORMAL(parallax, POWER(10, ((LOG10(parallax_error)+1)*1.3)-1)) AS parallax_obs 
-- This adds observational noise to the true parallaxes
FROM gedr3mock.main) AS sample
WHERE parallax_obs > 8
\end{lstlisting}

This retrieves a catalogue with 762,230 stars\footnote{The number of
stars retrieved will slightly change each time the query is run
because the random number generator does not accept seeds.}. Their
underlying true distance distribution is shown in
Fig.\,\ref{fig:prior_distance}.
The distribution of mock stars was inspected by comparing an in-plane,
$|$b$|<5^\circ$, and an out-of-plane, $|$b$|>65^\circ$. We found a
15\,\% deficiency of stars at 100\,pc distance for the out-of-plane
sample, as expected due to the stratification in the $z$ direction. When selecting
for specific stellar types, the directional dependence can increase
further, for instance for dynamically cold stellar populations. Here we ignored
these possibilities and used a distance prior independent of colour or
direction in the sky to let the exquisite data speak for themselves.

We sampled the posterior probability density function (PDF) using
Markov chain Monte Carlo methods \citep{2013PASP..125..306F}.  The
reported values, included in the online table\footnote{available from the CDS}, are the percentiles (from 1 to 99) of the stabilised
chain, that is, \texttt{dist\_50} represents the median of the
posterior distance estimation and \texttt{dist\_16}, \texttt{dist\_84}
the lower and upper 1$\sigma$ uncertainties. We also report
\texttt{mean\_acceptance\_fractions} and
\texttt{mean\_autocorrelation\_time} as quality indicators (but not
all sources have the latter).

\section{The \G\ Catalogue of Nearby Stars}
\label{sec:cat_overview}

\begin{table*}
 \tiny \caption{Content of the GCNS and rejected dataset with the first selected object as
 example.}  \label{GCNS_forpublication_cat.table} \begin{tabular}{lllll}
\hline                                                
Parameter & Unit & Comment & Example  \\
\hline
  source\_id                        &  ...     & Gaia EDR3 source ID                                                   &   2875125810310195712\\ 
  ra                                &  deg     & Right ascension (ICRS, epoch 2016.0)                               &      0.0157909\\        
  ra\_error                         &  mas     & Uncertainty                                                                   &    0.16\\               
  dec                               &  deg     & Declination (ICRS, epoch 2016.0)                                       &     34.1883005\\        
  dec\_error                        &  mas     & Uncertainty                                                                   &    0.13\\               
  parallax                          &  mas     & Gaia EDR3 parallax                                                    &    20.194\\             
  parallax\_error                   &  mas     & Gaia EDR3 parallax uncertainty                                               &   0.225\\               
  pmra*                             &  mas/yr  & Gaia EDR3 Proper motion in RA                                         &  -227.366\\             
  pmra*\_error                      &  mas/yr  & Gaia EDR3 RA proper motion uncertainty                                   &   0.206\\               
  pmdec                             &  mas/yr  & Gaia EDR3 Proper motion in  Dec                                       &   -56.934\\             
  pmdec\_error                      &  mas/yr  & Gaia EDR3 Dec proper motion uncertainty                                   &   0.159\\               
  phot\_g\_mean\_mag                &  mag     & Gaia  G Band magnitude                                        &   8.3483\\              
  phot\_g\_mean\_flux\_over\_error  &  mag     & Gaia  G flux to flux uncertainty ratio                                         &   6895.11\\             
  phot\_bp\_mean\_mag               &  mag     & Gaia BP Band magnitude                                        &   8.6769\\              
  phot\_bp\_mean\_flux\_over\_error &  mag     & Gaia BP flux to flux uncertainty ratio                                         &   3384.69\\             
  phot\_rp\_mean\_mag               &  mag     & Gaia RP Band magnitude                                        &   7.8431\\              
  phot\_rp\_mean\_flux\_over\_error &  mag     & Gaia RP flux to flux uncertainty ratio                                         &   3544.43\\             
  phot\_robust\_bp\_rp\_excess      &          & Ratio of the sum of the BP and RP flux to the G flux         &   1.2100\\              
  ruwe                              &          & Renormalised unit weight error                                         & 14.26\\                 
  ipd\_frac\_multi\_peak            &          & Fraction of windows with multiple peaks                 &   0\\                   
  adoptedRV                         &  km/s    & Adopted Radial Velocity from EDR3 or literature               &   -29.94\\              
  adoptedRV\_error                  &  km/s    & Uncertainty in adopted RV                                                    &     0.89\\              
  adoptedRV\_refname                &          & ADS Bibcode for RV                                                   &  2018A\&A...616A...1G\\  
  radial\_velocity\_is\_valid       &          & T/F Flag to indicate if RV is in eDR3                                & T\\
  GCNS\_prob                        &          & Probability 0 to 1 of having reliable astrometry                   & 1.00\\                  
  WD\_prob                          &          & Probability 0 to 1 of being a white dwarf                                 & 1.00\\                  
  dist\_1                           &  kpc     & 1st  percentile of the distance PDF, used in GCNS selection          &      0.04833\\          
  dist\_16                          &  kpc     & 16th percentile of the distance PDF, 1$\sigma$ lower bound              &      0.04901\\          
  dist\_50                          &  kpc     & 50th percentile of the distance PDF, the median distance              &      0.04952\\          
  dist\_84                          &  kpc     & 84th percentile of the distance PDF, 1$\sigma$ upper bound             &      0.05007\\          
  xcoord\_50                        &  pc      & x coordinate in the Galactic frame using dist\_50, median coordinate &    -15.72239\\          
  xcoord\_16                        &  pc      & x coordinate 1$\sigma$ lower bound                                         &    -15.55850\\          
  xcoord\_84                        &  pc      & x coordinate 1$\sigma$ upper bound                                     &    -15.89664\\          
  ycoord\_50                        &  pc      & y coordinate in the Galactic frame using dist\_50, median coordinate &     41.02444\\          
  ycoord\_16                        &  pc      & y coordinate 1$\sigma$ lower bound                                         &     40.59680\\          
  ycoord\_84                        &  pc      & y coordinate 1$\sigma$ upper bound                                     &     41.47911\\          
  zcoord\_50                        &  pc      & z coordinate in the Galactic frame using dist\_50, median coordinate &    -22.85814\\          
  zcoord\_16                        &  pc      & z coordinate 1$\sigma$ lower bound                                         &    -22.61987\\          
  zcoord\_84                        &  pc      & z coordinate 1$\sigma$ upper bound                                     &    -23.11148\\          
  uvel\_50                          &  km/s    & Velocity in the Galactic frame, direction positive x                   &   -61.07\\              
  uvel\_16                          &  km/s    & Velocity 1$\sigma$ lower bound                                       &   -61.69\\              
  uvel\_84                          &  km/s    & Velocity 1$\sigma$ upper bound                                       &   -60.43\\              
  vvel\_50                          &  km/s    & Velocity in the Galactic frame, direction positive y                   &    -5.58\\              
  vvel\_16                          &  km/s    & Velocity 1$\sigma$ lower bound                                       &    -6.39\\              
  vvel\_84                          &  km/s    & Velocity 1$\sigma$ upper bound                                       &    -4.88\\              
  wvel\_50                          &  km/s    & Velocity in the Galactic frame, direction positive z                   &    12.81\\              
  wvel\_16                          &  km/s    & Velocity 1$\sigma$ lower bound                                       &    12.43\\              
  wvel\_84                          &  km/s    & Velocity 1$\sigma$ upper bound                                       &    13.24\\              
  NAME\_GUNN                        &          & Name from the PanSTARRS/SDSS/SkyMapper survey                        &  1237663235523739680\\  
  REFNAME\_GUNN                     &          & ADS Bibcode Gunn bands                                               &  2017ApJS..233...25A\\  
  gmag\_GUNN                        &  mag     & GUNN G Band magnitude        (    SDSS:g, Skymapper:  g\_psf)         &  12.388\\               
  e\_gmag\_GUNN                     &  mag     & Uncertainty GUNN G Band magnitude  (SDSS:err\_g, Skymapper:e\_g\_psf)      &   0.007\\               
  rmag\_GUNN                        &  mag     & GUNN R Band magnitude        (    SDSS:r, Skymapper:  r\_psf)         &  12.293\\               
  e\_rmag\_GUNN                     &  mag     & Uncertainty GUNN R Band magnitude  (SDSS:err\_r, Skymapper:e\_r\_psf)      &   0.008\\               
  imag\_GUNN                        &  mag     & GUNN I Band magnitude        (    SDSS:i, Skymapper:  i\_psf)         &  12.445\\               
  e\_imag\_GUNN                     &  mag     & Uncertainty GUNN I Band magnitude  (SDSS:err\_i, Skymapper:e\_i\_psf)      &   0.008\\               
  zmag\_GUNN                        &  mag     & GUNN Z Band magnitude        (    SDSS:z, Skymapper:  z\_psf)         &   9.007\\               
  e\_zmag\_GUNN                     &  mag     & Uncertainty GUNN Z Band magnitude  (SDSS:err\_z, Skymapper:e\_z\_psf)      &   0.001\\               
  NAME\_2MASS                       &          & 2mass name                                                           &    00000410+3411189 \\ 
  j\_m\_2MASS                       &  mag     & 2MASS J band magnitude                                        &   7.249\\               
  j\_msig\_2MASS                    &  mag     & Uncertainty 2MASS J band magnitude                                     &   0.017\\               
  h\_m\_2MASS                       &  mag     & 2MASS H band magnitude                                        &   6.940\\               
  h\_msig\_2MASS                    &  mag     & Uncertainty 2MASS H band magnitude                                             &   0.016\\               
  k\_m\_2MASS                       &  mag     & 2MASS K band magnitude                                        &   6.885\\               
  k\_msig\_2MASS                    &  mag     & Uncertainty 2MASS K band magnitude                                             &   0.017\\               
  NAME\_WISE                        &          & WISE Name                                                             & J000003.81+341117.9 \\  
  w1mpro\_pm\_WISE                  &  mag     & CATWISE W1 Band magnitude                                               &   7.249\\               
  w1sigmpro\_pm\_WISE               &  mag     & Uncertainty CATWISE W1 Band magnitude                                     &   0.020\\               
  w2mpro\_pm\_WISE                  &  mag     & CATWISE W2 Band magnitude                                               &   6.922\\               
  w2sigmpro\_pm\_WISE               &  mag     & Uncertainty CATWISE W2 Band magnitude                                             &   0.008\\               
  w3mpro\_WISE                      &  mag     & ALLWISE W3 Band magnitude                                               &   6.883\\               
  w3sigmpro\_WISE                   &  mag     & Uncertainty ALLWISE W3 Band magnitude                                      &   0.016\\               
  w4mpro\_WISE                      &  mag     & ALLWISE W4 Band magnitude                                       &   6.824\\               
  w4sigmpro\_WISE                   &  mag     & Uncertainty ALLWISE W4 Band magnitude                                      &   0.085\\                         

\hline
\end{tabular}
\end{table*}

We now discuss the selection from the \NINEIGHTMAS\ objects with
$\hat\varpi > 8$\,mas for inclusion in the GCNS. As indicated in
Sect. \ref{sec:randomforest}, the optimal probability threshold
indicated by the ROC is $p = $\problimit. To enable a correct use of the
distance PDF
produced in Sect. \ref{sec:distance_estimation}, we retain all entries with a
non-zero probability of being inside 100\,pc, for which we used the distance with 1\% probability, \texttt{dist\_1}.  

Therefore the selection for inclusion in the GCNS is:
\begin{equation}
  p >= \problimit \rm{\ \&\&\ } \texttt{dist\_1} <= 0.1\,{\rm Kpc}.
\end{equation} This
selection resulted in \NFINAL\ objects that are listed in the online
table, an example of which
is reported in Table~\ref{GCNS_forpublication_cat.table}.
The \NREJECT\ objects from the full $\hat\varpi > 8$\,mas that did not
meet these criteria are provided in an identical table should they be
needed for characterisation\footnote{both tables available from the CDS}.

Our goal is to provide a stand-alone catalogue that will be useful
when observing or for simple exploratory studies. Following this goal,
we have retained minimal \gdrthree information, source ID, basic
astrometry, photometry, and a few of the quality flags used in this paper. In
keeping with the \G\ data release policy, we do not provide uncertainties on
the magnitudes but the \texttt{mean\_flux\_over\_error} for each passband. 
There
are \NNOGTOT\ objects in the full $ \hat\varpi > 8$\,mas sample that do
not have \G\ $G$ magnitudes, \NNOGGCNS\ of which meet our selection
criteria.  These \NNOGGCNS\ objects have on-board estimates of the 
$G$ magnitude in the 11--13
range, and we refer to the \gdrthree\ release
page\footnote{\url{https://www.cosmos.esa.int/web/gaia/early-data-release-3}} for their
values.

To the \gdrthree data we added the probability of reliable astrometry, $p$, calculated by the random forest classifier, as
detailed in Section \ref{sec:2Cataloggeneration}, which has a range of
0--1. We include four of the values from the posterior distance PDF
determined in Section \ref{sec:distance_estimation}: the median
distance \texttt{dist\_50}, its 1-$\sigma$ upper and lower bounds
(\texttt{dist\_16}, \texttt{dist\_84}) and the \texttt{dist\_1} value,
which is the 1\% distance probability and used in the selection of the
GCNS.

%
%
We include radial velocities included in \gdrthree \citep{EDR3-DPACP-128}, which are
\NDRTWORV\ entries; from the radial velocity
experiment \citep{2017AJ....153...75K}, which contributes \NRAVERV\ entries; and from a
5\arcsec cone search for each entry on the SIMBAD database\footnote{Set of Identifications, Measurements and Bibliography for Astronomical Data, \url{http://SIMBAD.u-strasbg.fr}}: \NSIMBADRV\
entries. From the RAVE and SIMBAD entries we removed 130 radial velocities that were 
$>~800$\,km\,s$^{-1}$ and 4937 objects without positive uncertainties or
without reference.  The total number of entries with a radial velocity
is \NGCNSRVS\ in the full sample, \NGCNSWITHRVS\ of which are in the GCNS.

We also provide magnitudes from external
optical, near-infrared, and mid-infrared catalogues. The optical
magnitudes are GUNN $g, r, z, i$ from, in preference order,
the Panoramic Survey Telescope and Rapid Response System first release
\citep[hereafter PS1,][]{2016arXiv161205560C}, the Sloan Digital Sky Survey 13th
data release \citep{2017ApJS..233...25A}, and the SkyMapper Southern
Survey \citep{2018PASA...35...10W}. The near-infrared magnitudes $J,
H, K$ are from the Two Micron All Sky
Survey \citep{2006AJ....131.1163S}, the mid-infrared magnitudes $W1 \text{ and } W2$ from the CATWISE2020 release \citep{2020ApJS..247...69E}, and
$W3 \text{ and } W4$ from the ALLWISE data
release \citep{2013wise.rept....1C}. All external matches came from
the \G\ cross-match tables \citep{2019A&A...621A.144M},
except for the
CATWISE2020 catalogue as it is not included for \gdrthree. For this
catalogue we used a simple nearest-neighbour cone search with a
5\arcsec limit. We emphasise that these magnitudes are provided to
have a record of the value we used in this paper and to enable a
simple direct use of the GCNS. If a sophisticated analysis is required
that wishes to exploit the external photometry, we recommend
to work directly with the external catalogues that also
have quality flags that should be consulted.

For analysis of the GCNS in a galactic framework, we require
coordinates  $(X,Y,Z)$  the coordinates in a barycentric rest frame
positive towards the Galactic centre, positive in the direction of
rotation, and positive towards the north Galactic pole, respectively. When
we ignore the low correlation between the equatorial coordinates, the
$(X,Y,Z)$  and their one-sigma bounds can be calculated using the distance
estimates from Sect.~\ref{sec:2Cataloggeneration}
and their Galactic coordinates.

We inferred space velocities in the Galactic reference frame
$U,V,W$ using a Bayesian formalism. Our model contains a top layer
with the parameters that we aim to infer (distances and space
velocities), a middle layer with their deterministic transformations
into observables (parallaxes, proper motions, and radial velocities),
and a bottom layer with the actual observations that are assumed to be
samples from multivariate (3D) Gaussian distributions with full
covariance matrices between parallaxes and proper motions, and an
independent univariate Gaussian for the radial velocity. We assumed the
classical deterministic relations that define space velocities in
terms of the observables (\G\,coordinates, parallaxes and proper
motions, and radial velocities), which we explicitly develop in
Appendix~\ref{app:UVW-priors}.  We neglect here for the sake of
simplicity and speed the uncertainties in the celestial coordinates
and their correlations with parallaxes and proper motions.  The full
covariance matrices are given by the catalogue uncertainties and
correlations.


We used the same empirical prior for the distance as described in
Section \ref{sec:distance_estimation} and defined three independent
priors for the space velocities $U$, $V,$ and $W$ (see
Appendix \ref{app:UVW-priors} for details). In all three cases we use
a modified Gaussian mixture model (GMM) fit to the space velocities
found in a local (140\,pc) simulation from the Besan\c con Galaxy
model \citep{2003A&A...409..523R}. The number of GMM components is
defined by the optimal Bayesian information criterion. The
modification consists of decreasing the proportion of the dominant
Gaussian component in each fit by 3\% and adding a new wide
component of equal size centred at 0\,km/s and with a standard
deviation of 120\,km/s to allow for potential solutions with high
speeds typical of halo stars that are not sufficiently represented in
the Besan\c con sample to justify a separate GMM component.
We then used {\sl Stan} \citep{JSSv076i01} to produce 2000 samples from
the posterior distribution and provide the median $U, V, W,$ and their
one-sigma upper and lower bounds in the output catalogue with suffixes
vel\_50, vel\_16, and vel\_84, respectively.


\section{Catalogue quality assurance}
\subsection{Sky variation}
\label{sec:completeness}

In this section we discuss the completeness of the GCNS in the context
of the full \gdrthree. In particular, we examine the changes in completeness
limit with the direction on the sky as a result of our distance
cut and as a result of separation of sources.

\subsubsection{\texorpdfstring{$G$}{G} magnitude limits over the sky}
\label{sec:gmag_lim}

\begin{figure}[!htb]
\center{\includegraphics[width=0.5\textwidth]{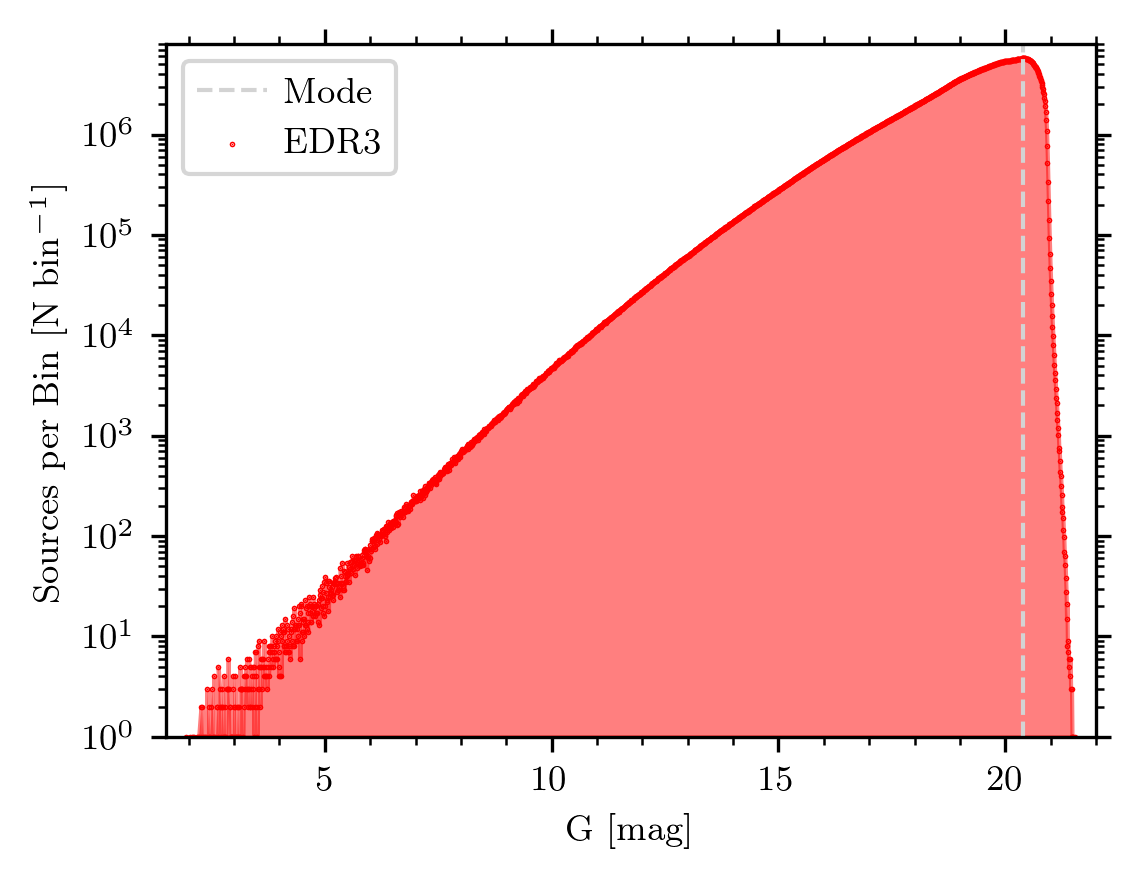}}  
\caption{$G$ magnitude distribution for \textbf{all sources} in \gdrthree
that have a $G$ magnitude and a parallax measurement (the bin size
is 0.01 mag). The mode is indicated as a grey dashed line at 20.41 mag.}
\label{fig:gdistribution}

\end{figure}

One of the main drivers of the completeness is the apparent
brightness of a source on the sky. It can be either too bright, such
that the CCDs are overexposed, or it can be too faint, such that it can hardly be picked up from background noise. For \gdrthree the $G$
magnitude distribution is depicted in Fig.\,\ref{fig:gdistribution}
for the sources that have both a $G$ and a parallax measurement. At
the bright end, we have a limit at about 3 mag, and at the faint end, the
magnitude distribution peaks at 20.41 mag {(the mode), which indicates
that not all sources at this magnitude are recovered by \G\, because
otherwise the source count would still rise}.

First- and second-order effects arise from the underlying source
density, for example, if there are too many sources for \G\ to process, $\sim10^6$ per deg$^2$ \citep{2012Ap&SS.341...31D}, then sources with brighter
on-board $G$ magnitude estimate are prioritised; and the scanning law,
for instance, expected scans per source, vary over the sky, which can improve
coverage for fainter sources. Because the latter effect is complex to
simulate \citep{2020arXiv200508983B}, we employed an empirical approach
using
the \texttt{gdr2\_completeness}\footnote{\url{https://github.com/jan-rybizki/gdr2_completeness}} \texttt{python}
package \citep{2018ascl.soft11018R}. We essentially focused on the $G$
magnitude distribution per HEALpix \citep{2005ApJ...622..759G}, but used
percentiles instead of the mode as an estimator of the limiting
magnitude because the mode is noisy in low-density fields and prone
to biases. Red clump stars towards the bulge or the
Magellanic clouds can produce a mode in the distribution at
brighter magnitudes, for example (cf. discussion in Sec. 3.2
of \citet{2020PASP..132g4501R}). 

\begin{figure}[!htb]
\center{\includegraphics[width=0.48\textwidth]{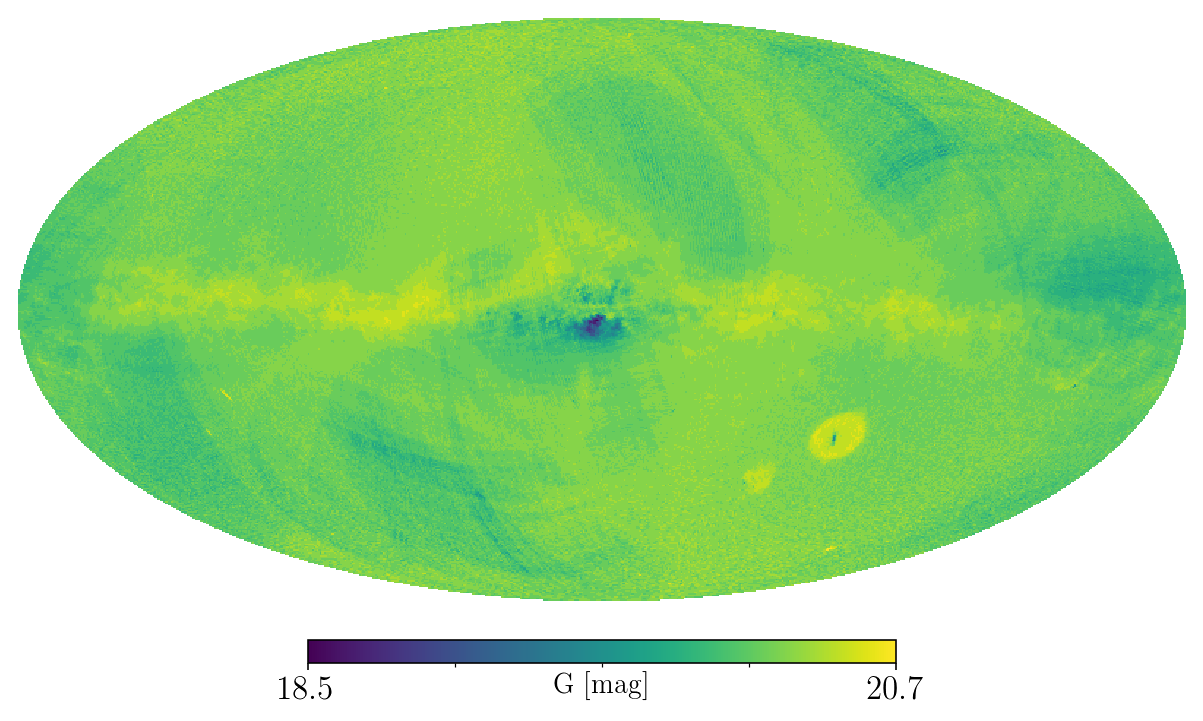}}
\caption{80th percentile of the $G$ magnitude distribution
per level 7 HEALpix over the sky as a Mollweide projection
in Galactic coordinates. The Galactic centre is in the middle, and the longitude increases to the left.}
\label{fig:maglim}
\end{figure}

We decided which percentile of
the magnitude distribution was used. Limits of $\mathrm{G}=20.28$ and
$\mathrm{G}=20.54$ encompass 80\,\% and 90\,\% of the sources,
respectively. These limits are approximately at the left and right
edge of the grey line in Fig.\,\ref{fig:gdistribution} denoting the
mode at $\mathrm{G}=20.41$, which includes 85\,\% of the sources. We
expect a reasonable cut for most lines of sight to be between these
values. We show the resulting empirical magnitude limit map in HEALpix
level 7 for sources with $G$ and parallax measurement in \gdrthree for
the 80th percentile in Fig.\,\ref{fig:maglim}. Scanning law patterns
as well as the high-density areas of the bulge and the Large
Magellanic Cloud can be seen. Sources with even fainter
magnitudes still enter the catalogue, but they do not represent the
complete underlying population of sources at these magnitudes. These sources instead enter the \gdrthree catalogue in a non-deterministic
fashion as a consequence of the imprecise on-board $G$ magnitude
estimate. We provide the empirical $G$ magnitude limit map including
all percentiles at the HEALpix fifth level as a supplementary
table\footnote{available from the CDS} because this is used in Sect.\,5.

\begin{figure}[!htb]
\center{\includegraphics[width=0.48\textwidth]{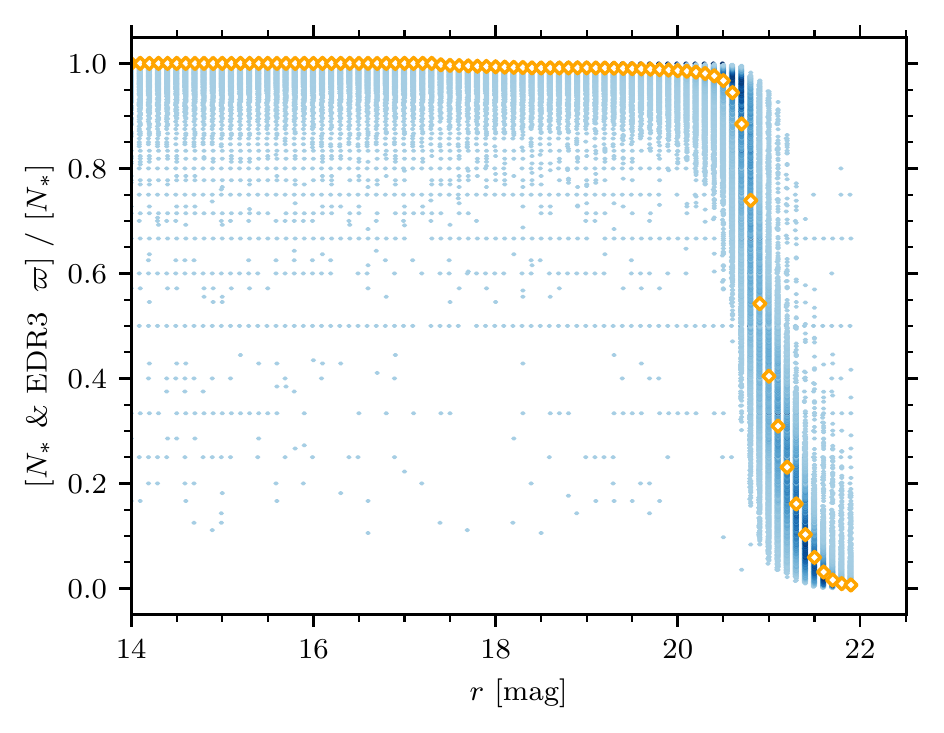}}
\caption{Each point is the ratio of objects classified as stellar in PanStarrs
with \gdrthree\ parallaxes to all objects classified as stellar per level 6 HEALpixel binned in $r$ magnitudes, as discussed in section \ref{sec:gmag_lim} The orange diamonds represent the median value for all HEALpixels. }
\label{fig:completeness_validation}
\end{figure}

An external validation of our usage of percentiles as a proxy for
completeness limits can be achieved by comparing \gdrthree\ results
cross-matched with PS1 sources classified as stars. A PS1 star is
defined as an object with a probability from \cite{2018PASP..130l8001T} greater than 0.5. We compare the full PS1 footprint
and make two assumptions: that the PS1 is complete in the relevant
magnitude range, and that $r \sim G$. This assumption means that the limit of the PS1 is
significantly fainter than the \G\ limit, and for all but the reddest
objects, the median $r - G$ is zero. When we assume this a simple cut at $G$ =
(19.9, 20.2, 20.5)\,mag, which is the mean magnitude limit of the 70th,
80th, and 90th percentile map, this results in a source-count averaged
completeness of 97\%, 95\%, and 91\%. Using our map at 70th, 80th, and 90th
percentiles, we find an 98\%, 97\%, and 95\% completeness.
Figure\,\ref{fig:completeness_validation} shows the ratio of PS1
stellar sources with \gdrthree\ parallaxes to all PS1 stellar sources
in bins of magnitudes in level 6 HEALpixels. The median ratio is 99\% until 19.5, drops to 95\% at 20.5 (slightly different to the above $G$ because it is averaged across the sky),  and quickly sinks to 50\%
at 21.0.

\subsubsection{Volume completeness with \texorpdfstring{M$_\mathrm{G}$}{MG}}
\label{sec:volcompleteness}

With regard to volume completeness per absolute magnitude, which needs to be corrected for when a luminosity function is constructed, as we do in Sects.\,\ref{sec:lf} and \ref{sec:wdlf}, we take into account (a) the apparent magnitude limits and (b) the distance probability distribution. 
For (a) we conservatively employed the 80th percentile apparent magnitude limit map from Sect.\,\ref{sec:gmag_lim} per level 5 HEALpix. All stars that are not within these limits were excluded from the analysis.
For (b) we used all of the 99 PDF samples with a distance estimate $\le$ 100\,pc instead of a single distance estimate per source, for example by the
median distance.

On the selected samples, we performed our analysis (e.g. used the respective
distance and $G$ magnitude to derive M$_\mathrm{G}$ and counted the
sources per absolute magnitude bin), finally dividing our resulting
numbers by 99 to recover the true stellar numbers. Objects that are close to the 100\,pc border only contribute partially to our analysis, down-weighted by the probability mass, which resides within 100\,pc. Similarly, owing to the distance PDF samples, individual sources can contribute to different absolute magnitude bins.



\subsubsection{Contrast sensitivity}
\label{sec:contrast_sensitivity}

The resolving power of the \G\ instrument of two sources that lie close together in the sky mainly depends on the angular
separation and the magnitude difference \citep{2015A&A...576A..74D} and is called \textit{\textup{contrast sensitivity}},
see \citet{2019A&A...621A..86B} for a \gdrtwo determination. In dense
regions we especially lose faint sources due to this effect \citep{2020PASP..132g4501R}, which directly affects our ability to resolve
binaries. We empirically
estimated this function using the distribution of close pairs from
the full \gdrthree.

\begin{figure}[!htb]
\center{\includegraphics[width=0.45\textwidth]{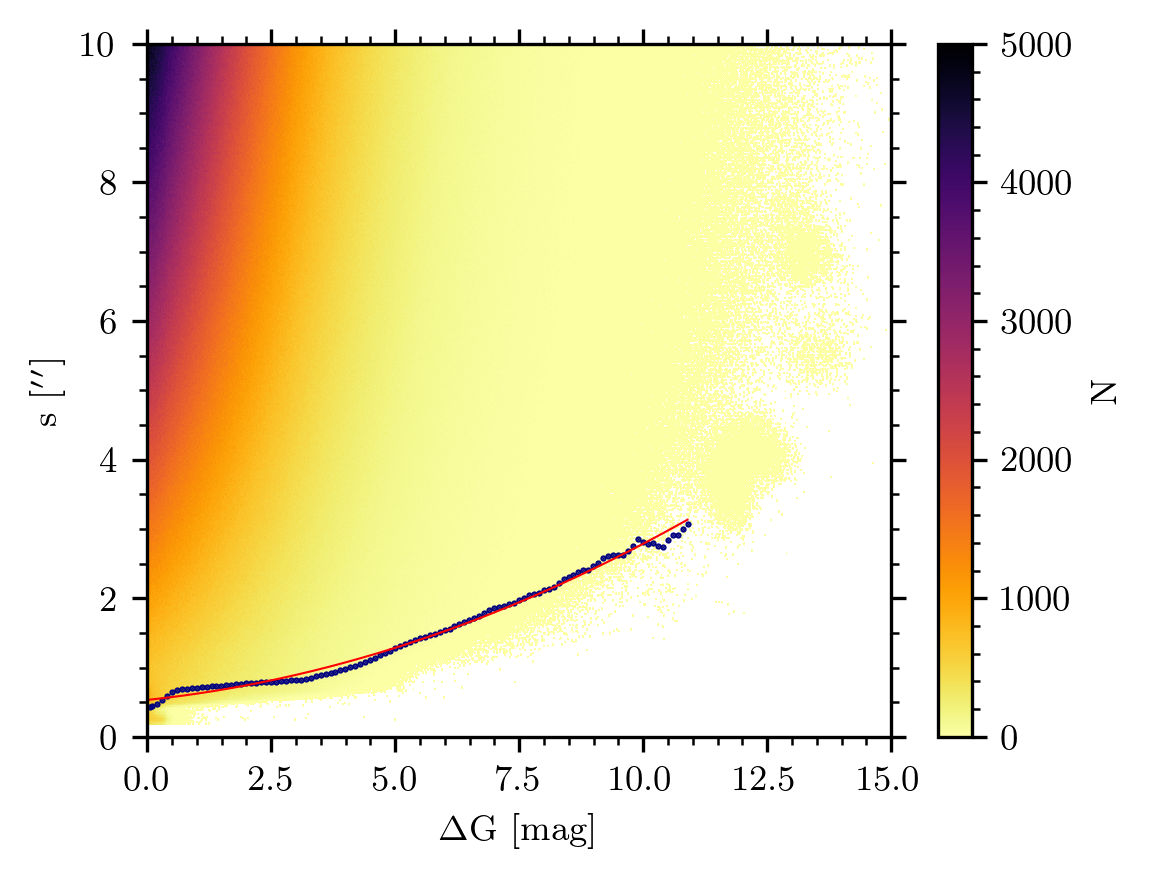}}  
\caption{Magnitude difference, $\Delta G$, vs. angular separation, $s$, of all objects in \gdrthree colour-coded by density in [0.02, 0.02] bins. The blue points represent the 99.5 percentiles of the separations, and the red line is a fit to these values and is reported in Eq.~\ref{eq:smin}.}
\label{fig:close_pairs}
\end{figure}

In Fig.\,\ref{fig:close_pairs} we plot the angular separation of
entries in the \gdrthree as a function of the magnitude
difference. The blue points are the 99.5 percentiles of the
separations binned in overlapping magnitude bins of 0.2\,mag in the
magnitude range 0--11\,mag. We adopted these percentiles as the minimum resolvable
separation, $s_{min}$, and therefore the dependence on the magnitude difference,
$\Delta G$, is approximated by the red line, 
\begin{equation}
    s_{min}= 0.532728 +  0.075526 \cdot \Delta G + 0.014981\cdot (\Delta G)^2
    \label{eq:smin}
.\end{equation}
The structure and over-densities in Fig.\,\ref{fig:close_pairs} after the
12th magnitude are due to the
gating and windowing effects for bright objects observed by \G.

\subsection{Comparison to previous compilations}
\label{sec:compprevcomp}

\begin{figure}
\includegraphics[width=0.4
\textwidth]{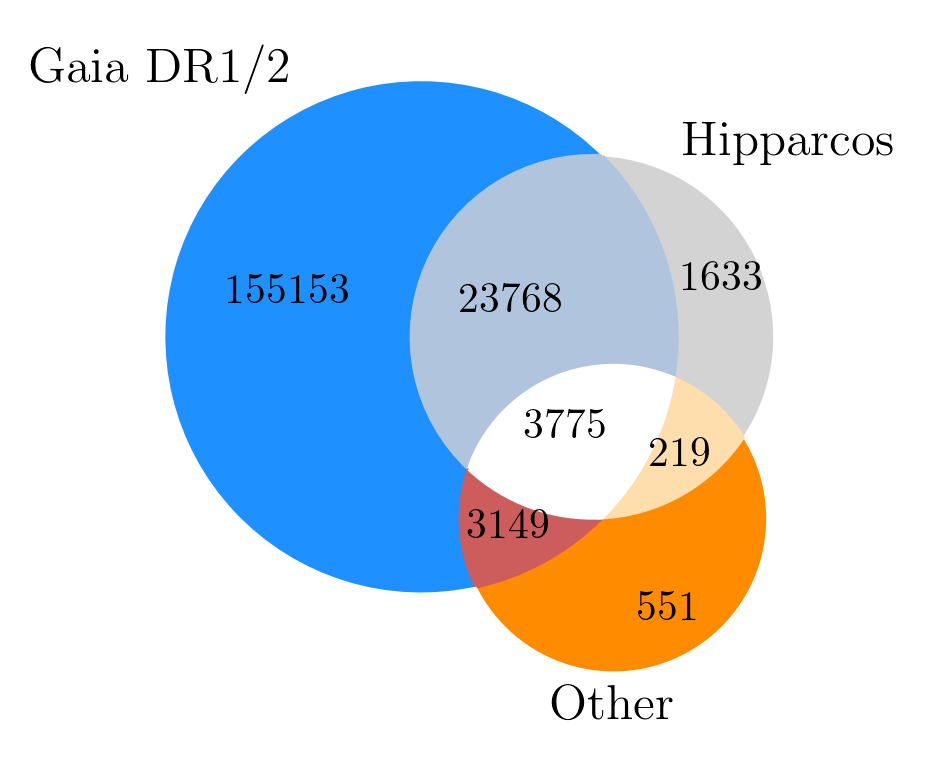}
\caption{Content of the 8\,mas sample from the SIMBAD query. The number of stars is given in different samples depending on the origin of their parallax. The ratio of the area for the three main primary circles is proportional to the ratio of the square root of the total number of objects per sample.  }
\label{fig:SIMBAD8mas}
\end{figure}

The Set of Identifications, Measurements and Bibliography for
Astronomical Data (SIMBAD) database provides information on
astronomical objects of interest that have been studied in scientific
articles. All objects in this database have therefore been individually
vetted by a professional in some way, and while the census is not complete because not
all objects have been studied, the contamination is low. From this
database we retrieved all stars with a parallax larger than 8 mas through
the following query performed with the TAP
service\footnote{\url{http://SIMBAD.u-strasbg.fr/SIMBAD/sim-tap}}:

\begin{lstlisting}
SELECT main_id, plx_value, plx_bibcode, string_agg(bibcode||';'||plx,';')
FROM basic LEFT JOIN mesPlx on oid=oidref
WHERE plx_value>8
GROUP BY main_id, plx_value, plx_bibcode
\end{lstlisting}

The 8\,mas limit, or 125\,pc, was chosen at this stage because we will
further cross-match with the GCNS and expect some of the sources to
have a new \G\ parallax above 10 mas, which means that they enter the 100
pc sample, or vice versa. This query returned 189\,096 objects. 
Eight hundred and thirty-nine objects in binary systems are duplicates: they have one entry
as a multiple system, plus one or two (or even three) entries for the
individual components (e.g. $\alpha$ Cen is listed three times, first as
a system, but then $\alpha$ Cen A and $\alpha$ Cen B are also listed
individually). Moreover, obvious errors are, for example, HIP\,114176,
2MASS\,J01365444-3509524, and 2MASS\,J06154370-6531528, which last case is a galaxy.

This leaves a sample of 188 248 objects.  Most of them, $\sim 98\%,$
have parallaxes from \gdrone \citep[566
objects;][]{2016A&A...595A...2G} and \gdrtwo \citep[184\,584
objects;][]{2018A&A...616A...1G}, and $\sim 2\%$
have parallaxes from Hipparcos\ \citep[2534
objects;][]{1997A&A...323L..49P,2007A&A...474..653V}. The few
remaining objects (564) are from other trigonometric parallax
programs \citep[e.g.][]{1995gcts.book.....V,
2013MNRAS.433.2054S, 2014ApJ...784..156D, 2018ApJ...867..109M}.

 SIMBAD does not systematically replace \texttt{plx$\_$value}
by the most recent determination, but prefers the value with the
lowest measurement uncertainty. In particular, for 693 very bright
stars from this query, the astrometric solution of Hipparcos\ is chosen over
that of \gdrtwo.

Our SIMBAD query also gives all existing trigonometric parallax
measurements (from the table \texttt{mesPlx}) for each
star. Fig.\,\ref{fig:SIMBAD8mas} shows the content of the SIMBAD query
in terms of the number of stars and the origin of their parallax. It shows
that the SIMBAD 8\,mas sample has mostly been fed by \G. For
this reason, we first compared GCNS with the compilation, excluding the
objects for which only a \G\ parallax is available (blue sample in
Fig.\,\ref{fig:SIMBAD8mas}). 

Next we compared GCNS with the full \gdrtwo
data (and not only with the stars listed in SIMBAD, which are about half
of the full \gdrtwo catalogue). 
Within the 100\,pc sphere, the total number of objects having an
astrometric parallax determination consequently increases from 26 536
stars prior to \G\, to \NDISTFIFTY\ stars in GCNS
with \texttt{dist\_50}$<0.1$, or 301\,797 stars when each source
with \texttt{dist\_1}$<0.1$ is counted and weighted by its probability
mass (see Sect.\,\ref{sec:volcompleteness})

\begin{figure}
\includegraphics[width=0.5\textwidth]{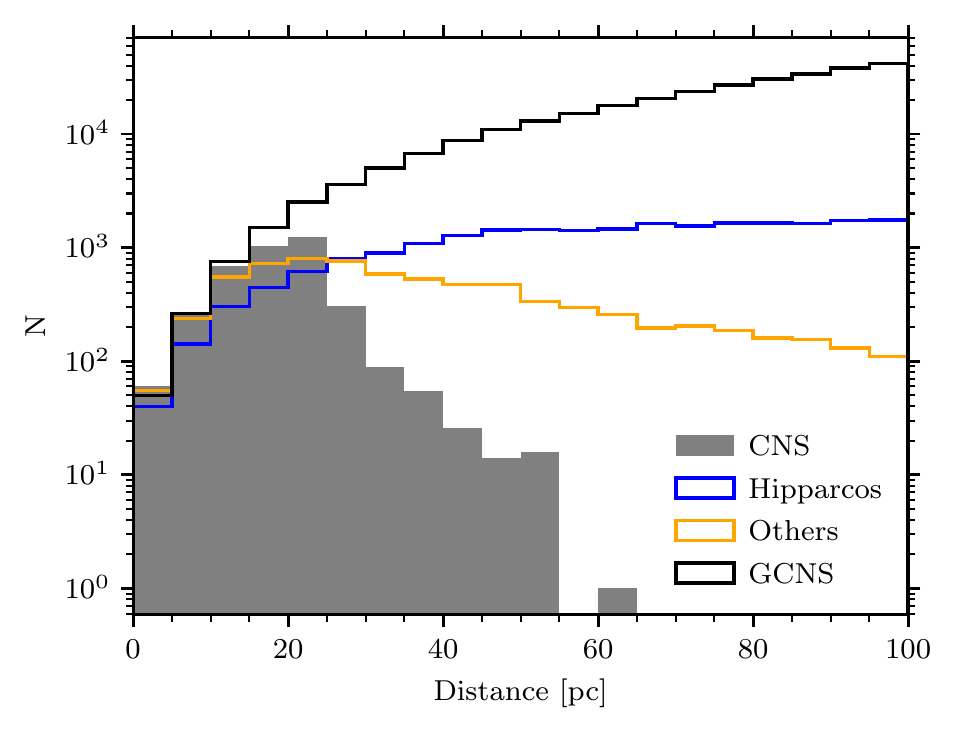}
\caption{Distance distributions of the GCNS compared to previous compilations. The distance is computed as the inverse of the parallax, taken from the respective catalogue. The y-axis is a log scale.}
\label{fig:hist_dist_previous}
\end{figure}

Fig.\,\ref{fig:hist_dist_previous} shows the distribution in distance
of the GCNS catalogue, of the Hipparcos catalogue, and of all objects
having a parallax from other programmes (mainly ground-based), prior to
\G. The Catalogue of Nearby Stars \citep{1991adc..rept.....G} is
also shown. Although CNS is based on ground-based programs, CNS contains more stars in some bins than are listed in SIMBAD. The main
reason is that CNS lists all the components in multiple systems,
whereas SIMBAD has only one entry for the systems with one parallax
measurement.

\begin{figure}
\includegraphics[width=0.5\textwidth]{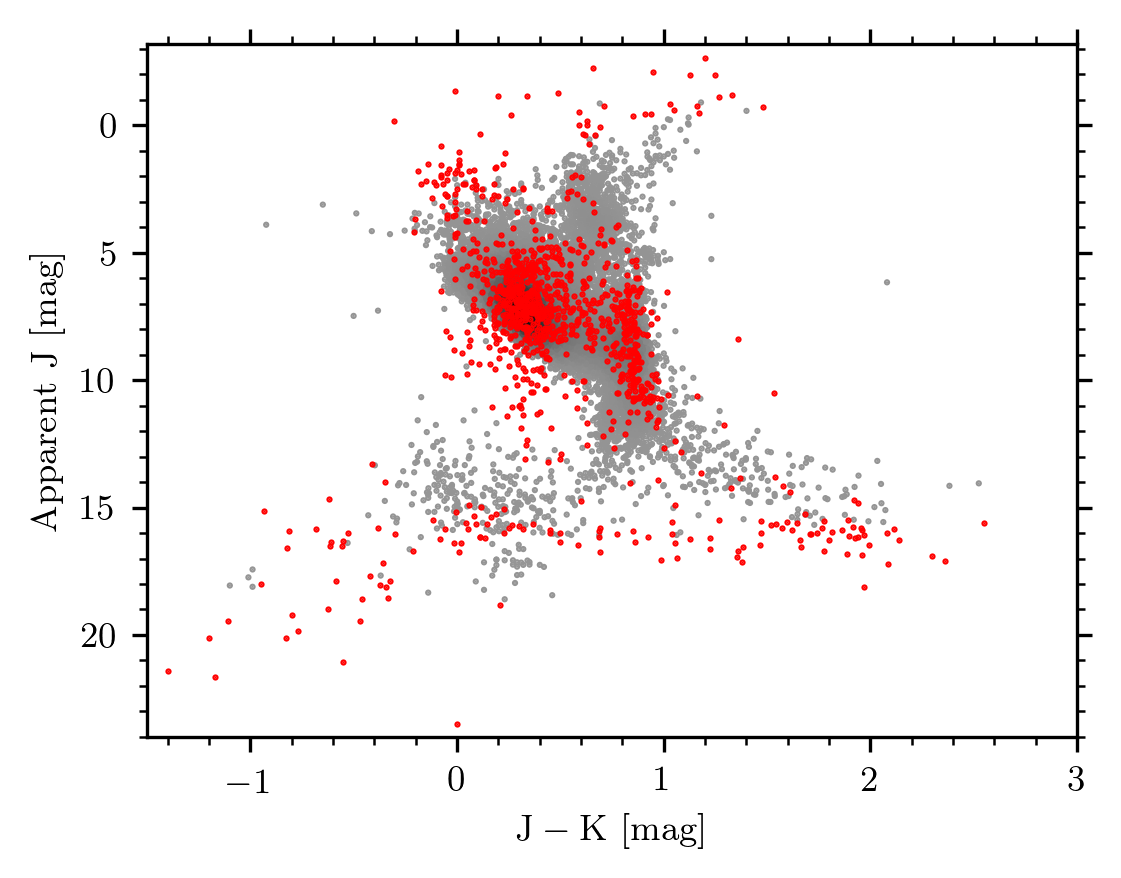}
\caption{$J$ vs. $J-K$ of the SIMBAD 100\,pc sample before \G.
Grey dots: Stars found in \gdrthree. Red dots: Stars not found in \gdrthree.}
\label{fig:SIMBADvsdr3}
\end{figure}

\begin{figure}
\includegraphics[width=0.5\textwidth]{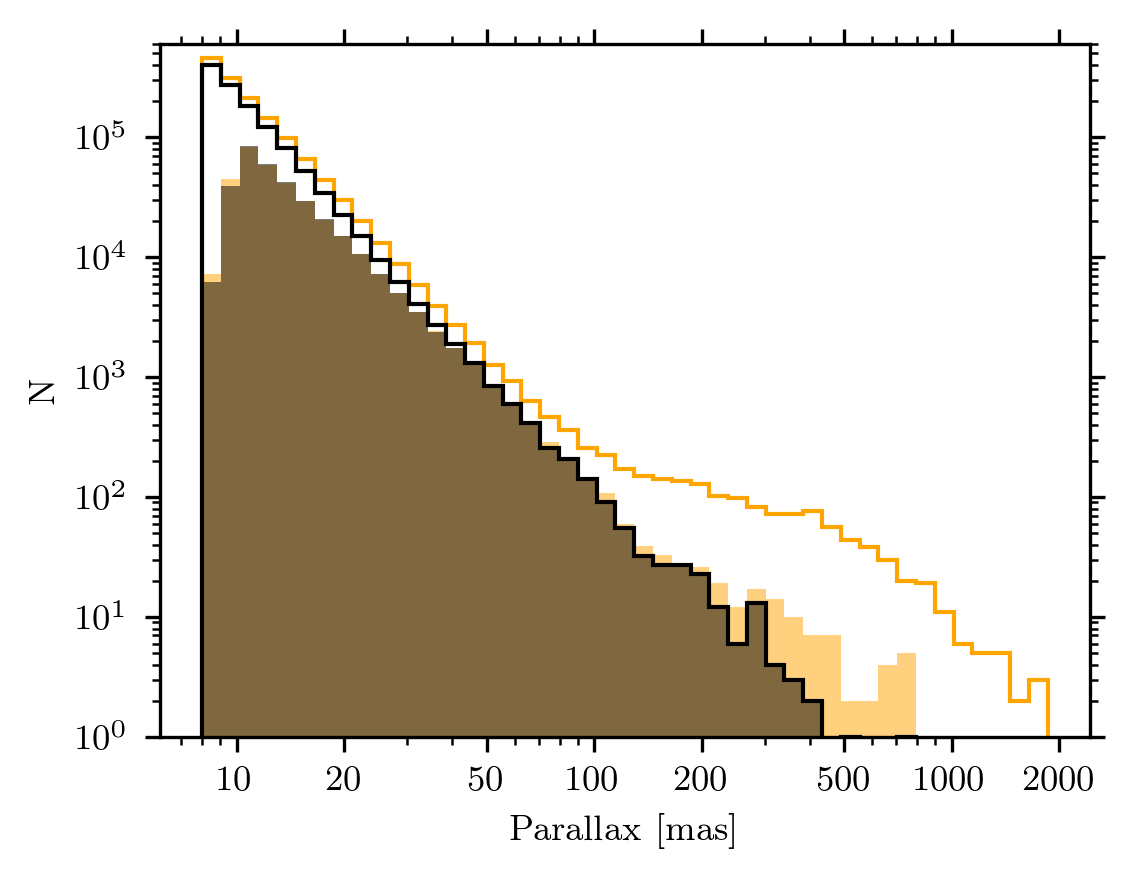}
\caption{Parallax distribution in \gdrtwo with $\hat\varpi > 8$\,mas (empty, orange), \gdrtwo with \texttt{dist\_1}$<0.1$ and $p>0.43$ (filled, orange), \gdrthree with $\hat\varpi > 8$\,mas (empty, black), and GCNS (filled, black).}
\label{fig:distdr2vsdr3}
\end{figure}

In what follows, we compare \gdrthree objects with a parallax
$\hat\varpi > 8$ mas and a probability $p>$\problimit\ (see
Sect.\,\ref{sec:cat-gen}) with previous compilations.
We first cross-matched \gdrthree with the SIMBAD sample (excluding exoplanets and
stars with only a \G\ parallax), and we retrieved 94\% of the
objects.
The \gdrthree adds 402
stars to the 100\,pc sphere and removes 318 stars. Some stars have
very different parallax determinations. For instance, HD\,215415 has a
parallax of $79.78 \pm 21.65$\,mas from Hipparcos and $10.32 \pm
0.05$\,mas from \gdrthree. This is a double star, which may question
the validity of the measurements.

SIMBAD contains 1\,245 objects with $\texttt{plx—value}\,>\,10$\,mas that are not
in \gdrthree. They are shown in Figure\,\ref{fig:SIMBADvsdr3}.
Some of them are too faint or bright or are binaries, but for some there is no clear
consistent reason why they are missing. In particular, half of
them are in \gdrtwo.  We provide an online table\footnote{available from the CDS} of these missing
objects, in which we also included 4 stars within 10\,pc and 9 confirmed
ultra-cool dwarfs with a parallax measurement from \gdrtwo\ that were not individually in SIMBAD, but were confirmed independently.


%



We next compared our sample with \gdrtwo, to which the same process of
training set construction and random forest classifier creation was applied for quality assurance (maintaining the same overall
choices, but adapting the feature space to those available
in \gdrtwo). We selected the stars with parallax $\hat\varpi > 8$\,mas and a
probability $p > 0.43$, which is the optimum threshold given by the ROC
for \gdrtwo. With this, we retrieved 95\% of the \gdrtwo stars 
in \gdrthree.
Figure\,\ref{fig:distdr2vsdr3}  shows the comparison in
the parallax distributions. It is
clear that \gdrtwo has significantly more false entries and spurious
large parallaxes. One reason that a GCNS was not attempted
with \gdrtwo was that the amount of false objects in the original data was
excessive; the selection procedure would have found 15 objects with
$\hat\varpi > 500$\,mas if we had made a GCNS with \gdrtwo.

\begin{figure*}
\begin{center}
\includegraphics[width=0.425\textwidth, height=6.5cm]{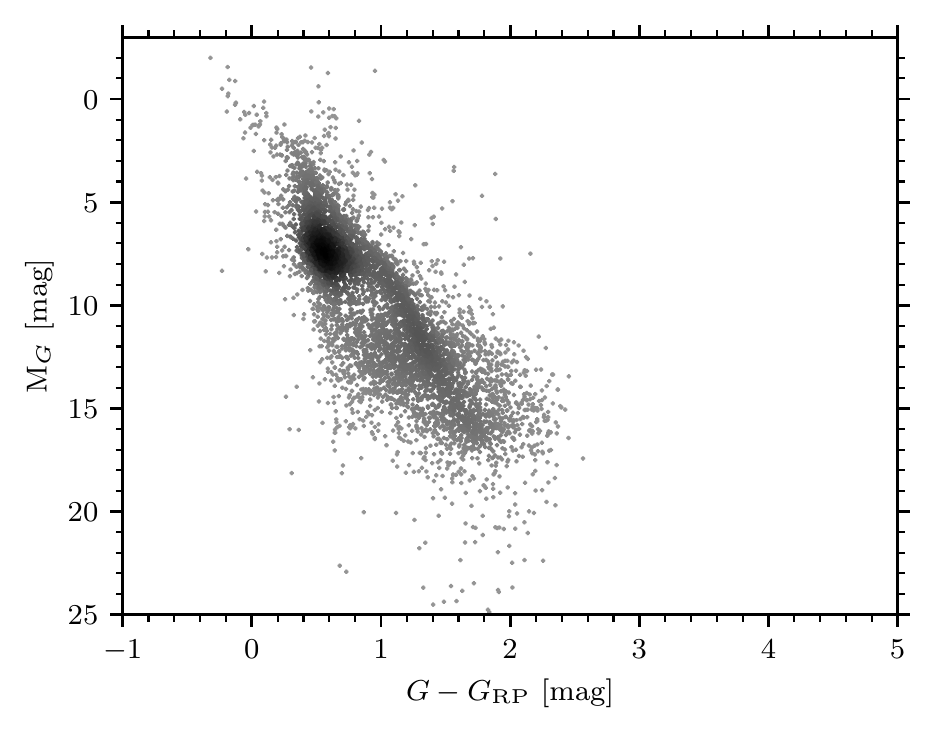}
\includegraphics[width=0.475\textwidth, height=6.5cm]{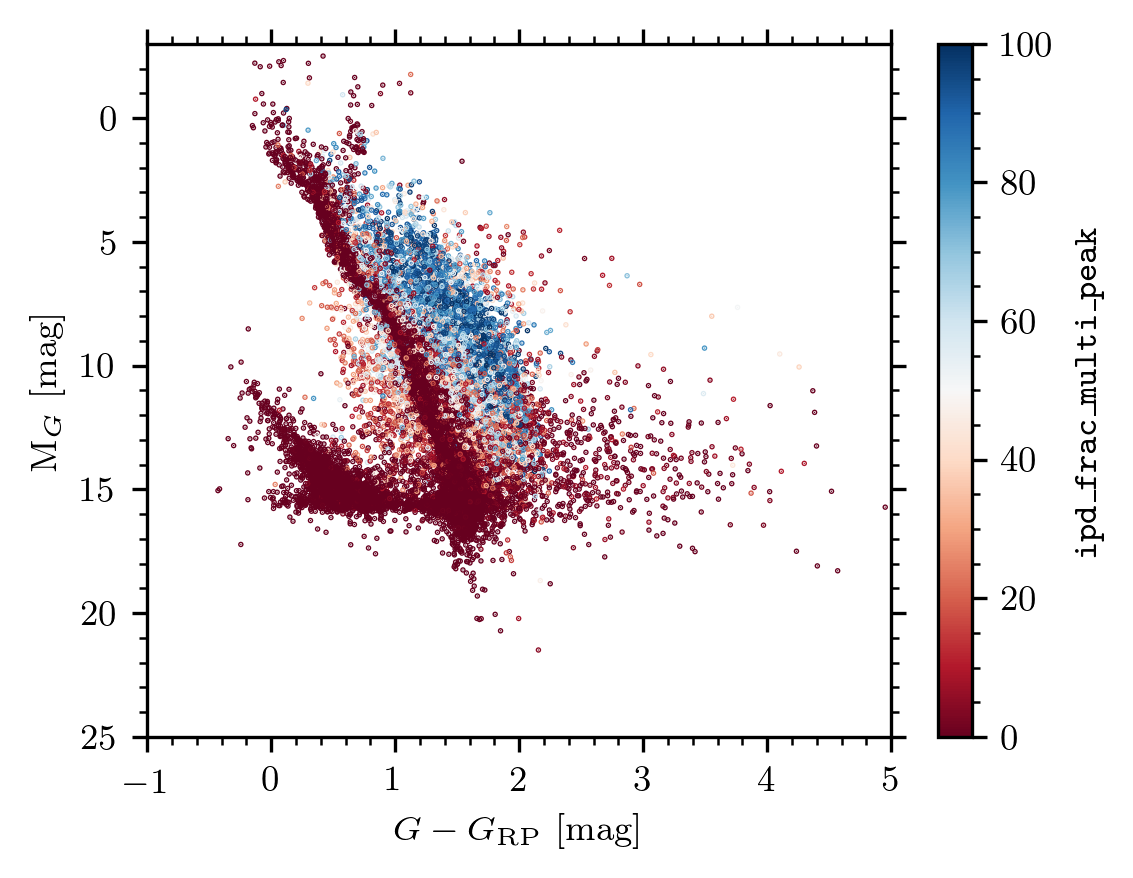}
\includegraphics[width=0.425\textwidth, height=6.5cm]{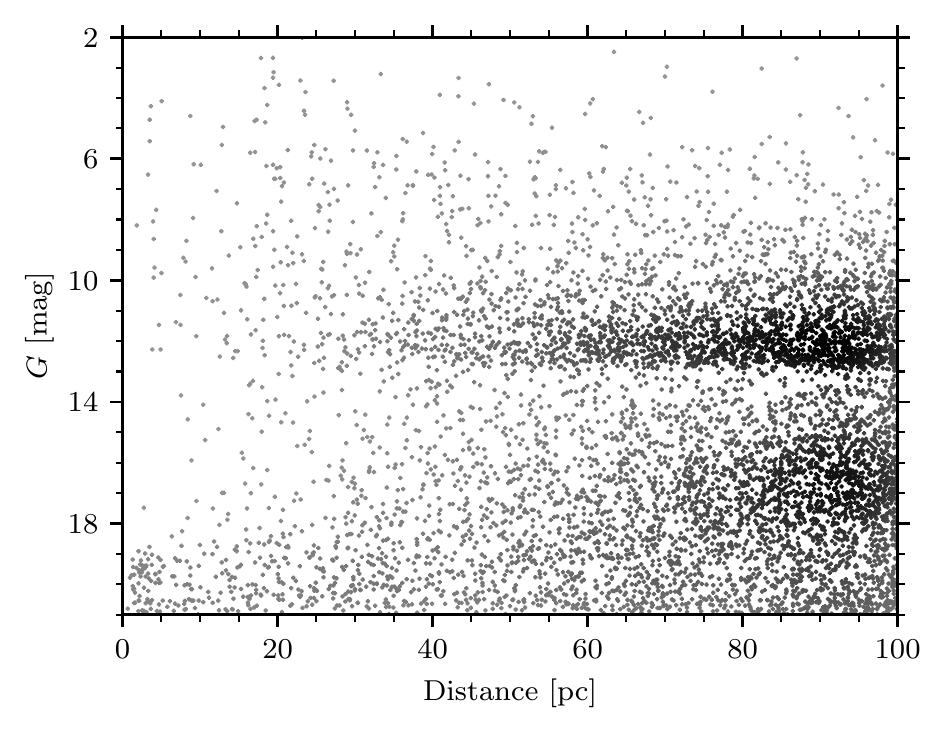}
\includegraphics[width=0.475\textwidth, height=6.5cm]{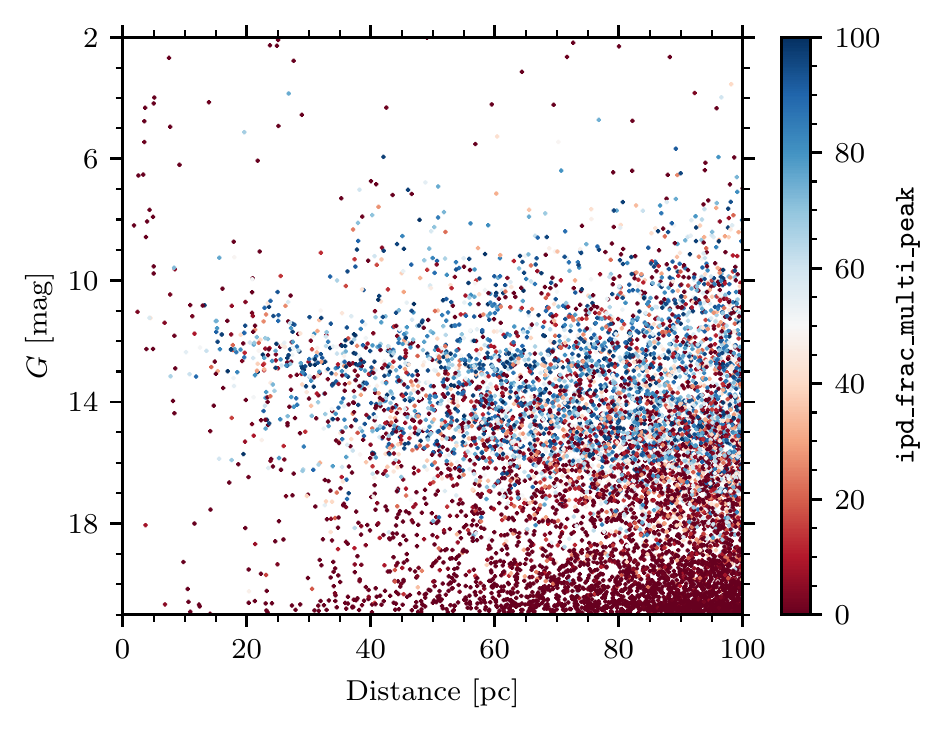}

\includegraphics[width=0.45\textwidth]{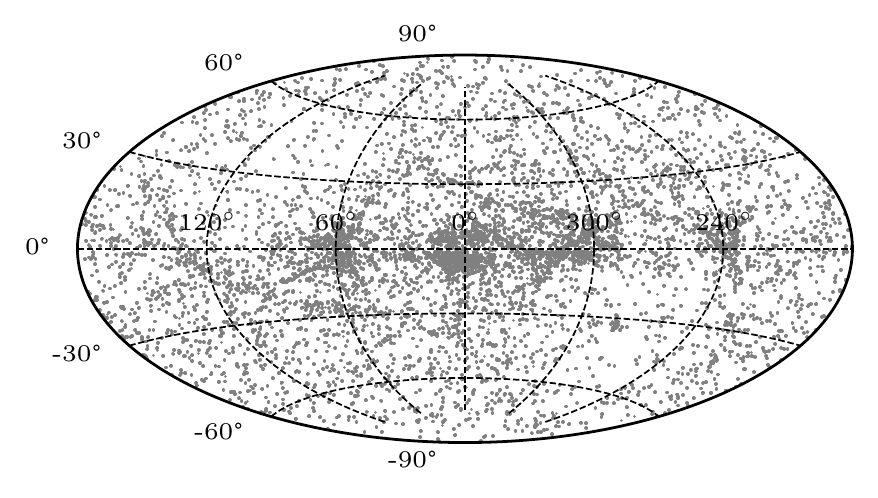}
\includegraphics[width=0.45\textwidth]{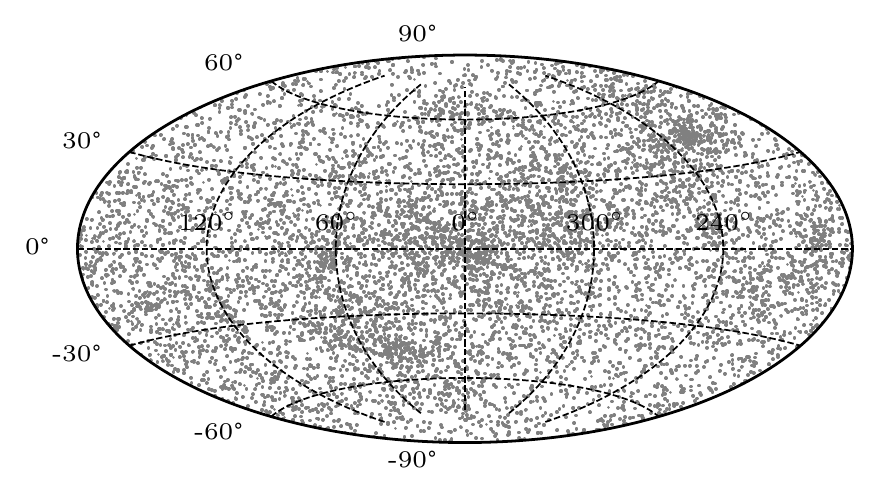}
\end{center}
\caption{Comparison between \gdrtwo and \gdrthree, from top to bottom, in the CAMD, in
a $G$ vs. distance diagram, and on the sky in galactic coordinates. Left: Stars in \gdrtwo not found in \gdrthree. Right: Stars in \gdrthree not in \gdrtwo. The upper and middle right panels are coloured with the \texttt{ipd$\_$frac$\_$multi$\_$peak}. This parameter, available in \gdrthree, provides the fraction of windows as percentage from 0 to 100 for which the detection algorithm has identified a "double peak", meaning that it was probably a visually resolved double star (either just a visual double or a real binary).}
\label{fig:dr2_not_edr3}
\end{figure*}

%
Within 100\,pc (i.e. \texttt{dist$\_$50} <0.1\,kpc), 7079 stars
published in \gdrtwo are not found in \gdrthree, and 8760 stars
in \gdrthree are not in \gdrtwo. Their position in the
CAMD, in
a $G$ versus distance diagram, and on the sky in Galactic coordinates is shown in Fig.\,\ref{fig:dr2_not_edr3}.
The left panels show the stars in \gdrtwo that are not \gdrthree.  Some of
them are very faint ($M_G>21$), very close ($\hat\varpi > 100$\,mas), and as
already noted, they are false entries in \gdrtwo. Stars in the left
part of the main sequence are also suspicious because it appears that they are also located
along scanning law patterns, as revealed by the lower left panel of
Fig.\,\ref{fig:dr2_not_edr3}. In particular, the clump around $M_G=8$
corresponds to the over-density of stars around $G=12$ in the $G$
versus distance diagram (see the middle left panel). We suspect
that the pile-up and gap at $G\sim13$ are related to the effects
in \gdrtwo of changing window class across this magnitude range,
see \citet{2018A&A...616A...4E, 2018A&A...616A...3R,
2016A&A...595A...7C}. Even given these known artefacts in the \G\ datasets,
we are still left with $\sim$3400  \gdrtwo\ stars that are located along
the main sequence for which there is no evident reason why they are not in \gdrthree.

In contrast, the $G$ versus distance diagram is smoother for the
stars that are found in \gdrthree but not in \gdrtwo\ (middle right panel). The
faint stars (at $G>20$) correspond to the WDs and low-mass
stars. Thousands of new candidates are thus expected for these faint
objects (see Sects.\,\ref{sec:wd} and \,\ref{sec:ucds}). The
CAMD in the top right panel is coloured as a function of the parameter \texttt{ipd$\_$frac$\_$multi$\_$peak}. It provides the
fraction of windows as percentage from 0 to 100 for which the
algorithm has identified a double peak, that is, a high value indicating that the
object is probably a resolved double star. Many stars lie at the right side
of the main sequence, where we expect over-luminous binary systems to lie, and a significant fraction have a consistently high probability to be binaries. 

Many objects with
low \texttt{ipd$\_$frac$\_$multi$\_$peak} values remain as outliers with red 
colours, probably due to inconsistent photometry (see
also Sect. \ref{sec:clusters}).  The sky map (lower right panel) shows regions (in
particular, $l\simeq 240^\circ$ and $b\simeq 45^\circ$) in which stars
are found in \gdrthree, but not \gdrtwo.


\subsection{ 10 pc sample}

As an illustration, we detail the 10\,pc sample.  The SIMBAD query
returns 393 objects (excluding exoplanets) 
From this list we removed 14
duplicates, one error (HIP 114176), and \gdrtwo 4733794485572154752, which
we suspect to be an artefact that lies in front of a globular
cluster. 
We added  the multiple brown dwarf Luhman 16 AB at 2 pc \citep{2013ApJ...767L...1L}. 2MASS
J19284155+2356016, a T6 at 6\,pc \citep{2019ApJS..240...19K}. The
resulting 10\,pc sample contains 378 objects, 307 
of which are in the GCNS.  The new \gdrthree parallax places \object{LP 388-55} outside the 10\,pc
sphere, and HD\,260655 enters this sphere. 

The GCNS lists the first individual parallax measurements for five stars in systems within the 10\,pc sample:
\object{HD\,32450B}, \object{CD-37 10765B},
the WD \object{$o$\,2\,Eri\,B}, \object{Wolf\,424\,B}, and one star in the
\object{$\mu$ Her} system, separated by 0.6\,\arcsec from $\mu$.02\,Her. This means that 312 stars are located within 10 pc in the GCNS.


 We removed all giants, WDs, and peculiar or uncertain types from the
 full set of SIMBAD spectral types, and we find a calibration between
 the median absolute magnitude, $M_G$, and each spectral class. With
 this calibration for the SIMBAD entries with spectral types, we
 predicted their apparent $G$ magnitudes. Ten of the objects missing
 in GCNS are stars that are too bright (\object{Sirius},
 \object{Fomalhaut}, \object{$\alpha$ Cen} A and B, \object{Vega},
 \object{Procyon}, \object{Altair}, \object{Mizar}, \object{$\chi$
   Dra}, and \object{HD\,156384}), 33 objects are T and Y brown dwarfs
 and are too faint, as are probably 2 late-L dwarfs. However, of the
 remaining 26 objects, 15 have \gdrtwo\ parallaxes. We note that 21 of
 these 26 objects are either spectroscopic binaries or in close binary
 systems that will give high residuals with a single-star solution,
 and for this reason, they may not have passed the \G\ five-parameter
 solution quality assurance tests. Five objects remain
 (\object{HD 152751}, \object{G 24-16}, \object{IRAS 21500+5903}, \object{SCR J1546-5534}, and \object{BPS CS
 22879-0089}) for which we do
 not have an obvious reason to explain the lack of a
 \gdrthree\ five-parameter solution.

The resulting 10\,pc sample contains 383 objects with a parallax
determination: 376 stars from SIMBAD minus LP 388-55, plus Luhman 16 AB, 2MASS J19284155+2356016, HD 260655, and the five companion stars with a first  parallax determination from \gdrthree. There are also known unresolved binary systems (Procyon,  $\eta$ Cas,  $\xi$\,UMa, etc.), and as there will undoubtedly be new systems \citep[e.g. see][]{2020arXiv200900121H}, we counted unresolved systems as one entry. 
 The T/Y types will not be complete in this list, for instance, the 16 T6 to Y2 brown dwarfs that are not included in this list have a parallax larger than 100 mas from \cite{2019ApJS..240...19K}, and more ultra-cool Y-dwarfs are expected to be discovered.
These 383 objects can be retrieved by selecting entries from the GCNS 
with \texttt{parallax}$ >= 100$\,mas (312 objects) and that in the file with missing objects that we provide have \texttt{plx\_value}$ >= 100$\,mas (71 objects). To provide a starting point for estimating the number of
objects expected within 100\,pc in the next section, we estimate that the
number of objects with $M_G<15.5$ within 10\,pc is \TENPC.


\subsection{Consistency check with the \texorpdfstring{10\,pc sample}{10pc}}
\label{sec:10pc_comparison}
In order to check for the plausibility of the total number of sources
that are classified as good by the random forest described in
Sect.\,\ref{sec:cat-gen}, we used the Einasto law with the maximum a posteriori values of the parameters inferred in
Sect.\,\ref{sec:verticalstrat} to produce synthetic samples of
sources with uniform densities in planes parallel to the Galactic
plane. We produced an arbitrary number of samples and then scaled
the numbers to match the observed number of sources within 10\,pc. As
the \G\ 10\,pc sample will be missing bright sources and to avoid
a possible circular reasoning, we used a census of known sources
inside 10\,pc with an absolute magnitude brighter than 15.5\,mag
regardless of whether the sources are detected by \G. 

In our simulations we assumed for the sake of simplicity that the
binary population properties are dominated by the M spectral type
regime. We set a binarity fraction of 25\% and a distribution of
binary separations ($a$) that is Gaussian in logarithmic scale, with
the mean and standard deviation equal to 1: $\log_{10}(a)\sim
\mathcal{N}(\mu=1,\sigma=1)$ \citep[see][and references
  therein]{2012A&A...543A.100R,FA-054}. We assumed that the orientation
of the orbital planes are random and uniform in space, giving rise to
the usual law for the inclinations $i$ with respect to the line of
sight given by $p(i)\sim \sin(i)$. Furthermore, we assigned a magnitude
difference between the two components (we did not include higher order
systems) based on the relative frequencies encountered in the GCNS and
discussed in Sect.\,\ref{sec:mult-resolved}. Based on the separations and
inclinations, we computed the fraction $f$ of the orbit where the
apparent angular separation of the binary components is larger than
the angular separation in Eq.\,\ref{eq:smin}. The probability of
detecting the binary system as two separate sources was then
approximated with the binomial distribution for a number of trials
equal to 22 (which is the mode of the distribution of the number of
astrometric transits in our dataset) and success probability $f$.
This is an optimistic estimate because it assumes that one single
separate detection suffices to resolve the binary system.

Using the procedure described above, we generated ten simulations with 40
million sources each, distributed in a cube of
$110\times110\times110$\,pc. From each simulation we extracted the
number of sources within 10 and 100\,pc ($N_{10}$ and $N_{100}$
, respectively) and the ratio between the two ($N_{100}/N_{10}$). The
average value of this ratio from our simulations is 878.2$\pm$28.2. When
we apply this scale to the observed number of sources within 10\,pc
(\TENPC\ sources), the expected number of sources in the GCNS selection
is $277\,511\pm8911$. This prediction has to be compared with the
number of sources in the GCNS catalogue with an absolute magnitude
brighter than 15.5 and within 100\,pc. In order to obtain this number,
we proceeded as described in Sect.\,\ref{sec:volcompleteness} and obtained
a total number of sources of 282\,652, which agrees well with
the prediction given the relatively large uncertainties and the fact
that the number of sources within 10\,pc (\TENPC) is itself a sample
from a Poisson distribution. It has to be borne in mind, however, that
the expected number (277\,511) does not take
incompleteness due to variations across the sky of the $G$ magnitude
level or due to the contrast sensitivity into account.
GeDR3mock simulations show that 1.8k sources
are fainter than the \gdrthree 85th percentile
magnitude limits (cf. Sect.\,\ref{sec:gmag_lim}). Additional 0.3k
sources are lost due to the contrast sensitivity, which will be a lower limit
because GeDR3mock
does not include binaries, so that this is only the contribution of chance
alignments in crowded regions.



\subsection{Contamination and completeness}
\label{sec:contamandcompl}

As described in \cite{2018A&A...616A...2L}, every solution in \G\ is the result of iteratively solving with
different versions of the input data and varying the calibration
models. The final solutions do not use all the observations and not
all solutions are published, many quality assurance tests are applied
to publish only high-confidence solutions. Internal parameter
tests that were applied to publish the five-parameter solution in \gdrthree\
were $G <= 21.0$; \texttt{astrometric\_sigma5d\_max} $<$\,1.2$\times10^{0.2 \mathrm{max}(6-G, 0, G-18)}$\,mas; \texttt{visibility\_periods\_used}\,$>$\,8;
\texttt{longestsemiMajorAxis} of the position uncertainty ellipse
$<=$100\,mas; and \texttt{duplicateSourceID}\,=0. The tests were 
calibrated to provide a balance between including poor solutions and
rejecting good solutions for the majority of objects, that is, distant,
slow-moving objects whose characteristics are different from those of the nearby
sample. In the current pipeline, the astrometric solution considers
targets as single stars, and for nearby unresolved or close binary
systems the residuals of the observed motion to the predicted motion
can be quite large, so that this causes some nearby objects to fail
the \texttt{astrometric\_sigma5d\_max} test.

For example, as we saw in Sec.\,\ref{sec:compprevcomp}, we expect 383
objects within the 10\,pc sample.  When the 35 L/T objects that we
consider too faint are removed, 348 objects remain that \G\ should see (we include the bright objects for the purpose of this exercise). Twenty-six of
these 348 objects do not have five-parameter solutions in
\gdrthree\ because they fail the solution quality checks. The fact that
many of the lost objects were in spectroscopic or close binary systems
is also an indication that the use of a single-star solution biases
the solutions for the nearby sample. If we take these numbers directly,
this loss is still relatively small: 26 of 348, or 7.4\%.  While this loss
is biased towards binary systems, it probably does not depend on direction
and the loss will diminish as the distance increases because the
effect of binary motion on the solutions decreases. The excess of
objects found in the GCNS compared to the prediction in
Sect.\,\ref{sec:10pc_comparison} supports this conclusion, and the
comparison of objects found in SIMBAD to those in the GCNS shows that only
6\% are missing, therefore we consider the 7.4\% as a worst-case estimate of the
GCNS stellar incompleteness.


Sect.\,\ref{sec:gmag_lim} showed that the mode, or peak, of the
apparent $G$ distribution is at $G = 20.41$\,mag, which includes
85\,\% of the sources. The median absolute magnitude of an M9 is
$M_G = 15.48$\,mag, which would translate into $G = 20.48$\,mag at 100\,pc;
therefore we should see at least 50\% of the M9-type stars at our catalogue
limit. Our comparison to the PS1 catalogue indicates that \gdrthree\ is 98\% complete at this magnitude. As discussed further in
Sect.\,\ref{sec:ucds}, the complete volume for later spectral types
becomes progressively smaller, but for spectral types up to M8, they are volume limited and not magnitude limited.

We lose small numbers of objects because we started with a sample that was selected with $\varpi > $ 8\,mas, which from the GDR3Mock is estimated
to be 55. We will lose objects that are separated by less than
0.6\arcsec due to contrast sensitivity
(Sect.\,\ref{sec:contrast_sensitivity}), which for chance alignments
from the GDR3Mock we have estimated to be 300 sources, but it will be
much higher for close binary systems and will bias our sample to not
include these objects. Finally, we lose objects that are incorrectly
removed because they have $p <$ \problimit. Based on
Table\,\ref{tab:confMatSel} and Sect.\,\ref{sec:2Cataloggeneration}, we
estimate this to be approximately 0.1\% of the good objects.

The incompleteness for non-binary objects to spectral type M8 is
therefore dominated by the 7.4\% of objects for which \G\ does not provide a parallax. For  objects later than M8, the complete volume decreases, as shown in Sect.\,\ref{sec:ucds}. We did not consider unresolved binary systems, which are considered in
Sects.\,\ref{sect:unresolved} and
\ref{sec:mult-resolved}.

We also considered the contamination of the GCNS. There are two types of contamination: objects that pass our probability cut but have poor
astrometric solutions, and objects that are beyond our 100\,pc
limit. The contamination of the good solutions is evident in the blue
points that populate the horizontal feature at $M_G$=15-16\,mag and between the main and WD sequence (Fig.\,\ref{fig:CAMDGminRP_KDE-MG}, right panel).  These are faint
objects ($G$>20.\,mag) that lie at the limit of our distance selection
(\dmed\~=~80-120\,pc), for example with a distance modulus of $\sim$5\,mag), and that therefore populate the $M_G >$15\,mag region.  These faint objects have the lowest
signal-to-noise ratio, and their parameters, used in the random
forest procedure, therefore have the largest uncertainties.  Because objects with poor astrometric solutions were accepted, we
estimate this contamination based on Table\,\ref{tab:confMatSel} to be $\sim$0.1\%, the false
positives. This means about 3000 objects for the GCNS.

The contamination by objects beyond the 100\,pc sphere can be
estimated by summing the number of distance probability quantiles inside
and outside 100\,pc. We find that 91,2\% of the probability mass lies
within 100\,pc and the rest outside. This means ~29k sources, or 9\%.
The use of the full distance PDF will allow addressing this possible source
of bias in any analysis.

These known shortcomings should be considered when the GCNS is used.
If the science case requires a clean 100\,pc sample, where no
contamination is a priority and completeness is of secondary importance, objects with a \dmed\ $<$0.1\,kpc should be selected from the GCNS. If the
science case requires a complete sample, all
objects with \dtol\ $<$ 0.1\,kpc should be selected and then weighted by the distance PDF. When a clean photometric sample is required, the
photometric flags should be applied, which we did not exploit to produce
this catalogue.  In the next section we investigate a number of
science questions, for which we apply different selection procedures to the
catalogue and use the distance PDF in different ways to illustrate
some optimal uses of the GNCS.


\section{GCNS exploitation}

\subsection{Vertical stratification}
\label{sec:verticalstrat}

In this section we study the vertical stratification as inferred from
the GCNS volume-limited sample. We did this using a relatively simple
Bayesian hierarchical model that we describe in the following
paragraphs. First we describe the data we used to infer the vertical
stratification parameters, however. The data consist of the latitudes, observed
parallaxes, and associated uncertainties of the sources in the GCNS with
observed parallaxes greater than 10\,mas.
In order to include the effect of the truncation in the observed
parallax, we also used the number of sources with observed parallaxes
between 8 and 10\,mas and their latitudes (but not their
parallaxes). The reasons for this (and the approximations underlying
this choice) will become clear after the inference model
specification.
The assumptions underlying the model listed below.
\begin{enumerate}
\item {\label{assump:trueDist} The data used for inference represent a sample of sources with true
parallaxes larger than 8\,mas. This is only an
approximation, and we know that the observed sample is incomplete and
contaminated. It is incomplete for several reasons, but in the context
of this model, the reason is that sources with true parallaxes greater than
8\,mas may have observed parallaxes smaller than this limit due to
observational uncertainties. It is also contaminated because the opposite is
also true: true parallaxes smaller than 8\,mas may be scattered in as a result of observational uncertainties as well. Because this effect is stronger than
the first reason and more sources lie at larger distances, we
expect fewer true sources with true parallaxes greater than
8\,mas (at distances closer than 125\,pc) than were
found in the GCNS.}
\item The source distribution in planes parallel to the Galactic plane
is isotropic. That is, the values of the true Galactic Cartesian
coordinates $x$ and $y$ are distributed uniformly in any such plane.
\item The measurement uncertainties associated with the observed Galactic
latitude values are sufficiently small that their effect on the
distance inference is negligible. Uncertainties in the measurement of
the Galactic latitude have an effect on the inference of distances because we
expect different distance probability distributions for different
Galactic latitudes. For example, for observing directions in the plane
that contains the Sun, the true distance distribution is only dictated
by the increase in the volume of rings at increasing true distances
(all rings are at the same height above the Galactic plane and
therefore have the same volume density of sources), while in other
directions the effect of increasing or decreasing volume densities due
to the stratification modifies the true distance distribution.
\item Galactic latitudes are angles measured with respect to a plane that
contains the Sun. This plane is parallel to the Galactic plane but offset with
respect to it by an unknown amount.
\item Parallax measurements of different sources are independent. This is
known to be untrue but the covariances amongst \G\ measurements are not
available and their effect is assumed to cancel out over the entire
celestial sphere.
\end{enumerate}

For a constant volume density $\rho$ and solid angle d$\Omega$ along a
given line of sight, the probability density for the distance $r$ is
proportional to $r^2$. In a scenario with vertical stratification,
however, the volume density is not constant along the line of sight
but depends on r through $z$, the Cartesian Galactic coordinate. For
the case of the Einasto stratification law
\citep{1979IAUS...84..451E} that is used in the Besan\c con Galaxy model 
\citep{2003A&A...409..523R}, the distribution of sources around the
Sun is determined by the $\epsilon$ parameter (the axis ratio) 
and the vertical offset of the
Sun , $Z_{\odot}$, with respect to the fundamental plane that defines the highest
density. The analytical expression of the Einasto law for
ages older than 0.15\,Gyr is

\begin{equation}
  \rho \propto \rho_0\cdot\exp{\left(-\left(0.5^2+\frac{a^2}{R_{+}^2}\right)^{\frac{1}{2}}\right)}- \exp{\left(-\left(0.5^2+\frac{a^2}{R_{-}^2}\right)^{\frac{1}{2}}\right)}
\label{eq:Einasto-dens}
,\end{equation}

\noindent where $a^2=R^2+\frac{z^2}{\epsilon^2}$, $R$ is the solar
galactocentric distance, $z$ is the Cartesian Galactic coordinate
(which depends on the Galactic latitude $b$ and the offset as
$z=r\cdot\sin(b)+Z_{\odot}$), $\epsilon$ is the axis ratio, and
we used the same values as in the Besan\c con model, $R_{+}=2530$\,pc
and $R_{-}=1320$\,pc. The value of $\epsilon$ 
 in general depends on age. We assumed a single value
for all GCNS sources independent of the age or the physical
parameters of the source such as mass, effective temperatures, and evolutionary state.

In our inference model we have the vertical stratification law
parameters ($\epsilon$ and $Z_{\odot}$) at the top.  We defined a prior
for the $\epsilon$ parameter given by a Gaussian distribution centred at
0.05 and with a standard deviation equal to 0.1, and a Gaussian prior
centred at 0 and with a standard deviation of 10\,pc for the offset of
the Sun with respect to the Galactic plane. Then, for a given source
with Galactic latitude $b$, the probability density for the true
distance $r$ is given by

\begin{equation}
p(r \mid \epsilon,Z_{\odot}) \propto \rho(z(r) \mid \epsilon,Z_{\odot})\cdot r^2
\label{eq:latDepDistPrior}
.\end{equation}

Equation\,\ref{eq:latDepDistPrior} is the natural extension of the
constant volume density distribution of the distances. Finally, for $N$
observations of the parallax $\hat{\varpi}_i$ with associated
uncertainties $\sigma_{\varpi_i}$, the likelihood is defined as

\begin{equation}
\mathcal{L} = \prod_1^N p(\hat{\varpi_i} \mid r_i,\epsilon,Z_{\odot}) = 
\prod_1^N  \mathcal{N}(\hat{\varpi_i} \mid r_i,\sigma_{\varpi_i})
\label{eq:likelihood_observed}
,\end{equation}

\noindent where $\mathcal{N}(\cdot \mid \mu,\sigma)$ represents the Gaussian 
(or normal) distribution centred at $\mu$ and with standard deviation $\sigma$,
and we have introduced the assumption that all parallax measurement
are independent. The model is defined by the stratification parameter
$\epsilon$, the (also) global parameter $Z_{\odot}$, and the $N$ true
distances to individual sources $r_i$. With this, the posterior
distribution for the full forward model can be expressed as

\begin{equation}
p(\epsilon,Z_{\odot},\vec{r} \mid \vec{\hat{\varpi}}) \propto \prod_1^N
p(\hat{\varpi_i} \mid r_i,\epsilon,Z_{\odot})\cdot
p(r_i\mid \epsilon,Z_{\odot})\cdot p(\epsilon)\cdot p(Z_{\odot})
,\end{equation}

\noindent where bold symbols represent vectors. For the sake of
computational efficiency, we marginalised over the $N$ individual
distance parameters $r_i$ and inferred only the two global parameters
$\epsilon$ and $Z_{\odot}$,

\begin{align}
\begin{split}
& p(\epsilon,Z_{\odot} \mid \vec{\hat{\varpi}}) \propto
\int p(\epsilon,Z_{\odot},\vec{r} \mid \vec{\hat{\varpi}})\cdot\vec{{\rm d}r}\\
& = \prod_1^N \int_0^{r_{max}} p(\hat{\varpi_i} \mid r_i,\epsilon,Z_{\odot})\cdot
p(r_i \mid \epsilon,Z_{\odot})\cdot{\rm d}r_i \cdot p(\epsilon)\cdot p(Z_{\odot}),\\
\end{split}
\end{align}

\noindent where $r_{max}$ represents the assumed maximum true 
distance in the sample of sources that defines the dataset.

The model described so far relies on the assumption that the dataset
used for the evaluation of the likelihood is a complete and uncontaminated set of the sources with true distances between 0 and
$r_{max}$. The selection of this dataset from the observations is impossible, however. On the one hand, the posterior distances derived
in Sect.\,\ref{sec:distance_estimation} assume an isotropic prior, and a selection based on it would therefore be (mildly) inconsistent. The
inconsistency is minor because the directional dependence of the prior
is a second-order effect with respect to the dominant $r^2$ factor. It
is also problematic because a source with a posterior median slightly
greater than 100\,pc would be left out of the sample even though it has
a relatively high probability to be inside, and {\sl \textup{vice versa}} for
sources with posterior medians slightly smaller than 100\,pc. On the
other hand, a selection based on the observed parallax (e.g.
defined by $\hat{\varpi} > 10\,mas$) is different from the sample assumed by
the model (which is defined by all true distances being within the 100\,pc
boundary). We decided to modify the model to account
for a truncation in the space of observations for illustration
purposes. It exemplifies an imperfect yet reasonable way to deal
with such truncations.

We inferred the model parameters from the set of sources with
observed parallaxes $\hat{\varpi} > 10$\,mas, but we modified the
likelihood term in order to include the truncation of observed
parallaxes. The dataset upon which our model infers the
stratification parameters was defined by all sources classified as
good astrometric solutions with the random forest described in
Sect.\,\ref{sec:cat-gen}, for example with $p >= 0.38$. We assumed that the total number of
sources with true distances smaller than 125\,pc is the same as that
with observed parallaxes greater than or equal to 8\,mas. This is an
approximation because we know that in general, the true number will be
smaller due to the effect of the measurement uncertainties scattering
more external sources in than internal sources out. However, the true number cannot be estimated without knowing the stratification
parameters. It is possible to infer the total number as another model
parameter, but that is beyond the scope of this demonstration paper. We modified the likelihood term to include the effect of the
truncation as follows.  The likelihood term was divided into two
contributions that distinguish sources with $\hat{\varpi} > 10$\,mas
and sources with smaller parallaxes ($8 < \hat{\varpi} < 10$\,mas). For
the former, the likelihood term was exactly as described by
Eq.\,\ref{eq:likelihood_observed}. For the latter we only retained
their Galactic latitudes, the fact that their observed parallaxes are
smaller than 10\,mas, and the total number of sources, but not the
parallax measurements themselves. The new likelihood term is

\begin{align}
\begin{split}
    \mathcal{L}=&p(\vec{\hat{\varpi}} \mid r,\epsilon,Z_{\odot}) \\ =
    & \prod_1^{N_{obs}} p(\hat{\varpi_i} \mid r_i,\epsilon,Z_{\odot})
    \cdot \prod_1^{N_{miss}} \int_{-\infty}^{10}
    p(\hat{\varpi_i} \mid r_i,\epsilon,Z_{\odot}) \cdot{\rm d}\hat{\varpi}_i,
\end{split}
\end{align}

\noindent where 
$N_{obs}$ is the number of sources with observed
parallaxes $\hat{\varpi}\geq 10$\,mas, and $N_{miss}$ is the number of
sources with observed parallaxes in the range from 8 to 10\,mas. With
this new likelihood expression, we can proceed to calculate posterior
probability densities for a given choice of priors. 

\begin{figure}
    \centering
    \includegraphics[width=0.45\textwidth]{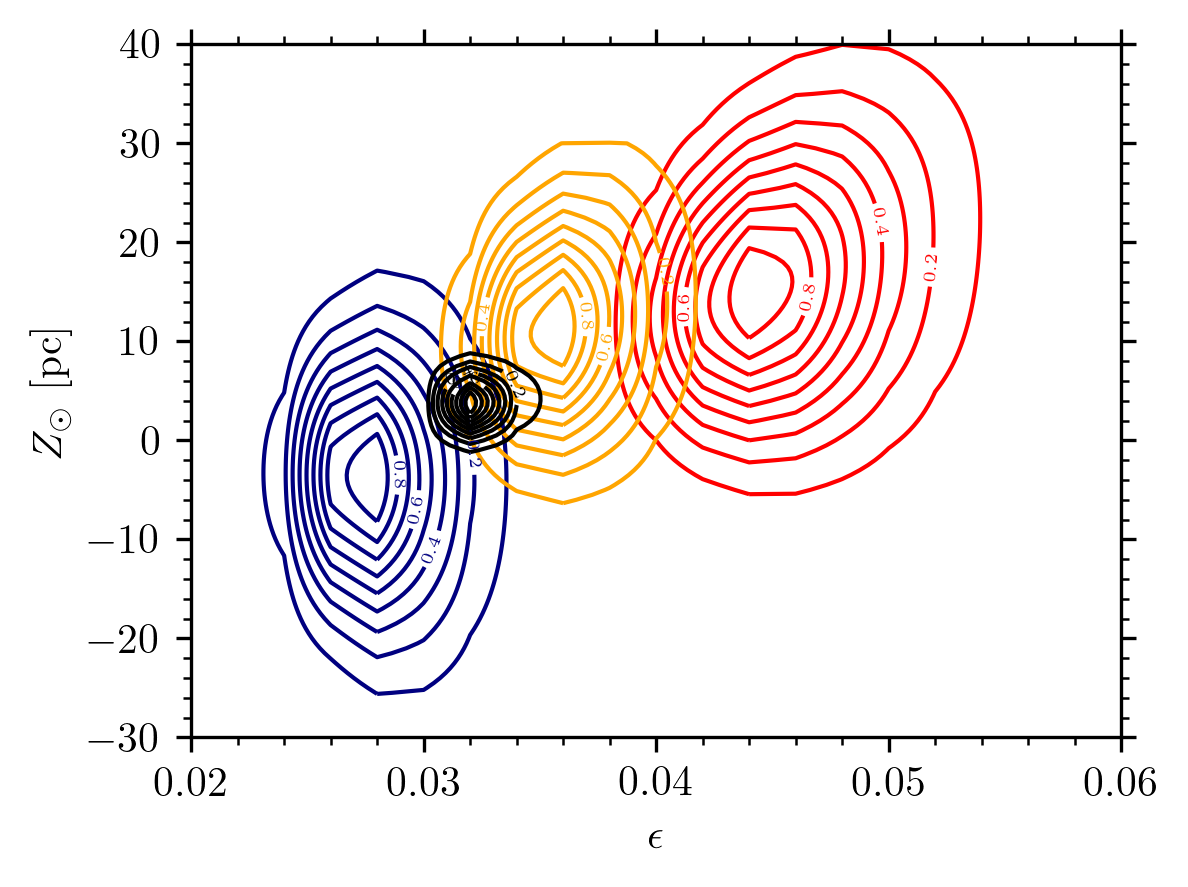}  
    \caption{Posterior
    probability density for the Einasto law $\epsilon$ parameter and
    the solar $z$ coordinate for the entire GCNS (black) and for three
    segments along the main-sequence distributions from left to right: early spectral types before the
    turn-off point (blue), spectral types G and early K (orange), and
    M-type stars (red).}
    \label{fig:posterior-epsilon}
\end{figure}

Figure \ref{fig:posterior-epsilon} shows the posterior density
contours for the $\epsilon$ parameter and the solar coordinate $z=Z_{\odot}$
under the Einasto model described above for the entire sample, small contours in the middle, and for three separate subsamples along the main sequence. The maximum a posteriori 
values of the model parameters for the full GCNS sample are $\epsilon
= 0.032$ and $Z_{\odot}=4$\,pc. Figure \ref{fig:posterior-epsilon}
shows that the hot population (defined as the main-sequence segment
brighter than $M_G=4$) seems characterised 
by a smaller $\epsilon$ parameter (with a maximum a posteriori
value of 
0.028) and a vertical coordinate $Z_{\odot}=-3.5,$ whereas the middle 
($4 < M_G < 7$) and cool ($12 < M_G < 15$)
segments of the main sequence are characterised by higher values of
$\epsilon$ (0.036 and 0.044, respectively) and $Z_{\odot}$ coordinates
larger than the inferred value for the full sample (11.5 and 15,
respectively). For comparison, the values used in the Besan\c con
Galaxy model \citep{2003A&A...409..523R} range between 0.0268 for
stars younger than 1\,Gyr and 0.0791 for those with ages between 7 and
10\,Gyr. The parameters inferred in this section are fully consistent
with these values given that the data samples are not characterised by
a single age but contain sources with a continuum of ages (younger on
average for the hot segment, and increasingly older for the middle or
cool segments) determined by the local star formation history and
kinematical mixing.

The values discussed in the previous paragraph did not take into account that the objects used to infer the stratification parameters include sources that do not belong to the thin disc. In order to assess the effect of the presence of thick disc stars in our dataset on the inferred parameter values, we applied the same method to an augmented dataset with an additional 6.6\% of sources (see Sect. \ref{sec:stellar-pops-orbits} for a justification of this value) distributed uniformly in the three Galactic Cartesian coordinates. This was an upper limit to the effect because thick-disc stars are also vertically stratified. It results in a small shift of the inferred parameters characterised by a maximum a posteriori value of $\epsilon$ and of the vertical coordinate $Z_{\odot}$ of 0.034 and 3.5\,pc, respectively.

\begin{figure}
    \centering 
    \includegraphics[width=0.45\textwidth]{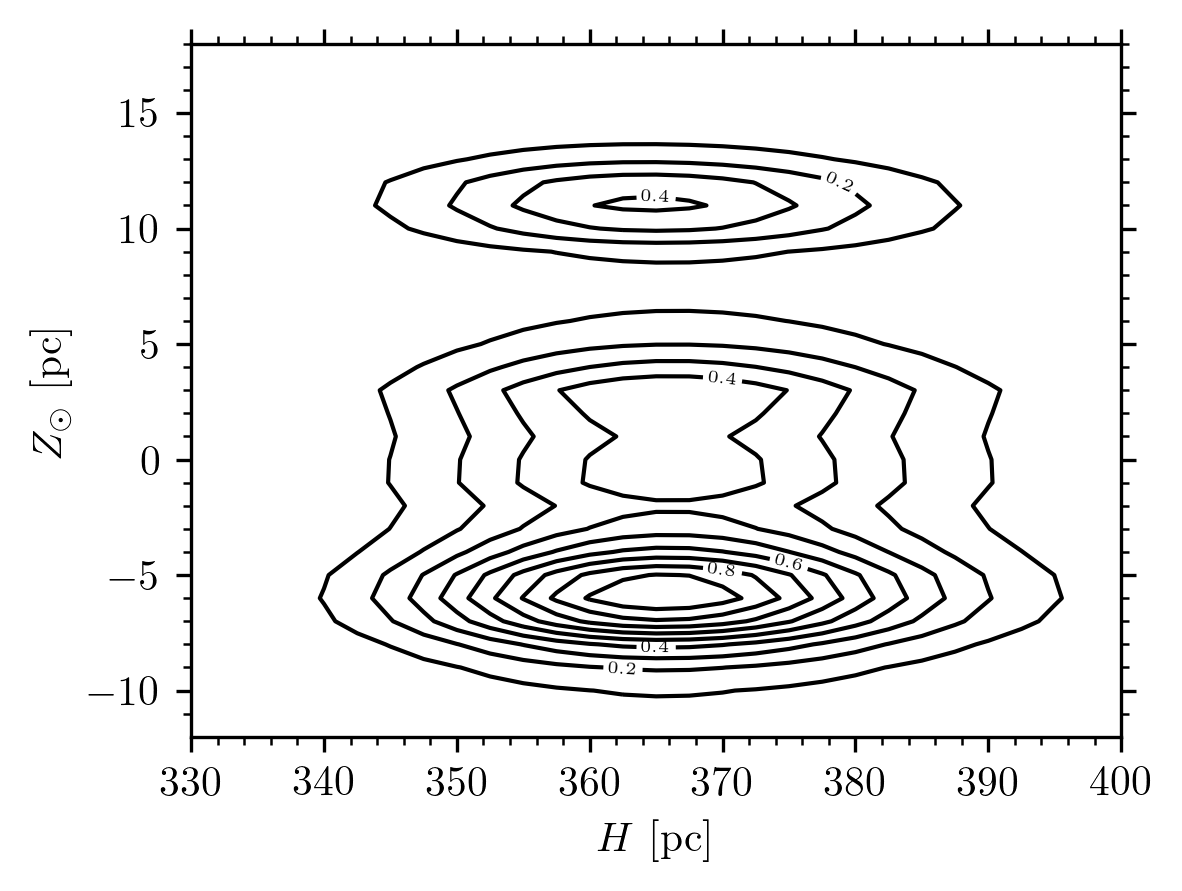}  
    \caption{Posterior
    probability density for the scale height $H$ and solar $z$
    coordinate with respect to to the Galactic plane inferred from the
    hierarchical Bayesian model.}
    \label{fig:HBM-posterior}
\end{figure}

We also applied the formalism described above to the alternative
stratification model defined by the exponential decay with scale
height $H$ \citep[see][for a recent application of Bayesian 
techniques to a set of analytical stratification laws that includes 
the exponential model]{2020OJAp....3E...5D}. The 
prior probability for the scale height is defined as
an exponential distribution with scale 1000\,pc, and that for the
offset is defined as a Gaussian distribution centred at 0 and with a
standard deviation of 10\,pc (the same as in the case of the Einasto
law). Figure \ref{fig:HBM-posterior} shows the un-normalised posterior
for the model described above and parameters $H$ and $Z_{\odot}$. The
maximum a posteriori value of the vertical scale height is 365\,pc,
above the value of 300\,pc commonly accepted in the
literature \citep[see][and references therein]{2013A&ARv..21...61R}, 
and certainly greater than more recent estimates such as those of 
\cite{2020OJAp....3E...5D}. {We interpret this difference as due to
the discontinuity of the derivative of the exponential distribution at
Z=0 and the limited range of distances of the sample used here. 
If the true density distribution is smooth at that point (i.e. if the likelihood term that includes the exponential decay is not
a good model of the data), then it is to be expected that the
inference model favours values of $H$ that are higher than would be
inferred over larger volumes. The value of $Z_{\odot}$ is} less
constrained by the data and the marginal distribution is multi-modal,
with a maximum at -6 and several local minima at positive
coordinates. Given the sharp peak of the exponential distribution, we
interpret the various maxima as the result of local
over-densities. The negative maximum a posteriori value of $Z_{\odot}$ is surprising
because the values found in the literature range between 5 and 60\,pc,
with most of the recent measurements concentrated between 5 and
30\,pc \cite[see Table 3 of][and references
therein]{2017MNRAS.465..472K}. A direct comparison of the values is
difficult, however, because each measurement defines the Galactic plane
in a different way. In our case, we measured the vertical position of
the Sun with respect to the $z$ coordinate of the local (within
100\,pc of the Sun) maximum volume density. This does not need to coincide
with the Galactic plane defined by the distribution of star counts of
different stellar populations (e.g. Cepheids, Wolf-Rayet stars, or
OB-type), the distribution of clusters, or the distribution
of molecular gas, especially if these distributions are not local but
averaged over much larger fractions of the Galactic disc.

\subsection{Luminosity function}
\label{sec:lf}

The GCNS is an exquisite dataset from which to derive the local luminosity
function. This is possible for the first time using volume-limited
samples with parallaxes not derived from photometric
measurements that are affected by related biases (Eddington or
Malmquist), and
homogeneously throughout the   HR diagram, from bright stars down to white
dwarfs and the substellar regime.
In this section, we present the luminosity function of main-sequence
and giant stars. 

We first removed all objects with a probability higher than 50\% to be a WD as defined in Section\,\ref{sec:wd}. 
The giant branch is well separated from the main sequence in the $M_G$
versus $\gbp-\grp$ diagram. 
Our giant star selection follows the two conditions
$M_G<3.85$ and $\gbp-\grp>0.91$ and gives 1573 stars, which is a
significant sample even given the small volume.  The remaining stars
are considered to belong to the main sequence.  At this stage, we did
not attempt to correct the luminosity function for binarity
effects. We thus defined a subsample of the main sequence keeping
only stars with \texttt{ipd\_frac\_multi\_peak} $=0$
corresponding to 81\% of
the main-sequence stars. As already mentioned, this parameter reflects
the probably of being a visually resolved binary star. This filter
decreases the binarity contribution and at the same time removes some
of the outliers with $G-\grp$ colour excess whose photometry is suspected to be incorrect (see Sect.\,\ref{sec:clusters}).

We determined the luminosity function using the generalised form of the
$V_{max}$ classical technique \citep{1968ApJ...151..393S}.  We computed the maximum
volume probed at a given absolute magnitude, and corrected it to take the decrease in stellar density with increasing
distance above the Galactic
plane \citep{1976ApJ...207..700F,1993ApJ...414..254T} into account,
\begin{equation}
\label{eq:vgen}
V_{max}=\Omega \frac{H^3}{\sin^3|b|}[2-(\xi^2+2\xi+2)\exp(-\xi)]
,\end{equation}
with 
\begin{equation}
\xi=\frac{d_{max}\sin |b|}{H}    
,\end{equation} 
where $H$ is the thin-disc scale height and
$d_{max}$ is the maximum distance of the detection. $b$ and $\Omega$ are the Galactic latitude and the area of the HEALpix to which the star belongs.
We assumed $H=365$\,pc as derived in Sect.\,\ref{sec:verticalstrat}.

Usually, $d_{max}$ was estimated for each object. Thus the object was
counted as the inverse of the maximum volume $V_{max}$ in which it is
observed. The luminosity function is the sum over all objects within
an absolute magnitude bin, 
\begin{equation}
\Phi(M)=\sum\frac{1}{V_{max}}.
\end{equation}
We followed this scheme, but as explained in Sect.\,\ref{sec:volcompleteness},
instead of using a single distance for each star, we considered its whole distance probability distribution, adding a
contribution of 1/99th of the sum of those probabilities within 100\,pc. The use of the Bayesian framework allowed us to avoid the
Lutz–Kelker bias for a volume-limited
sample \citep{1973PASP...85..573L}.

Finally, we took the 80th percentile $G$ magnitude sky distribution 
at HEALpix level\,5 (corresponding to an
angular resolution of three square degrees per HEALpix) to apply
sensitivity cuts and reach the highest completeness limit depending on
the sky position. These limits have a minimum, mean, and maximum $G$
limit of~18.7,~20.2, and~20.7 respectively.

\begin{figure*}
\sidecaption
\includegraphics[width=12cm]{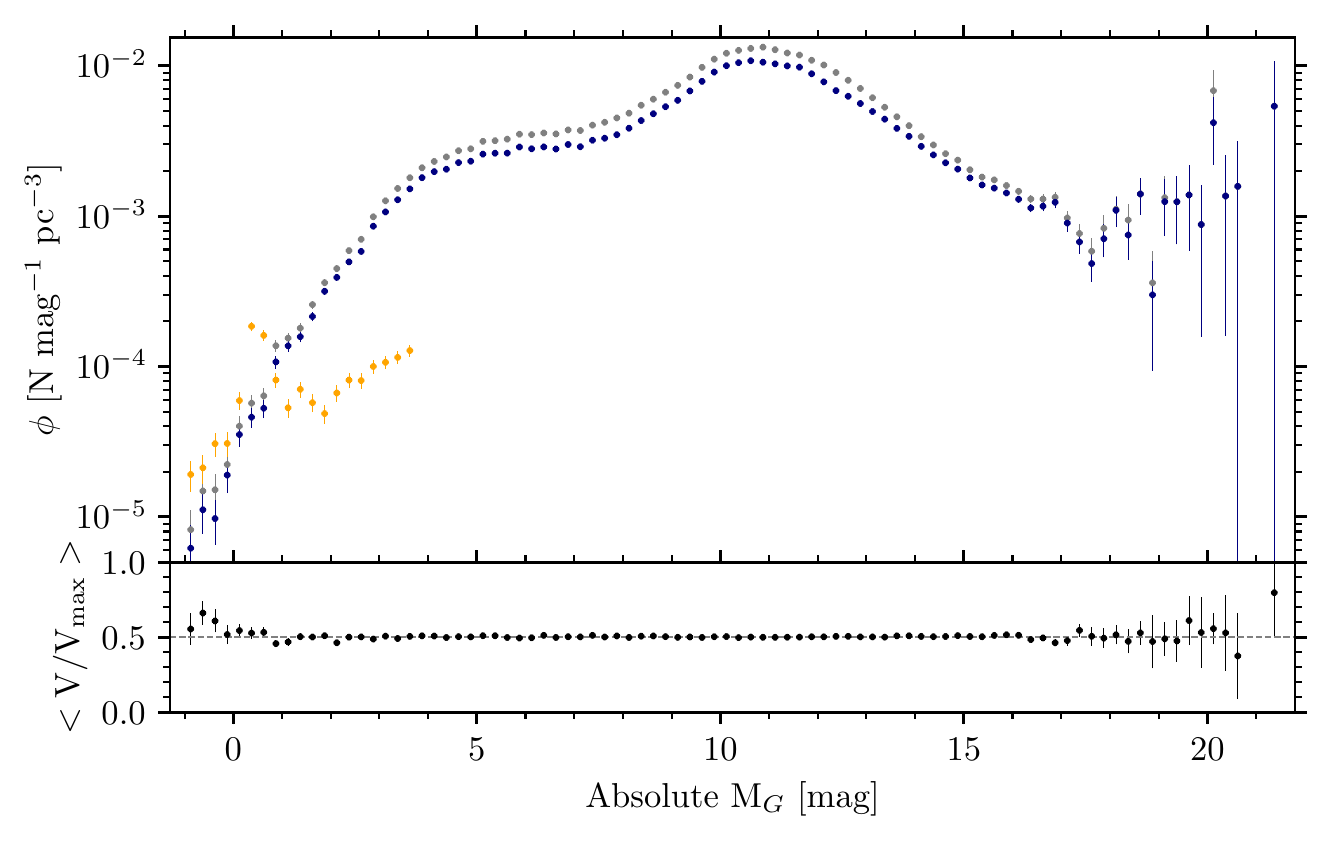}
\caption{Upper panel: Luminosity function of the GCNS, with a 0.25 bin, in log scale. The upper full curve plotted in grey shows main-sequence stars. The lower full curve plotted in blue points represents main-sequence stars with \texttt{ipd\_frac\_multi\_peak} $=0$, that is, probably single stars. The small lower partially orange curve shows giants stars. The confidence intervals reflect the Poisson uncertainties. Lower panel: $<$$V/V_{max}$$>$ vs. $M_G$. The expectation
value for the statistic is 0.5 for a uniform sample within the survey volume. \label{fig:LF_ms}} 
\end{figure*}

We also computed the mean value of $V/V_{max}$ in the bins of absolute magnitude, where $V$ is the (generalised) volume in which each object is discovered, that is,~the volume within the distance~$d$ to each object. This statistic $<$$V/V_{max}$$>$ should approach~0.5 for
a uniformly distributed sample with equal counts in each volume.
As shown in the lower panel of
Fig.\,\ref{fig:LF_ms},  this
is the case of our sample from $M_G=2$ to $\sim20.5$.\\

The luminosity function is shown in the upper panel of Fig.\,\ref{fig:LF_ms}.  The luminosity function of the giant sample is shown in
red. The red clump is clearly visible by the peak at $M_G=0.4$ with
$\Phi=1.9\pm0.1\times10 ^{-4}$ stars pc$^{-3}$ mag$^{-1}$. However, this
is underestimated here because objects brighter than
$G\simeq3$ are not included in \G.  The local
luminosity function of fainter giants, on the red giant branch, is
reliable, however. We compared our result with those obtained
by \cite{2015MNRAS.451..149J} using a sample of 2660 giants from \hip\
and the CNS up to 200\,pc. They found a value of $\Phi=8.3\times10 ^{-5}$
stars pc$^{-3}$ mag$^{-1}$ at $M_K=1$, which roughly corresponds to
$M_G=3,$ where we find a consistent $11.0\pm1.1\times10^{-5}$ stars
pc$^{-3}$ mag$^{-1}$.
 
 
The luminosity function of the main-sequence sample illustrates the
very high precision offered by the unprecedented quality of the
GCNS. The confidence intervals reflecting Poisson uncertainties are very small
even at the low-mass end down to $M_G\simeq18$\,mag, corresponding to L3-L4 
based on the spectral type versus $M_G$ relation derived for SIMBAD entries 
as described in 
Sect.\,\ref{sec:compprevcomp}. The overall density is
$0.081\pm0.003$ stars pc$^{-3}$. 

This can be compared with the previous efforts made to determine the
luminosity function within 25 pc based on \hip\,
CNS \citep[e.g. ][]{2015MNRAS.451..149J}, and ground-based
observations \citep[e.g. ][and references
therein]{2002AJ....124.2721R}.  By using a combination of \hip\
and astrometric and spectroscopic observations, \cite{2002AJ....124.2721R}
were able to derive the solar neighbourhood (25\,pc) luminosity
function from bright to low-mass stars, including 
the contribution from companions. There is an overall agreement with
our determination, in particular within their confidence intervals, that can be
20 times larger than in the GCNS luminosity function (see their Fig.
8). One main difference is the double-peaked shape in their
luminosity function, with one maximum at $M_V=12.5$\,mag (corresponding to
our maximum at $M_G=10.5$\,mag) and a higher one at $M_V=15.5$\,mag that should
stand at $M_G=14-14.5$\,mag and does not appear in the GCNS luminosity
function. This second peak is poorly defined: it has a
large confidence interval, for instance, and
does not appear in the 8\,pc luminosity function determined
by \cite{2003AJ....125..354R}.

\begin{figure*}[!htb]
\begin{center}
\includegraphics[height=6.4cm, width=0.45\textwidth]{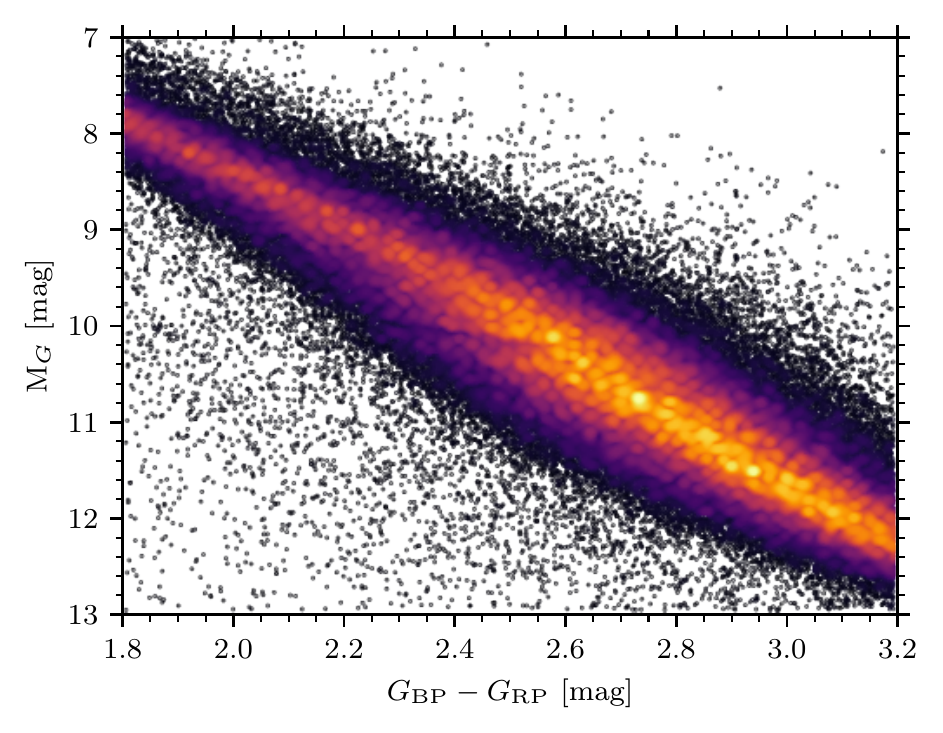}
\includegraphics[height=6.4cm, width=0.45\textwidth]{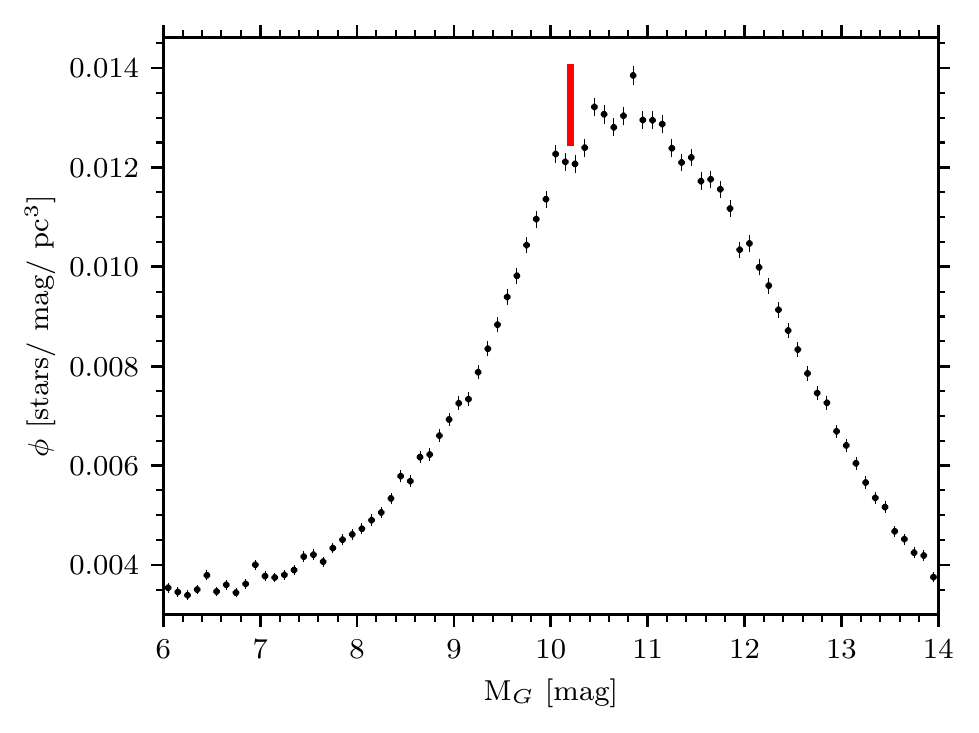}

\caption{Left panel: So-called Jao gap in the $M_G$ vs. $\grp-\gbp$ diagram. Right panel: Zoom of the luminosity function from Fig.\,\ref{fig:LF_ms} computed in 0.1 magnitude bins, in linear scale. The red line indicates the position of the Jao gap. 
\label{fig:LF_jaogap_hrd}}
\end{center}
\end{figure*}

The high precision of the luminosity function enables searching
for signatures of structures in the CAMD, such as the Jao gap.
Using \gdrtwo, \cite{2018ApJ...861L..11J} discovered this narrow gap
($\sim$0.05\,mag) in the lower main sequence, which is hypothesised to
be the result of a dip in the luminosity function associated with
complex evolutionary features of stars with mass
$\sim$0.35\,$M_\odot$ \citep{2018MNRAS.480.1711M,2018A&A...619A.177B}.
Therefore, we first inspected the lower main sequence of the complete
GCNS catalogue to verify the presence of this feature, and find that
the gap stands out distinctly, as depicted in the left panel of
Fig.\,\ref{fig:LF_jaogap_hrd}.
By breaking down the GCNS sample into $\gbp-\grp$ colour and magnitude
bins according to \cite{2018ApJ...861L..11J}, their Table 1, we also
confirm the largest decrement of counts around $M_G=10.14$, or
$M_{RP}=9.04$. The effects of this gap are reflected in the
luminosity function, as shown in the right panel of Fig.\,\ref{fig:LF_jaogap_hrd} by the red line.  The main sequence also shows an inflection  close to the gap that is very likely the effect noted by \cite{1998ApJ...496..352C}. Other structure is apparent in the luminosity
function that may be connected to the main-sequence structure found
in \cite{2020AJ....160..102J} as well as the more classical variations \citep{1983nssl.conf..163W, 1990MNRAS.244...76K}, but this is
beyond the scope of this contribution.




Recent works have been made to derive the luminosity function at the
stellar to substellar boundary \citep{2019ApJ...883..205B} and for L to
Y brown dwarfs \citep{2019ApJS..240...19K}.  Although the statistical
noise increases in the brown dwarf regime, the luminosity function can
be derived down to $M_G=20.5$ (translating into $\sim$L9 spectral
type). Several features can still be seen: a dip at $M_G=16.4$\,mag or
L0; a dip at $M_G=18.9$\,mag or L5; a peak at $M_G=20.1$\,mag or L8, but
with a large confidence interval. A better investigation of the possible
contamination by red objects with potentially inaccurate photometry,
in particular in the late-L regime, should be made before any
conclusions are drawn as to the reality of these features.

%
%

However, the clear dip at $M_G=17.6$ in
Fig.~\ref{fig:LF_ms}, which is also seen at $M_{RP}=16.11$ in the $M_{RP}$
luminosity function, corresponding to the L3 spectral type, probably is a real
feature. This can be seen in Fig.\,26
from \cite{2019ApJ...883..205B}, for instance, where a plateau appears at
$M_J\simeq13$ to 14 (corresponding to $M_G\simeq16$ to 18, M9 to L4),
followed by an increase in luminosity function. This absolute
magnitude region lies at the edge of other studies
(\cite{2007AJ....133..439C,2019ApJ...883..205B} for M7 to L5
and \citep{2010A&A...522A.112R} for L5 to T0) using different
samples. In contrast, the GCNS offers a homogeneous sample that
gives confidence to the physical significance of that dip.  This minimum probably is a signature of the stellar to substellar boundary because
brown dwarfs are rapidly cooling down with time. Models predict that they pass through several spectral types within 1 Gyr \citep[see
e.g.][]{2015A&A...577A..42B}. Thus they depopulate earlier spectral
types to go to later ones.  The homogeneous dataset offered by GCNS
will allow us to refine the locus of this boundary.

\subsection{Kinematics}
\label{sec:Kinematics}

We explored the kinematics of the GCNS catalogue by restricting the sample
to 74\,281 stars with a valid radial velocity in \gdrthree. We used the
\texttt{vel$\_$50} Cartesian velocities $(U,V,W)$ as determined in
Sect.\,\ref{sec:cat_overview}.

\begin{figure}
\center{\includegraphics[width=0.45\textwidth]{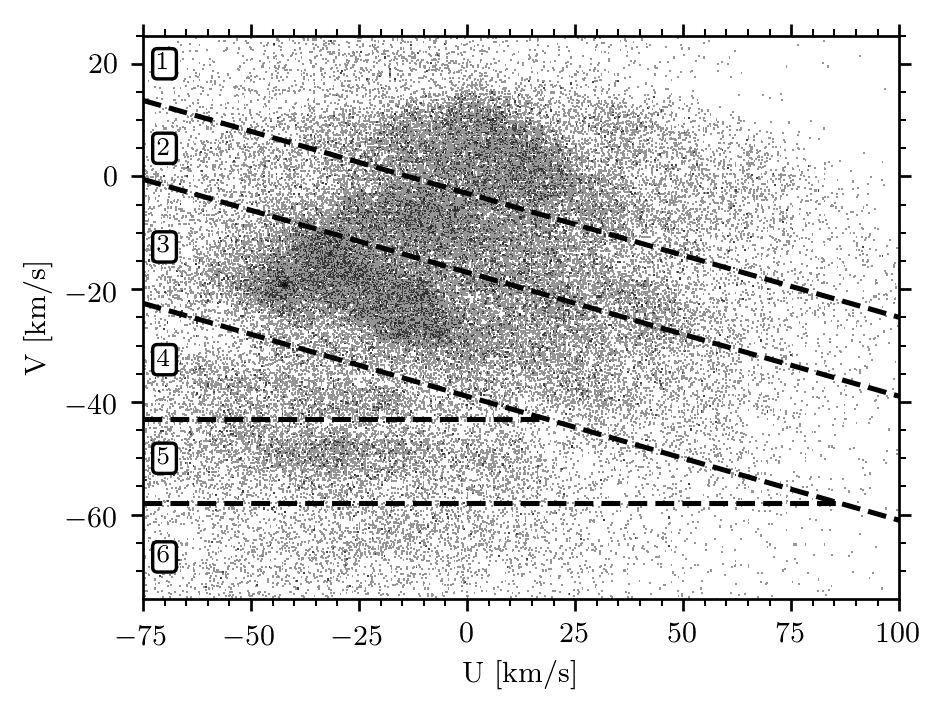}
\label{fig:kine-uv-plane}
\caption{GCNS stars in the $(U,V)$ plane. We identify substructures, labelled strip 1 to 6, from top to bottom, separated by indicative straight lines: $V=0.22*U-3$, $V=0.22*U-17$, $V=0.22*U-39$, $V=-43$, and $V=-58$.
}}
\end{figure}

\subsubsection{Structures in the \texorpdfstring{$(U,V)$}{(U,V)} plane}

The sample in the $(U,V)$ plane shows several substructures, as already pointed out in early studies by Eggen, who
identified numerous groups or
superclusters \citep{1958Obs....78...21E,1971PASP...83..251E}, then
from Hipparcos data in \cite{1999MNRAS.308..731S}
and \cite{1997ESASP.402..545C}. See \cite{2010LNEA....4...13B} for a
detailed historical review. These substructures were confirmed
in \gdrtwo
\citep{2018A&A...616A..11G}.

In this local sample, the $(U,V)$ plane also appears to be highly
structured, as shown in fig~\ref{fig:kine-uv-plane}, where we show the approximate structuralisation of the
velocity space by straight line divisions. The three top strips have been largely studied from
Hipparcos data, particularly by \cite{1999MNRAS.308..731S} from
wavelet transform, who labelled them as 1) the Sirius branch at the
top (where the Sirius supercluster identified by Eggen is located), 2)
the middle branch, which is less populated, and 3) the Pleiades branch, which is most
populated, where the Hyades and the Pleiades groups are located.
A significant gap lies just below this strip 3; it has been presented in previous studies and is nicely visible in \gdrtwo\
data. The strips below the gap are nearly parallel to the $U$
axis. Strips 4 and 5, seen at $V\approx -35$\kms\ and $\approx -45$\kms\ , are most probably associated with
the Hercules stream that was identified by \cite{1958Obs....78...21E}, where the high-velocity star $\zeta$
Hercules is located. The
Hercules stream was identified at $V \approx -50$\kms\
by \cite{1999MNRAS.308..731S}, but appears itself to be
substructured \citep{1998AJ....115.2384D} when the sample is
sufficiently populated and has 
accurate velocities. The \cite{2018A&A...616A..11G} suggested that the strip at
$V \approx -70$\kms\ (strip 6) might also be linked to the 
Hercules stream.

The strips are not related to cluster disruptions, as pointed out by
different
studies \citep[e.g.][]{1998AJ....115.2384D,2010LNEA....4...13B}, they
cover wide age ranges \citep{2008A&A...483..453F}. Some studies argued that
they are due to resonances from either the bar, the spiral arms, or
both.  As a verification test, we investigated whether there is
evidence of an age difference between different strips by examining
their turn-off colour.

\begin{figure}
\center{
 \includegraphics[width=0.4\textwidth]{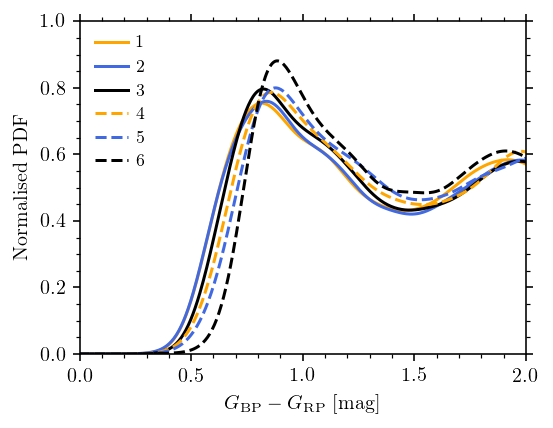}
\caption{KDE distribution of $\gbp-\grp$ colour for different strips of the
$(U,V)$ plane of fig~\ref{fig:kine-uv-plane}. Turn-off colours are
similar for the three strips above the gap (strips 1 to 3), but older
and older in strips below it (strips 4 to 6).}
\label{fig:bprp-KDE-strips}}
\end{figure}




We show in Fig.~\ref{fig:bprp-KDE-strips} the KDE distribution of
colours for the different strips. There is a clear colour shift of the
turn-off of the different strips for those that are below the main
gap (strips 4, 5 and 6 in fig~\ref{fig:kine-uv-plane}), indicating that they
are increasingly older when the asymmetric drift is stronger and extends farther, as
expected from secular evolution. However, this global trend is
superimposed on a structure that is probably due to resonances that
several studies attributed to the outer Lindblad resonance of the
bar \citep{2012MNRAS.426L...1A,
2014A&A...563A..60A,2017MNRAS.466L.113M} , while others linked it to the spiral
structure \citep{ 2018MNRAS.481.3794H,2018ApJ...863L..37M}. There is no indication of an age
dependence in our study for the three strips above the gap. They all appear to have the same turn-off
colour. Therefore the structure of the velocities can be
due solely to dynamical effects, such as resonances of the bar and/or the
spiral \citep{2009ApJ...700L..78A}.

 \begin{figure*}
 \begin{center}
\includegraphics[height=6cm]{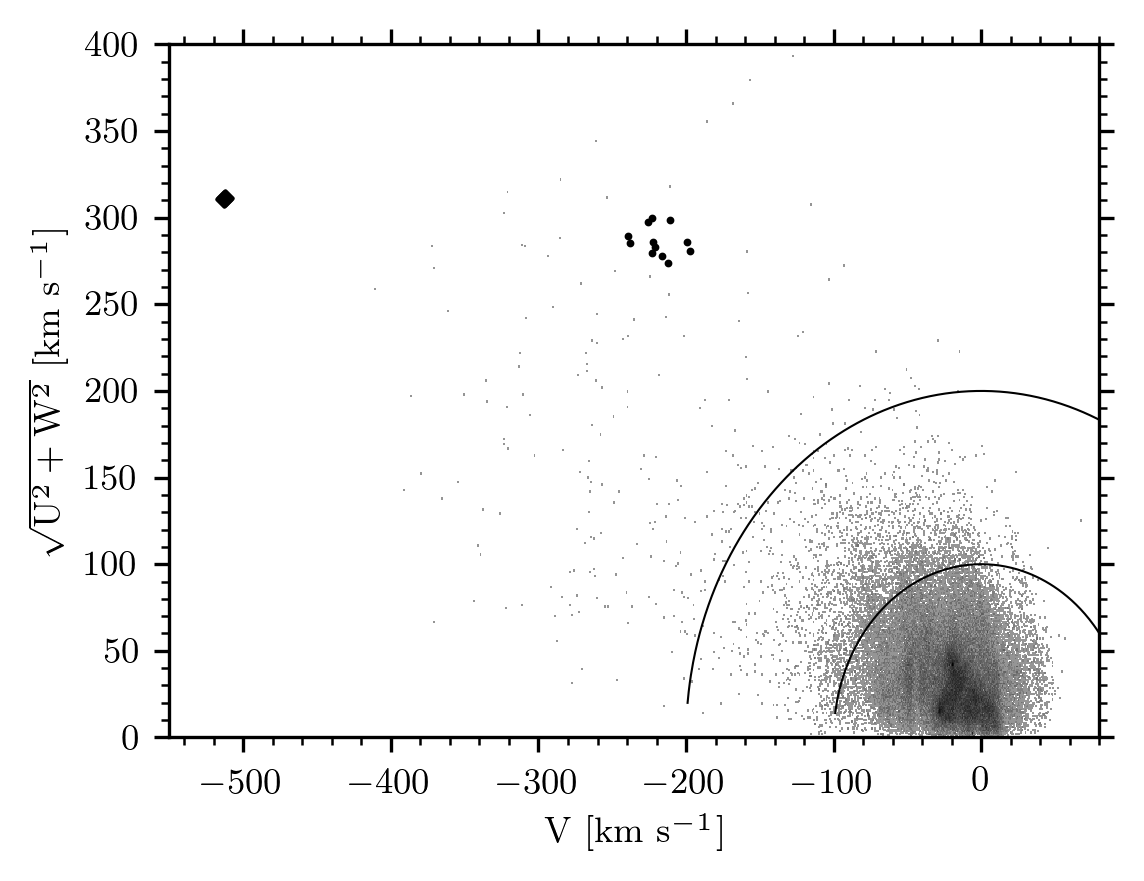}
\includegraphics[height=6cm]{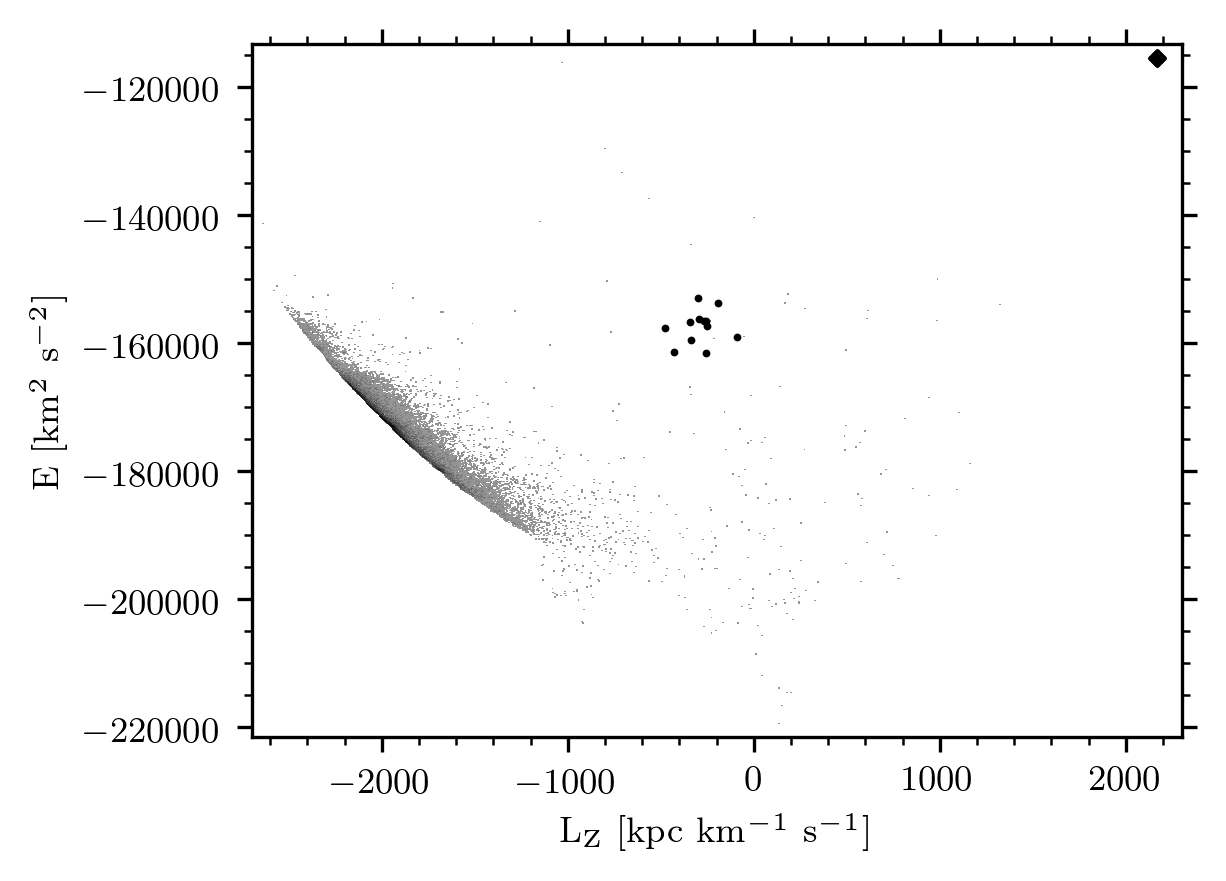}
\caption{Left panel: Toomre diagram for all the GCNS entries. The diamond symbols are the binary HD\,134439/HD\,134440, the circles are the  \G\--Enceladus group members. Right panel: Energy vs. angular momentum for the GCNS. The symbols are the same same as the left panel.}\label{fig:toomre}
\end{center}
\end{figure*}




\begin{figure}
    \centering
    \includegraphics[width=0.5\textwidth]{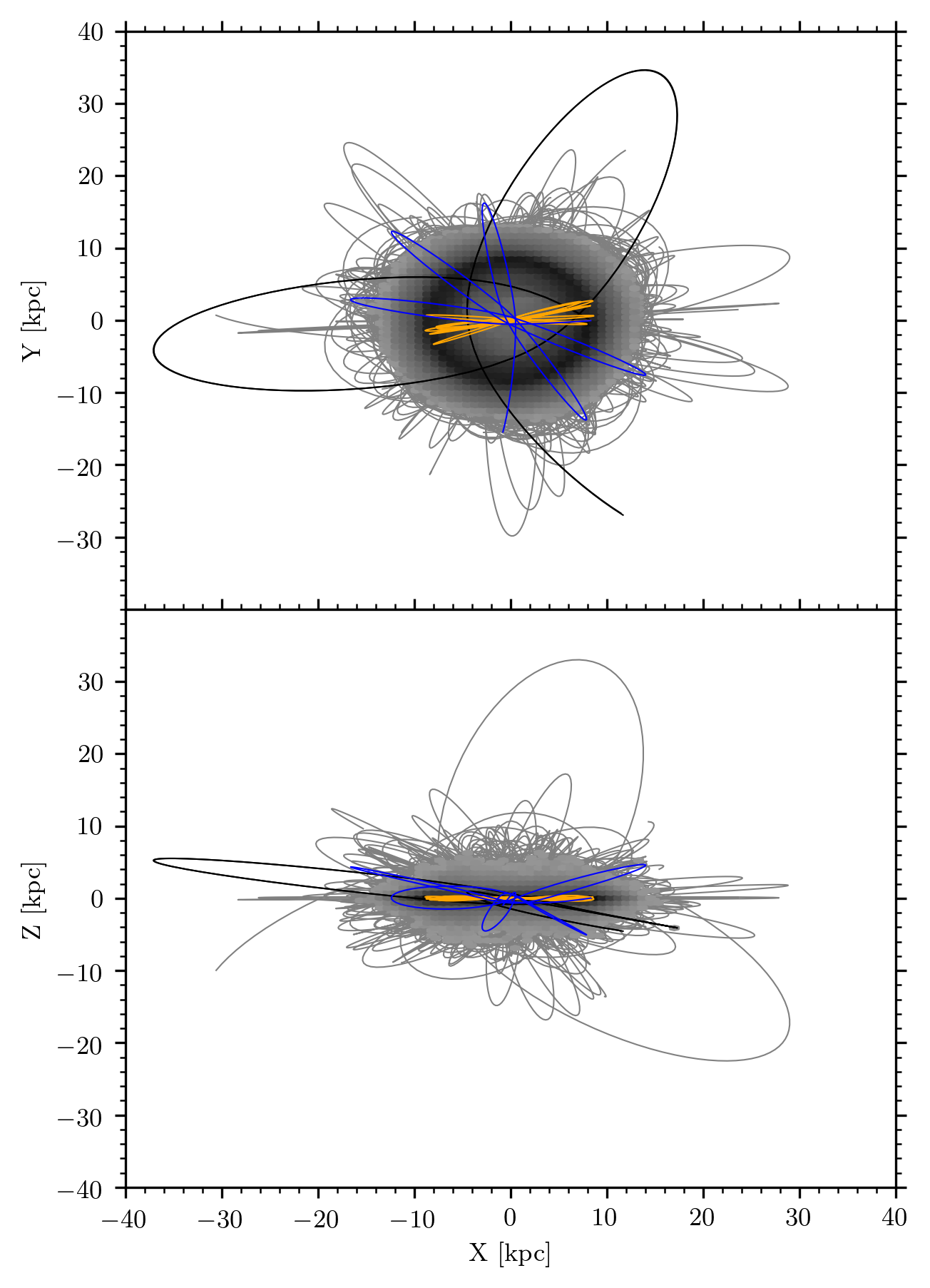}
    \caption{Orbits of the GCNS sample. The orbits are
computed over 1 Gyr and plotted in the referential system of the
Galaxy. Orbits are highlighted for a few stars: the halo pair HD\,134439
and HD\,134440 (black), a star from the Enceladus group (blue), and a star with
minimum energy (orange) coming from the central regions of the
Galaxy.\label{fig:orbit}}
\end{figure}

\subsubsection{Stellar populations and orbits} 
\label{sec:stellar-pops-orbits}
The left panel in Fig.~\ref{fig:toomre}
shows the Toomre diagram of the sample. The circles with
 100 and 200\kms\ radii delineate thin-disc, thick-disc, and halo
 stars. Using these limits, we estimate that 95\% of the stars belongs
 to the thin disc, 6.6\% to the thick disc, and 0.4\% to the
 halo. However, we show in what follows that the GCNS also contains
 tens of stars that visit from the central part of the Milky Way.
 
 As discussed before, the disc kinematics is not smooth, even in the
 100 pc sphere. The Toomre diagram  shows many structures, but
 not only in the disc. The nearby halo is clumpy as well.

 
The diamond, with the highest retrograde velocity, shows the twin
pair \object{HD\,134439} and \object{HD\,134440}. They are known to be chemically
anomalous stars. Their chemical compositions are close to those
observed in dwarf galaxies such as Draco and Fornax, indicating an
extragalactic origin \citep{2018MNRAS.475.3502R}, and are consistent
with the kinematics study of \cite{1996AJ....112..668C}.

The orbital parameters were computed using the online tool
Gravpot16\footnote{\url{https://gravpot.utinam.cnrs.fr/}}. The
Galactic potential we used is a non-axisymmetric potential including
the bar, developed by
\cite{phdthesisFT}\footnote{\url{http://theses.fr/s108979}} and used
in \cite{2018A&A...616A..12G} to derive orbital parameters of globular
clusters and dwarf galaxies from \gdrtwo\ data. We assumed a bar mass
of $10^{10}$ M$_\odot$, with a pattern speed of
43\,km\,s$^{-1}$\,kpc$^{-1}$ and a bar angle of 20$\degree$.

The orbits integrated forward over 1 Gyr are shown in
Fig.~\ref{fig:orbit} in the $(X,Y,Z)$ referential system of the
Galaxy. The most numerous disc stars populate the circular orbits in
the Galactic plane $(Z=0)$. Halo stars have higher eccentricities and
inclinations. 
The central part of the $(X,Y)$ plane is populated by the
orbits of stars coming from the central regions of the Galaxy.

The 12 circled dots in Fig.~\ref{fig:toomre} are the stars that we
identified as related to \object{\textit{Gaia}--Enceladus} (they are out of the panel of
Fig.~\ref{fig:kine-uv-plane}, but their velocities are centred on
$U=267 \pm 10$\,km s$^{-1}$ and $V=-221\pm 11$\,km s$^{-1}$.)  We
selected them based on their orbital parameters, in particular, in the
total energy versus angular momentum (Fig.~\ref{fig:toomre}, right panel),
where they are concentrated at $E=-156 000$ km$^2$s$^{-2}$ with a
dispersion of 2660 km$^2$s$^{-2}$ and $L_z=-273$\,kpc\,km s$^{-1}$
with a dispersion of 94\,kpc\,km s$^{-1}$. These values can be
compared with the last large merger event experienced by the Milky Way
discovered by \cite{2018Natur.563...85H}, who identified the so-called
\G\--Enceladus substructure with a selection of $-1
500$\,kpc\,km\,s$^{-1} < $Lz $< $150\,kpc\,km s$^{-1}$ and $E > -180
000$\,km$^2$ s$^{-2}$. In our local sample, the number of stars that
can be attributed to Enceladus is much smaller than in the discovery
paper. However, thanks to the exquisite accuracy of the parallaxes and
proper motions, the structure is concentrated in orbital
elements as well as velocity space. The pericentres are close
to the Galactic centre ($<2$\,kpc) and apocentres between 16 and
20\,kpc (see Fig.~\ref{fig:orbit}, blue orbit).

The solar neighbourhood is also visited by stars from the central
region of the Galaxy. The apocentres of about 40 stars in our sample
lie close to the Sun, and the pericentre distance of the stars is
smaller than 1\,kpc (see Fig.~\ref{fig:orbit}, orange orbit). They
have high eccentricities, a minimum energy $E \approx -200
000$\,km$^2$ s$^{-2}$ , and small angular momentum
$|L_z|$<$400$\,kpc\,km s$^{-1}$), some are slightly retrograde.

\subsubsection{Solar motion}
\begin{figure*}
\center{
\center \includegraphics[width=0.9\textwidth]{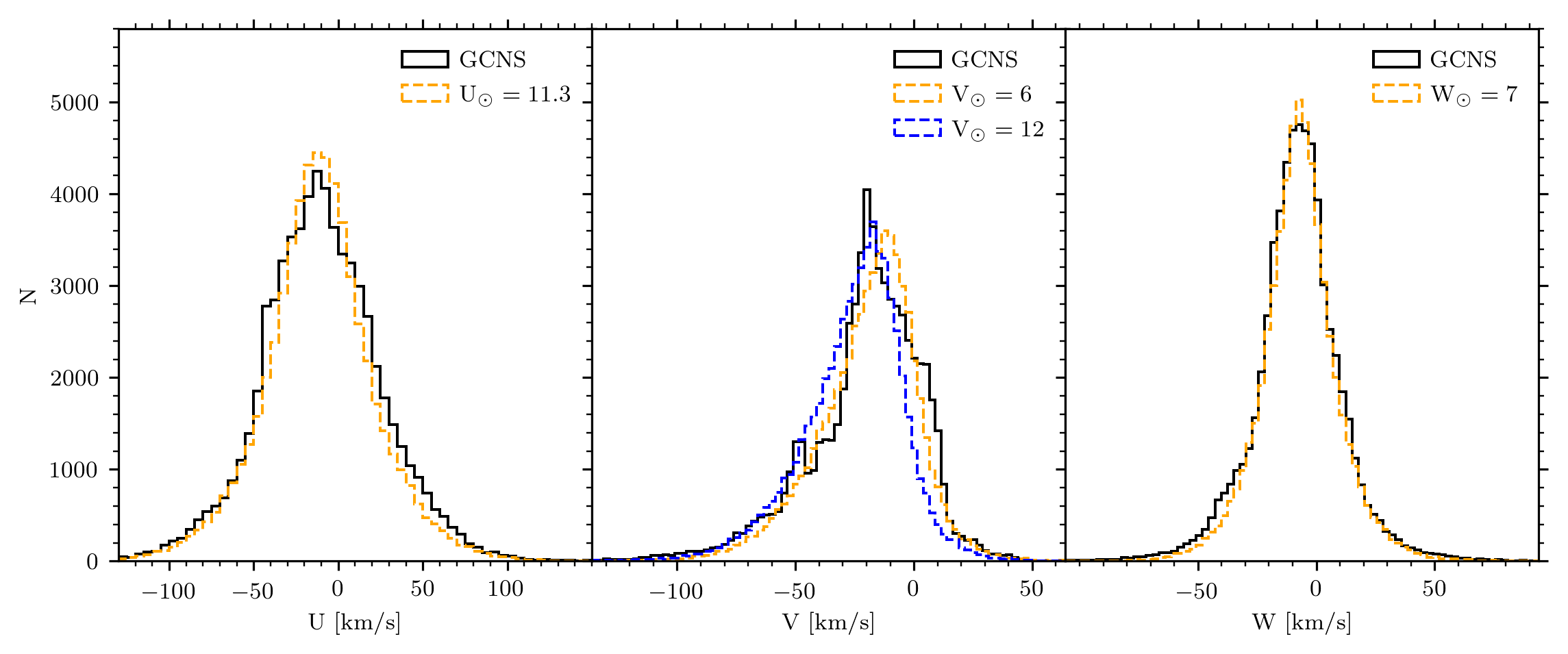}
\caption{Histograms of $U$, $V,$ and $W$ velocity in GCNS catalogue with $G<13$\,mag
(black lines) compared with simulations with distance $<$ 100\,pc and
$G<13$ (dashed lines). The simulations assume solar velocities:
$U$=11.3, $V$=6\kms\ (dashed orange line), and 12\kms\ (dashed blue line),
$W$=7\kms. The cluster members have not been removed from the data.}
\label{fig:hist_UVW}}
\end{figure*}

\begin{figure}
\center{
\includegraphics[width=0.5\textwidth]{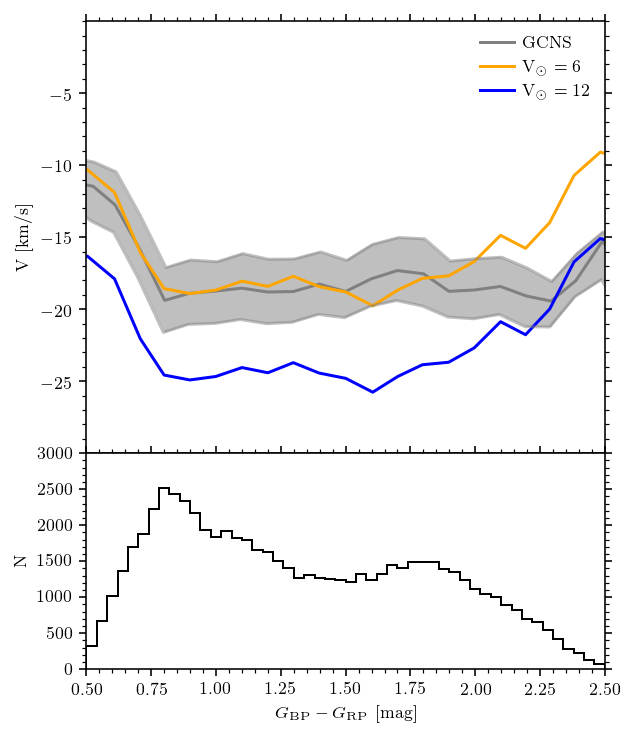}
\caption{Upper panel: Median velocity as a function of $\gbp-\grp$ colour for the GCNS sample with $G<13$. The data with quantiles 0.45 and 0.55 are plotted in grey and the median is shown in black. The simulation was made with \Vsun\, of 6\kms\ (orange) and 12\kms\ (blue). Lower panel: Histogram of the colour distribution in \gdrthree.} 
\label{fig:V_vs_BPRP}}
\end{figure}
The GCNS is well suited for measuring the solar motion relative to the
local standard of rest (LSR). \USun\, and \WSun\, velocities are easy
to measure, while \VSun\, is subject to controversies, with values
varying from 1 to 21\kms
\citep[e.g.][]{1998MNRAS.298..387D,2010MNRAS.403.1829S,2012ApJ...759..131B,2017A&A...605A...1R}.
The traditional way of computing the $V$ solar motion is to
extrapolate the distribution of $V$ as a function of $U^2$ to $U$=0
for a given sample. It is expected that at $U$=0 the mean $V$
corresponds to the solar velocity because as a function of age the
rotation of stars experiences a lag induced by the secular evolution
(stars become more eccentric and the motion is less circular). The
youngest stars have a lowest U velocity also because of secular
evolution. Depending on the mean age of the sample, on its location
(close to or farther away from the solar neighbourhood), the
literature values of \VSun\ have been disputed. In this sense, the
younger the stars are in the sample, the better the sample is for
measuring \Vsun.  However, the youngest stars are also those that
experience clumping because their kinematics are far from relaxed, so
they do not represent the LSR well in a general manner. Therefore we
considered the whole GCNS sample as representative of the LSR, and
compared the median $V$ velocity of the sample with a simulation of
the Besan\c{c}on Galaxy model. The simulation is the same as we used
for unresolved stellar multiplicity (see
Sect.\,\ref{sect:unresolved}).  The kinematics was computed from a
self-consistent dynamical solution using an approximate Staeckel
potential \citep{2015A&A...581A.123B,2018A&A...620A.103B}, while the
kinematic parameters of the disc populations (mainly the age-velocity
dispersion relation) were fitted to \G\ DR1 and RAVE data as described
in \cite{2017A&A...605A...1R}. We used of the latest determination of
the rotation curve from \cite{2019ApJ...871..120E} based on
\gdrtwo. In contrast to the \G\ object generator (GOG;
\citealt{2014A&A...566A.119L}), we added observational noise from a
simplified model of the parallax and photometric uncertainties to
these simulations using the equations given on the ESA web
site \footnote{\url{https://www.cosmos.esa.int/web/gaia/science-performance}}
as a function of magnitude and position on the sky.

To study the solar motion, we considered the GCNS sample, selecting only
stars having $G<13$ as a clear cut for stars with good radial velocities that
can then be applied simply to the simulation.
We first plot
histograms of the distribution in $U$, $V,$ and $W$ for the sample in
Fig. ~\ref{fig:hist_UVW}. We over-plot the simulation
where the solar velocities are assumed to be (11.3 ,6 ,7)\kms. While
the $U$ and $W$ velocities are well represented by the simulation, this
is not the case for the $V$ velocity distribution, which shows
significant non-Gaussianity. The clusters were not
removed from the observed sample.
For comparison we also over-plot a simulation assuming
alternative \Vsun\, of 12\kms \citep{2010MNRAS.403.1829S}.

The non-Gaussianity of the $V$ distribution was already known and is
partly due to secular evolution and asymmetric drift, as expected even
in an axisymmetric galaxy, and partly due to substructures in the
($U,V$) plane that are probably associated with resonances due to the
bar and the spiral arms. It is beyond the scope of this paper to
analyse and interpret the detailed features. However, we explored the $V$ velocity distribution slightly more and plot the median $V$ as a
function of $\gbp-\grp$. The blue stars are expected to be younger in
the mean, while redder stars cover all disc
ages. Fig~\ref{fig:V_vs_BPRP} shows that at $\gbp-\grp>0.8,$ the median
$V$ velocity is constant at about $V=-20$\kms\ , while at
$\gbp-\grp<0.7$ there is a shift of the median $V$. We over-plot a
simulation with \Vsun\,=6\kms\ and 12\kms. The data agree
well with the simulation when a solar \Vsun\ of 6\kms\ is assumed, especially for 
$\gbp - \grp < 1.5$\,mag, which dominate the sample. For redder stars the mean \Vsun\ varies more with colour, it is therefore less secure to define the mean solar motion
in the region. However, the median $V$ velocity even for a local sample is a
complex mix of substructures with different mean motions and of
the expected asymmetric drift. With these solar velocities, the solar
apex is towards l=31.5\degree, b=27.2\degree.

We also considered the vertex deviation concept defined as the apex of
the velocity ellipsoid when it is slightly rotated and does not point
towards the Galactic centre, as would be expected in an axisymmetric
disc. It has long been seen that young populations experience a vertex
deviation, at least locally, which has been interpreted as an effect
of the spiral perturbation, for instance, some theoretical analysis can be found  
in \cite{1970A&A.....6...60M} and \cite{1973A&A....27..281C}. The
distribution in the $(U,V)$ plane clearly shows that strip 1 shows an
inclination of the ellipsoid in the mean. However, this might also be
due to the substructures in the ($U,V$) plane rather than a true
deviation of the vertex because this strip appears to be made of the
superposition of at least three superclusters or groups.


\subsection{Stellar to substellar boundary}
\label{sec:ucds}

\begin{figure*}[!htb]
    \centering

     \includegraphics[width=0.96\textwidth]{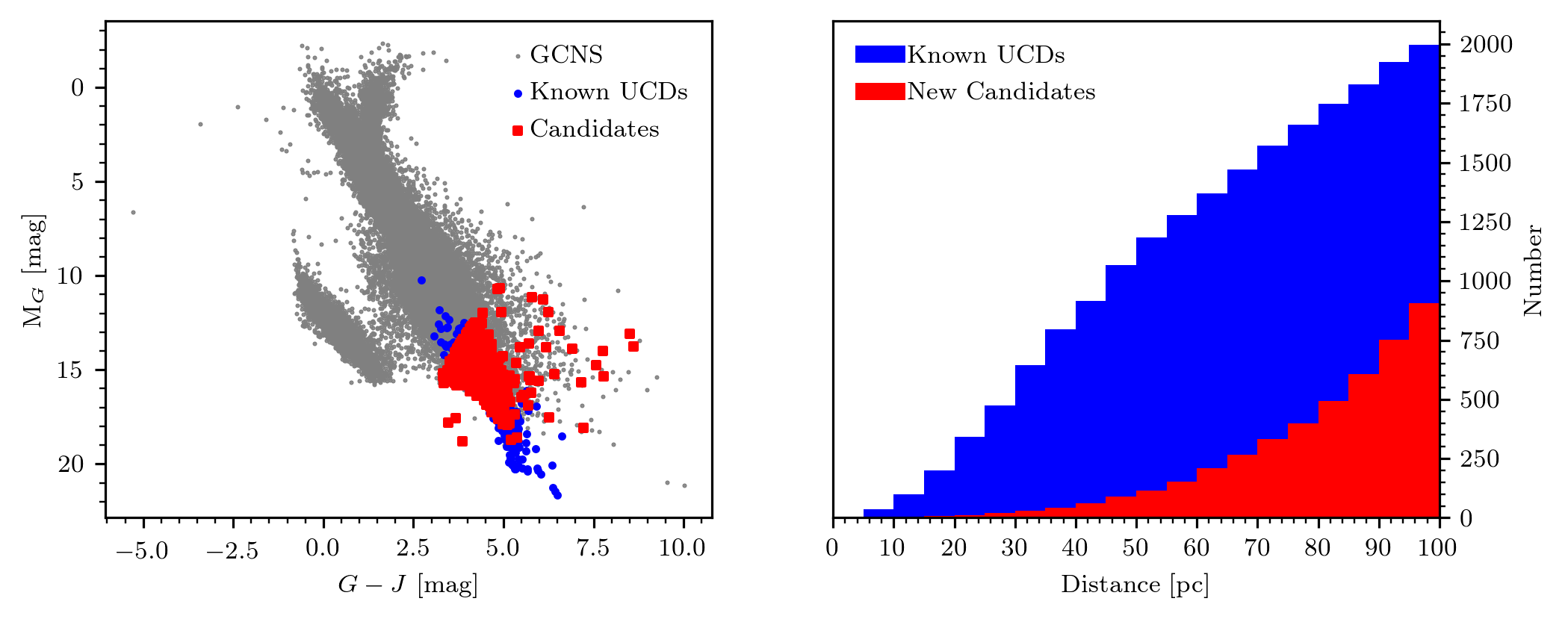}
    \caption{    \label{fig:candidates_ucds} Left: $M_G$ vs. $G-J$ diagram of
    stars in GCNS that are not found in \gdrtwo. The red dots are new UCD
    candidates, the blue points are known UCDs (spectral types between M7 and T8), and the grey points are
    the full GCNS sample. The new candidates are selected following
    the condition $M_G>-3\times(G-J)+25$, after removing stars whose
    probability of being a WD is higher than 20\%. Right: Distance distribution of the new candidates in the GCNS (red) and the
    known UCDs (blue).}
\end{figure*}

The nearby sample is particularly important for the ultra-cool dwarfs
(UCDs), which are the lowest-mass, coldest, and faintest products of
star formation, making them difficult to detect at large distances. They
were defined by \cite{1997AJ....113.1421K} as objects with spectral
types M7 and later, through L, T, and Y types, have masses $M < 0.1
M_\odot$, and effective temperatures $< 2700 K$. UCDs are of
particular interest because they include both very low-mass stars that
slowly fuse hydrogen, and brown dwarfs, which have insufficient mass
(below about 0.075 $M_\odot$) to sustain hydrogen fusion in their
cores, and slowly cool down with time.

The full sky coverage and high-precision
observations of \G\ offer the means of uncovering nearby UCDs through
astrometric rather than purely photometric
selection \citep{2018A&A...619L...8R,2019MNRAS.485.4423S,2020A&A...637A..45S}. Gaia provides a large homogeneous sample. The capability of \G\
to study the stellar to substellar boundary is illustrated in
Sect.\,\ref{sec:lf}, where the luminosity
function can be computed for the first time with one unique dataset throughout the main
sequence down to the brown dwarf regime. It nicely shows a dip in the
space density at spectral type L3, defining the locus of the
stellar to substellar boundary.\\

\subsubsection{New UCD candidates in \gdrthree}
As mentioned in Sect.\,\ref{sec:compprevcomp}, GCNS contains thousands
of faint stars (WDs and low-mass stars) that have no parallax
from \gdrtwo. We investigate the potential new UCD
candidates in GCNS in more detail. Following the selection procedure
from \cite{2018A&A...619L...8R}, we selected UCD candidates from the
$M_G$ versus $G-J$ diagram (Fig.\,\ref{fig:candidates_ucds}, left
panel). GCNS contains 2879 additional candidates compared to \gdrtwo, 1016 of which have a median distance inside 100\,pc. This is a valuable
contribution to complete the solar neighbourhood census in the region
of the stellar to substellar boundary, as shown in the right panel of Fig.\,\ref{fig:candidates_ucds}.

\begin{figure}[!htb]
    \center{ \includegraphics[width=0.49\textwidth]{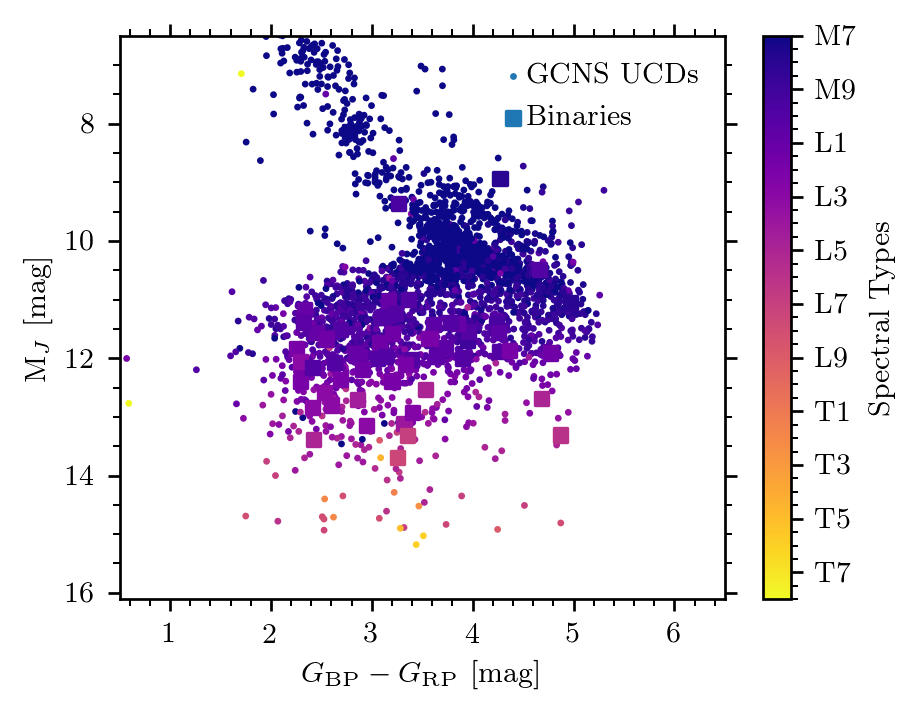}} \caption{\label{fig:ucds_bprp}
    CAMD of \gbp\ - \grp\ [mag] against
    $\mathrm{M_{J}}$ [mag]. The full sample is from the GUCDS, and known
    binaries are over plotted as squares. Points are coloured by their
    published spectral types.}
\end{figure}

In Fig.\,\ref{fig:ucds_bprp} we examine \gbp\ - \grp\ versus $M_J$ for 
known UCDs taken from the \G\ Ultra-cool Dwarf Sample
(\citeauthor[GUCDS,][]{2017MNRAS.469..401S}, \citeyear{2017MNRAS.469..401S}, \citeyear{2019MNRAS.485.4423S}). The non-monotonic
decrease of $M_J$ with \gbp\ - \grp\ indicates that \gbp\ is unreliable in the
UCD regime, in agreement with the conclusions
in \cite{2019MNRAS.485.4423S}. For a full discussion and explanation of
the limits on \gbp, see \cite{EDR3-DPACP-117}.

\subsubsection{GCNS completeness in the UCD regime}

\begin{figure}[!htb]
    \center{
    \includegraphics[width=\columnwidth]{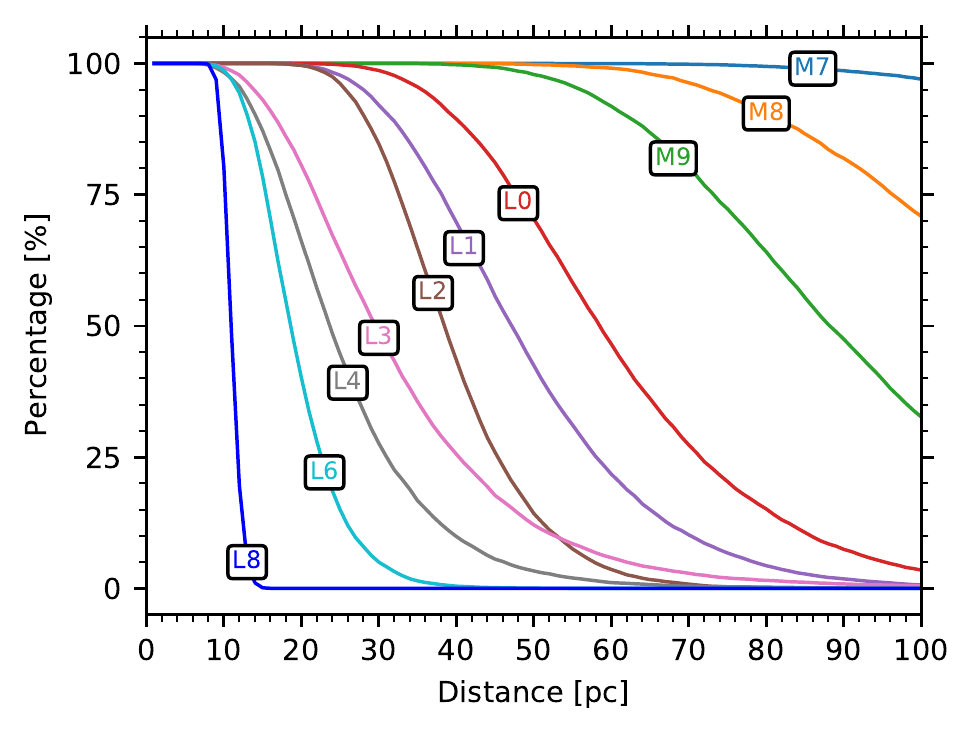}}
    \caption{\label{fig:ucd_completeness_theo}Simulated completeness per parsec for each spectral type.
    Each spectral type from M7-L8
    (right to left) is labelled next to its respective simulated
    completeness level. We skip L5 and L7 for better readability.
    }
\end{figure}

We show the simulated completeness for M7-L8
in 
Fig.\,\ref{fig:ucd_completeness_theo}. This was calculated using
median absolute magnitudes $M_{\text{G}}$ and standard deviations for
each spectral type derived from the GCNS sample (in
Sect.\,\ref{sec:compprevcomp}) and assuming a sky-isotropic $\mathrm{G}$
apparent magnitude limit of
20.4\,mag with Monte Carlo sampling.
Fig.\,\ref{fig:ucd_completeness_theo} indicates that an
incompleteness begins at spectral type M7 and increases until L8,
where the catalogue is only complete for the first 10\,pc. The
standard deviations of absolute magnitudes per spectral type bin are
large (0.5 to 1\,mag) and often have small sample sizes; therefore the noise in
these simulations was quite large, which explains the crossing of the mean
relation for some sequential spectral types.

\subsubsection{UCD empirical completeness exceptions}


We considered the simulated completeness from
Fig.~\ref{fig:ucd_completeness_theo} with respect to a known sample,
objects in the GUCDS identified in one of the \G\ releases, and spectral
type from M7 to T6. This corresponds to $2925$ sources. We find that
98 objects were not included in the GCNS that are in \gdrthree, but they
either do not have parallaxes (34) or failed our probability selection
(25), and 39 had parallaxes $<8$ mas.  Of the 34 objects that did not
have parallaxes, 21 did have parallaxes in \gdrtwo\ but the five-parameter solutions in \gdrthree\ were not published because
their \texttt{astrometric\_sigma5d\_max} $>$ 1.2\,mas. This
could be because these objects are non-single or simply because they
are very faint and at the limit of our precision. 
%

An example of a system that we would expect to be in the GCNS is the
nearby L/T binary  \object{Luhman  16} AB;
\gdrtwo~5353626573555863424 and 5353626573562355584 for
A and B, respectively, with $\pi = 496 \pm 37$ mas
(\cite{2013ApJ...767L...1L}) and $G=16.93$ \& $G=16.96$\,mag. The
primary is in \gdrtwo and \gdrthree (without complete astrometric
solution in either), whilst the secondary is only in \gdrtwo. This is a
very close binary system with a short period, so that the use of a single-star astrometric solution may result in significant residuals that
may have resulted in its exclusion in the current release.

\subsection{Clusters within 100~pc}
\label{sec:clusters}

\begin{figure*}[!htb]
\sidecaption
\includegraphics[width=12cm]{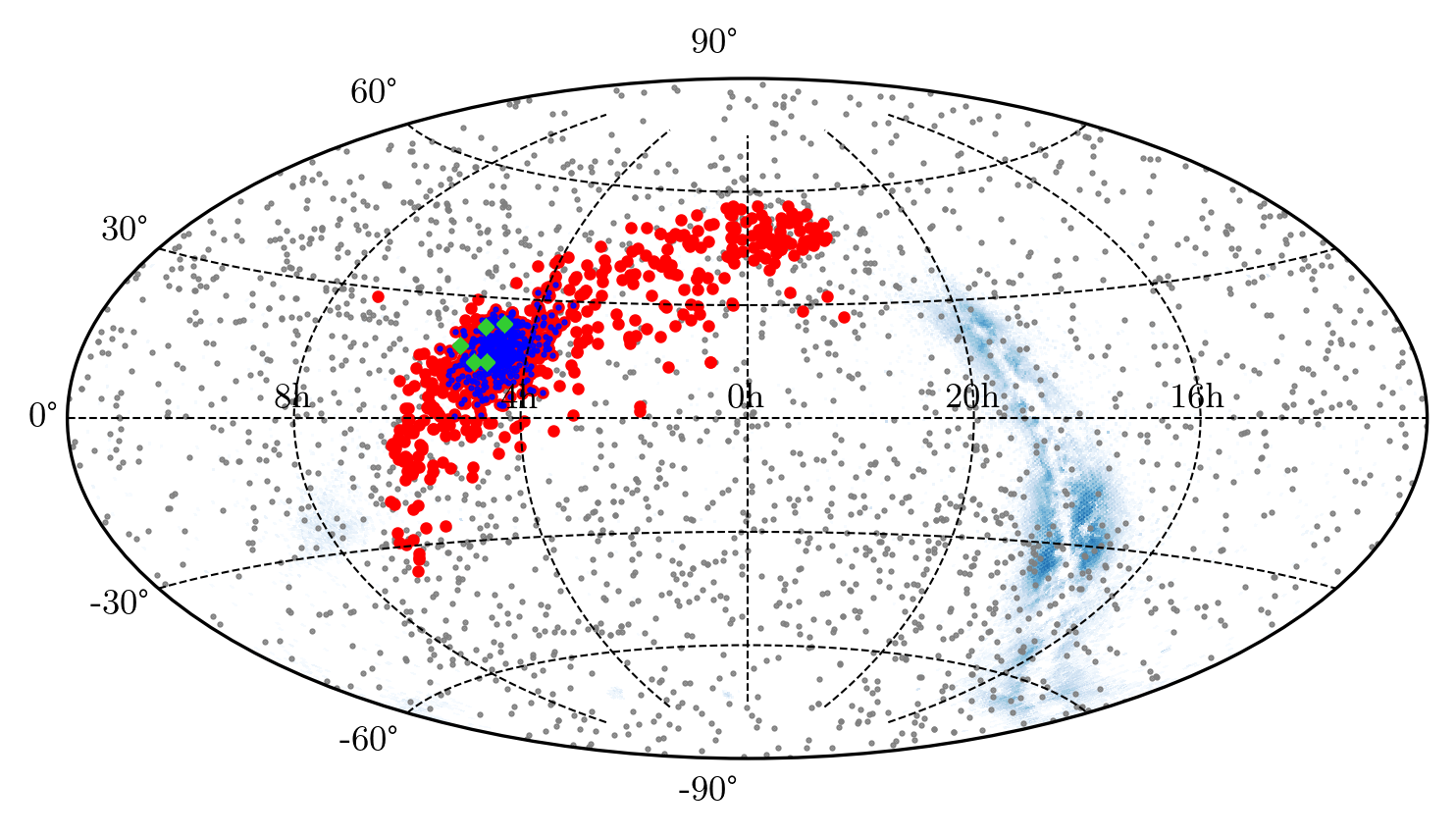}  
\caption{Sky projection, in equatorial coordinates, of candidate
Hyades members. The cloudy light blue structure in the background
denotes the densest part of the Galactic plane in the direction of the Galactic centre. Grey dots denote all 3055 candidates, and filled red circles
indicate the 920 sources that survived our ad hoc density filter aimed
at suppressing contamination and bringing out the classical cluster
and its tidal tails. Small blue dots denote 510 of the 515
\citet{2018A&A...616A..10G} members that are confirmed
by \gdrthree, and the green diamonds denote the five deprecated \gdrtwo
members.\label{fig:Hyades_sky_map}\newline ~ \newline }

\sidecaption
\includegraphics[width=12cm]{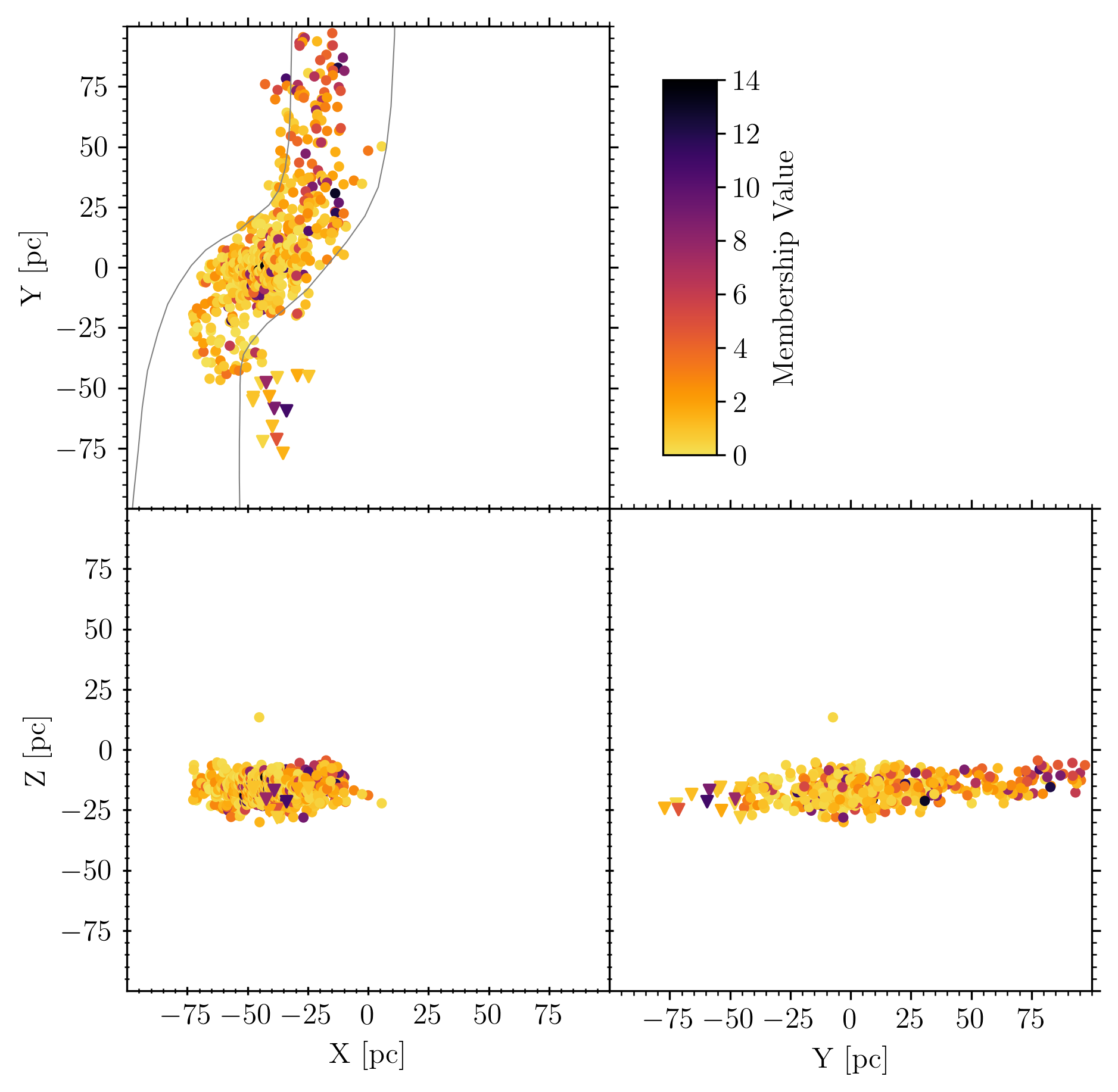}  
\caption{Projections of the Hyades and its tidal tails in Galactic Cartesian
coordinates $(X,Y,Z)$ with the Sun at the origin. The grey lines
(courtesy of Stefan Meingast) denote the approximate contours of the
Hyades tidal tails as simulated by \cite{2005AstL...31..308C}. The
trailing tail shows a peculiar bend (triangles) where stars deviate
from the simple $N$-body model prediction. These stars are well
compatible with the cluster's space motion and agree well
with the remaining cluster population based on their location
in the CAMD (Fig.\,\ref{fig:Hyades_HRD}). We have no reason to assume that they do not belong to
the cluster. \newline ~ \newline 
\label{fig:Hyades_3D_map}}
\end{figure*}

\begin{figure*}[!htb]
\center{\includegraphics[width=16cm]{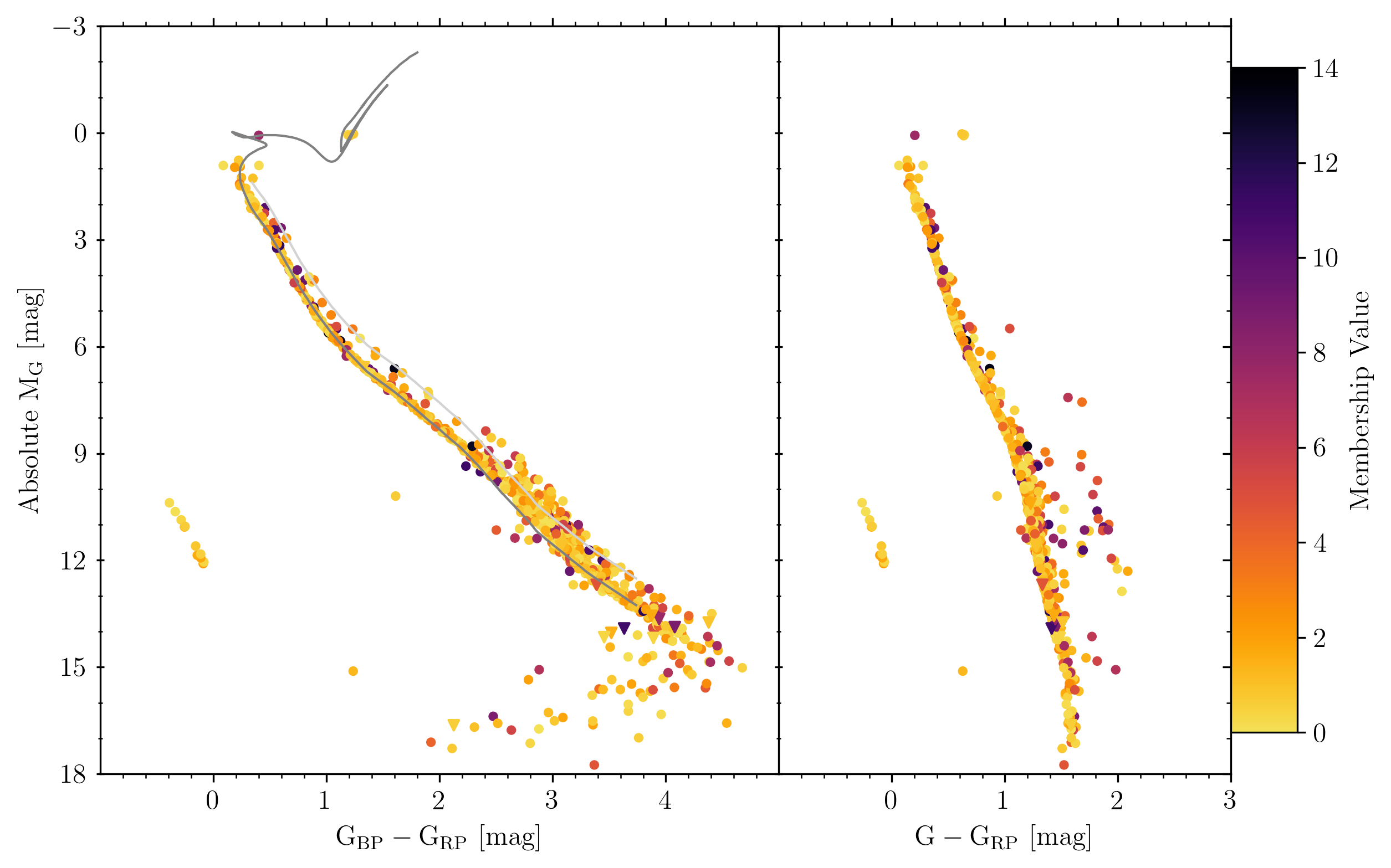}}  
\caption{CAMDs for the 920 Hyades
candidate members using \gbp-\grp (left) and $G$-\grp\
(right). Extinction and reddening are not included but are
generally negligible for the Hyades. Absolute magnitudes have been
computed using \texttt{dist\_50} as distance estimate. The hook at
the faint end of the \gbp\ -\grp\ main sequence (left panel) is a
known artificial feature of \gdrthree caused by spurious \gbp\
magnitude estimates for very faint intrinsically red sources. The outliers to the right above the $G$-\grp\ main sequence (right panel) have biased \grp\ and \gbp\ magnitudes, as indicated by the high BP/RP flux excess values. Because their \gbp\ and \grp\ magnitudes are biased by the same amount, the \gbp-\grp\ value (left panel) is fairly accurate. The colour of the symbols
encodes our membership probability, with low $c$ values (yellow)
indicating highly probable members. The grey curves in the left panel
denote a 800-Myr PARSEC isochrone and its associated equal-mass binary
sequence (both based on \gdrtwo passbands); they are not a best fit,
but are only meant to guide the eye. Fourteen stars are marked with
triangles; they are nearly indistinguishable from the remaining stars
in the two CAMDs. These correspond to a group of stars that deviates from
a simple $N$-body prediction for the development of the tidal tails,
see the $(x, y)$-diagram in
Figure\,\ref{fig:Hyades_3D_map}.\label{fig:Hyades_HRD}}
\end{figure*}

The 100\,pc sample contains two well-known open clusters, the \object{Hyades}
(Melotte 25, at $\sim$47\,pc) and Coma Berenices (\object{Melotte 111}, at
$\sim$86\,pc). Both clusters stand out as density concentrations in 3D
configuration as well as in 3D velocity space.

\subsubsection{Membership}

In order to identify candidate members, we largely followed the approach
of \cite{2018MNRAS.477.3197R}. Their method uses astrometry
(positions, parallaxes, and proper motions), combined with radial
velocity data when present, to compute 3D space motions and select
stars as candidate members of each cluster. We slightly adopted the
original approach and added an iterative loop in order to remove the
dependence on the assumed initial conditions of the cluster. After
convergence, the method attributes a membership probability to each
star, expressing the statistical compatibility of the computed space
motion of the star with the mean cluster motion, taking the full
covariance matrix of the measurements as well as the cluster velocity
dispersion into account \citep[for details,
see][]{2018MNRAS.477.3197R}. In contrast
to \cite{2018MNRAS.477.3197R}, who used the method on a limited-size
field on the sky centred on the Hyades, we used the full all-sky GCNS
catalogue. It is worth noting that the method only uses observables
such as proper motions and radial velocities and does not depend on
other parameters, in particular, on the GCNS probability ($p$) or the
renormalised astrometric unit weight error (ruwe).


\subsubsection{Hyades}

For the Hyades, using the approach outlined above but limiting the
radial velocities to those present in \gdrthree (i.e. excluding
ground-based values in GCNS), we identify 3055 candidate
members. Their distribution on the sky
(Figure\,\ref{fig:Hyades_sky_map}) shows three main features: (i) a
dense concentration at the location of the cluster core, (ii) clear
signs of two tidal tails, and (iii) a uniform spread of interlopers
throughout the sky.

The tidal tails were discovered independently, based on \gdrtwo data,
by \cite{2019A&A...621L...3M} and \cite{2019A&A...621L...2R}. These
studies have noted the need to remove contamination, and both
adopted a spatial density filter with subjective limits to highlight
spatial over-densities: \citeauthor{2019A&A...621L...3M} eliminated all
sources with fewer than three neighbours within 20\,pc,
while \citeauthor{2019A&A...621L...2R} first drew a sphere with a
10\,pc radius around each star, then counted the number of stars that
fell into this sphere, subsequently selected all spheres that were
filled by six stars or more, and finally selected all stars that
belonged to at least one of these spheres. In our sample, 920
candidate members remain after a density filter was adopted that somewhat arbitrarily accepted all stars with eight or more
neighbours in a 10\,pc sphere. By construction, the resulting set is
strongly concentrated towards the classical cluster (630 stars are
within two tidal radii, i.e. 20\,pc of the centre) and the two tidal
tails. The interpretation of the filtered sample clearly requires
care, if only because edge effects are expected to be present close to
the GCNS sample border at 100\,pc.

The cluster was studied with \gdrtwo data
by \cite{2018A&A...616A..10G}. We confirm 510
objects and refuse 5 of their 515 members.
When the GCNS sample is extended to include the rejected stars with low
probabilities ($p<0.38$; Section\,\ref{sec:randomforest}) no members
within 20\,pc from the cluster centre are added. The closest 'new' candidate
member is found at 20.6~pc distance from the cluster centre and has 
$p=0.01.$ It also has an excessive $\textrm{ruwe} = 2.73$. This 19th$^{\rm }$ magnitude object is most
likely a partially resolved binary with suspect astrometry because a
nearby polluting secondary star was detected but not accounted for in
the \G\ DR3 data processing in about one-third of the
transits of this object. We conclude that our membership, at least
within two tidal radii, is not affected by the astrometric cleaning
that underlies the GCNS sample definition.

Beyond the cluster core and surrounding corona, the candidate members
show clear signs of tidal tails (Figure\,\ref{fig:Hyades_3D_map}). In
the \gdrtwo data, these tails were found to extend out to at least
170\,pc from the cluster centre for the leading tail (which extends
towards positive $y$ into the northern
hemisphere). \cite{2020arXiv200702969O} suggested that tail lengths of
$\sim$400-800\,pc can be expected. The two \gdrtwo discovery studies used
a sample out to 200\,pc from the Sun. Because the GCNS sample is by
construction limited to a distance of 100\,pc from the Sun, we cannot
use GCNS to study the full extent of the tails. The GCNS does
confirm, however, the \gdrtwo-based observation that the trailing tail
(at southern latitudes and negative $y$ values) is less pronounced,
deviates (triangles in Fig.\,\ref{fig:Hyades_3D_map}) from the
expected S-shape predicted by $N$-body
simulations \citep[e.g.][]{2005AstL...31..308C,2009A&A...495..807K},
and is currently dissolving \citep[see also][]{2020arXiv200702969O}.

The classical \gbp\ -\grp\ CAMD that is displayed in the left panel of
Fig.\,\ref{fig:Hyades_HRD} shows a narrow main sequence that
extends the \cite{2018A&A...616A..10G} sequence based on \gdrtwo by
$\sim$2\,magnitudes towards fainter objects, a well-defined white
dwarf sequence, and a clear sign of an equal-mass binary
sequence that extends to the faintest objects ($M_G \sim
15$\,mag). Further noticeable features in the CAMD include the
broadening of the main sequence for M dwarfs, caused by radius
inflation \citep[e.g.][]{2019ApJ...879...39J}, and a
hook at the faint end comprising $\sim$50 low-mass objects.  The
latter feature has been present as an artefact in \gdrtwo
\citep[e.g.][]{2019A&A...623A..35L} and is caused by spurious mean
\gbp\ magnitudes exhibited by faint red targets for which negative
\gbp\ transit fluxes that remain after background subtraction were not
accounted for while forming the mean published \gbp\ magnitudes. This
hook entirely consists of objects \gbp\ $<20.3,$ which would
therefore be cut had we applied the photometric quality
filter suggested in \cite{EDR3-DPACP-117}.

As expected, the $G$-\grp\ CAMD in the right panel of
Fig.\,\ref{fig:Hyades_HRD} shows a continuous, smooth main sequence
all the way down to $M_G \sim 17$\,mag.  This CAMD shows another cloud
of $\sim$20 outliers to the right above the main sequence. These
objects have problematic \gbp\ and \grp\ magnitudes, as indicated by
their non-nominal BP/RP flux excess values. These sources can be
identified using the blended fraction $\beta$ as described in
\cite{EDR3-DPACP-117}. Because \gbp\ and \grp\ are biased in the same
way for these sources (because more than one source lies within the
BP/RP windows, which has not been accounted for in the
\gdrthree\ processing), the difference \gbp-\grp\ is fairly accurate.
The CAMD outliers in both the \gbp-\grp\ version (hook) and the
$G$-\grp\ version are fully explained by known features of the
\gdrthree photometry and are not correlated to the position of the
stars in the cluster or tidal tails.  All in all, both CAMDs
demonstrate the overall exquisite (and improved) quality of the
\gdrthree astrometry and photometry.

\subsubsection{Coma Berenices}

The 100\,pc sample contains a second open cluster, Coma Berenices. It
has similar age ($\sim$800\,Myr) and tidal radius ($\sim$7\,pc) as the
Hyades, but is twice as distant, close to the GCNS sample limit. The
cluster has been studied with \gdrtwo data
by \cite{2018A&A...616A..10G}, who found 153 members in a limited-size
field of view. We used the \gdrthree astrometry and repeated the same
procedures as outlined above. Within the central 14\,pc we confirm 146
of the \gdrtwo members and add 15 new candidate
members. \cite{2018ApJ...862..106T} noted that the cluster is
elongated along its orbit towards the Galactic plane, and subsequently
reported tidal
tails \citep{2019ApJ...877...12T}. Our \gdrthree candidate members
show very clear signs of tidal tails beyond two tidal radii from the
cluster centre, but their precise shape and membership depends
sensitively on the spatial density filter that is needed to remove
contamination from the all-sky GCNS sample. Moreover, a study of the
cluster and its tidal tails based on the GCNS sample is complicated because it lies close to the sample border.


\subsection{Stellar multiplicity: Resolved systems}
\label{sec:mult-resolved}


\begin{figure*}
\centering

\begin{tabular}{cc}  
\includegraphics[width=0.48\textwidth]{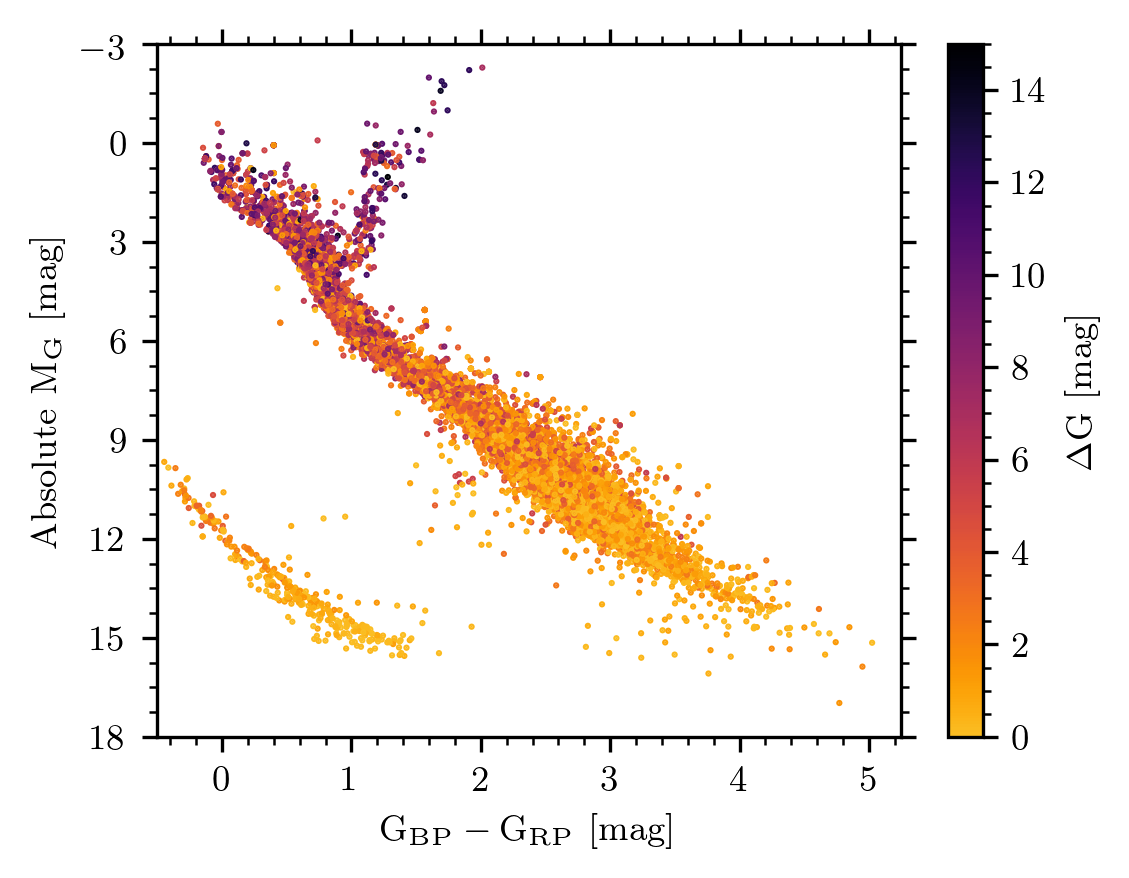} &
\includegraphics[width=0.48\textwidth]{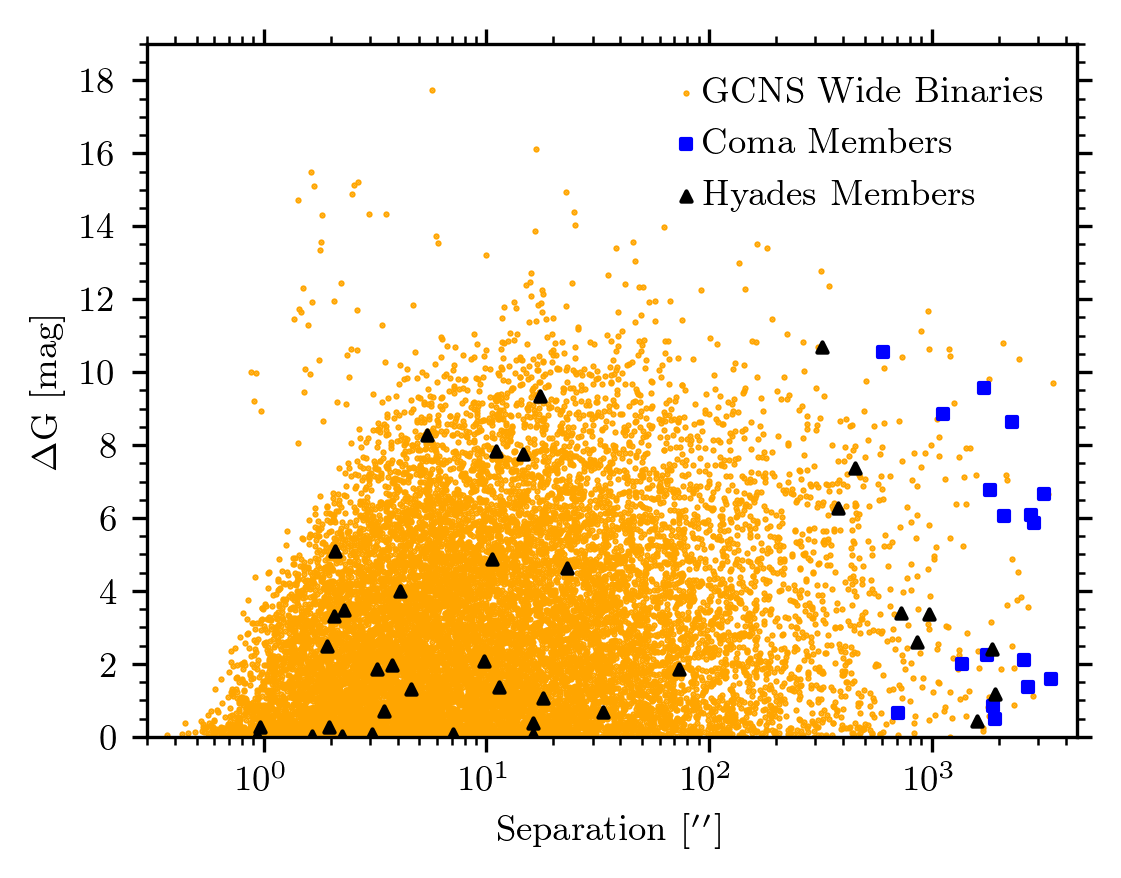} \\
\includegraphics[width=0.48\textwidth]{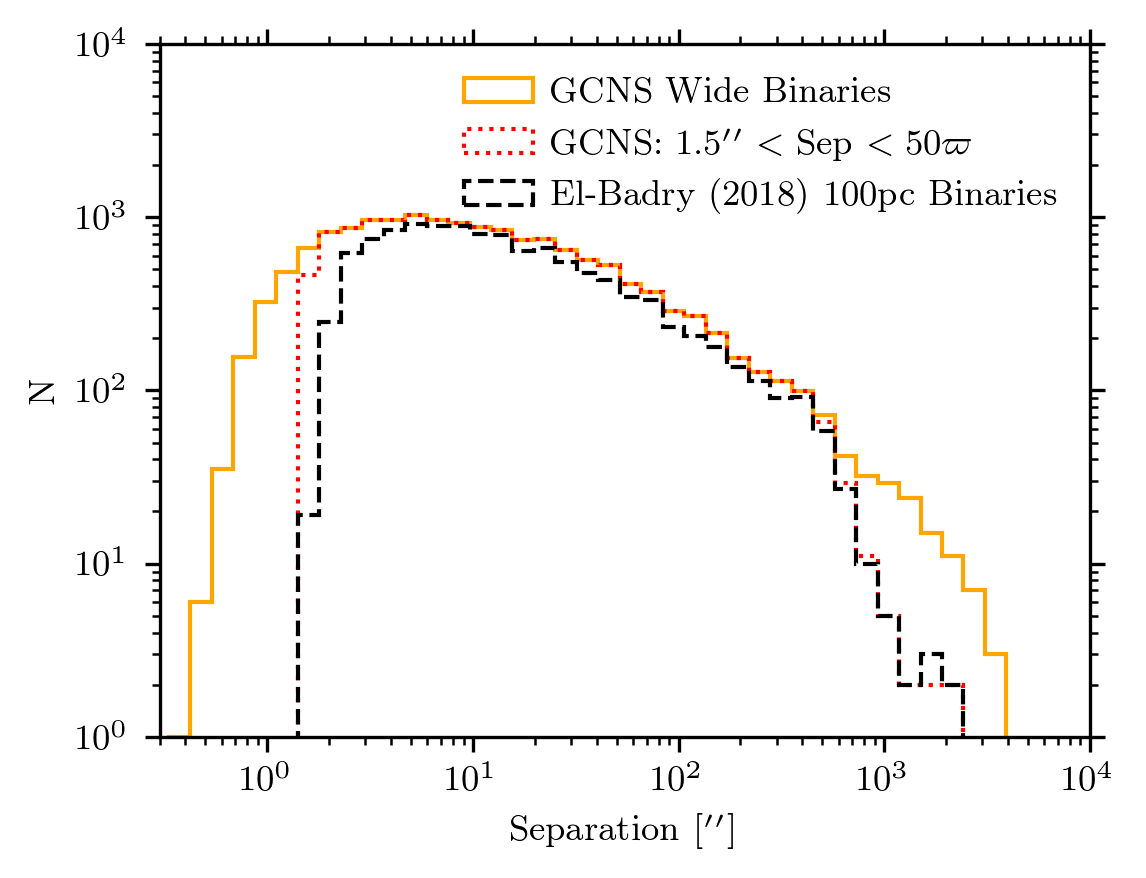} & 
\includegraphics[width=0.48\textwidth]{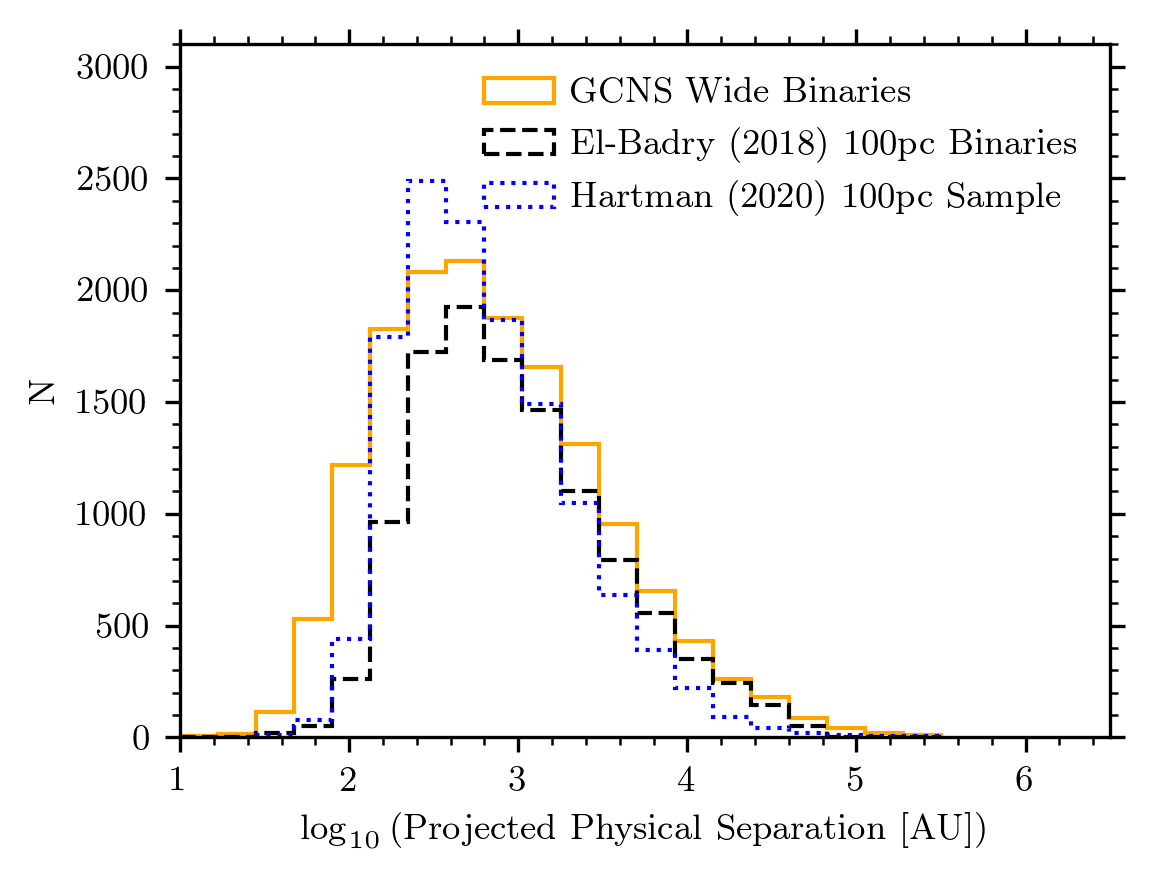} \\

\end{tabular}
\caption{\label{fig:GCNS_binaries} Top left: CAMD for the GCNS
  systems colour-coded by the magnitude differences of the binary
  components.  Top right: Separation vs. $G$ mag difference for the
  resolved stellar systems in GCNS (orange points). Known members of
  the Hyades (black triangles) and Coma Ber (blue squares) clusters are
  highlighted. Bottom left: Histogram (solid orange) of separations for wide
  binaries in the GCNS sample compared to the DR2-based catalogue (dashed black) from \citet{2018MNRAS.480.4884E}. The dotted red histogram corresponds to the separation distribution of GCNS wide-binary candidates adopting the exact boundaries in \citet{2018MNRAS.480.4884E}. Bottom right: Physical projected separation distribution for the wide-binary candidates identified in this work (solid orange) compared to those from \citet{2018MNRAS.480.4884E} (dashed black) and \citet{2020ApJS..247...66H} (dotted blue), restricted to systems within 100\,pc.}
\end{figure*}

\begin{figure}
\includegraphics[width=0.48\textwidth,scale=0.5]{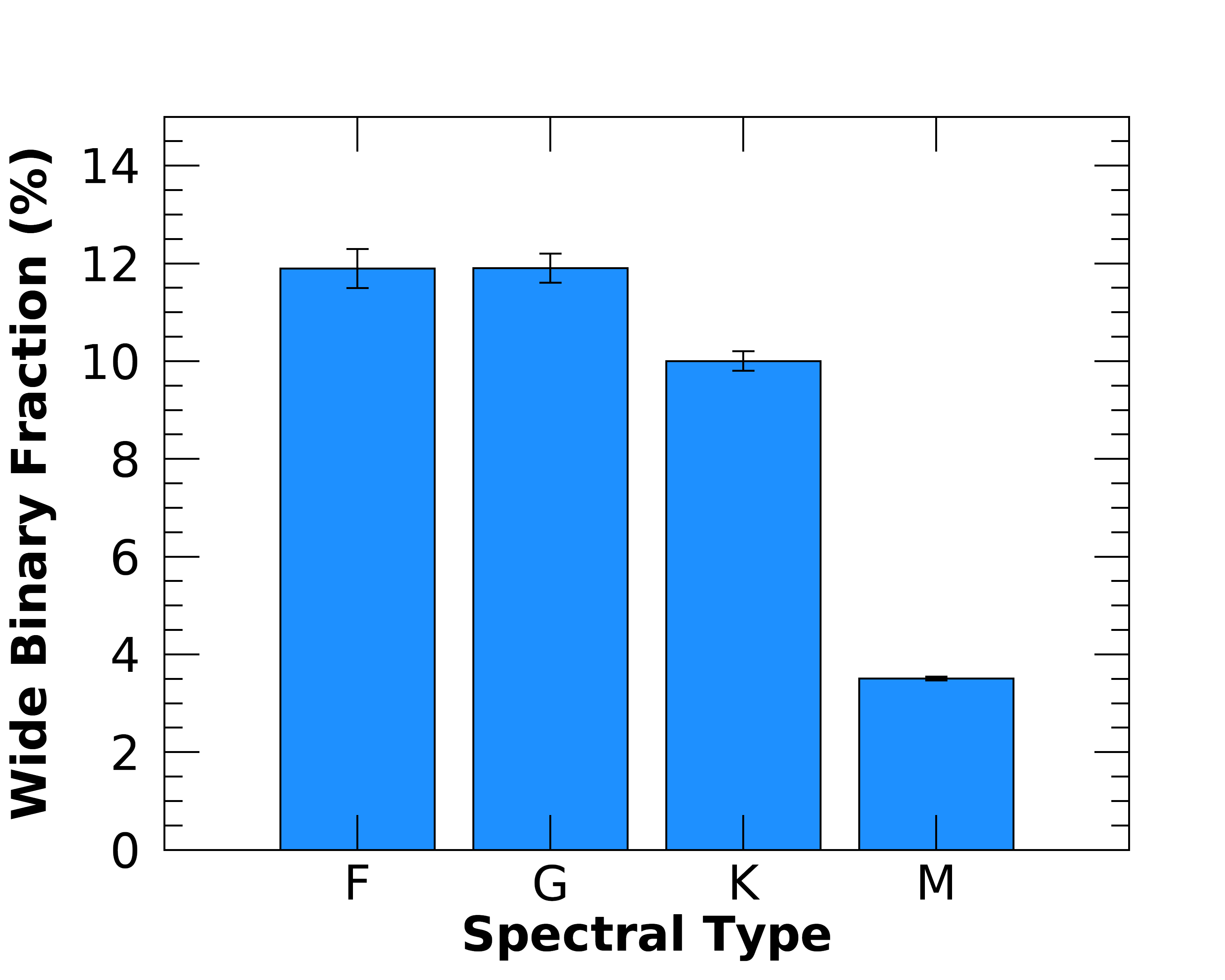} 
\caption{\label{fig:GCNS_binfrac} Histogram of the wide-binary fraction within 100\,pc as recovered from the GCNS catalogue. Error bars, representing $1\-sigma$ confidence intervals, are derived from a binomial distribution. }
\end{figure}

Statistical studies of stellar multiplicity are key to a proper
understanding of many topics in modern-day astrophysics, including
star formation processes, the dynamics of dense stellar environments,
the various stages of stellar evolution, the formation and evolution
of planetary systems, the genesis of extreme high-energy phenomena
(supernovae, gamma ray bursts, and gravitational waves), and the formation
of the large-scale structure of the Universe (e.g.
\citealt{2020MNRAS.496.1922B}, and references therein).  Early
investigations of the statistical properties of stellar systems in the
solar neighbourhood based on small sample sizes (hundreds of
stars, e.g. see \citealt{1991A&A...248..485D,2010ApJS..190....1R}, and
reference therein) have  revealed not only a high binary
(and higher multiplicity) fraction, but also trends in stellar
multiplicity with spectral type, age, and metallicity. With the
improvements in the host of techniques used to search for stellar
binaries, which include photometry, spectroscopy, astrometry,
high-contrast imaging, and interferometry
\citep[e.g.][]{2017ApJS..230...15M}, such trends are now being placed on
solid statistical grounds based on typical sample sizes of tens of
thousands of systems
\citep[e.g.][]{2018MNRAS.480.4884E,2018ApJS..235....6T,2019ApJ...875...61M,2020A&A...635A.155M,2020ApJ...895....2P}.
A new revolution is in the making, however, with the \G\ mission bound
to provide a further order-of-magnitude increase in known stellar
systems across all mass ratios and orbital separations
\citep{2004ASPC..318..413S,2005ESASP.576...97S}, detected based on\ the  astrometric, spectroscopic, photometric, and spatial resolution
information of \textit{Gaia}.

The first two \G\ data releases (DR1 and DR2) have
permitted detailed investigations with unprecedented precision of
the regime of spatially resolved intermediate- to wide-separation
binaries (e.g.
\citealt{2017MNRAS.472..675A,2017AJ....153..257O,2017AJ....153..259O,2018MNRAS.480.4884E,2019ApJ...875...61M,2019AJ....157...78J,2020ApJS..247...66H},
and references therein). Such systems are of particular interest
because of their low binding energies, they can be used as probes of
the dynamical evolution history of the Galaxy and of the mass
distribution and number density of dark objects in the Milky Way
(e.g. \citealt{2018MNRAS.480.4884E}, and references therein;
\citealt{2020ApJS..247...66H}, and references therein). They were
born at the same time and in the same environment but evolved in an
entirely independent way, therefore  they are very useful tools for testing
stellar evolutionary models, they can be used as calibrators for age
and metallicity relations (e.g., \citealt{2019AJ....157...78J}, and
references therein). Because they are common in the field
\citep{2010ApJS..190....1R}, they are natural laboratories in which to
study the effect of stellar companions on the
formation, architecture, and evolution of planetary systems
\citep[e.g.
][]{2007A&A...462..345D,2016MNRAS.455.4212D,2016AJ....152....8K,2019AJ....157..252K}.

Using the updated astrometric information in the GCNS catalogue, we
performed a new search for wide binaries within 100\,pc of the Sun. We
first identified neighbouring objects with an angular separation in the sky
$<1$ deg (which implies a non-constant projected separation in au),
similarly to \citet{2020ApJS..247...66H}. We did not impose a lower
limit on the projected separation in order to characterise the loss in
efficiency in detecting pairs when the resolution limit of \gdrthree\ is approached.  We then followed \citet{2019MNRAS.485.4423S} and adopted
standard criteria to select a sample of likely bound stellar systems:
1) scalar proper motion difference within 10\% of the total proper
motion ($\Delta\mu< 0.1\mu$), and 2) parallax difference within either
3$\sigma$ or 1\,mas, whichever is greater ($\Delta\hat\varpi < max[1.0,
  3\sigma]$). We further refined the selection with a second pass based
on the requirement of boundedness of the orbits, following
\citet{2018MNRAS.480.4884E}, but placing the more stringent constraint
$\Delta\mu < \Delta\mu_\mathrm{orbit}$, with
$\Delta\mu_\mathrm{orbit}$ defined as in Eq. 4 of
\citet{2018MNRAS.480.4884E}.

\begin{table*}
\centering
\caption{Summary data on binary pairs in the GCNS catalogue of wide binaries.}            
\begin{minipage}[t]{15.0cm} 
\setlength{\tabcolsep}{2.0mm}
\hskip-1.5cm\begin{tabular}{r r c c c c c c c}        
\hline\hline
\noalign{\smallskip}
 SourceId 1               & SourceId 2                & Separation      & $\Delta G$              & Proj. Sep.   & Bound & Hyades & Coma & Binary \\
\noalign{\smallskip}
 (Primary)               & (Secondary)                & (arcsec)   &              & (au)   &  &  &  &  \\
\hline
\noalign{\smallskip}
 83154862613888 & 83154861954304 & 3.8353 & 3.2631  &  244.7406 & true & false & false & true \\ 
554329954689280 & 554329954689152  &  3.7164 & 0.4823 &  358.7470 & true & false & false & true  \\
1611029348657664 & 1611029348487680 &  6.1252 & 0.8744  &  513.6358 & true & false & false & true \\
1950331764866304 & 1962117155125760 &  9.3117 & 6.8372  &  810.1125 & false & false & false & true \\
 \dots & \dots & \dots & \dots & \dots & \dots & \dots & \dots & \dots \\
\noalign{\smallskip}
\hline\hline       
\end{tabular}
\end{minipage}
\label{tab:GCNS_WB_cat}
\end{table*}

The application of our selection criteria to the GCNS catalogue allowed
us to identify a total of 16\,556 resolved binary candidates (this
number increases to 19\,176 when we do not impose the
bound orbit criterion). The relevant information is reported in
Table \ref{tab:GCNS_WB_cat}\footnote{full table available from the CDS}. The selection by
construction contains objects that are co-moving because they are
members of rich open clusters (Hyades and Coma Berenices), more
sparsely populated young moving groups
\citep[e.g. ][]{2018ApJ...863...91F}, as well as higher-order resolved
multiples in which more than one companion is identified to either
member of a pair.  In Table \ref{tab:GCNS_WB_cat} we flag
both higher-order multiples and cluster members (1758 and 286,
respectively) based on the updated cluster membership list
in Sect.\,\ref{sec:clusters}.

The upper left panel of Fig. \ref{fig:GCNS_binaries} shows the
colour-magnitude diagram for the primaries in the 100 pc wide-binary
candidate sample.  The plot is colour-coded by magnitude difference
with the secondary. A small number of objects are removed as they do
not have full colour information in \gdrthree. The diagram is almost
free of spurious objects located in between the main sequence and the
WD cooling sequence, which amount to no more than 0.2\% of the
sample. These objects are likely misclassified due a variety of
reasons that are summarised in \citet{2020ApJS..247...66H}.  Similarly
to \citet{2018MNRAS.480.4884E} and \citet{2020ApJS..247...66H}, for
instance, the diagram also displays an indication of a secondary main
sequence, offset upward by $\sim0.5$ mag particularly in the $1.0
\lesssim (G_\mathrm{BP} - G_\mathrm{RP})\lesssim 2.0$\,mag range. This
is the unresolved binary sequence composed of hierarchical systems in
which one or both of the resolved components is itself a spatially
unresolved binary with a typical mass ratio $q \gtrsim 0.5$
(e.g. \citealt{2018ApJ...857..114W}, and references therein).

Further evidence of the presence of the photometric binary main-sequence  
is found by colour-coding the plot in the upper left panel of Fig. \ref{fig:GCNS_binaries} using the value of the ruwe, which exhibits a notable excess in this region (plot not shown, but see e.g. \citealt{2020MNRAS.496.1922B}). Overall, $\sim24\%$ of the objects in our catalogue have ruwe $\gtrsim 1.4$ (indicative of an ill-behaved astrometric solution), and in $\sim2\%$ of the cases, both components of a binary have a high ruwe value. These numbers might be explained based on the combined effects from higher-order multiples with short-period components (this number is difficult to derive as it entails understanding the selection function of short-period binaries with wide-separation stellar companions) and larger samples of intermediate-separation binaries that become unresolved or partially resolved as a function of increasing distance (preliminary estimates indicate that this percentage is about $15-20\%$).

The upper right panel of Fig. \ref{fig:GCNS_binaries} shows the $G$
mag difference $\Delta G$ of our wide-binary candidates as a function
of angular separation. The sample of objects flagged as Hyades and
Coma Ber cluster members, as determined in Sect. \ref{sec:clusters}, is also reported. The slope of increasingly
lower $\Delta G$ at separations $<10$\,\arcsec\ is the footprint of the
\G\  sensitivity loss, which nicely follows the behaviour in
contrast sensitivity shown in Fig. 9. 
At separations $\gtrsim 10$\,\arcsec\ , the interval of $\Delta G$ is
essentially independent of separation. Interestingly, all Coma Ber
bona fide cluster members flagged as candidate binaries reside at very
wide separations (at the distance of Coma Ber, the typical projected
separation $\gtrsim 4\times10^4$ au). Even when the requirement of
formally bound orbits is enforced, a significant fraction of the very
wide binaries could still be a result of chance alignment. We
estimated the contamination rate of our sample of wide binaries using
the GeDR3 mock catalogue \citep{2020PASP..132g4501R}.  The catalogue does
not contain any true binaries: an adoption of our selection criteria
for pair identification in the mock catalogue provides a direct
measurement of the number of false positives (pairs due to chance
alignment) in our sample, particularly in the regime of very wide
separations. When we applied our two-pass search criteria, the mock
catalogue returned five pairs, all at a separation $\gtrsim 1000$ au.  This means that the contamination level in our sample probably is
$0.05-0.1\%$.

A comparison with the recent DR2-based catalogue of wide binaries
produced by \citet{2018MNRAS.480.4884E} shows good agreement with our
selection in the overlapping regime of separations when cluster members and higher-order multiples are
excluded (see Fig.
\ref{fig:GCNS_binaries}, bottom left panel). When we restrict ourselves
to the regime of projected physical separations defined by
\citet{2018MNRAS.480.4884E}, our candidates match$\text{}$ 
those found by \citet{2018MNRAS.480.4884E} within 100\,pc to 78.1\%. The
discrepancy is likely due to significantly revised values of parallax
and proper motion in \gdrthree  with respect to the DR2 values. We also
note an overall increase of $\sim$20\% in detected pairs with
respect to the \citet{2018MNRAS.480.4884E} 100\,pc sample. The larger
number of close pairs identified in the GCNS samples of candidates is
a possible indication of a moderate improvement in sensitivity in the
$\approx 1$\arcsec\ $-$ 3\arcsec\ regime with respect to DR2-based
estimates \citep[e.g. ][]{2019A&A...621A..86B}.

The bottom right panel of Fig. \ref{fig:GCNS_binaries} shows the projected
physical separation of our wide-binary candidates, compared to the
distributions of the same quantity in the \citet{2018MNRAS.480.4884E}
and \citet{2020ApJS..247...66H} catalogues, both restricted to $d< 100$\,pc (and the latter with Bayesian binary probability $>99\%$). All distributions peak
around $10^{2.5}$ au, and they all exhibit the same exponential decay at
wider separations. Finally, we retrieve 25 of the 63 very wide
binaries within 100\,pc in the \citet{2019AJ....157...78J} catalogue
(this number increases to 41 if we lift the requirement on formally
bound orbits). Similarly to the searches performed by
\citet{2018MNRAS.480.4884E} and \citet{2020ApJS..247...66H}, we find
no evidence of bi-modality in the distribution of projected physical
separations due to a second population of binaries with companions at
$>100\,000$\,au, as had been previously suggested.  As a matter of
fact, such a feature is instead clearly seen (plot not shown) in our
first-pass sample selected without imposing that the orbits be
physically bound.

Using the spectral type and median $M_G$ calibration found in Sect.\,\ref{sec:compprevcomp}
and the list of binary candidates cleaned for cluster members and
higher-order multiples, we briefly comment  on the wide-binary fraction
$f_\mathrm{WB}$ in the 100\,pc sample. For instance, we obtain
$f_\mathrm{WB}=4.8_{-0.3}^{+0.4}\%$ (with $1-\sigma$ errors derived using the binomial distribution, 
e.g. see \citealt{2003ApJ...586..512B,2009ApJ...697..544S}) for M dwarfs
within 25\,pc in the regime of separations $>2$\,\arcsec\ (171 wide
systems in a sample of 3555 M-type stars). This differs at the
$\sim3.5\sigma$ level from the $7.9\pm0.8$ multiplicity rate reported by
\citet{2019AJ....157..216W}, although subtracting the approximately
25\% of higher-order multiples from their sample the results become
compatible within $1.6\sigma$.  As highlighted by the histogram in
Fig. \ref{fig:GCNS_binfrac}, in the volume-limited 100\,pc sample the
wide-binary fraction appears constant for F- and G-type dwarfs, with a
measured rate entirely in line with previous estimates
($f_\mathrm{WB}\simeq 10-15\%$) in the literature (e.g.,
\citealt{2019ApJ...875...61M}, and references therein).  The hint of a
decline in wide-binary fraction for K-dwarfs is likely real, as based
on the spectral type versus $M_G$ relation provided in Sect. 4.2 we
are complete for all K types. The clear decline in
$f_\mathrm{WB}$ for the M dwarf sample is real, and only mildly affected by incompleteness at the latest sub-spectral types ($>$ M7).


\subsection{Stellar Multiplicity: Unresolved systems}
\label{sect:unresolved}

\begin{figure}
\centering
\includegraphics[width=0.45\textwidth]{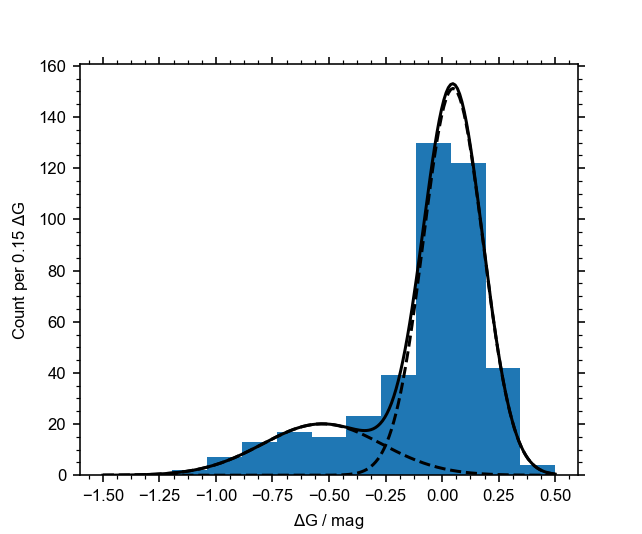}

\includegraphics[width=0.45\textwidth]{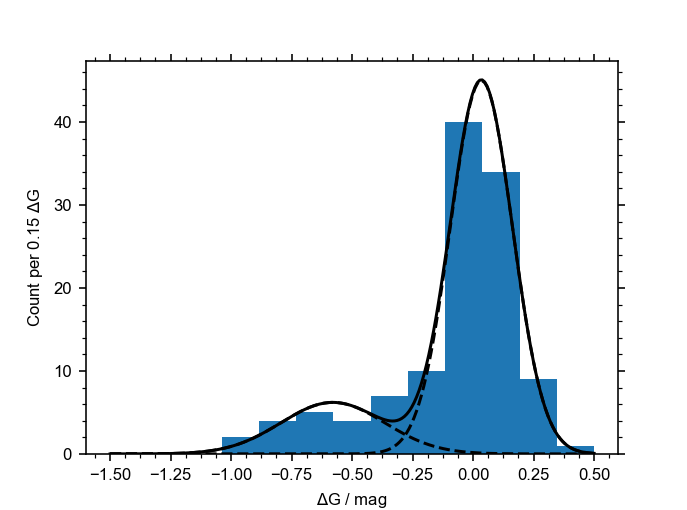} 
\caption{\label{fig:hrds} Two-component Gaussian mixture model of the distributions of star counts per 0.15\,mag $\Delta G$ bin for the Hyades cluster in the upper panel and Coma\,Ber in the lower panel, as described in the text.}
\end{figure}

\begin{figure*}
\centering
\begin{tabular}{cc}
\includegraphics[width=0.45\textwidth,scale=0.5,trim=0 0 0 39,clip]{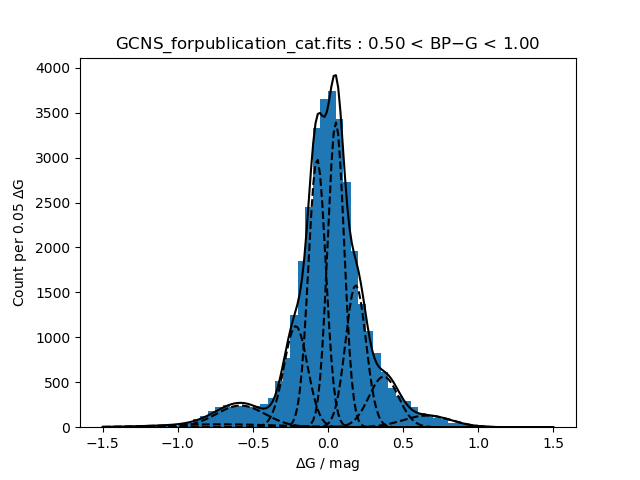} &
\includegraphics[width=0.45\textwidth,scale=0.5,trim=0 0 0 39,clip]{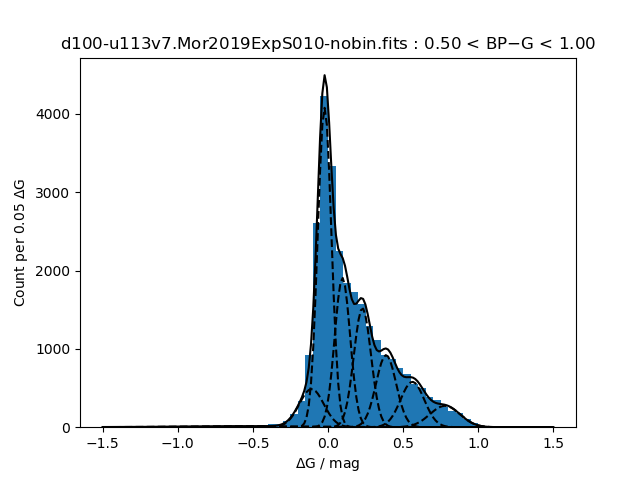} \\
\includegraphics[width=0.45\textwidth,scale=0.5,trim=0 0 0 39,clip]{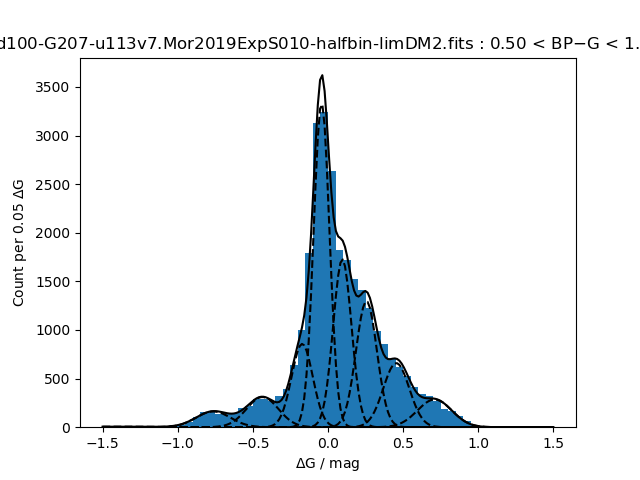} &
\includegraphics[width=0.45\textwidth,scale=0.5,trim=0 0 0 39,clip]{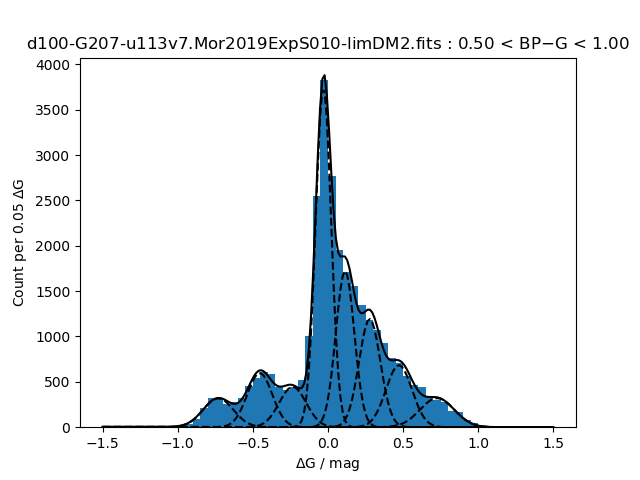}\\
\end{tabular}
\caption{\label{fig:fieldgmm}
Gaussian mixture models of the distribution of star counts per
0.1\,mag $\Delta G$ bin in the range $0.5< \gbp - G <1.0$
after the slope of the main sequence is subtracted in the CAMD. From top
to bottom, the first panel shows the GCNS, the next three panels show
simulations employing no binarity, half--fiducial, and fiducial
binarity according to the prescription
of ~\citet{2011AIPC.1346..107A}.} 
\end{figure*}

The advent of precise photoelectric photometry in the latter half of
the last century contributed to the discovery of a significant level
of close binarity amongst the stellar populations. Photometric
observations in coeval populations with very low dispersion in
chemical composition (e.g.~rich clusters) reveal a faint
sequence parallel to the locus of the main sequence in
the CAMD --
e.g.~\cite{1984ApJ...280..189S} and references therein.
Given sufficient precision in the photometric measurements a single star
sequence is often accompanied by a second sequence at brighter
magnitudes and redder colours.  The reason is that in a
significant fraction of spatially unresolved binaries, twins
(i.e.~equal-mass binaries) show a vertical elevation of 0.75\,mag in
the CAMD, while extreme mass ratio binaries exhibit a significantly
redder colour close to the same brightness as a single
star (see previously in Section~\ref{sec:mult-resolved} and Fig.~\ref{fig:GCNS_binaries}), and 
an elevated locus of
unresolved binaries is populated by all mass ratios between these two
extremes. The high-precision photometry of \G\ leads to particularly
fine examples of this phenomenon in clean astrometric samples of
cluster stars (e.g.~\citealt{2018A&A...616A..10G}).

Two large coeval populations of stars overlap in the GCNS sample: the Hyades and Coma Berenices clusters
(Section\,\ref{sec:clusters}). The Hyades in particular present a rich
sequence of photometrically unresolved binaries that is evident in Fig.\,\ref{fig:Hyades_HRD}. Using the cluster
members derived in Section\,\ref{sec:clusters}, and limiting our selection to within\,two tidal
radii of the respective cluster centres, we made subsamples of the
GCNS catalogue for the Hyades and Coma\,Ber.  The only additional
filtering on photometric quality applied in this case was as defined
in~\citet{2018A&A...616A..10G}, namely $\sigma_{\rm G} < 0.022$ and $\sigma_{\rm
BP,RP} < 0.054,$ along with their photometric quality cut (via {\tt
phot\_bp\_rp\_excess\_factor}; see Appendix~B in \citealt{2018A&A...616A..10G}).
We traced the locus
of the single-star sequence using a low-order polynomial fit that allowed
us to subtract the slope in the CAMD. This yielded a set of $\Delta G$ versus
colour. Marginalising over the whole range in colour and employing a two-component Gaussian mixture resulted in the models for the star counts
versus $\Delta G$ shown in Fig.\,\ref{fig:hrds}.


The difference in $\Delta G$ between these two components is $\sim0.7$\,mag, as expected for a 
dominantly single-star population along with a subordinate population of near equal--mass
but unresolved binaries. According to this simple model the binary fraction, measured as the
ratio of weights of the subordinate to dominant component and counting one star in the latter and 
two stars in the former, is~34\% for the Hyades and~31\% in Coma\,Ber. This is for the range
$0.5< \gbp - G <2.5,$ which corresponds roughly to main-sequence masses in
the range~1.4M$_\odot$ down to~0.2M$_\odot$ (according to simulations -- see below).

The general field population sampled by the GCNS is neither coeval nor
chemically homogeneous.  However, an analogous procedure can be applied
in its CAMD, noting that the ghostly signature of unresolved binarity
is easily visible at intermediate colours
(e.g. the top right panel in\,Fig.\,\ref{fig:dr2_not_edr3} ).  Furthermore, it
is instructive to apply the same procedure to \G\ CAMD simulations
generated without binaries, a fiducial level of binarity, and a few
mid-fractions.

\begin{table*}
    \footnotesize
    \centering
    \begin{tabular}{cccccc}\hline
    & Observations & Arenou (2011) & $\times0.8$ & $\times0.5$ & $\times0.0$ \\\hline
    \multicolumn{6}{c}{ }\\
    Component weights   & $-0.716$~:~0.0180 & $-1.440$~:~0.0044 & $-1.331$~:~0.0054 & $-2.157$~:~0.0008 & $-2.000$~:~0.0019 \\
    above main sequence & $-0.580$~:~0.0579 & $-0.730$~:~0.0527 & $-0.713$~:~0.0476 & $-1.393$~:~0.0028 & $-0.557$~:~0.0105\\
    $0.5<\gbp - G<1.0$  & $-0.214$~:~0.1301 & $-0.454$~:~0.0823 & $-0.424$~:~0.0871 & $-0.763$~:~0.0322 & \\
    ($\Delta G$~:~weight) &                  & $-0.235$~:~0.0660 &                   & $-0.436$~:~0.0573 & \\
    \multicolumn{6}{c}{ }\\
    TOTAL WEIGHT: & ~~~~~~~~~~0.2060 & ~~~~~~~~~~0.2054 & ~~~~~~~~~~0.1401 & ~~~~~~~~~~0.0931 & ~~~~~~~~~~0.0124 \\
    \multicolumn{6}{c}{ }\\\hline
    \end{tabular}
    \caption{Component weights contributing to the Gaussian mixture models 
    for $\Delta G$ indicating photometrically unresolved binarity. Columns\,3 to\,6 are from simulations using the prescription of\,\cite{2011AIPC.1346..107A} in full and at
    fractional reductions of\,0.8,\,0.5, and\,0.0.}
    \label{tab:fieldgmm}
    \normalsize
\end{table*}

In GOG, in particular, those provided as part of \gdrthree
\citep{EDR3-DPACP-130}, binary stars are generated but the unresolved
binaries do not have the fluxes of the combined components. They have
the flux of the primary. Therefore the sequence of twin binaries is
not present in GOG, and we used another set of simulations to
analyse unresolved binaries. For this we used the last version of the
Besan\c{c}on Galaxy model, 
where the initial mass
function and star formation history were fitted to \G\ DR2 data
\citep[see][and references
  therein]{2018A&A...620A..79M,2019A&A...624L...1M} where the star
formation history of the thin disc is assumed to decrease exponentially.
The stellar evolutionary tracks we used are the
new set from the STAREVOL library \citep[see][and
references therein]{2017A&A...601A..27L}. The complete scheme of the model is described
in \cite{2003A&A...409..523R}, while the binarity treatment is
explained in \cite{2014A&A...564A.102C}. The generation of
binaries, probability, separation, and mass ratio is the same as in the GOG
simulations. However, in contrast to the GOG simulations, unresolved
binaries are treated such that the magnitude and colours reflect the
total flux and energy distribution of the combined components.

This allows a comparison with the cluster results above and also provides a confirmation of the level of
realism in the simulations. The binary angular separation\,$s$ in arcseconds at which simulated pairs become unresolved was assumed to follow Equation\,\ref{eq:smin} in Section\,\ref{sec:contrast_sensitivity}.
The fiducial level of binarity follows\,\citet{2011AIPC.1346..107A} (see also~\citealt{FA-054}), which is
near 60\% for solar mass stars and decreases to about 10\% for stars of mass~0.1\,$M_{\odot}$. 
Figure\,\ref{fig:fieldgmm} shows Gaussian mixture models for histograms of~$\Delta G$ for the stellar main sequence in the colour range
$0.5< \gbp -  G <1.0$ after subtracting the sloping locus in the CAMD, this time tracing the latter using the dominant component in a
Gaussian mixture model in colour bins of~0.05mag; a final mixture model was again employed for the resulting marginal distributions of star counts
versus $\Delta G$. 
Table\,\ref{tab:fieldgmm} quantifies the measured level of photometrically unresolved binarity by summing the weights of all components that significantly contribute to the counts for the range in $\Delta G$ that is affected by unresolved binarity according to the simulations. The observations appear to match the fiducial binarity level simulation by this metric very well.

Given the approximations and assumptions made in this simple analysis,
the agreement between simulations and observations is gratifying. A
significant source of uncertainty on the observational side is the
assumed angular resolution. In reality, the extant processing pipeline
(at Data Reduction Cycle\,3 corresponding to \gdrthree) does not
deblend close pairs observed in single transit windows
\citep{EDR3-DPACP-130}. The effects on photometry and astrometry
depend on the scan angle with respect to the position angle of the
binary in a given transit, as well as on the angular separation. Another
limitation on the simulation side is the treatment of the distribution
in metallicity, as shown by the sharper features in the histogram
counts in Fig.\,\ref{fig:fieldgmm}.  Because of these complications, we draw
no more quantitative conclusions as to the true level of binarity in
the GCNS sample.  Further and more detailed studies of the effects of
binarity (e.g.~\citealt{2020MNRAS.tmp.1659B};
\citealt{2020arXiv200611092L}) are clearly warranted but are beyond
the scope of this demonstration work.

\subsection{White dwarfs}
\label{sec:wd}

\subsubsection{White dwarf selection}

To recover the WD population in our catalogue, we started analysing all
$1\,040\,614$ sources with $\hat\varpi > 8$\,mas for which the three \G\
photometric passbands are available. We used the $29\,341$ sources from this larger sample that are in common with three
different catalogues of known WDs
(\citealt{GentileFusillo2019}, \citealt{Torres2019},
and \citealt{JimenezEsteban2018}) to build training and test
datasets for our WD random forest classification algorithm. We
selected $20\,000$ of them to constitute the WD sample in the training
dataset, and the other $9\,341$ sources became the test dataset.

\begin{figure}[!htb]
\center{
\includegraphics[width=\columnwidth]{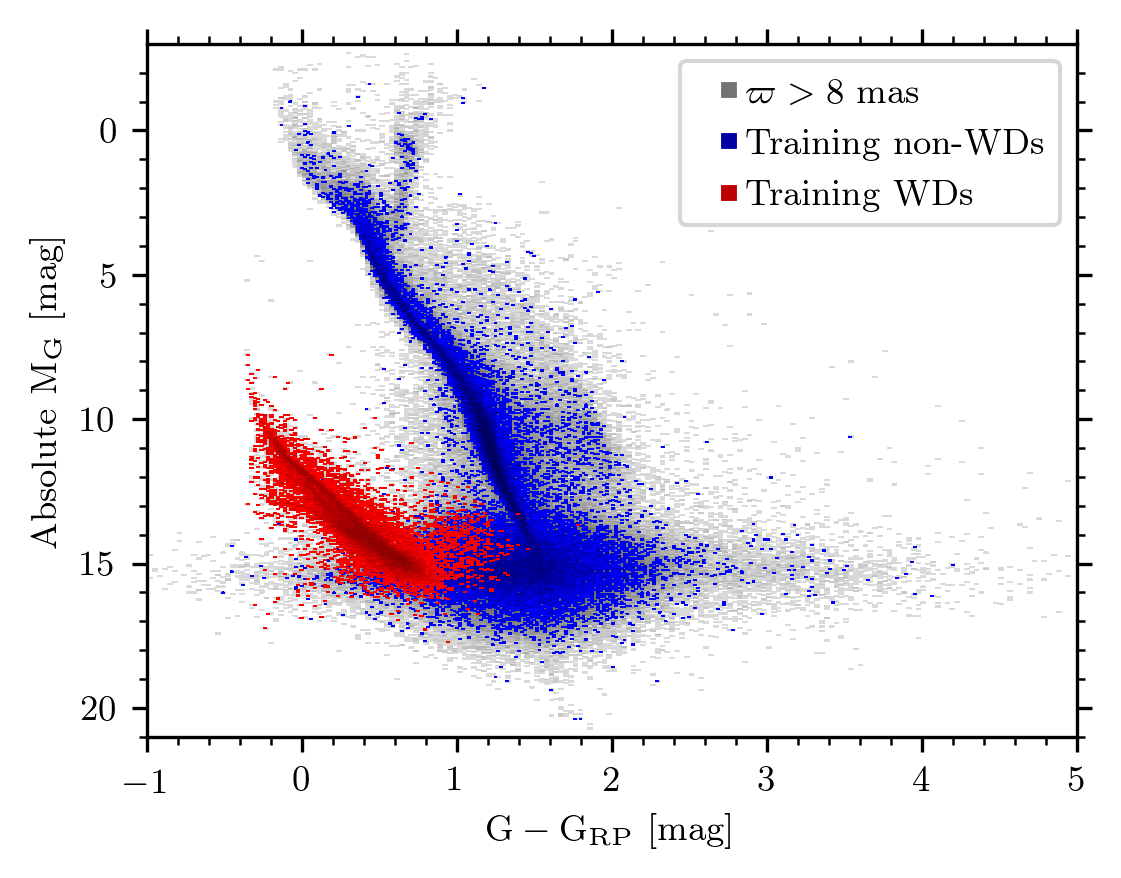}
}
\caption{CAMD for the training 
dataset based on which we classified the WDs. Red 
data points represent the WD population.
Grey points are all sources in the whole $\hat\varpi > 8$\,mas sample from which these samples were extracted.
The appearance of the test dataset is quite similar to the one plotted here.
\label{fig:HRdiagramTrainTestWDs}
} 
\end{figure}

After these known WDs were excluded from the whole set of $1\,040\,614$
sources with $\hat\varpi > 8$\,mas and \G\ photometry, we randomly selected
$40\,000$ and $37\,364$ sources to constitute the training and test
dataset of non-WD sources, respectively. We chose these particular
numbers of sources in order to use a training dataset with twice the
number of non-WDs with respect to the number of WDs, but four times
for the test dataset. This is useful to detect whether the ratio of
WDs in the sample analysed affects the classification.  This random
selection of non-WDs was made with the aim of maintaining the colour distribution
of the whole sample and better populating our sample with non-WDs with
blue colours that might be confused with WDs. Thus, we
selected 9.6\% of the sources having $G-\grp<0.75$\,mag and the rest
with larger $G-\grp$ values.
This selection resulted in a set of $60\,000$ training
and $46\,705$ test sources.
Their distribution in the CAMD for the training dataset is shown in 
Fig.\,\ref{fig:HRdiagramTrainTestWDs} (the CAMD diagram for the test dataset looks very similar). These
figures show the concentration of WDs in $G-\grp<1$~mag,
 increasing the normalised colour distribution in this
range. As explained, this was the main reason to better populate blue
colours in the non-WD dataset to avoid confusion with the WD
sample.





Using these datasets, we trained the random forest algorithm with the purpose of classifying WDs. We used the Python random forest classifier, performing a cross-validation to obtain the most appropriate hyperparameters\footnote{This cross-validation process returned the following best values for the hyperparameters: bootstrap= True, max\_depth= 20, max\_features= 'sqrt', min\_samples\_leaf= 4, min\_samples\_split= 10, n\_estimators= 30.}. As information for the classification of WDs we considered the three photometric \G\ magnitudes ($G, \gbp$  , and \grp), proper motions ($\mu_\alpha$ and $\mu_\delta$), and parallaxes ($\hat\varpi$), and also their uncertainties. 

The most important features that help in the WD classification are the red magnitude (24.7\% of the total weight) and the parallax uncertainty (14.5\%). The parallax uncertainty is more important than the parallax itself, which is, in fact, one of the least important parameters because WDs are well separated from non-WDs in a CAMD inside 125 pc.


After the algorithm was trained, we evaluated how well the test
dataset was classified. It correctly classified 9160 WDs (98.1\% of
the total) and $37\,214$ non-WDs (99.6\% of the total) in the test
dataset. The resulting list of WD candidates is contaminated by 147
non-WDs (representing 1.6\% of the list of WD candidates derived from
the test dataset). We then applied the classification algorithm to the
whole $\hat\varpi > 8$\,mas dataset with three passband
\G\ photometry. The random forest algorithm outputs a value
representing the probability of each source of being a WD.

\begin{figure}[!htb]
\center{
\includegraphics[width=\columnwidth]{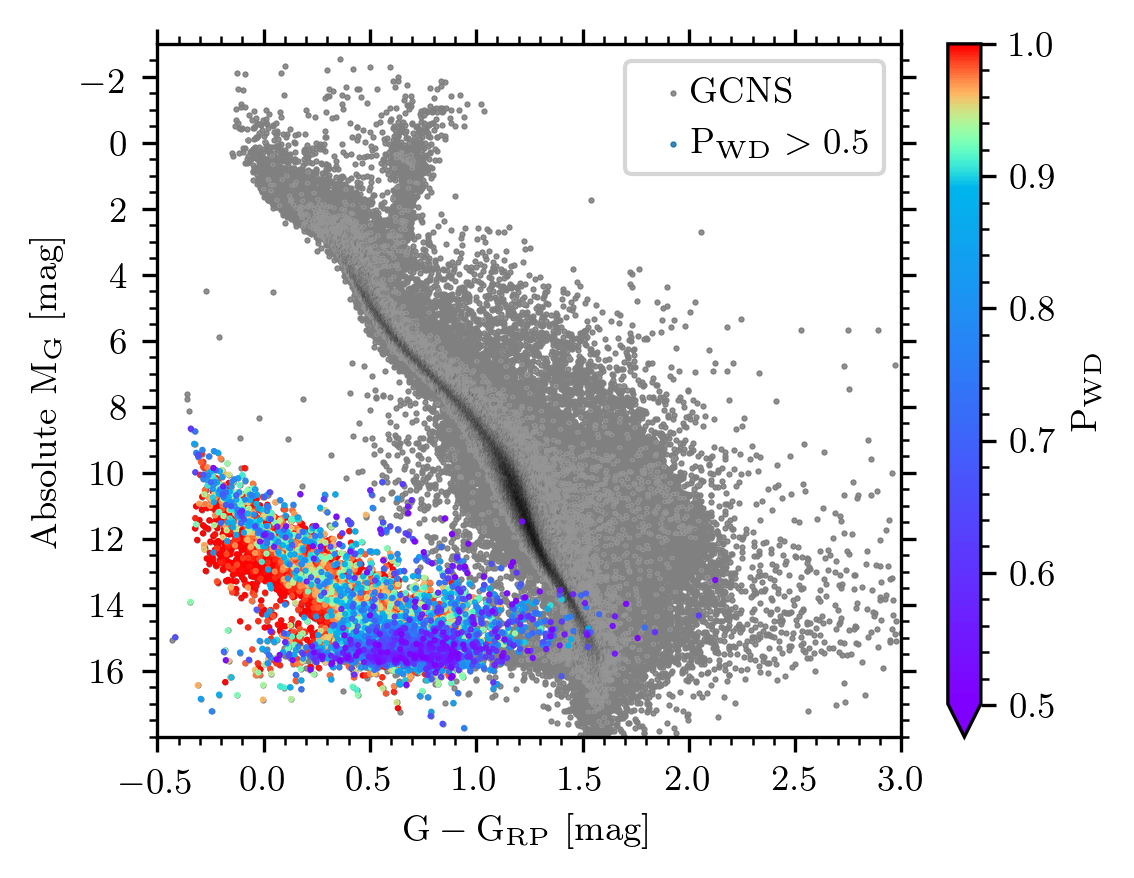}  
}
\caption{CAMD of sources included in the GCNS (grey) and the WD candidates obtained with the random forest classification algorithm having $P_{\rm WD}>0.5$ (with the value of $P_{\rm WD}$ shown with the colour index).
\label{fig:HRWDGCNS}
} 
\end{figure}

From the total of $1\,040\,614$ sources we found $32\,948$ sources
with a probability of being a WD ($P_{\rm WD}$) higher than
$0.5$\footnote{full table available from the CDS}. After the selection
in Sec.~\ref{sec:2Cataloggeneration}, $21\,848$ of these sources were
included in GCNS dataset\footnote{There are 7005 sources in GCNS that do not have all three \G\ photometric passbands. These
sources were not assigned any value for the probability of being
a WD.}. The CAMD of these WD candidates in
GCNS is shown in Fig.\,\ref{fig:HRWDGCNS}.  We verified that the
distribution in the sky of the WD candidates is homogeneous, as
it is expected to be in the 100\,pc bubble.
Of the 11\,106 sources with $P_{\rm WD}>0.5$ having $\hat\varpi > 8$\,mas
that are not included in GCNS catalogue, only 815 with
dist\_1<0.1 fail the GCNS $p$ criteria. They are at the red side of the WD locus, where more
contamination is expected, and it is therefore possible that they are not 
real WDs.

\begin{figure}[!htb]
\center{
\includegraphics[width=\columnwidth]{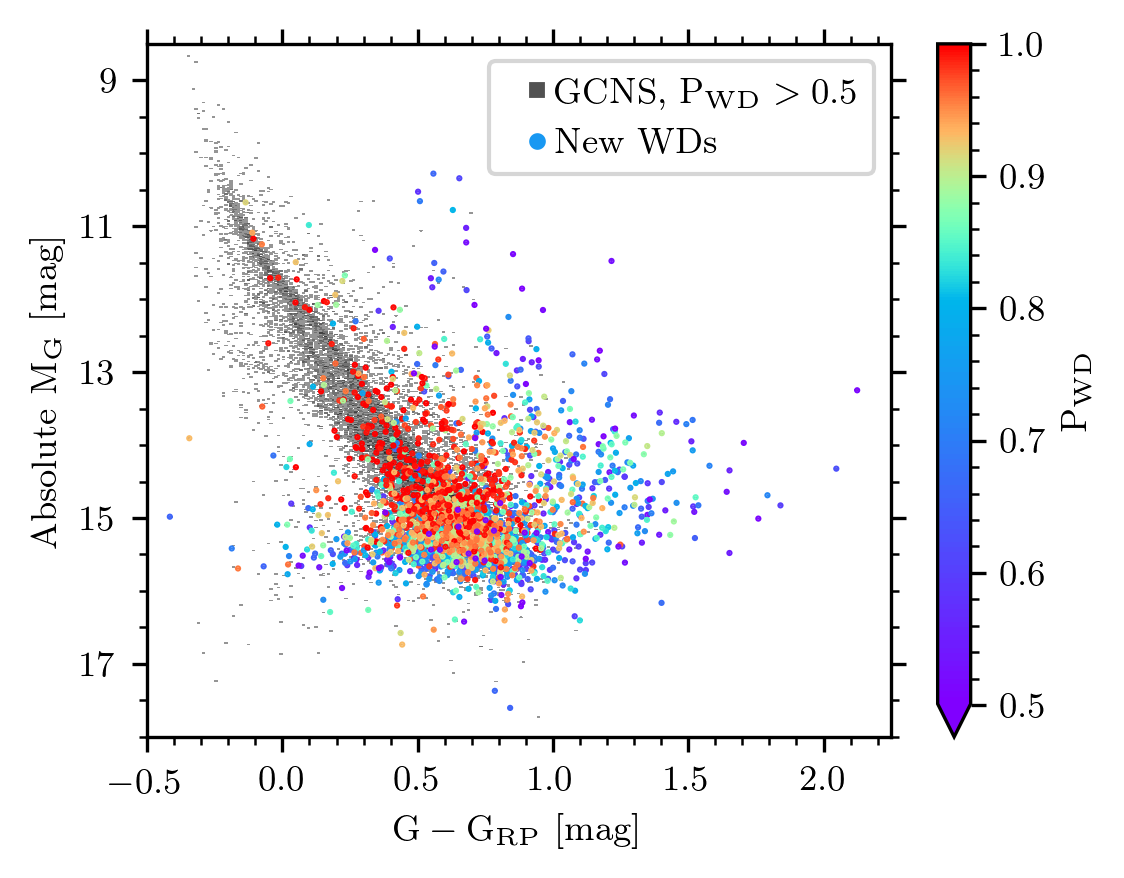}  
\includegraphics[width=\columnwidth]{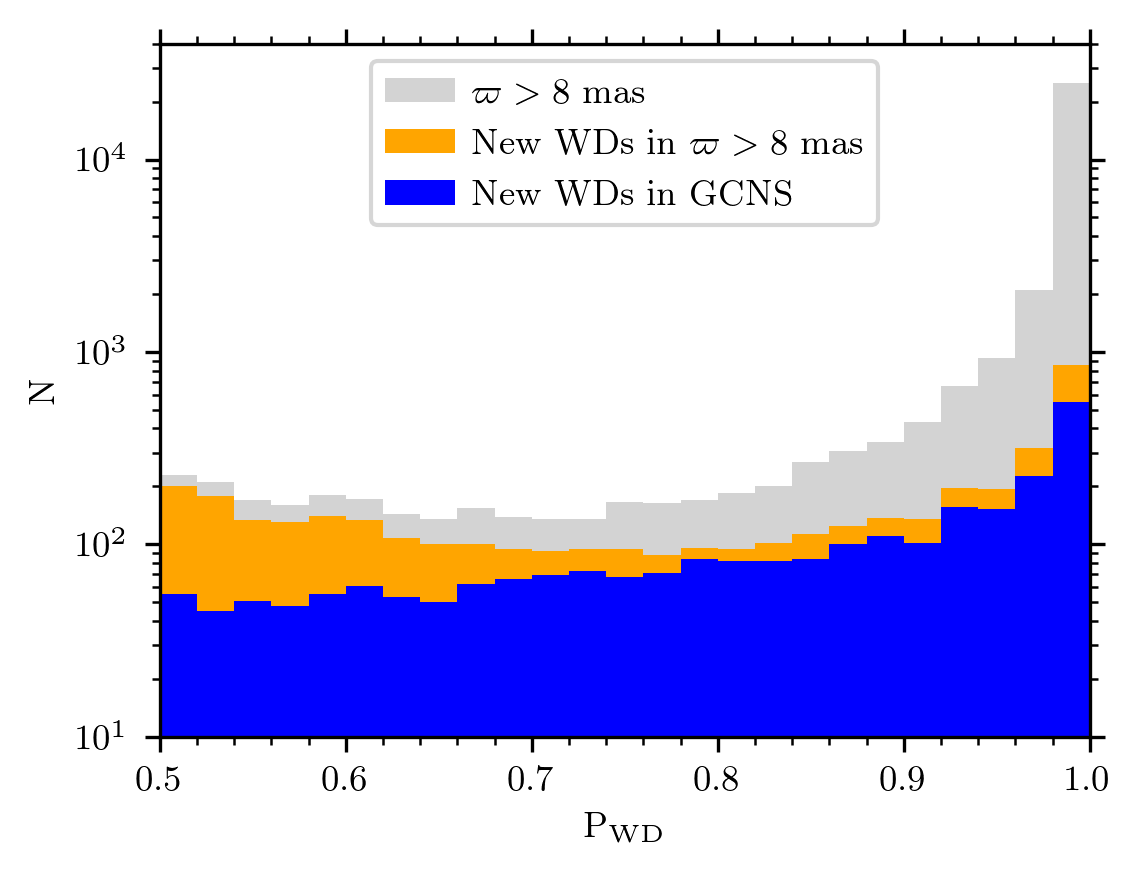}  
}
\caption{Top: CAMD with the new WDs found here that were not included in \cite{GentileFusillo2019}, \cite{Torres2019}, or \cite{JimenezEsteban2018}. Bottom: Probability distribution of the new WD candidates.
\label{fig:newWDs}
} 
\end{figure}

\begin{figure}
\center{
\includegraphics[width=0.45\textwidth]{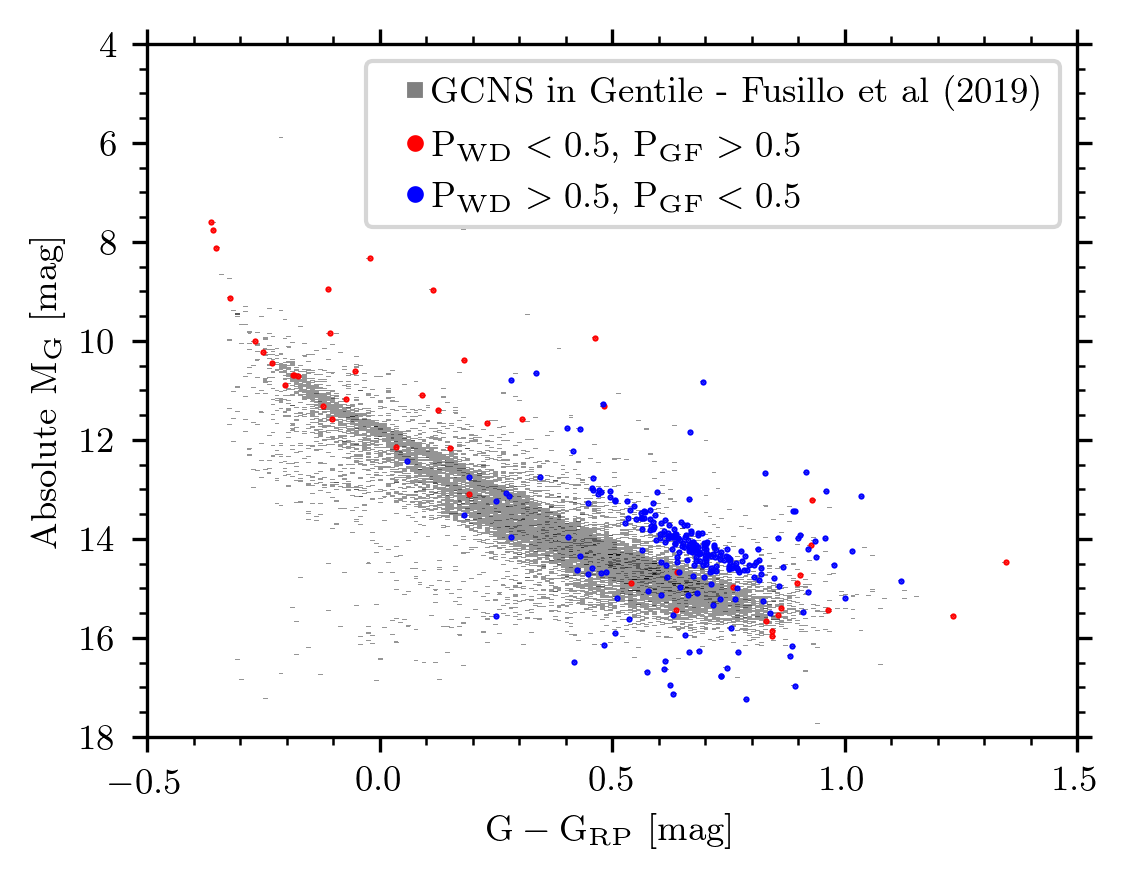}  
}
\caption{Position in the CAMD of the WD candidates in \cite{GentileFusillo2019}, $P_{\rm GF}$ , and our candidate $P_{\rm WD}$. Coloured points indicate contradictory conclusions between the two studies.\label{fig:knownWDs}
} 
\end{figure}

When we compared our list of WD candidates with the $29\,341$ WDs used
for training and testing the algorithm (extracted
from \citealt{GentileFusillo2019}, \citealt{Torres2019},
and \citealt{JimenezEsteban2018}), we detected 2553 new WD candidates
that were not included in the referenced bibliography (see
Fig.\,\ref{fig:newWDs} to see their position in the CAMD and
their probability distribution of being a WD). These new candidates
are mostly located in the red region of the WD locus, where the
contamination is expected to be higher.
In Fig.\,\ref{fig:knownWDs} we plot our WD candidates, $P_{\rm WD}$, and
those of \cite{GentileFusillo2019}, $P_{\rm GF}$. 

There are 250 sources assigned $P_{\rm WD}>0.5$ in this work, but which have a
low probability in \cite{GentileFusillo2019}. Based on their position in
the CAMD (blue points in Fig.\,\ref{fig:knownWDs}), we suggest
that this is probably due to very restrictive filtering
in \cite{GentileFusillo2019} because these sources are mostly concentrated
in the upper red part of the WD locus.  On the other hand,
45 sources with $P_{\rm WD}<0.5$ are in our work, but are present
in \cite{GentileFusillo2019} with $P_{\rm GF}>0.5$ (red points in
Fig.\,\ref{fig:knownWDs}). These red points include some sources that are
located in the upper blue region of the WD locus, where our algorithm
appears to fail to recognise these extreme sources as WDs. Some of these sources that are not recognised as WDs have very bright magnitudes compared with the training dataset we used (Sirius B and LAWD 37 are two examples of this). Because they are very few and are already contained in previous catalogues, they are easily recognisable. For completeness,
we decided to include these 45 sources as a supplementary table\footnote{available from the CDS}
including $P_{\rm GF}$ values.

\subsubsection{White dwarf luminosity function}
\label{sec:wdlf}

The white dwarf luminosity function (WDLF) 
tracks the collective
evolution of all WDs since they were formed.  Stars with
masses up to $\rm \sim8M_\odot$ \ will become WDs, 
and their individual luminosity is determined by a relatively simple
cooling law because all energy-generation processes have ceased, unless
they are part of a binary system where later mass transfer can heat
the envelope.  In simple terms, the stored energy in the isothermal
degenerate core of a WD is radiated into space through its
surface.  Therefore the rate of energy loss is determined to first
order by the core temperature and the surface area. Higher mass WDs
cool more slowly at a given temperature because their
radii are smaller. In reality, cooling rates are modified by the core composition,
which determines the core heat capacity, and by the composition and
structure of the envelope.  Several research groups have published
detailed evolutionary models that provide cooling curves for a range
of remnant masses (arising from the progenitor evolution) and core and
envelope compositions (see e.g. \citealt{2019ApJ...876...67B} and
references therein).  In principle, the shape of the WDLF reflects
historical star formation rates moderated by the distribution of main-sequence lifetimes and subsequent WD cooling
times. Furthermore, as the age of the galaxy exceeds the
combined main-sequence and cooling lifetimes of the oldest white
dwarfs, the cutoff at the highest absolute magnitude (lowest
luminosity) can provide a low limit to the age of the disc for
comparison with determinations from other methods.  The WDLF also
provides insight into physical processes in WD interiors. For
example, phase changes such as crystallisation release latent heat,
which delays the cooling for a time. Conversely, energy loss through
postulated dark matter particles (e.g. axions) might produce a
detectable enhancement in cooling, if they exist.

\begin{figure}
    \center{
    \includegraphics[width=0.45\textwidth]{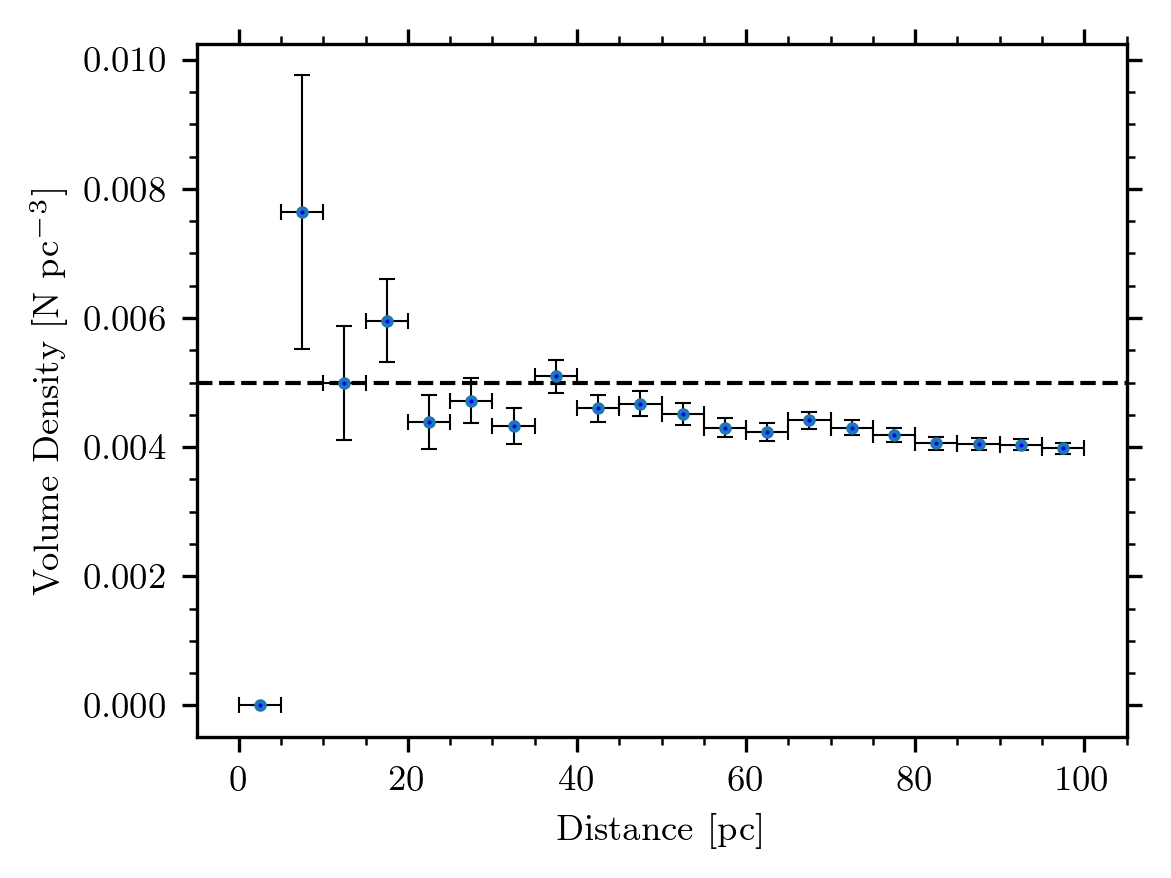}}
    \caption{Volume density of WDs as a function of distance. Values are computed for a 5\,pc wide spherical shell.}
    \label{fig:wd_num_dist}
\end{figure}

The GCNS WD catalogue presents an opportunity to derive an WDLF without
recourse to the considerably complex
corrections\,\citep{2019MNRAS.482..715L} required when treating
kinematically biased samples, especially those derived from reduced
proper motion\,\citep{2006AJ....131..571H,2011MNRAS.417...93R}.  The
GCNS sample is highly reliable and complete within a well--defined
survey volume. In principle, it thus enables a straightforward
derivation of the WDLF.  However, the relatively low luminosity of white
dwarfs compared to the apparent magnitude limit of the \G\ catalogue
leads to some incompleteness within 100\,pc.
We calculated  the WD volume density as a function
of distance (fig.\, \ref{fig:wd_num_dist}). The values shown were
calculated based on the number of WDs in a spherical shell with a
width of 5\,pc for each distance point. Within a distance of 40\,pc, the WD
volume density measurements show scatter from statistical number count
fluctuations, but then show a clear decline by approximately 15\% \ of
the value between 40\,pc and 100\,pc, likely a consequence of the combined
effect of the \G\ magnitude limit and a vertical decline in density
in the disc. 

\begin{figure*}
    \sidecaption
    \includegraphics[width=12cm]{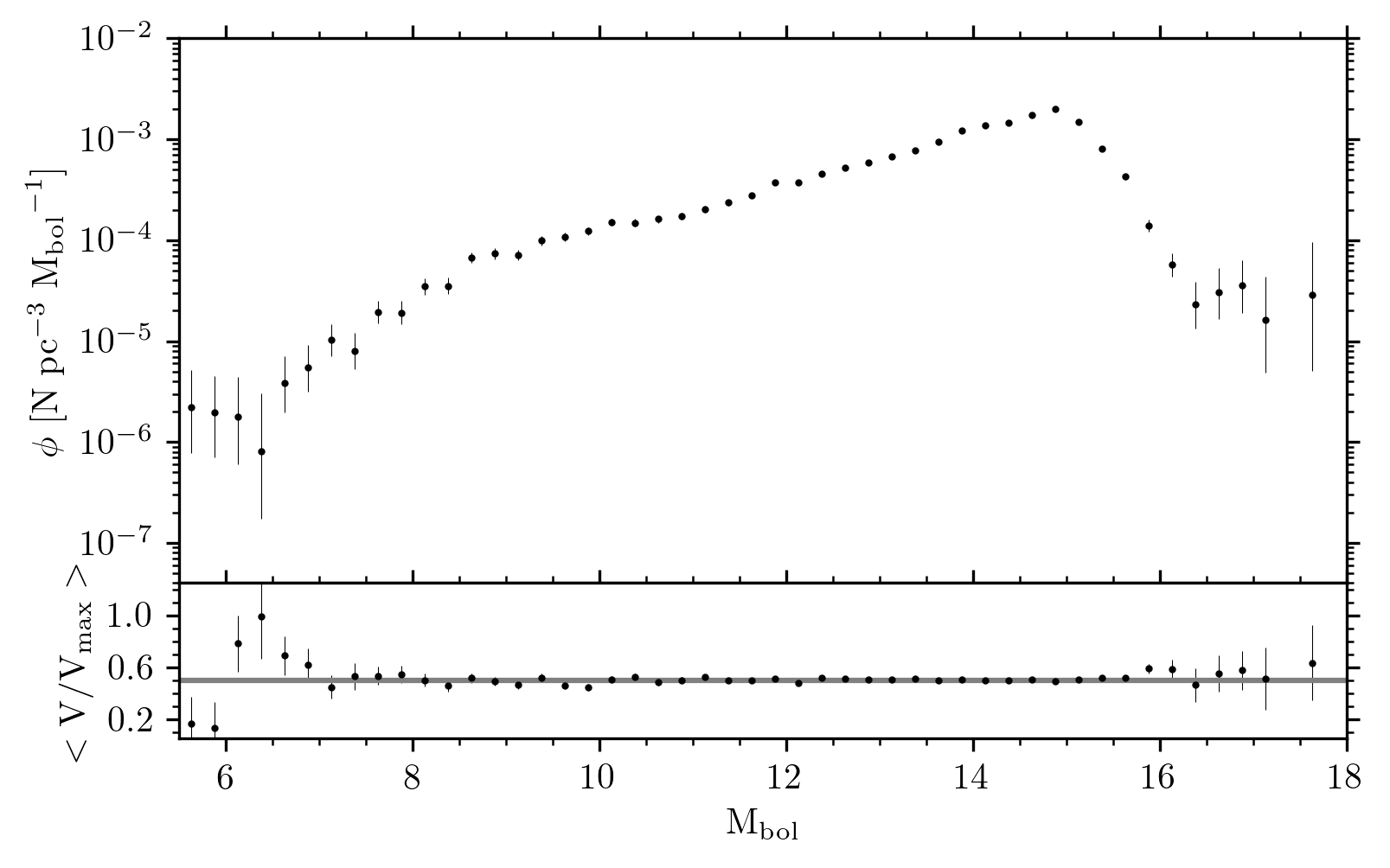}
    \caption{Upper panel: WDLF for the 100\,pc sample. The bin width is 0.25M$_{\rm bol}$ and the confidence intervals are Poisson uncertainties~\citep{1986ApJ...303..336G}. The structure and features in the WDLF are statistically significant for all but the first and last few bins. \newline ~ \newline ~ \newline 
    Lower panel: V/$V_{Max}$  statistic for the WDLF sample plotted in the upper panel. The expectation value for the statistic is~0.5 for a uniform sample within the survey volume (see the main text for further details).\newline ~ \newline ~ }
    \label{fig:wdlfmain_vvmax}
\end{figure*}

We again employed the 1/$V_{Max}$  technique as detailed in Section\,\ref{sec:lf} in bins of bolometric magnitude.
An advantage in
deriving the WDLF from \G\ data lies in measuring bolometric magnitudes for the fainter WDs. The white-light G passband measures a large
fraction of the flux of the cooler WDs where bolometric corrections
are only a weak function of effective temperature. We made the
simplifying assumption that bolometric corrections $({\rm M}_{\rm bol}
- {\rm M}_{\rm G})$ can be taken from pure hydrogen models\,(we
employed those of the Montreal group: \cite{2019ApJ...876...67B} and references therein) for all WDs,
ignoring the effects of varying the H/He atmospheric composition and
surface gravity. We interpolated amongst the model tabulated
values to look up G-band bolometric corrections as a function of
$(G-G_{\rm RP})$ to correct ${\rm M}_{\rm G}$ to ${\rm
M}_{\rm bol}$. For effective temperatures $T_{\rm eff} < 4500$\,K, we
assumed the bolometric correction in $G = 0$ because the pure--DA model bolometric--correction--colour
relationship is non--monotonic due to the effects of
collisionally induced opacities in the high-pressure atmospheres. The
model grids indicate that inaccuracies introduced by these
simplifying assumptions are never more than a few tenths of a
magnitude and are limited to the intermediately hot and very cool
effective temperature ranges of the scale.
Our resulting WDLF is displayed
in Fig.\,\ref{fig:wdlfmain_vvmax} ,{ where we also} show the
mean V/$V_{Max}$  statistic as a function of bolometric magnitude for the
sample. 
As described in~\citet{2011MNRAS.417...93R}, for a sample
uniformly distributed within the (generalised) survey volume, the
expectation value of this statistic is~$0.5\pm1/\surd12N$ for~$N$
stars.  For the WDLF sample we find overall
V/$V_{Max}$ \,$=0.5050\pm0.0023$, with no obvious indications of systematic
effects as a function of luminosity or position.

The statistical power of the GCNS sample is evident in Fig.\,\ref{fig:wdlfmain_vvmax}. At the peak of the WDLF, nearly~1900 WDs contribute in the bin range $14.75 < {\rm M}_{\rm bol} < 15.00$. There appears to be evidence of 
a series of features in the WDLF at high confidence: the feature around M$_{\rm bol}=10.5$ has been noted previously\,\citep{2015ApJS..219...19L,2006AJ....131..571H} but those at fainter levels (e.g.~around M$_{\rm bol}=14.25$) have not been so apparent. The segments between these features are linear and consistent in gradient, resulting in an apparent series of steps. 
The high signal-to-noise ratio and detail in this WDLF will facilitate derivation of star formation histories with inversion techniques~\citep{2013MNRAS.434.1549R}. 
The peak itself appears
broader than some recent determinations, and especially so with respect to simulations, although this may be the result of simplifying assumptions in such population synthesis codes as noted by~\citet{2015ApJS..219...19L}. Furthermore, the peak appears to be slightly brighter than M$_{\rm bol}=15,$ whereas several recent determinations have reported the peak to be slightly fainter than this level. This may be an age effect, where the greater volumes studied in deep proper-motion-selected samples will net larger fractions of older thick-disc and spheroid WDs.



\section{Conclusions}

We have provided a well-characterised catalogue of objects within 100\,pc
of the Sun. In this catalogue we inferred a distance probability density function  for all
sources using the parallaxes and a single distance prior
that takes the observational parallax cut at 8\,mas and
the distribution of parallax uncertainties in \gdrthree\ into account.  We provide all-sky maps
at HEALpix level 5 of empirical magnitude limits, which we generated
using all \gdrthree entries with a $G$ magnitude and parallax
measurement. We base our magnitude limit estimator on the G magnitude
distribution per HEALpix and advocate a limit between the 80th
(conservative) and 90th (optimistic) percentile. 

The GCNS catalogue has an estimated \NDISTTEN\ entries within 100\,pc. This is an increase of
an order of magnitude with respect to the most complete nearby star
census prior to the \G\ mission. A comparison with \gdrtwo shows that the last
release contained more contamination than \gdrthree, but also that
a few percent of real objects are still not included in
\gdrthree. The overall completeness of the GCNS to M8 at 100\,pc is probably better than 95\%.
An examination of the 10\,pc sample finds that we provide the
first direct parallax of five stars in multiple systems. 

%

The GCNS was used to undertake a number of investigations into local
populations, structures, and distributions. We list this below. 
\begin{itemize}
     \item We computed the luminosity function from the brightest main-sequence 
       stars ($M_G=2$), including part of giant stars, to the late-L
       brown dwarfs ($M_G=20.5$). We found an overall density of
       $0.081\pm0.003$ (main sequence) stars pc$^{-3}$. The high
       signal-to-noise ratio of the luminosity function indicates
       features such as the Jao gap \citep{2018ApJ...861L..11J} and the
       drop in object counts at the stellar to substellar boundary
       \citep{2019ApJ...883..205B}.
    \item We explored the kinematical plane for the GCNS stars that have a
      \G\ radial velocity (74\,281 stars).  We show that even in
      the local sample, the kinematical plane shows substructures in
      the disc that are associated with several streams and superclusters, such
      as Sirius and Hercules, and in the halo, where we identified
      12 stars from the \G\ Enceladus.
    \item We provide orbits for the sample. As expected, most of the
      stars have circular in-plane orbits similar to the
      Sun. However, the solar neighbourhood is also visited by several
      tens of stars with eccentric orbits that come from the Galactic
      central regions, as well as stars coming from external
      regions, for example the Enceladus objects. 
    \item We briefly investigated the value for the solar motion,
      proposed a revision of the \Vsun value to 7 km/s, and
      discussed the vertex deviation.    
    \item We find 2879 new UCD candidates compared to the \gdrtwo,\ but
    we also note that the very nearby binary brown dwarf Luhman 16 AB does not have a 
     five-parameter solution in the \gdrthree.
    \item We provided a revised catalogue of 16\,556 high-probability
      resolved binary candidates. We confirmed the absence of
      bimodality in the physical projected separations distribution,
      placing previous DR2-based results on more solid ground. We
      refined the wide-binary fraction statistics as a function of
      spectral type, quantifying the decline in $f_\mathrm{WB}$ for
      later (K and M) spectral types.
    \item We re-examined the Hyades cluster and produced a list of
      candidate members using a procedure that did not use the GCNS selection criteria. We found only one candidate that would not have have made the GCNS.
    \item Using a random forest algorithm, we identified 21\,848
      sources with a high probability of being a WD, 2553 of
      which are new WD candidates. 
     \item We derived a white dwarf luminosity function of
       unprecedented statistical power. Several features are clearly
       present that appear as a series of steps in the function. These may
       be indicative of variations in the historical star formation
       rate in the 100\,pc volume and can be examined further by
       direction inversion techniques or comparison with population
       synthesis calculations.
\end{itemize}

In these investigations we have illustrated different ways of using
the GCNS: the direct use of the astrometric parameters
(Sects.\,\ref{sec:verticalstrat} and \ref{sec:clusters}), the use of
derived distance PDFs (Sect.\,\ref{sec:lf}), and derived quantities
(Sect.\,\ref{sec:Kinematics}). We indicated other quality cuts
that can be made to clean the catalogue using photometric flags
(Sect.\,\ref{sec:wd}) and indicators of binarity
(Sect.\,\ref{sec:lf}). Finally, we have shown that even though we know that the
catalogue 
volume is incomplete, useful
conclusions and constraints can be drawn (Sect.\,\ref{sec:ucds}).

We expect the next releases of the \G\ mission to improve the
GCNS in particular with the inclusion of unresolved companions and
with the application of non-single star solutions in the
\G\ processing chain where the current single-star solution will often
result in erroneous astrometric parameters. In addition, the \G~DR3,
due to be released in 2021, will provide astrophysical
parameters for nearly all the stellar sources in the \G\ Catalogue of Nearby Stars.


\section{Acknowledgements\label{sec:acknowl}}
We thank the anonymous referee for comments and suggestions that improved this article. This work presents results from the European Space Agency (ESA) space
mission \gaia. \gaia\ data are being processed by the \gaia\ Data
Processing and Analysis Consortium (DPAC). Funding for the DPAC is
provided by national institutions, in particular the institutions
participating in the \gaia\ MultiLateral Agreement (MLA). The
\gaia\ mission website is \url{https://www.cosmos.esa.int/gaia}. The
\gaia\ archive website is \url{https://archives.esac.esa.int/gaia}.

The \gaia\ mission and data processing have financially been supported by, in alphabetical order by country:
\begin{itemize}
\item the Algerian Centre de Recherche en Astronomie, Astrophysique et G\'{e}ophysique of Bouzareah Observatory;
\item the Austrian Fonds zur F\"{o}rderung der wissenschaftlichen Forschung (FWF) Hertha Firnberg Programme through grants T359, P20046, and P23737;
\item the BELgian federal Science Policy Office (BELSPO) through various PROgramme de D\'eveloppement d'Exp\'eriences scientifiques (PRODEX) grants and the Polish Academy of Sciences - Fonds Wetenschappelijk Onderzoek through grant VS.091.16N, and the Fonds de la Recherche Scientifique (FNRS);
\item the Brazil-France exchange programmes Funda\c{c}\~{a}o de Amparo \`{a} Pesquisa do Estado de S\~{a}o Paulo (FAPESP) and Coordena\c{c}\~{a}o de Aperfeicoamento de Pessoal de N\'{\i}vel Superior (CAPES) - Comit\'{e} Fran\c{c}ais d'Evaluation de la Coop\'{e}ration Universitaire et Scientifique avec le Br\'{e}sil (COFECUB);
\item the National Science Foundation of China (NSFC) through grants 11573054 and 11703065 and the China Scholarship Council through grant 201806040200;  
\item the Tenure Track Pilot Programme of the Croatian Science Foundation and the \'{E}cole Polytechnique F\'{e}d\'{e}rale de Lausanne and the project TTP-2018-07-1171 'Mining the Variable Sky', with the funds of the Croatian-Swiss Research Programme;
\item the Czech-Republic Ministry of Education, Youth, and Sports through grant LG 15010 and INTER-EXCELLENCE grant LTAUSA18093, and the Czech Space Office through ESA PECS contract 98058;
\item the Danish Ministry of Science;
\item the Estonian Ministry of Education and Research through grant IUT40-1;
\item the European Commission’s Sixth Framework Programme through the European Leadership in Space Astrometry (\href{https://www.cosmos.esa.int/web/gaia/elsa-rtn-programme}{ELSA}) Marie Curie Research Training Network (MRTN-CT-2006-033481), through Marie Curie project PIOF-GA-2009-255267 (Space AsteroSeismology \& RR Lyrae stars, SAS-RRL), and through a Marie Curie Transfer-of-Knowledge (ToK) fellowship (MTKD-CT-2004-014188); the European Commission's Seventh Framework Programme through grant FP7-606740 (FP7-SPACE-2013-1) for the \gaia\ European Network for Improved data User Services (\href{https://gaia.ub.edu/twiki/do/view/GENIUS/}{GENIUS}) and through grant 264895 for the \gaia\ Research for European Astronomy Training (\href{https://www.cosmos.esa.int/web/gaia/great-programme}{GREAT-ITN}) network;
\item the European Research Council (ERC) through grants 320360 and 647208 and through the European Union’s Horizon 2020 research and innovation and excellent science programmes through Marie Sk{\l}odowska-Curie grant 745617 as well as grants 670519 (Mixing and Angular Momentum tranSport of massIvE stars -- MAMSIE), 687378 (Small Bodies: Near and Far), 682115 (Using the Magellanic Clouds to Understand the Interaction of Galaxies), and 695099 (A sub-percent distance scale from binaries and Cepheids -- CepBin);
\item the European Science Foundation (ESF), in the framework of the \gaia\ Research for European Astronomy Training Research Network Programme (\href{https://www.cosmos.esa.int/web/gaia/great-programme}{GREAT-ESF});
\item the European Space Agency (ESA) in the framework of the \gaia\ project, through the Plan for European Cooperating States (PECS) programme through grants for Slovenia, through contracts C98090 and 4000106398/12/NL/KML for Hungary, and through contract 4000115263/15/NL/IB for Germany;
\item the Academy of Finland and the Magnus Ehrnrooth Foundation;
\item the French Centre National d’Etudes Spatiales (CNES), the Agence Nationale de la Recherche (ANR) through grant ANR-10-IDEX-0001-02 for the 'Investissements d'avenir' programme, through grant ANR-15-CE31-0007 for project 'Modelling the Milky Way in the Gaia era' (MOD4Gaia), through grant ANR-14-CE33-0014-01 for project 'The Milky Way disc formation in the Gaia era' (ARCHEOGAL), and through grant ANR-15-CE31-0012-01 for project 'Unlocking the potential of Cepheids as primary distance calibrators' (UnlockCepheids), the Centre National de la Recherche Scientifique (CNRS) and its SNO Gaia of the Institut des Sciences de l’Univers (INSU), the 'Action F\'{e}d\'{e}ratrice Gaia' of the Observatoire de Paris, the R\'{e}gion de Franche-Comt\'{e}, and the Programme National de Gravitation, R\'{e}f\'{e}rences, Astronomie, et M\'{e}trologie (GRAM) of CNRS/INSU with the Institut National Polytechnique (INP) and the Institut National de Physique nucléaire et de Physique des Particules (IN2P3) co-funded by CNES;
\item the German Aerospace Agency (Deutsches Zentrum f\"{u}r Luft- und Raumfahrt e.V., DLR) through grants 50QG0501, 50QG0601, 50QG0602, 50QG0701, 50QG0901, 50QG1001, 50QG1101, 50QG1401, 50QG1402, 50QG1403, 50QG1404, and 50QG1904 and the Centre for Information Services and High Performance Computing (ZIH) at the Technische Universit\"{a}t (TU) Dresden for generous allocations of computer time;
\item the Hungarian Academy of Sciences through the Lend\"{u}let Programme grants LP2014-17 and LP2018-7 and through the Premium Postdoctoral Research Programme (L.~Moln\'{a}r), and the Hungarian National Research, Development, and Innovation Office (NKFIH) through grant KH\_18-130405;
\item the Science Foundation Ireland (SFI) through a Royal Society - SFI University Research Fellowship (M.~Fraser);
\item the Israel Science Foundation (ISF) through grant 848/16;
\item the Agenzia Spaziale Italiana (ASI) through contracts I/037/08/0, I/058/10/0, 2014-025-R.0, 2014-025-R.1.2015, and 2018-24-HH.0 to the Italian Istituto Nazionale di Astrofisica (INAF), contract 2014-049-R.0/1/2 to INAF for the Space Science Data Centre (SSDC, formerly known as the ASI Science Data Center, ASDC), contracts I/008/10/0, 2013/030/I.0, 2013-030-I.0.1-2015, and 2016-17-I.0 to the Aerospace Logistics Technology Engineering Company (ALTEC S.p.A.), INAF, and the Italian Ministry of Education, University, and Research (Ministero dell'Istruzione, dell'Universit\`{a} e della Ricerca) through the Premiale project 'MIning The Cosmos Big Data and Innovative Italian Technology for Frontier Astrophysics and Cosmology' (MITiC);
\item the Netherlands Organisation for Scientific Research (NWO) through grant NWO-M-614.061.414, through a VICI grant (A.~Helmi), and through a Spinoza prize (A.~Helmi), and the Netherlands Research School for Astronomy (NOVA);
\item the Polish National Science Centre through HARMONIA grant 2018/06/M/ST9/00311, DAINA grant 2017/27/L/ST9/03221, and PRELUDIUM grant 2017/25/N/ST9/01253, and the Ministry of Science and Higher Education (MNiSW) through grant DIR/WK/2018/12;
\item the Portugese Funda\c{c}\~ao para a Ci\^{e}ncia e a Tecnologia (FCT) through grants SFRH/BPD/74697/2010 and SFRH/BD/128840/2017 and the Strategic Programme UID/FIS/00099/2019 for CENTRA;
\item the Slovenian Research Agency through grant P1-0188;
\item the Spanish Ministry of Economy (MINECO/FEDER, UE) through grants ESP2016-80079-C2-1-R, ESP2016-80079-C2-2-R, RTI2018-095076-B-C21, RTI2018-095076-B-C22, BES-2016-078499, and BES-2017-083126 and the Juan de la Cierva formaci\'{o}n 2015 grant FJCI-2015-2671, the Spanish Ministry of Education, Culture, and Sports through grant FPU16/03827, the Spanish Ministry of Science and Innovation (MICINN) through grant AYA2017-89841P for project 'Estudio de las propiedades de los f\'{o}siles estelares en el entorno del Grupo Local' and through grant TIN2015-65316-P for project 'Computaci\'{o}n de Altas Prestaciones VII', the Severo Ochoa Centre of Excellence Programme of the Spanish Government through grant SEV2015-0493, the Institute of Cosmos Sciences University of Barcelona (ICCUB, Unidad de Excelencia ’Mar\'{\i}a de Maeztu’) through grants MDM-2014-0369 and CEX2019-000918-M, the University of Barcelona's official doctoral programme for the development of an R+D+i project through an Ajuts de Personal Investigador en Formaci\'{o} (APIF) grant, the Spanish Virtual Observatory through project AyA2017-84089, the Galician Regional Government, Xunta de Galicia, through grants ED431B-2018/42 and ED481A-2019/155, support received from the Centro de Investigaci\'{o}n en Tecnolog\'{\i}as de la Informaci\'{o}n y las Comunicaciones (CITIC) funded by the Xunta de Galicia, the Xunta de Galicia and the Centros Singulares de Investigaci\'{o}n de Galicia for the period 2016-2019 through CITIC, the European Union through the European Regional Development Fund (ERDF) / Fondo Europeo de Desenvolvemento Rexional (FEDER) for the Galicia 2014-2020 Programme through grant ED431G-2019/01, the Red Espa\~{n}ola de Supercomputaci\'{o}n (RES) computer resources at MareNostrum, the Barcelona Supercomputing Centre - Centro Nacional de Supercomputaci\'{o}n (BSC-CNS) through activities AECT-2016-1-0006, AECT-2016-2-0013, AECT-2016-3-0011, and AECT-2017-1-0020, the Departament d'Innovaci\'{o}, Universitats i Empresa de la Generalitat de Catalunya through grant 2014-SGR-1051 for project 'Models de Programaci\'{o} i Entorns d'Execuci\'{o} Parallels' (MPEXPAR), and Ramon y Cajal Fellowship RYC2018-025968-I;
\item the Swedish National Space Agency (SNSA/Rymdstyrelsen);
\item the Swiss State Secretariat for Education, Research, and Innovation through
the Mesures d’Accompagnement, the Swiss Activit\'es Nationales Compl\'ementaires, and the Swiss National Science Foundation;
\item the United Kingdom Particle Physics and Astronomy Research Council (PPARC), the United Kingdom Science and Technology Facilities Council (STFC), and the United Kingdom Space Agency (UKSA) through the following grants to the University of Bristol, the University of Cambridge, the University of Edinburgh, the University of Leicester, the Mullard Space Sciences Laboratory of University College London, and the United Kingdom Rutherford Appleton Laboratory (RAL): PP/D006511/1, PP/D006546/1, PP/D006570/1, ST/I000852/1, ST/J005045/1, ST/K00056X/1, ST/K000209/1, ST/K000756/1, ST/L006561/1, ST/N000595/1, ST/N000641/1, ST/N000978/1, ST/N001117/1, ST/S000089/1, ST/S000976/1, ST/S001123/1, ST/S001948/1, ST/S002103/1, and ST/V000969/1.
\end{itemize}

The \gaia\ project, data processing and this contribution have made use of:
\begin{itemize}
\item the Set of Identifications, Measurements, and Bibliography for Astronomical Data \citep[SIMBAD,][]{2000A&AS..143....9W}, the 'Aladin sky atlas' \citep{2000A&AS..143...33B,2014ASPC..485..277B}, and the VizieR catalogue access tool \citep{2000A&AS..143...23O}, all operated at the Centre de Donn\'ees astronomiques de Strasbourg (\href{http://cds.u-strasbg.fr/}{CDS});
\item the National Aeronautics and Space Administration (NASA) Astrophysics Data System (\href{http://adsabs.harvard.edu/abstract_service.html}{ADS});
\item the software products \href{http://www.starlink.ac.uk/topcat/}{TOPCAT}, and \href{http://www.starlink.ac.uk/stilts}{STILTS} \citep{2005ASPC..347...29T,2006ASPC..351..666T};
\item Astropy, a community-developed core Python package for Astronomy \citep{2018AJ....156..123A};
\item data products from the Two Micron All Sky Survey \citep[2MASS,][]{2006AJ....131.1163S}, which is a joint project of the University of Massachusetts and the Infrared Processing and Analysis Center (IPAC) / California Institute of Technology, funded by the National Aeronautics and Space Administration (NASA) and the National Science Foundation (NSF) of the USA;
\item the first data release of the Pan-STARRS survey \citep{2016arXiv161205560C}
The Pan-STARRS1 Surveys (PS1) and the PS1 public science archive have been made possible through contributions by the Institute for Astronomy, the University of Hawaii, the Pan-STARRS Project Office, the Max-Planck Society and its participating institutes, the Max Planck Institute for Astronomy, Heidelberg and the Max Planck Institute for Extraterrestrial Physics, Garching, The Johns Hopkins University, Durham University, the University of Edinburgh, the Queen's University Belfast, the Harvard-Smithsonian Center for Astrophysics, the Las Cumbres Observatory Global Telescope Network Incorporated, the National Central University of Taiwan, the Space Telescope Science Institute, the National Aeronautics and Space Administration (NASA) through grant NNX08AR22G issued through the Planetary Science Division of the NASA Science Mission Directorate, the National Science Foundation through grant AST-1238877, the University of Maryland, Eotvos Lorand University (ELTE), the Los Alamos National Laboratory, and the Gordon and Betty Moore Foundation;
\item data products from the Wide-field Infrared Survey Explorer (WISE), which is a joint project of the University of California, Los Angeles, and the Jet Propulsion Laboratory/California Institute of Technology, and NEOWISE, which is a project of the Jet Propulsion Laboratory/California Institute of Technology. WISE and NEOWISE are funded by the National Aeronautics and Space Administration (NASA);
\item the fifth data release of the Radial Velocity Experiment \citep[RAVE DR5,][]{2017AJ....153...75K}. Funding for RAVE has been provided by the Australian Astronomical Observatory, the Leibniz-Institut f\"ur Astrophysik Potsdam (AIP), the Australian National University, the Australian Research Council, the French National Research Agency, the German Research Foundation (SPP 1177 and SFB 881), the European Research Council (ERC-StG 240271 Galactica), the Istituto Nazionale di Astrofisica at Padova, The Johns Hopkins University, the National Science Foundation of the USA (AST-0908326), the W. M. Keck foundation, the Macquarie University, the Netherlands Research School for Astronomy, the Natural Sciences and Engineering Research Council of Canada, the Slovenian Research Agency, the Swiss National Science Foundation, the Science \& Technology Facilities Council of the UK, Opticon, Strasbourg Observatory, and the Universities of Groningen, Heidelberg, and Sydney. The RAVE website is at \url{https://www.rave-survey.org/};
\item the thirteenth release of the Sloan Digital Sky Survey \citep[SDSS DR13,][]{2017ApJS..233...25A}. Funding for SDSS-IV has been provided by the Alfred P. Sloan Foundation, the United States Department of Energy Office of Science, and the Participating Institutions. SDSS-IV acknowledges support and resources from the Center for High-Performance Computing at the University of Utah. The SDSS web site is \url{https://www.sdss.org/}. SDSS-IV is managed by the Astrophysical Research Consortium for the Participating Institutions of the SDSS Collaboration including the Brazilian Participation Group, the Carnegie Institution for Science, Carnegie Mellon University, the Chilean Participation Group, the French Participation Group, Harvard-Smithsonian Center for Astrophysics, Instituto de Astrof\'isica de Canarias, The Johns Hopkins University, Kavli Institute for the Physics and Mathematics of the Universe (IPMU) / University of Tokyo, the Korean Participation Group, Lawrence Berkeley National Laboratory, Leibniz Institut f\"ur Astrophysik Potsdam (AIP),  Max-Planck-Institut f\"ur Astronomie (MPIA Heidelberg), Max-Planck-Institut f\"ur Astrophysik (MPA Garching), Max-Planck-Institut f\"ur Extraterrestrische Physik (MPE), National Astronomical Observatories of China, New Mexico State University, New York University, University of Notre Dame, Observat\'ario Nacional / MCTI, The Ohio State University, Pennsylvania State University, Shanghai Astronomical Observatory, United Kingdom Participation Group, Universidad Nacional Aut\'onoma de M\'exico, University of Arizona, University of Colorado Boulder, University of Oxford, University of Portsmouth, University of Utah, University of Virginia, University of Washington, University of Wisconsin, Vanderbilt University, and Yale University;
\item the second release of the SkyMapper catalogue \citep[SkyMapper DR2,][Digital Object Identifier 10.25914/5ce60d31ce759]{2019PASA...36...33O}. The national facility capability for SkyMapper has been funded through grant LE130100104 from the Australian Research Council (ARC) Linkage Infrastructure, Equipment, and Facilities (LIEF) programme, awarded to the University of Sydney, the Australian National University, Swinburne University of Technology, the University of Queensland, the University of Western Australia, the University of Melbourne, Curtin University of Technology, Monash University, and the Australian Astronomical Observatory. SkyMapper is owned and operated by The Australian National University's Research School of Astronomy and Astrophysics. The survey data were processed and provided by the SkyMapper Team at the the Australian National University. The SkyMapper node of the All-Sky Virtual Observatory (ASVO) is hosted at the National Computational Infrastructure (NCI). Development and support the SkyMapper node of the ASVO has been funded in part by Astronomy Australia Limited (AAL) and the Australian Government through the Commonwealth's Education Investment Fund (EIF) and National Collaborative Research Infrastructure Strategy (NCRIS), particularly the National eResearch Collaboration Tools and Resources (NeCTAR) and the Australian National Data Service Projects (ANDS).
\end{itemize}

%
%
 \bibliographystyle{aa}
 \bibliography{refs} 

\begin{appendix}
\section{Details of the Random Forest classifier parameters and training set.}

\subsection{Colour-Absolute Magnitude Diagrams}

\begin{figure}[!htb]
\center{\includegraphics[width=0.5\textwidth]{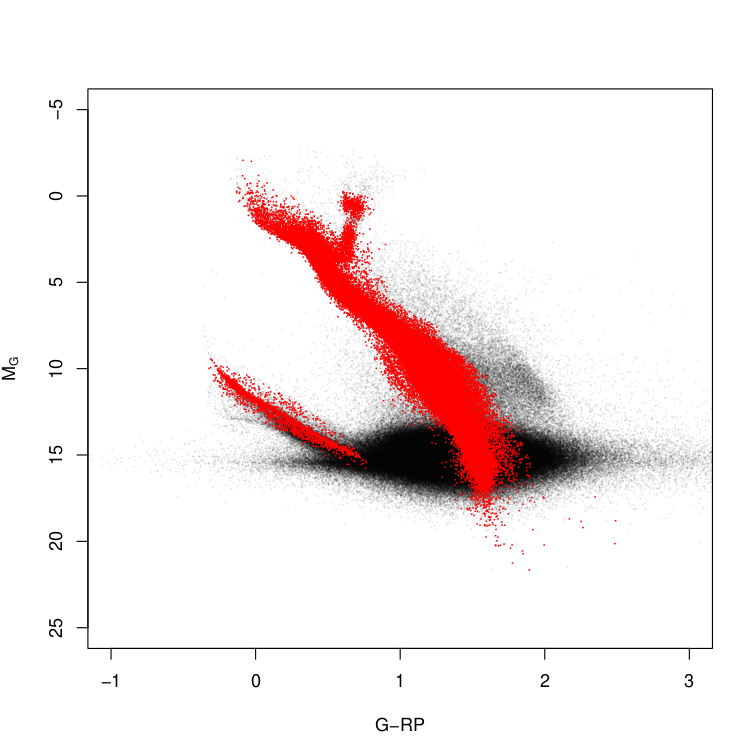}}
\caption{CAMD with colour $G$-\grp\ and absolute magnitude $M_G$ of the set
  of sources with parallaxes greater than or equal to 8 mas (black)
  and those used as examples of good astrometry (red points) in the
  random forest training set.\label{fig:app-HRD-Gaia}}
\end{figure}

\begin{figure}[!htb]
\center{\includegraphics[width=0.5\textwidth]{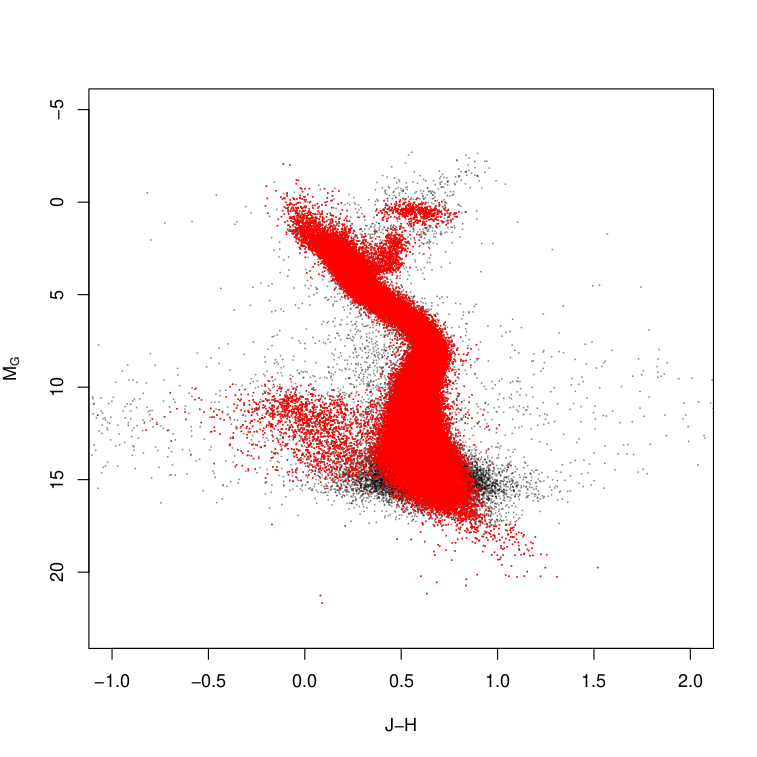}}
\caption{As Figure \ref{fig:app-HRD-Gaia} for $M_G$ and the 2MASS colour index $J$-$H$.
\label{fig:app-HRD-JHG}}
\end{figure}

\begin{figure}[!htb]
\center{\includegraphics[width=0.5\textwidth]{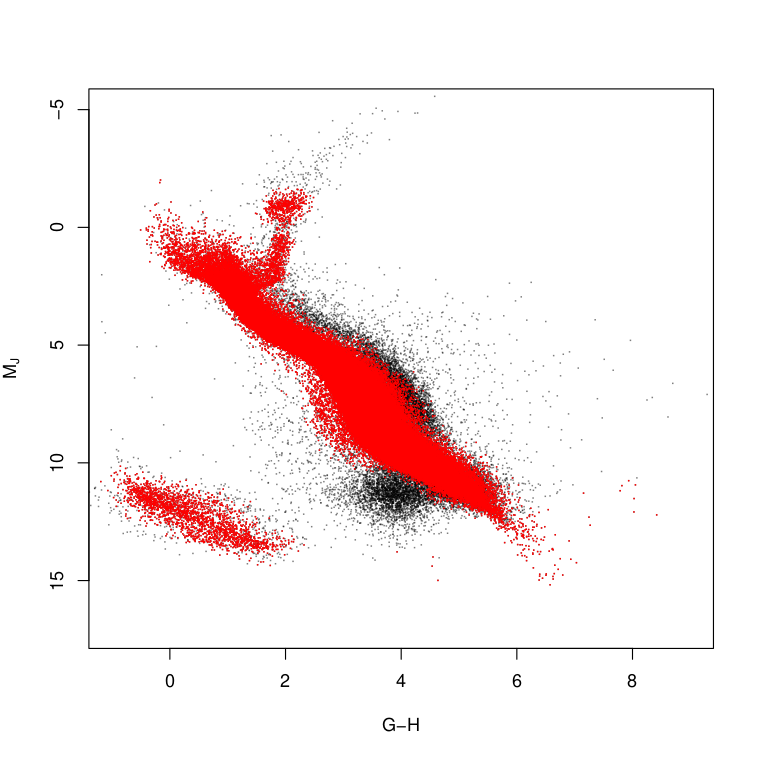}}
\caption{As Figure \ref{fig:app-HRD-Gaia} for $M_J$ and the $G$-$H$ colour index.
\label{fig:app-HRD-GHJ}}
\end{figure}

Figures \ref{fig:app-HRD-Gaia}-\ref{fig:app-HRD-GHJ} show the position
in several colour-absolute magnitude diagrams of the sources selected
as examples of good astrometry in the training set (red) superimposed
on the full distribution of sources with observed parallaxes greater
than or equal to 8\,mas. Figures \ref{fig:app-HRD-GHJ} and
\ref{fig:app-HRD-JHG} show that the requirement to have a 2MASS
counterpart to the \G\ source already removes most of the sources with
spurious observed parallaxes greater than 8 mas..

\subsection{Parameters tested for relevance}

Table \ref{tab:importance-all} lists all catalogue columns tested for
relevance in the classification problem of separating good astrometric
solutions from spurious ones. We did not check for the relevance of
astrometric\_primary\_flag, astrometric\_weight\_ac",
nu\_eff\_used\_in\_astrometry, pseudocolour, and pseudocolour\_error
due to the high fraction of missing values in the training set.

\begin{table*}
\begin{center}
\begin{tabular}{l l l l}

Feature name &  Mean Decrease & Feature name & Mean Decrease\\
             & accuracy &                    & Gini index \\
\hline\hline
\cellcolor{gray!25} parallax\_error  &  0.125  &  \cellcolor{gray!25} parallax\_error  &  33821  \\
\cellcolor{gray!25} parallax\_over\_error  &  0.087  &  \cellcolor{gray!25} parallax\_over\_error  &  27713  \\
\cellcolor{gray!25} pmra  &  0.056  &  \cellcolor{gray!25} astrometric\_sigma5d\_max  &  24035  \\
\cellcolor{gray!25} astrometric\_sigma5d\_max  &  0.052  &  \cellcolor{gray!25} pmra\_error  &  20226  \\
\cellcolor{gray!25} pmdec  &  0.047  &  \cellcolor{gray!25} pmdec\_error & 14866  \\
\cellcolor{gray!25} pmdec\_error  &  0.027  &  \cellcolor{gray!25} astrometric\_excess\_noise  & 12737    \\
\cellcolor{gray!25} pmra\_error  &  0.025  &    \cellcolor{red!25} astrometric\_params\_solved &    7677 \\
\cellcolor{gray!25} astrometric\_excess\_noise  &  0.013  & \cellcolor{gray!25} ipd\_gof\_harmonic\_amplitude   & 5628   \\
\cellcolor{gray!25} visibility\_periods\_used  &  0.01  &  \cellcolor{gray!25} ruwe  &  3383    \\
\cellcolor{gray!25} ruwe  &  0.008   &  \cellcolor{gray!25} visibility\_periods\_used  &  2371  \\
\cellcolor{gray!25} astrometric\_gof\_al  &  0.005 &  \cellcolor{gray!25} pmdec  &  2263  \\
astrometric\_n\_obs\_ac  &  0.005   &  \cellcolor{gray!25} pmra  &  2039  \\
\cellcolor{gray!25} ipd\_gof\_harmonic\_amplitude  &  0.004  &  \cellcolor{gray!25} ipd\_frac\_odd\_win  &  1566  \\
astrometric\_excess\_noise\_sig  &  0.003 & \cellcolor{gray!25} ipd\_frac\_multi\_peak  &  1006    \\
\cellcolor{gray!25} ipd\_frac\_odd\_win  &  0.002    & \cellcolor{gray!25} astrometric\_gof\_al  &  801      \\
astrometric\_chi2\_al  &  0.002 &  scan\_direction\_strength\_k2  &  694\\
parallax\_pmdec\_corr  &  0.002 &  parallax\_pmdec\_corr  &  522    \\
\cellcolor{gray!25} ipd\_frac\_multi\_peak  &  0.002   &  astrometric\_excess\_noise\_sig  &  413\\
scan\_direction\_strength\_k2  &  0.001  & astrometric\_n\_good\_obs\_al  &  394   \\
astrometric\_n\_good\_obs\_al  &  0.001  &  astrometric\_chi2\_al  &  275   \\
\cellcolor{red!25} astrometric\_params\_solved  &  0.001  &  astrometric\_n\_obs\_al  &  244 \\
astrometric\_n\_obs\_al  &  0.001  & astrometric\_n\_obs\_ac  &  224   \\
astrometric\_matched\_transits  &  0.001  &    dec\_parallax\_corr  &  208 \\
dec\_parallax\_corr  &  0.001  & astrometric\_matched\_transits  &  165  \\
dec\_pmdec\_corr  &  0.001  &    dec\_pmdec\_corr  &  157 \\
scan\_direction\_mean\_k2  &  0.001  &   ra\_dec\_corr  &  65   \\
ra\_parallax\_corr  &  0  &  scan\_direction\_strength\_k1  &  59\\
scan\_direction\_strength\_k4  &  0  &   scan\_direction\_mean\_k2  &  50  \\
ra\_dec\_corr  &  0  &  scan\_direction\_strength\_k4  &  50  \\
scan\_direction\_strength\_k1  &  0  &  parallax\_pmra\_corr  &  49  \\
scan\_direction\_mean\_k4  &  0  &  ra\_parallax\_corr  &  48  \\
scan\_direction\_strength\_k3  &  0  &  ra\_pmdec\_corr  &  44  \\
parallax\_pmra\_corr  &  0  &  scan\_direction\_mean\_k4  &  42  \\
astrometric\_n\_bad\_obs\_al  &  0  &  scan\_direction\_strength\_k3  &  41  \\
ra\_pmdec\_corr  &  0  &  astrometric\_n\_bad\_obs\_al  &  38  \\
scan\_direction\_mean\_k3  &  0  &  scan\_direction\_mean\_k3  &  30  \\
ipd\_gof\_harmonic\_phase  &  0  &  ipd\_gof\_harmonic\_phase  &  29  \\
pmra\_pmdec\_corr  &  0  &  ra\_pmra\_corr  &  28  \\
scan\_direction\_mean\_k1  &  0  &  pmra\_pmdec\_corr  &  27  \\
ra\_pmra\_corr  &  0  &  scan\_direction\_mean\_k1  &  24  \\
dec\_pmra\_corr  &  0  &  dec\_pmra\_corr  &  22  \\

\end{tabular}
\end{center}
\caption{\label{tab:importance-all} Importance of all features tested
  for classification by the Random Forest classifier ordered according
  to the mean decrease in accuracy (two leftmost columns) and by the
  mean decrease in the Gini index (two rightmost columns).}
\end{table*}

\if

\subsection{Distributions of features}
\label{app:tsbad}
\begin{figure}[!htb]
\center{\includegraphics[width=0.5\textwidth]{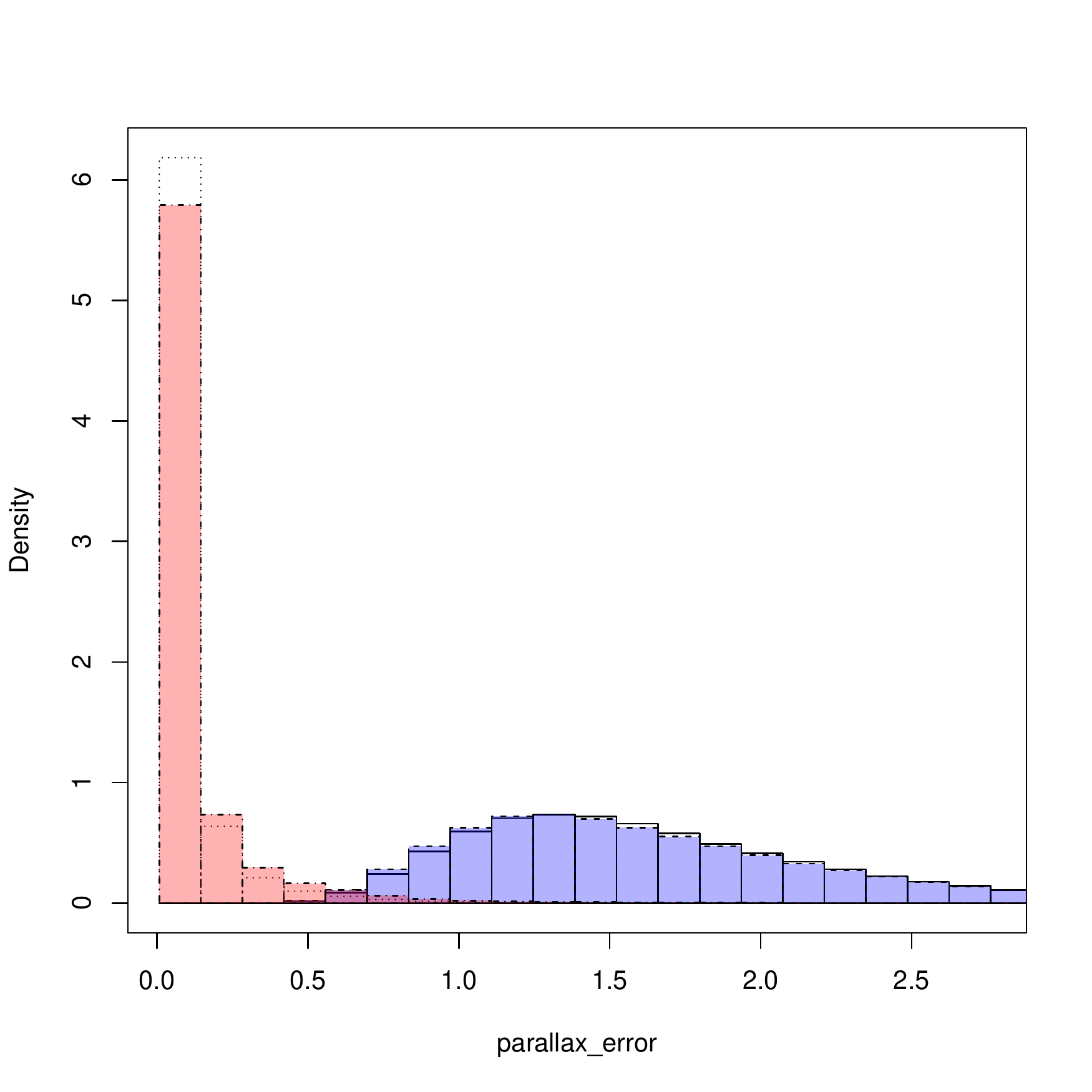}}
\caption{\label{fig:app-hists-parallaxerror} Distribution of values of
  the parallax\_error feature in the set of training examples of the
  bad category (continuous line, white filling); in the set of
  training examples of the good category (dotted line, white filling);
  the set of sources classified as bad astrometric solutions (dashed
  line, blue transparent filling); and the set of sources classified
  as good astrometric solutions (dash-dotted line, red transparent
  filling). }
\end{figure}

\begin{figure}[!htb]
\center{\includegraphics[width=0.5\textwidth]{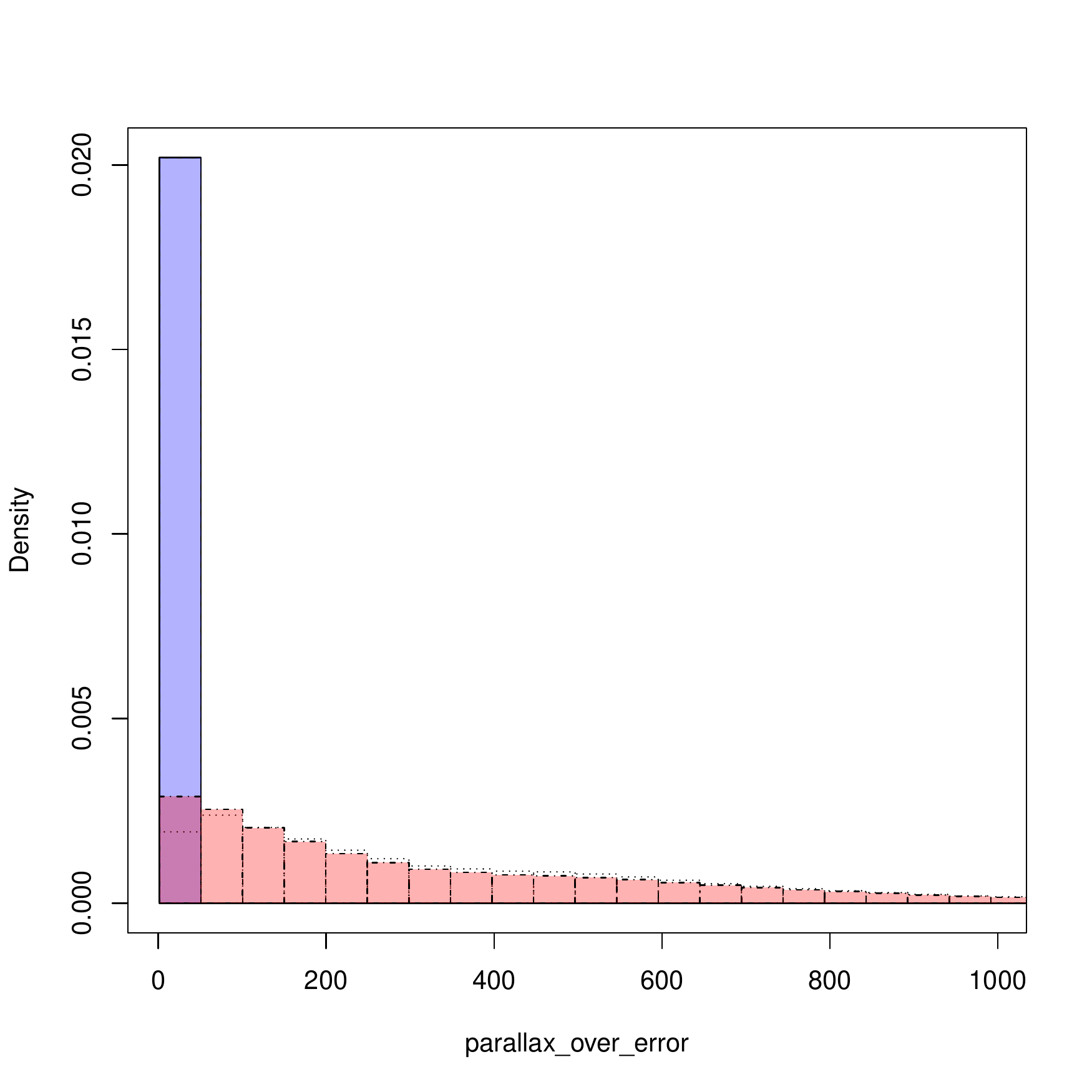}}
\caption{\label{fig:app-hists-parallaxovererror} Distribution of
  values of the parallax\_over\_error feature in the set of training
  examples of the bad category (continuous line, white filling); in
  the set of training examples of the good category (dotted line,
  white filling); the set of sources classified as bad astrometric
  solutions (dashed line, blue transparent filling); and the set of
  sources classified as good astrometric solutions (dash-dotted line,
  red transparent filling). }
\end{figure}

\begin{figure}[!htb]
\center{\includegraphics[width=0.5\textwidth]{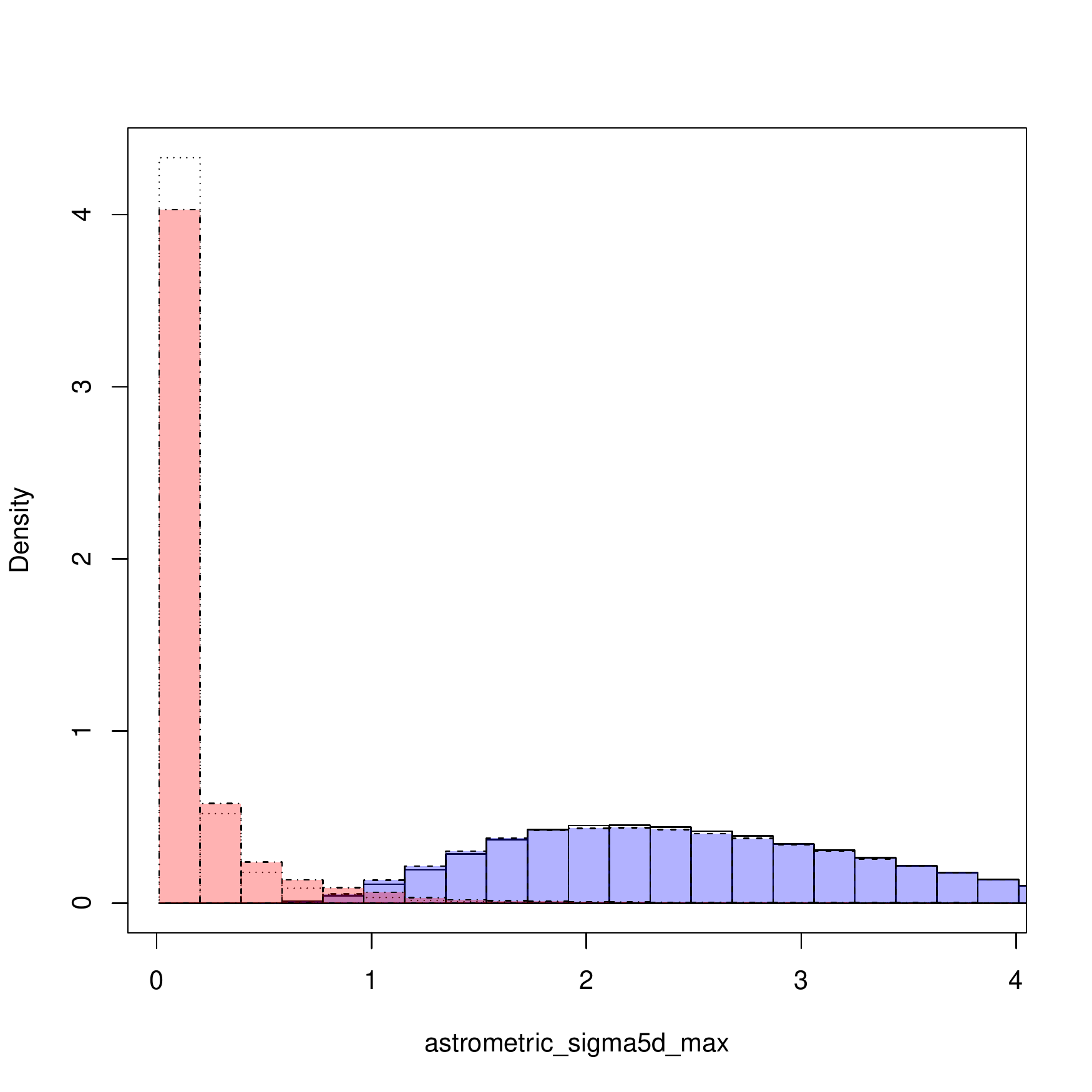}}
\caption{\label{fig:app-hists-astrometric_sigma5d_max} Distribution of values of the astrometric\_sigma5d\_max feature in the set of training examples of the bad category (continuous line, white filling); in the set of training examples of the good category (dotted line, white filling); the set of sources classified as bad astrometric solutions (dashed line, blue transparent filling); and the set of sources classified as good astrometric solutions (dash-dotted line, red transparent filling). }
\end{figure}

\begin{figure}[!htb]
\center{\includegraphics[width=0.5\textwidth]{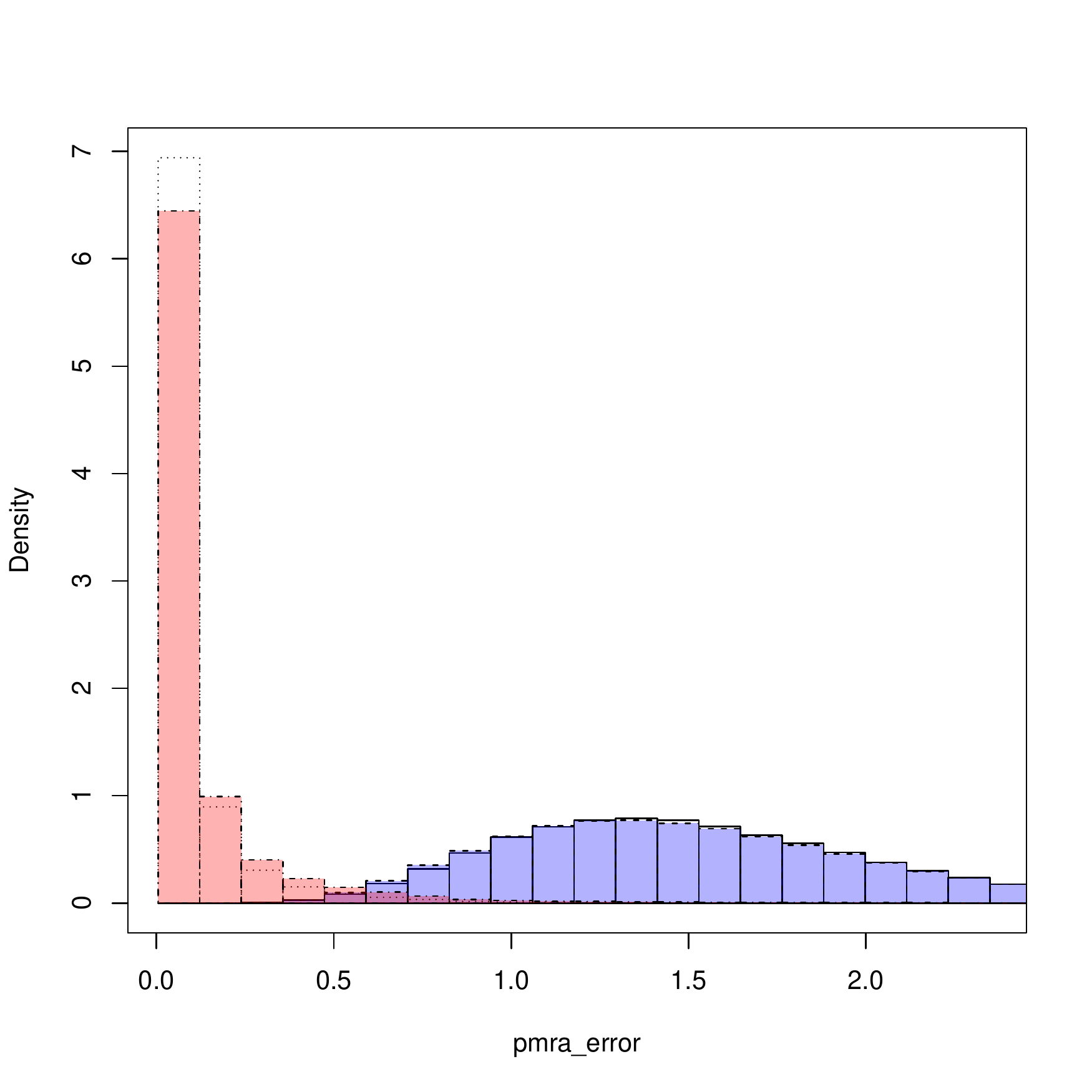}}
\caption{\label{fig:app-hists-pmra_error} Distribution of values of the pmra\_error feature in the set of training examples of the bad category (continuous line, white filling); in the set of training examples of the good category (dotted line, white filling); the set of sources classified as bad astrometric solutions (dashed line, blue transparent filling); and the set of sources classified as good astrometric solutions (dash-dotted line, red transparent filling). }
\end{figure}

\begin{figure}[!htb]
\center{\includegraphics[width=0.5\textwidth]{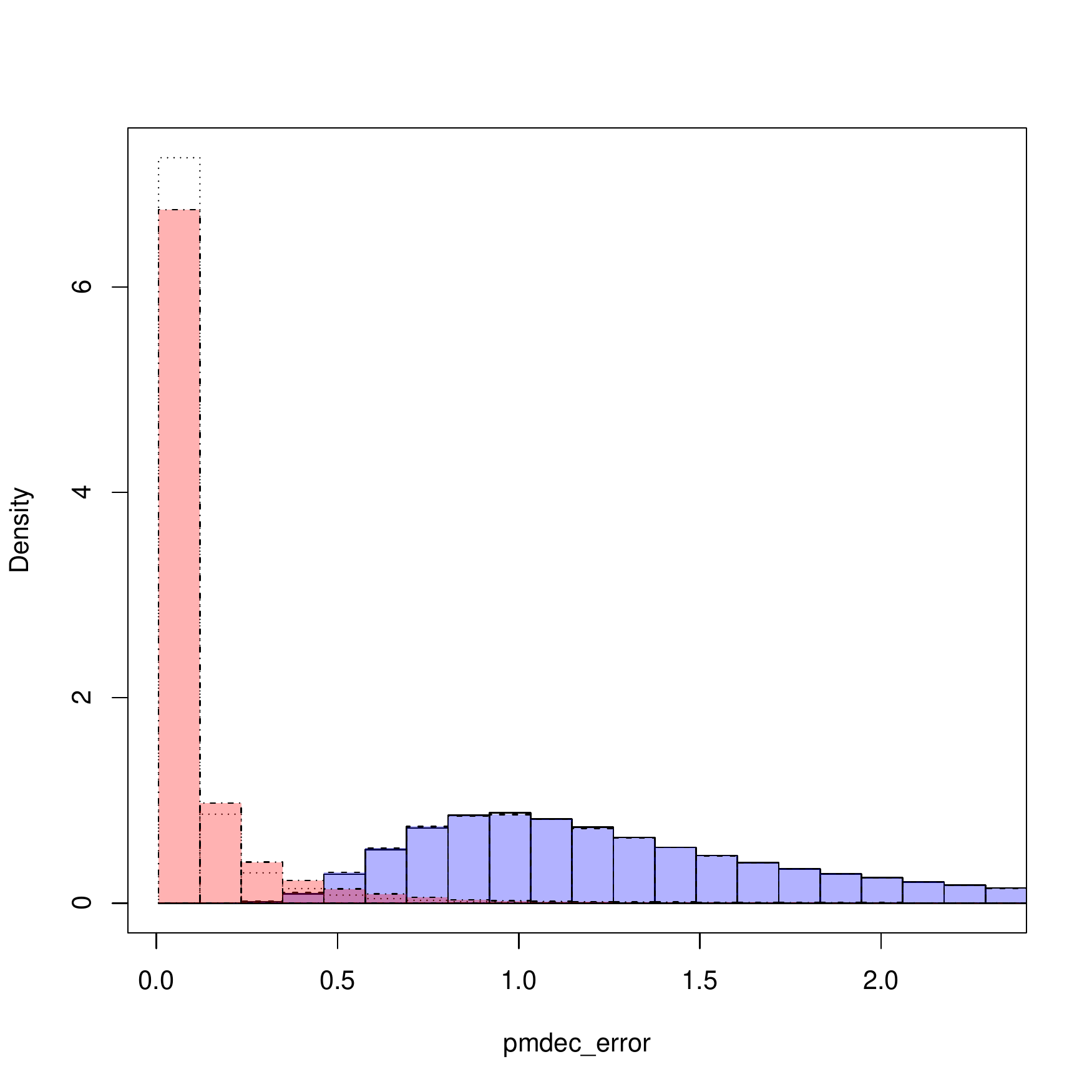}}
\caption{\label{fig:app-hists-pmdec_error} Distribution of values of the pmdec\_error feature in the set of training examples of the bad category (continuous line, white filling); in the set of training examples of the good category (dotted line, white filling); the set of sources classified as bad astrometric solutions (dashed line, blue transparent filling); and the set of sources classified as good astrometric solutions (dash-dotted line, red transparent filling). }
\end{figure}

\begin{figure}[!htb]
\center{\includegraphics[width=0.5\textwidth]{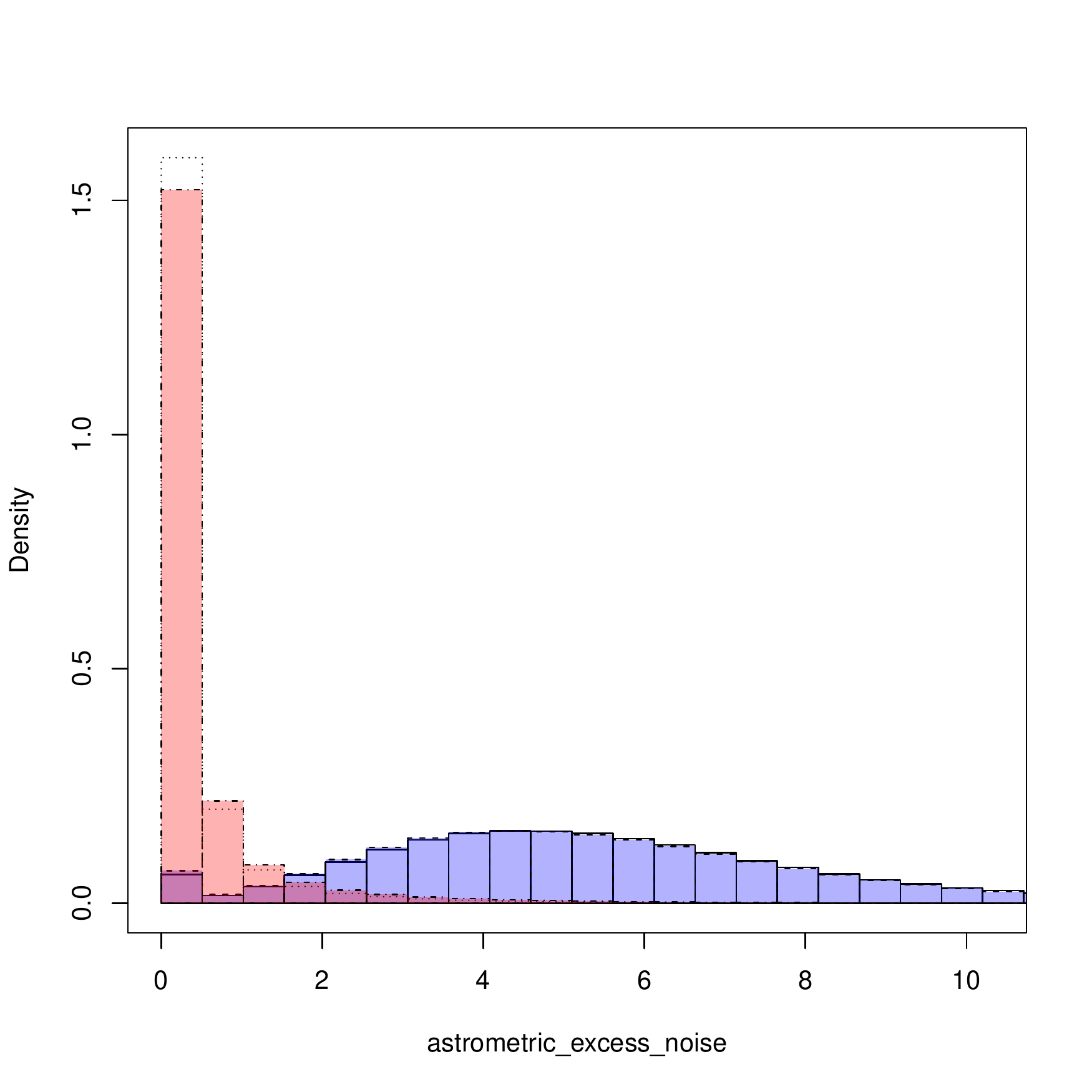}}
\caption{\label{fig:app-hists-astrometric_excess_noise} Distribution of values of the astrometric\_excess\_noise feature in the set of training examples of the bad category (continuous line, white filling); in the set of training examples of the good category (dotted line, white filling); the set of sources classified as bad astrometric solutions (dashed line, blue transparent filling); and the set of sources classified as good astrometric solutions (dash-dotted line, red transparent filling). }
\end{figure}

\begin{figure}[!htb]
\center{\includegraphics[width=0.5\textwidth]{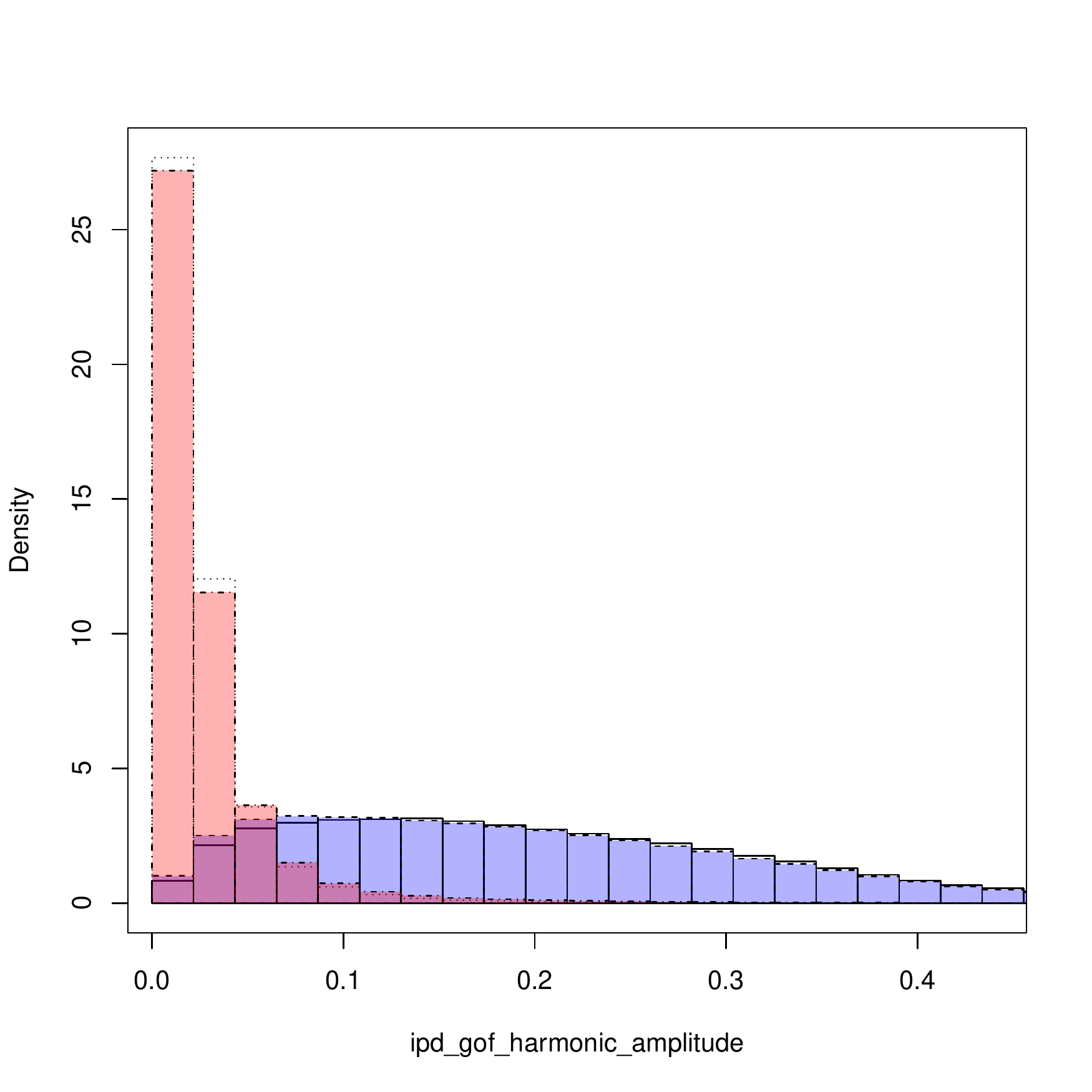}}
\caption{\label{fig:app-hists-ipd_gof_harmonic_amplitude} Distribution of values of the ipd\_gof\_harmonic\_amplitude feature in the set of training examples of the bad category (continuous line, white filling); in the set of training examples of the good category (dotted line, white filling); the set of sources classified as bad astrometric solutions (dashed line, blue transparent filling); and the set of sources classified as good astrometric solutions (dash-dotted line, red transparent filling). }
\end{figure}

\begin{figure}[!htb]
\center{\includegraphics[width=0.5\textwidth]{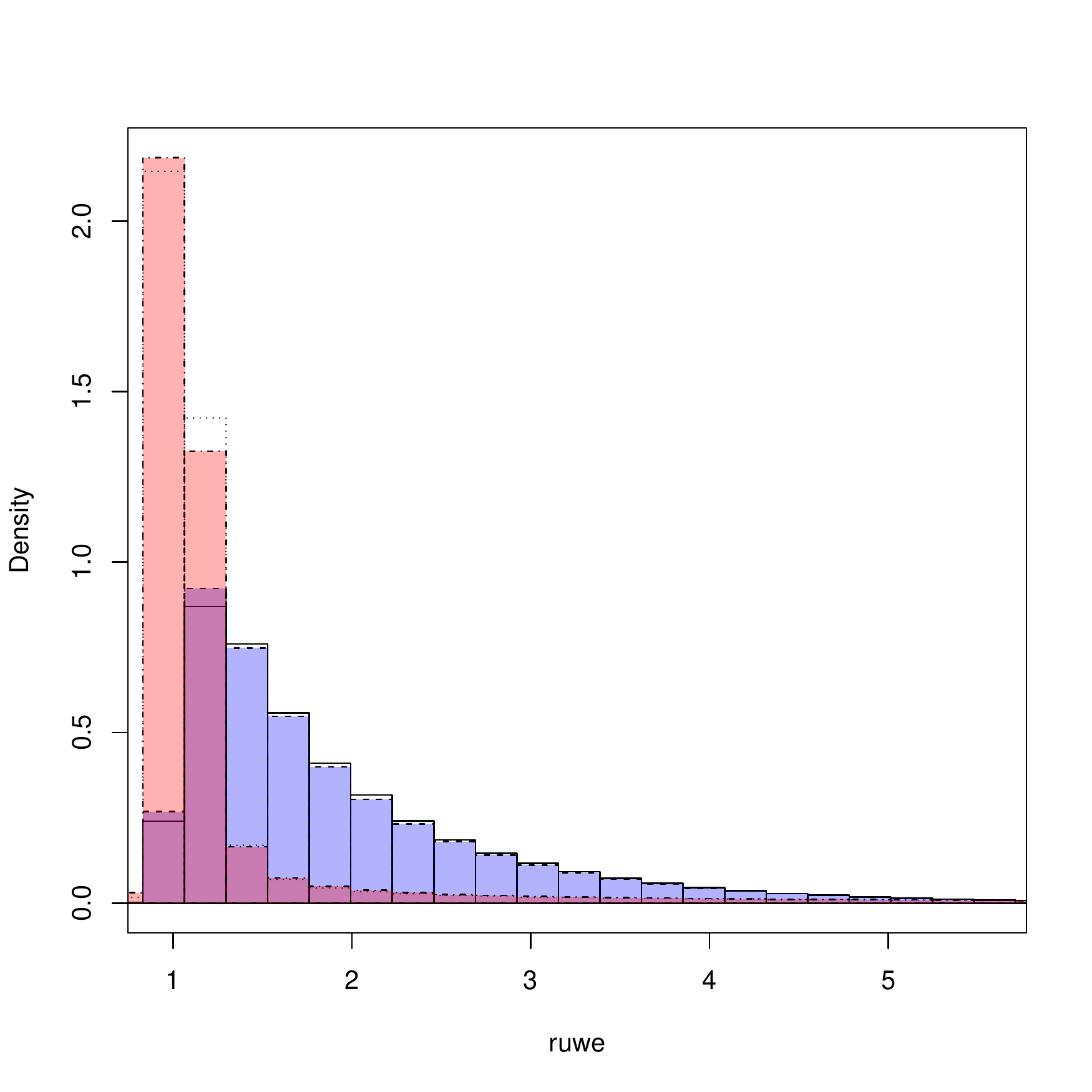}}
\caption{\label{fig:app-hists-ruwe} Distribution of values of the ruwe feature in the set of training examples of the bad category (continuous line, white filling); in the set of training examples of the good category (dotted line, white filling); the set of sources classified as bad astrometric solutions (dashed line, blue transparent filling); and the set of sources classified as good astrometric solutions (dash-dotted line, red transparent filling). }
\end{figure}

\begin{figure}[!htb]
\center{\includegraphics[width=0.5\textwidth]{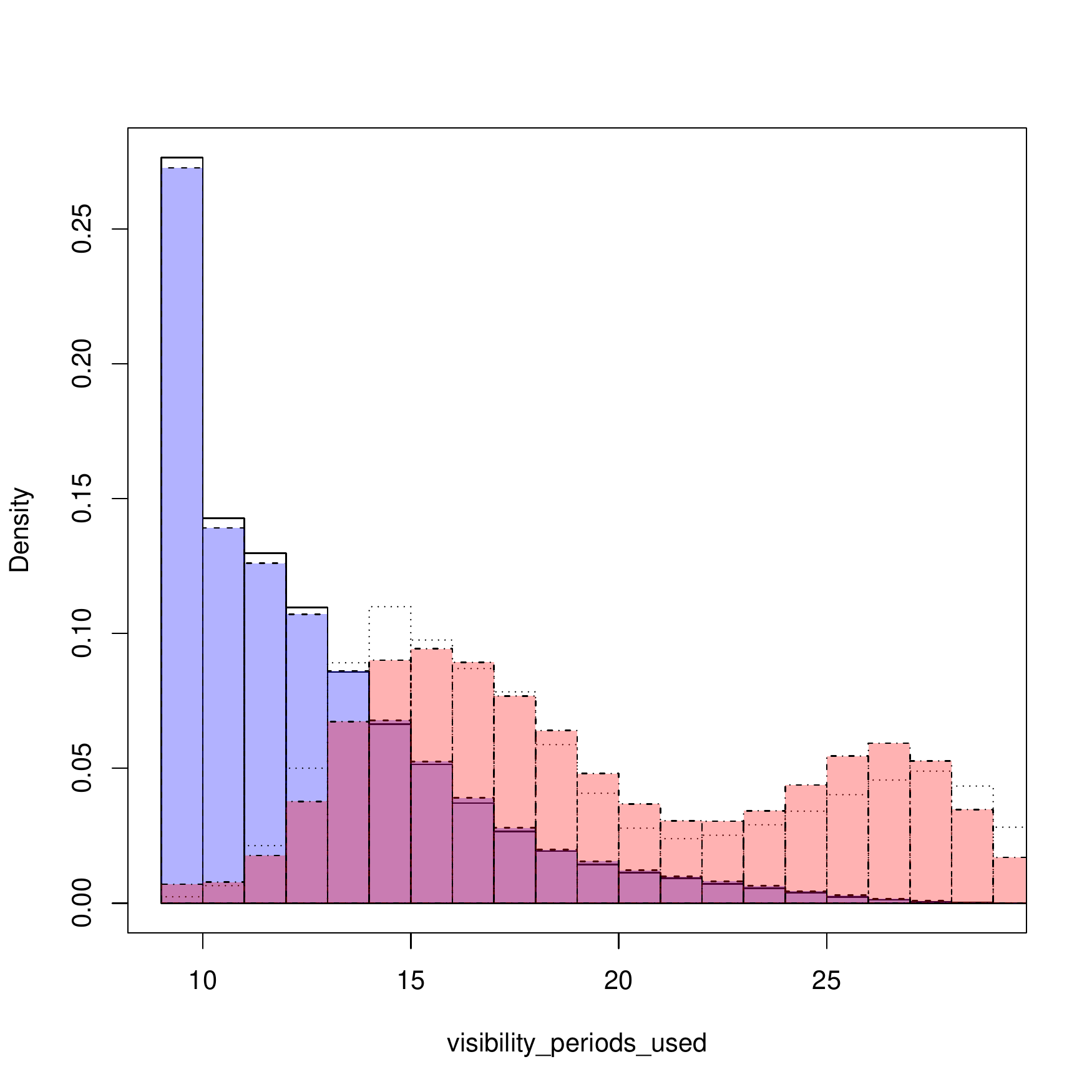}}
\caption{\label{fig:app-hists-visibility_periods_used} Distribution of values of the visibility\_periods\_used feature in the set of training examples of the bad category (continuous line, white filling); in the set of training examples of the good category (dotted line, white filling); the set of sources classified as bad astrometric solutions (dashed line, blue transparent filling); and the set of sources classified as good astrometric solutions (dash-dotted line, red transparent filling). }
\end{figure}

\begin{figure}[!htb]
\center{\includegraphics[width=0.5\textwidth]{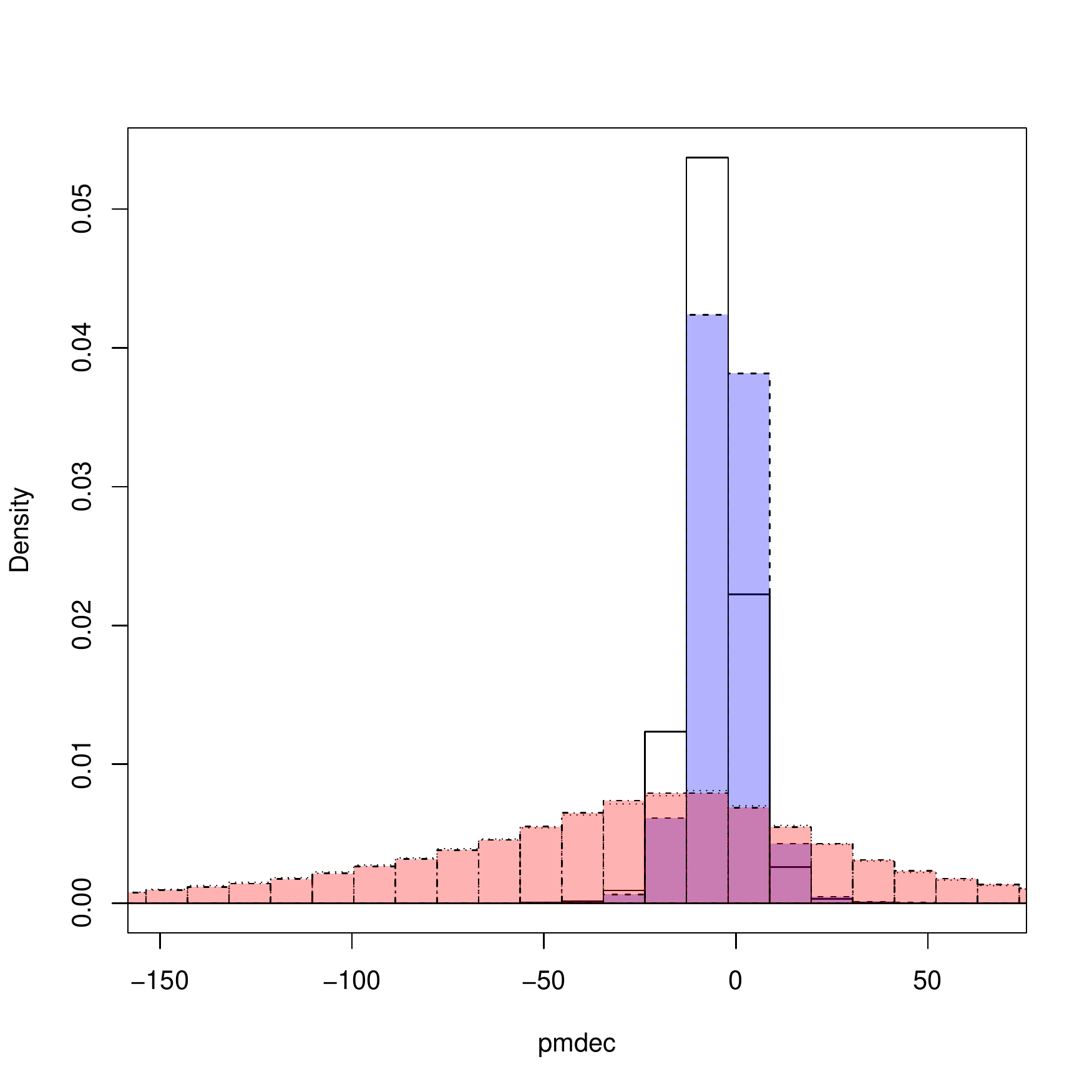}}
\caption{\label{fig:app-hists-pmdec} Distribution of values of the pmdec feature in the set of training examples of the bad category (continuous line, white filling); in the set of training examples of the good category (dotted line, white filling); the set of sources classified as bad astrometric solutions (dashed line, blue transparent filling); and the set of sources classified as good astrometric solutions (dash-dotted line, red transparent filling). }
\end{figure}

\begin{figure}[!htb]
\center{\includegraphics[width=0.5\textwidth]{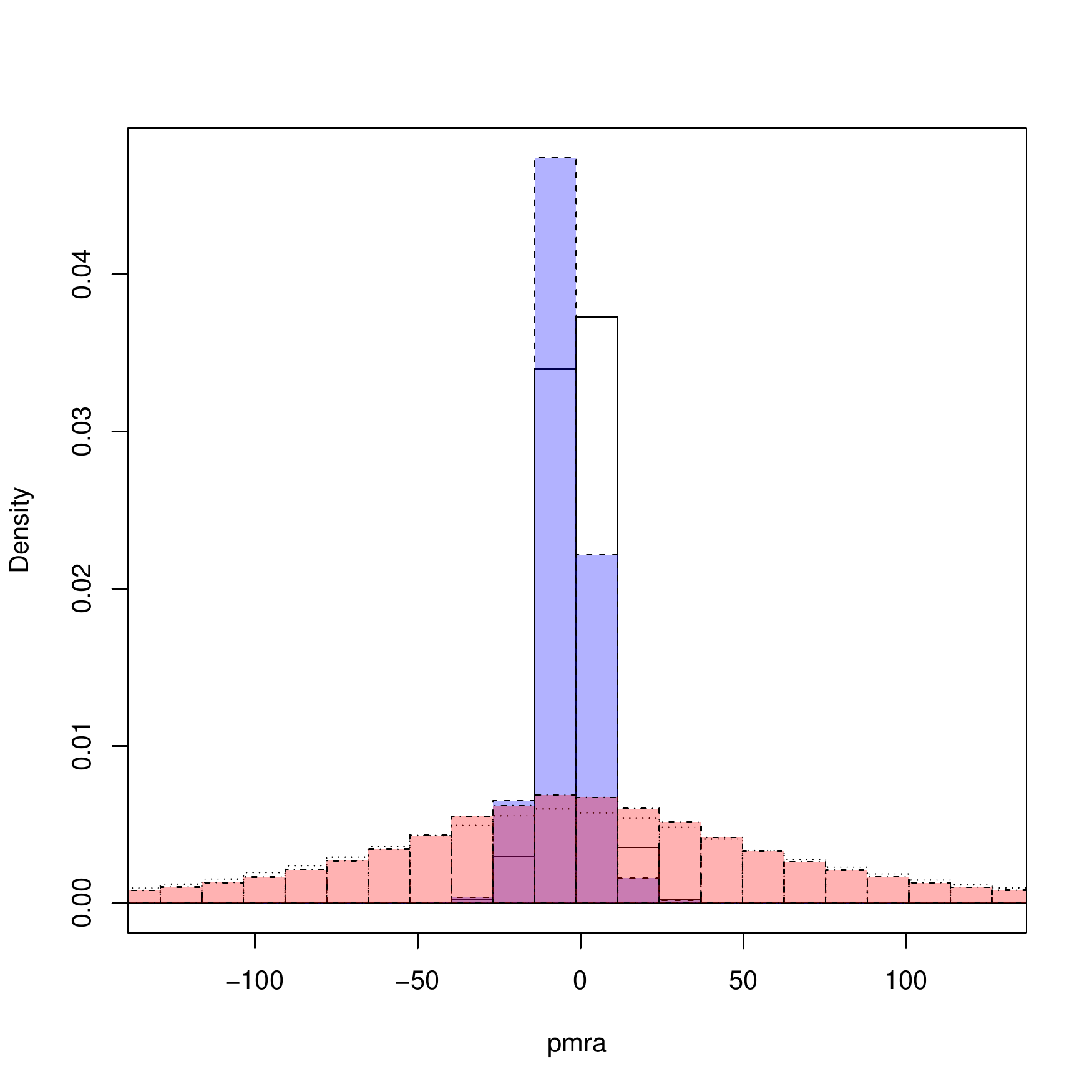}}
\caption{\label{fig:app-hists-pmra} Distribution of values of the pmra feature in the set of training examples of the bad category (continuous line, white filling); in the set of training examples of the good category (dotted line, white filling); the set of sources classified as bad astrometric solutions (dashed line, blue transparent filling); and the set of sources classified as good astrometric solutions (dash-dotted line, red transparent filling). }
\end{figure}

\begin{figure}[!htb]
\center{\includegraphics[width=0.5\textwidth]{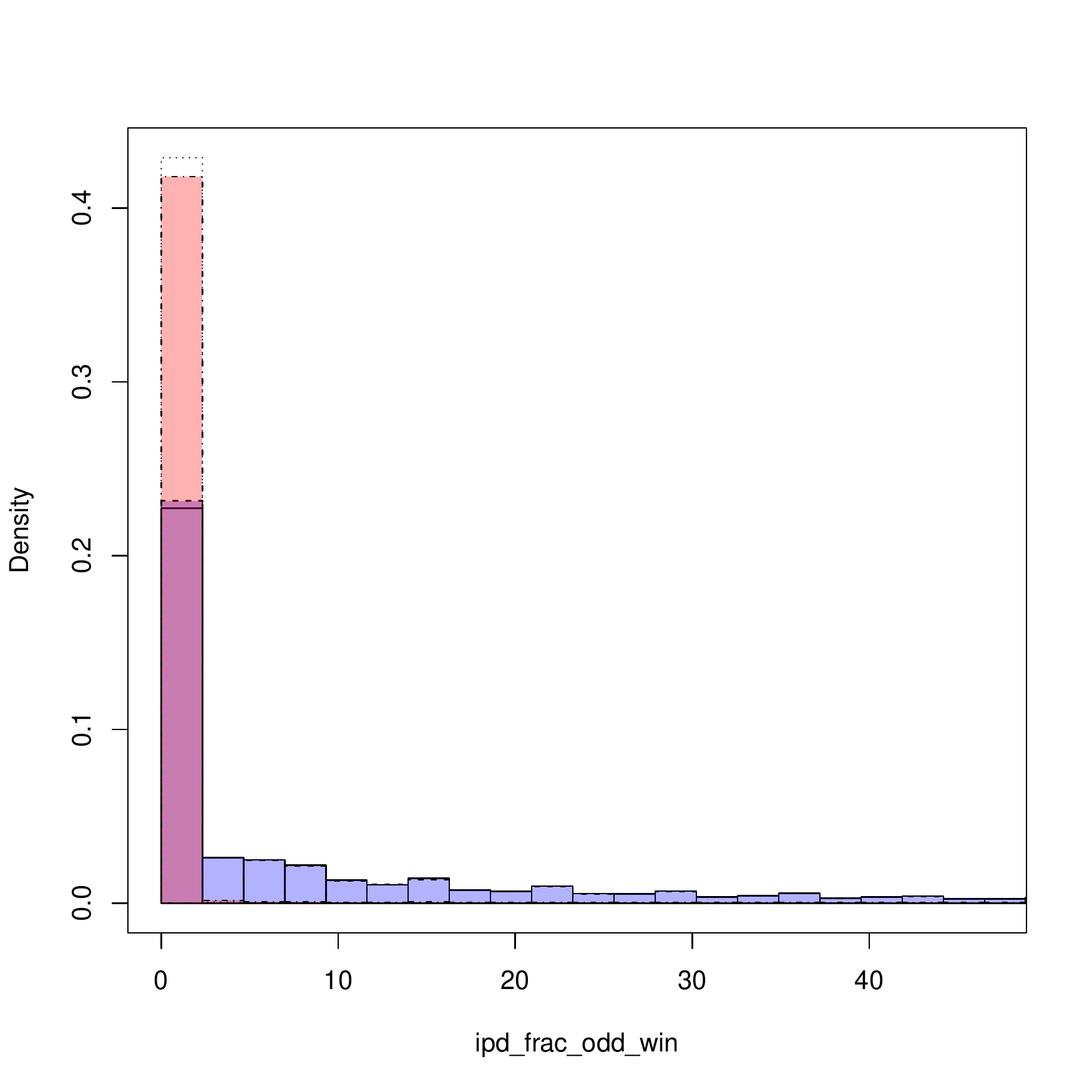}}
\caption{\label{fig:app-hists-ipd_frac_odd_win} Distribution of values of the ipd\_frac\_odd\_win feature in the set of training examples of the bad category (continuous line, white filling); in the set of training examples of the good category (dotted line, white filling); the set of sources classified as bad astrometric solutions (dashed line, blue transparent filling); and the set of sources classified as good astrometric solutions (dash-dotted line, red transparent filling). }
\end{figure}

\begin{figure}[!htb]
\center{\includegraphics[width=0.5\textwidth]{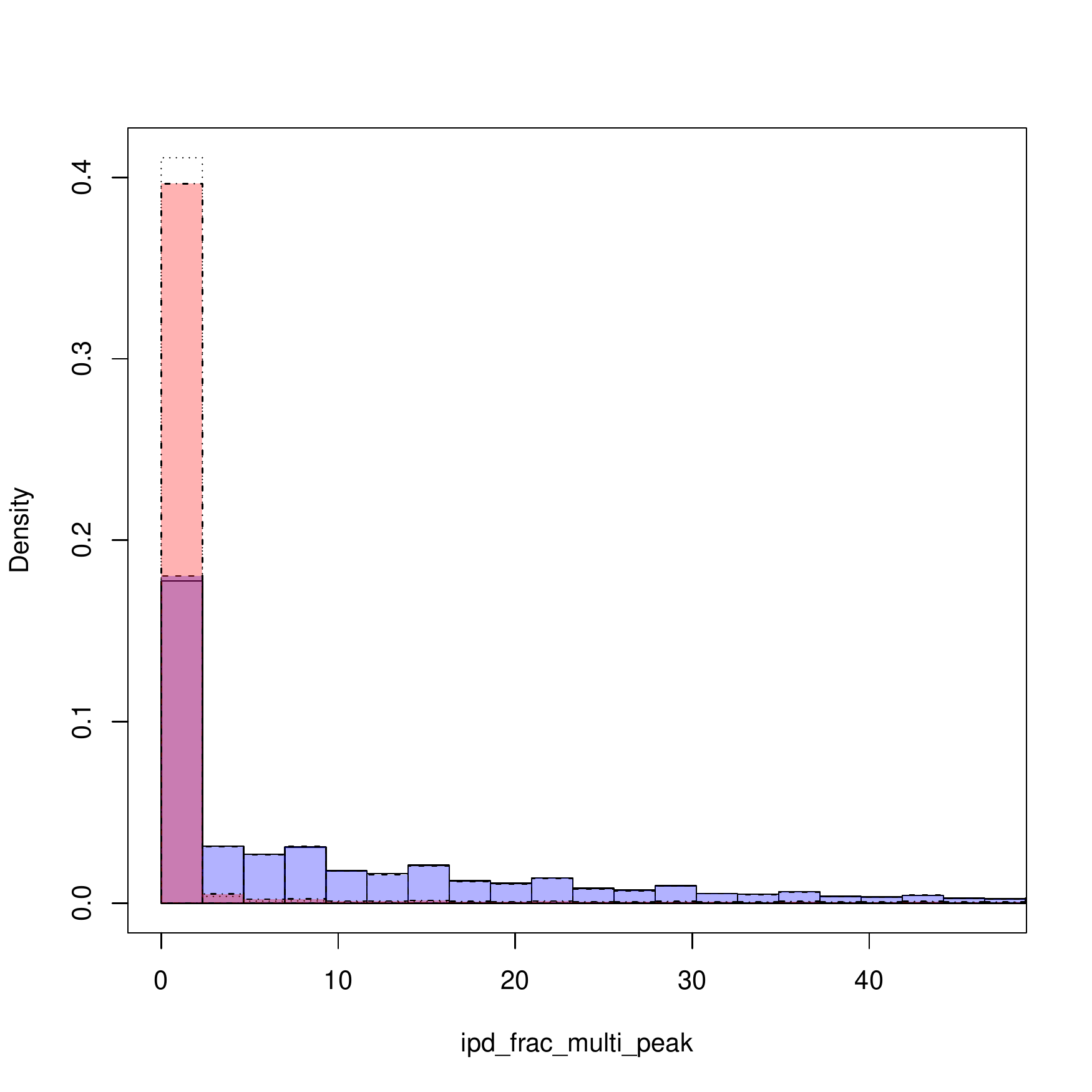}}
\caption{\label{fig:app-hists-ipd_frac_multi_peak} Distribution of values of the ipd\_frac\_multi\_peak feature in the set of training examples of the bad category (continuous line, white filling); in the set of training examples of the good category (dotted line, white filling); the set of sources classified as bad astrometric solutions (dashed line, blue transparent filling); and the set of sources classified as good astrometric solutions (dash-dotted line, red transparent filling). }
\end{figure}

\begin{figure}[!htb]
\center{\includegraphics[width=0.5\textwidth]{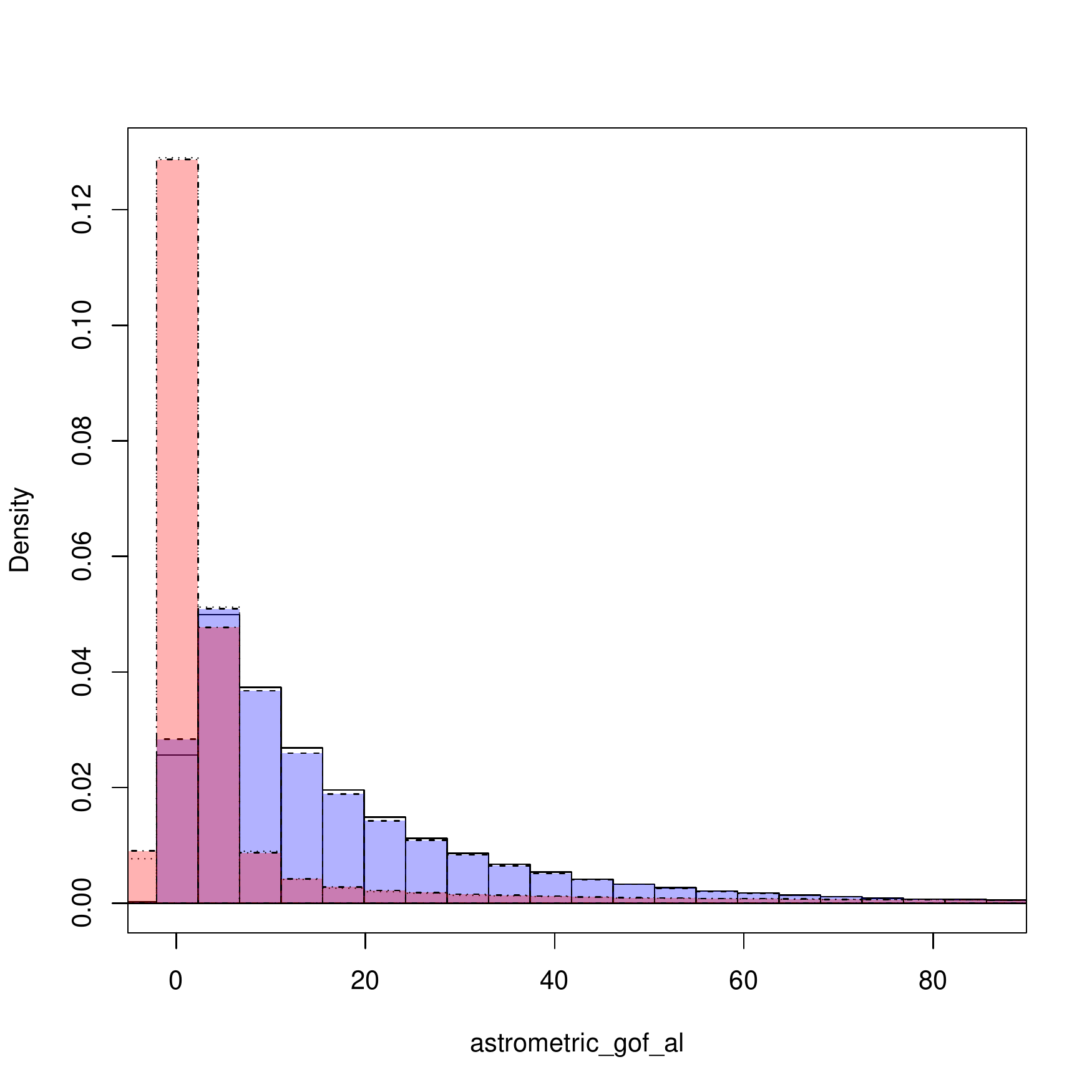}}
\caption{\label{fig:app-hists-astrometric_gof_al} Distribution of values of the astrometric\_gof\_al feature in the set of training examples of the bad category (continuous line, white filling); in the set of training examples of the good category (dotted line, white filling); the set of sources classified as bad astrometric solutions (dashed line, blue transparent filling); and the set of sources classified as good astrometric solutions (dash-dotted line, red transparent filling). }
\end{figure}

\begin{figure}[!htb]
\center{\includegraphics[width=0.5\textwidth]{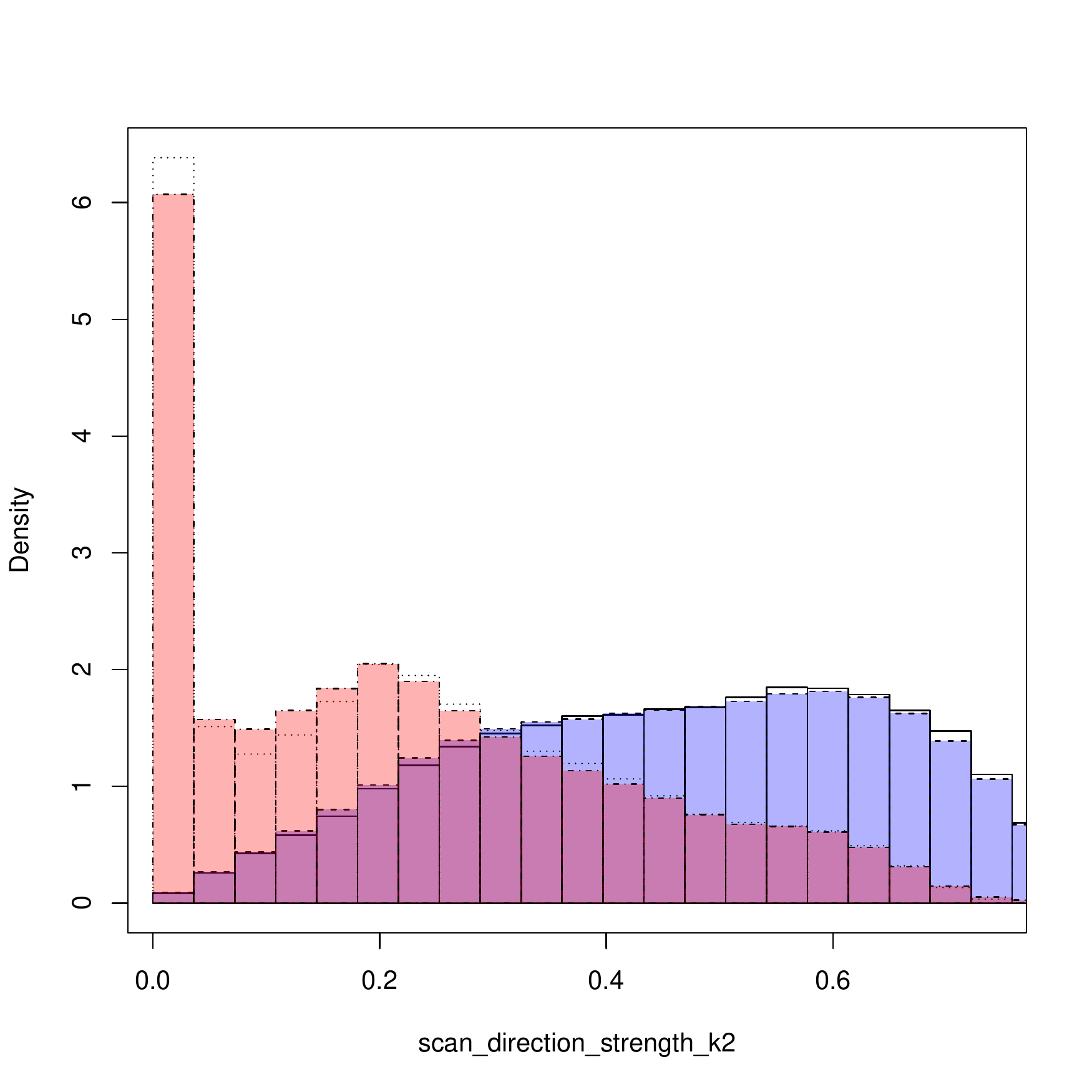}}
\caption{\label{fig:app-hists-scan_direction_strength_k2} Distribution of values of the scan\_direction\_strength\_k2 feature in the set of training examples of the bad category (continuous line, white filling); in the set of training examples of the good category (dotted line, white filling); the set of sources classified as bad astrometric solutions (dashed line, blue transparent filling); and the set of sources classified as good astrometric solutions (dash-dotted line, red transparent filling). }
\end{figure}

\clearpage
\section{The relations and Gaussian Mixture Model priors used for the determination of space velocities in Galactic coordinates}
\label{app:UVW-priors}

The relations that define space velocities in
terms of the observables, \G\,coordinates, parallaxes and proper
motions, and radial velocities are: 

\begin{equation}
\begin{pmatrix}
u\\ 
v\\ 
w
\end{pmatrix} = A_G^T \cdot A \begin{pmatrix}
4.74047~ {\mu_{\alpha^*}}/{\varpi}\\ 
4.74047~ {\mu_{\delta}}/{\varpi}\\
v_r 
\end{pmatrix}
\label{eq:det-rel-uvw}
\end{equation}

where $A_G$ is the transformation matrix to Galactic coordinates from
the introduction to the Hipparcos catalogue \citep{1997ESASP1200.....E}
and matrix $A$ is obtained from the components of the normal triad
at the star as:

\begin{equation}
A = \begin{pmatrix}
-\sin \alpha & -\sin \delta \cos \alpha  & \cos \delta \cos \alpha \\
 \cos \alpha & -\sin \delta \sin \alpha  & \cos \delta \sin \alpha\\
     0       &        \cos \delta        &          \sin \delta\\ 
\end{pmatrix}
\end{equation}

The Bayesian model used to infer posterior probabilities for the space
velocities requires the definition of priors for the model
parameters. As described in Section\,\ref{sec:cat_overview}, we fit
Gaussian Mixture Models to a local (140\,pc) simulation from the
Besan\c con Galaxy model \citep{2003A&A...409..523R} and modify the
result by adding a wide non-informative component. The resulting
priors used in the inference process are defined in Equations \ref{eq:priorU}-\ref{eq:priorW} using the notation $\mathcal{N}(\cdot\mid\mu,\sigma)$ to denote the Gaussian distribution centred at $\mu$ and with standard deviation $\sigma$.

\begin{align}
\begin{split}
    \pi(U) = & 0.52\cdot\mathcal{N}(U\mid -11.3,23.2)+\\
    & 0.45\cdot\mathcal{N}(U\mid -11,44)+\\
    & 0.03\cdot\mathcal{N}(U\mid 0,120)
    \label{eq:priorU}
\end{split}
\end{align}

\begin{align}
\begin{split}
    \pi(V) = & 0.588\cdot\mathcal{N}(V\mid -26.1,23.7)+\\
    & 0.375\cdot\mathcal{N}(V\mid -13,11.3)+\\
    & 0.03\cdot\mathcal{N}(V\mid 0,120)+\\
    & 0.007\cdot\mathcal{N}(V\mid -115.8,114.3)
    \label{eq:priorV}
\end{split}
\end{align}

\begin{align}
\begin{split}
    \pi(W) = &0.53\cdot\mathcal{N}(W\mid -7.3,19.4)+\\
    &0.2\cdot\mathcal{N}(W\mid -10,9.2)+\\
    &0.21\cdot\mathcal{N}(W\mid -4.1,10.1)+\\
    &0.03\cdot\mathcal{N}(W\mid -7,43.3)+\\
    &0.03\cdot\mathcal{N}(W\mid 0,120)
    \label{eq:priorW}
\end{split}
\end{align}

\fi

\end{appendix}

\end{document}